# UNIVERSITÀ DEGLI STUDI DI MILANO

**Scuola di Dottorato in Fisica, Astrofisica e Fisica Applicata**

Dipartimento di Fisica
Corso di Dottorato di Ricerca in Fisica, Astrofisica e Fisica Applicata

Ciclo XXXI

# Random Combinatorial Optimization Problems: Mean Field and Finite-Dimensional Results

Settore Scientifico Disciplinare FIS/02

**Coordinator**
Prof. Francesco Ragusa

**Candidate**
Enrico Maria Malatesta

**Advisors**
Prof. Sergio Caracciolo
Prof. Giorgio Parisi

Thesis defended on 21 December 2018
in front of a Board of Examiners composed by:

Paolo Biscari
Sergio Caracciolo
Luca Dall'Asta

External Referees:

Florent Krzakala
Federico Ricci-Tersenghi

*to Chiara*
*for all the wonderful years together*

## Acknowledgements

These 3 years of PhD were intense and beautiful; I owe a lot to the people I attended and with whom I have worked during these years. First of all I want to thank Prof. Sergio Caracciolo, not only for his precious guidance, for his advices, and for the long discussions we had during those 3 years, but also for his kindness and complete availability. In the same way I also acknowledge Prof. Giorgio Parisi for all his advices, and for his availability to every doubt or question. I thank also Prof. Riccardo Zecchina and Carlo Baldassi, for giving me the opportunity to work together in the last year of PhD and Luca Guido Molinari, for the beautiful lessons on random matrix theory. A special acknowledgement goes to the "mythological" Tommaso Rizzo and to the "legendary" Andrea Maiorano, for the always warm hospitality, and for all the discussions (on physics, of course, but also on politics) we made. I am grateful to Carlo Lucibello, and Gabriele Sicuro, for their hospitality respectively in Turin and Rome, as long as the many and stimulating discussions. I also thank the others Post-Docs in Rome: Valerio Astuti, Fernanda Benetti and Giacomo Gradenigo. In the same way I also mention all the Milan people: Bruno Bassetti, Francesco Borra, Riccardo Capelli, Matteo Cardella, Vittorio Erba, Riccardo Fabbricatore, Marco Gherardi, Andrea Di Gioacchino, Alessandro Montoli, Mauro Pastore, Pietro Rotondo and German Sborlini. The office would have been boring without them. I thank all the other people with whom I had the pleasure to interact with or to attend conference and lectures together: Ada Altieri, Francesco Concetti, Gianpaolo Folena, Cosimo Lupo, Jacopo Rocchi, Francesco De Santis, Valerio Volpati. A special mention goes also to the infn secretary: without pain that's no fun.

Then I want thank my parents for giving me the opportunity to study and all the rest of my family: my uncles Paola, Bob, Marina, Enrico, my sister Francesca, Francesco, my nieces Sveva and Greta, my cousins Davide and Diego. A special thank to my old friends Andrea, Luca and Matteo for all the good time evenings together and to Fabrizio for the incredible fencing journeys around Italy. I thank also my old university friends, Federico and Eleonora, Francesco and Roberta, Luca M., Luca S., Claudia, Margherita and many others. I finally want to give a very special thank to Chiara, for her patience and for her presence every time I needed.

# Contents

# Introduction

Since the introduction of the first spin glass model by Edwards and Anderson [EA75] in 1975, the research area of disordered systems has undergone a huge progress, thanks to the introduction of new analytical techniques and numerical tools, as long as the development of novel concepts and ideas. In particular, the rich phenomenology found by the extensive study of mean field spin glass models, not only proved to be the basis for an explanation of many different physical phenomena and permitted to strengthen the traditional relationship between physics and mathematics, but also it allowed physicist to apply those concepts to research areas that were thought to be completely disconnected from physics before, such as theoretical computer science, biology, engineering, economics and finance. Therefore, on the analytical level, disordered systems techniques proved to be an excellent tool in order to give a physical insights on different areas of knowledge and many times they were able to solve long-standing issues of the field paving the way to the development of rigorous proofs by the mathematician community. Those continuous interactions and exchange of ideas between different communities guaranteed, on the numerical level, the birth of new, performing algorithms.

In this thesis we will analyze *combinatorial optimization problems*, that are one of those multidisciplinary applications we were alluding, from a physics point of view. This thesis is organized as follows

- Part I: Preliminaries
  - Chapter 1: Phase transitions in pure and disordered systems.

    Here we give some basic definitions in order to set up the notation that will be used throughout all this thesis work. In particular we remind some essential definitions of graph and random graph theory and we introduce basilar concepts and techniques of statistical physics such as mean-field, Bethe-Peierls approximation and Belief Propagation. We will also deal with preliminary definitions in spin glass theory. In the final section of this chapter the main topic of the thesis, that is combinatorial optimization, is introduced.
  - Chapter 2: The Sherrington-Kirkpatrick model.

    In this chapter we will develop the replica formalism for the Sherrington-Kirkpatrick model, which historically was the first analyzed using replicas. This can be an useful introduction to whom has not ever seen replicas in action. We will describe the results coming from the replica

symmetric solutions and how it fails in the spin glass phase. We will explain how and when replica symmetry breaking occurs, studying both the negative entropy problem and performing a stability analysis of the solution. Then we will introduce Parisi's correct scheme of replica symmetry breaking, discussing the phase diagram of the model, as long as physical consequences.

- Part II: Mean Field
  - Chapter 3: Finite-size corrections in random matching problems.
    This is the first original chapter of this thesis work. We will mainly deal with the computation of the finite-size corrections in the random assignment problem, which was one of the first combinatorial optimization problems that was studied by means of the replica technique. Here we present how the choice of the disorder distribution affects the finite-size corrections to the average cost. We have found that corrections are smaller in the case of a pure power law probability distribution, i.e. $\rho(w) \sim w^r$. In this case, only analytical corrections are present, that is in inverse powers of the number of points. On the contrary, and interestingly enough, the exponent of the leading correction changes as a function of $r$ whenever $\rho(w) \sim w^r(\eta_0 + \eta_1 w + \dots)$ with $\eta_1 \neq 0$. These results were published in [Car+17]. Similar conclusions can be derived for the random matching problem case. The form of the finite-size correction is quite much complicated because of a non-trivial determinant contribution. This contribution can be interpreted as the sum over odd loops of the graph, a form which is reminiscent of the finite-size corrections of many other diluited models. We remark here that analogous equations arise when computing Gaussian fluctuations around instantons for determining the large order behavior of perturbation theory in quantum field theory [MPR17]. Combined with cavity arguments [Par77], instantons could be an useful approach in computing finite-size corrections in diluted models without resorting to replicas.
  - Chapter 4: The random fractional matching problem.
    We study here the random fractional matching problem, which is a relaxation of the random matching problem analyzed in Chapter 3. We show how this slight modification to the problem does not modify the average optimal cost but only finite-size corrections [Luc+18]. In addition we show that the determinant contribution of the random matching problem completely disappears. We will also consider a "loopy" variation of the model which was extensively studied by the mathematical community.
  - Chapter 5: Replica-cavity connection
    Here shall investigate the key connection between replica and cavity quantities in random-link combinatorial optimization problems. This is of interest because for both the random-link 2-factor and the random-link traveling salesman problem one encounters some technical prob-

lems when dealing with replicas, whereas there are no such issues using the cavity method. Since the replica method is the best-suited approach for computing finite-size corrections to average quantities of interest, one still does not know how to compute them in such two models. We prove here a relation that has been mentioned in [MP86b] and we show that it is valid for the random-link matching, the 2-factor and traveling salesman problem. Our approach is based on the use of Belief Propagation equations.

- Part III: Finite Dimension
  - Chapter 6: Euclidean TSP in one dimension.

    Encouraged by some recent results on the bipartite Euclidean matching problems in 2 dimensions we started studying a much more challenging problem from the point of view of complexity theory: the Traveling Salesman Problem (TSP). We started by studying it in one dimension, both on the fully connected [Car+18b] and complete bipartite topologies [Car+18a]. Here both problems will be solved when the cost is a convex and an increasing function of the Euclidean distance between points unveiling that for every realization of the disorder the solution is always of the same "shape". This shows also that in one dimension, the TSP is polynomial. In the fully connected case we will also completely solve the problem in the increasing but concave case and in the decreasing case. The scaling of the cost will be showed to be the same of previously studied combinatorial optimization problems. In particular in the bipartite case, where the an anomalous scaling of the cost is present due to the local difference of the number of red and blue points, we establish a very general connection with the assignment problem, namely the average optimal cost of the one-dimensional TSP tends, for large number of points, to 2 times the average optimal cost of the assignment.
  - Chapter 7: The Euclidean 2-factor problem in one dimension.

    From the analytical findings of Chapter 6 many results follow for the Euclidean 2-factor problem in one dimension [CDGM18a]. This problem can be seen as a relaxation of the TSP since one does not have the constraint of having only one loop in a configuration. We will show how even if it is a one-dimensional model, it is not a completely trivial one, since, when the cost is a convex and increasing function with a power $p = 2$ of the Euclidean distance between points, there appears an exponential number of possible solutions, differently to what happens in previously studied models. We will derive some upper bounds to the average optimal cost and we compare the analytical results with numerical simulations. In Appendix E we will sketch how to derive in the $p > 1$ generic case, not only those upper bounds, but also the average optimal cost in the assignment and TSP using Selberg integrals [CDGM18b]. Their use in the context of combinatorial optimization were substantially ignored and there were no explicit formula

in the generic $p > 1$ case even in the well-studied matching case.

- Chapter 8: Going to higher dimension.

  In this Chapter the analogy between the bipartite Euclidean TSP and the assignment problem found in Chapter 6 will be extended in 2 dimensions and discussed in higher ones [Cap+18]. In those cases the Euclidean TSP is really an "hard" problem from the point of view of complexity theory. The results will be justified by a scale argument and confirmed by extensive numerical simulations. Also the 2-factor problem will be discussed.

• Part IV: Conclusions

- Chapter 9: Conclusions and perspectives

  In this final Chapter we will discuss the results obtained in this thesis and we will highlight several possible directions of future work.

# Part I

# Preliminaries

# Chapter 1

# Phase Transitions in pure and disordered systems

In this chapter we shall give a very basic introduction to concepts and techniques used in statistical physics that are useful to treat disordered systems. The same physical and mathematical ideas turn out to have numerous multidisciplinary applications. After a brief list of definitions of graph theory (section 1.1) which has the task of fixing the notation once and for all, I will discuss the concepts of mean field and Bethe-Peierls approximations in section 1.2, then I will recall some basic notions of spin glass theory in section 1.3. In the last section 1.4 I will finally introduce the main topic of my thesis work: combinatorial optimization.

## 1.1 Graph Theory preamble

It is essential to recall some basic definitions of graph theory. A *graph* is an ordered pair $\mathcal{G} = (\mathcal{V}, \mathcal{E})$ where $\mathcal{V}$ is a set of elements called *vertices* (or *nodes*) and $\mathcal{E}$ is a set of pairs of vertices called *edges*. We will denote the cardinality of the set of vertices by $|\mathcal{V}| = N$ and with $(i, j)$ an edge that connects vertex $i$ with vertex $j$. A graph is said to be *undirected* if the edges have no orientation, i.e. the edge set $\mathcal{E}$ is a set of *unordered* pairs of vertices and *directed* otherwise. A graph is *simple* if there are no *multiple edges* between two vertices and there are no *self-loop* i.e. edges that connects a vertices to itself $(i, i)$. From now on we will focus on simple and undirected graphs if not otherwise stated. A graph is *weighted* if it is assigned, to every edge of the graph, a real number. We will say that two vertices $i$ and $j$ are *adjacent* (or *neighbors*) is there is an edge $e \in \mathcal{E}$ such that $e = (i, j)$, i.e. it connects $i$ with $j$. The *neighborhood* of a vertex $i$ is the set $\partial i$ that contains all the vertices adjacent to $i$ i.e. $\partial i = \{j \in \mathcal{V}; (i, j) \in \mathcal{E}\}$. The *connectivity* (or *degree*) of a vertex $i$ is the cardinality of its neighborhood $|\partial i|$. Note that for every graph $\mathcal{G}$ it holds $\sum_{i=1}^{N} |\partial i| = 2\,|E|$. A vertex which has degree zero is called *isolated* vertex whereas a vertex with degree one is called a *leaf*. A graph is *k-regular* if $|\partial i| = k$ for every $i \in \mathcal{V}$. A graph is *bipartite* if its vertex set can be partitioned into two set $\mathcal{V}_1$ and $\mathcal{V}_2$ such that for every edge $e = (i, j) \in \mathcal{E}$ one has $i \in \mathcal{V}_1$ and $j \in \mathcal{V}_2$. The *adjacency*

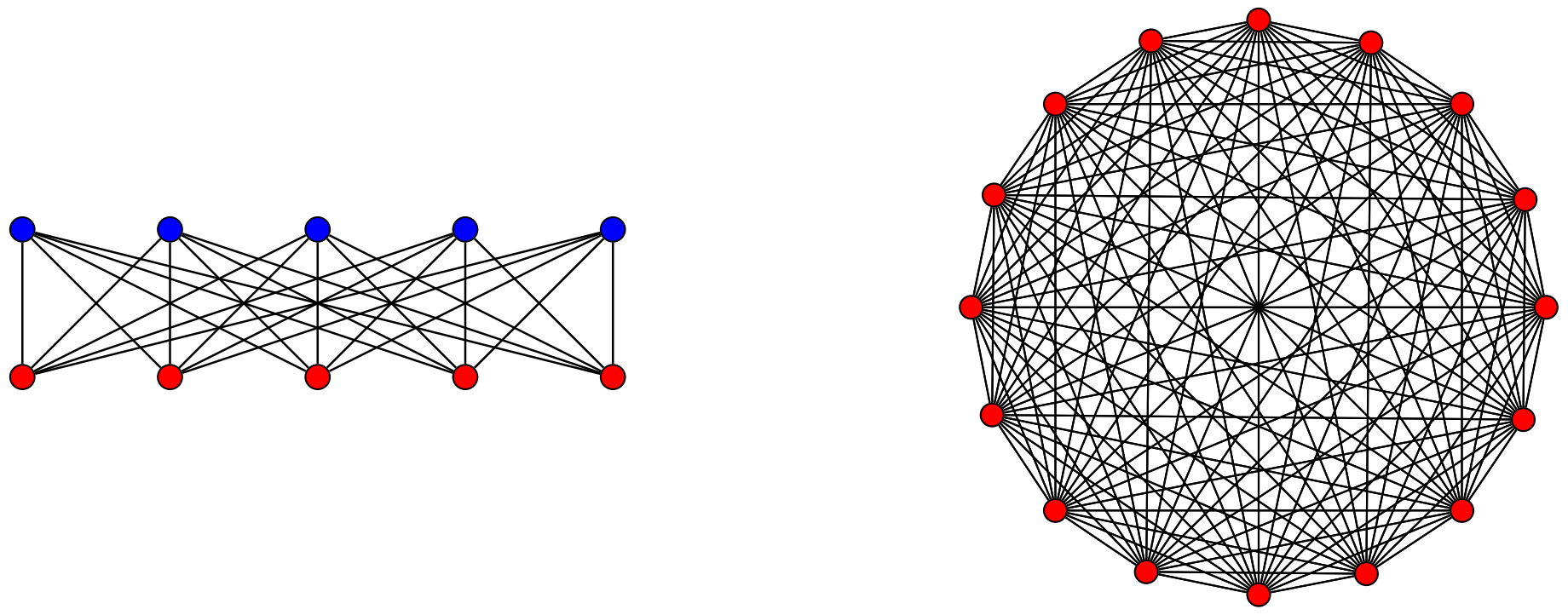

**(a)** The complete bipartite graph $\mathcal{K}_{N,M}$ with $N = M = 5$.

**(b)** The complete graph $\mathcal{K}_N$ with $N = 16$.

**Figure 1.1**

*matrix* of a graph $\mathcal{G}$ is defined as

$$A_{ij} = \begin{cases} 1 & \text{if } (i,j) \in \mathcal{E} \\ 0 & \text{otherwise}. \end{cases} \tag{1.1}$$

The adjacency matrix is symmetric by construction $A^T = A$ and has the property

$$\sum_{j=1}^{N} A_{ij} = |\partial i| \ . \tag{1.2}$$

If the graph is bipartite with, for simplicity, the same cardinality $N$ of the two sets of vertices the adjacency matrix can be written in the block form

$$A = \begin{pmatrix} \mathbf{0} & B \\ B^T & \mathbf{0} \end{pmatrix} . \tag{1.3}$$

where $B$ is a $N \times N$ matrix containing the only non-zero entries corresponding to edges connecting the two different types of points. Of course $B$ uniquely identifies the adjacency matrix $A$.

At this point it is interesting to introduce two types of graphs that will be used extensively throughout this thesis. The *complete graph* of $N$ vertices $\mathcal{K}_N$ is a graph such that which every pair of vertices is connected by an edge (the adjacency matrix is composed by all 1 beside the zeros on the diagonal). The *complete bipartite graph* $\mathcal{K}_{N,M}$ is a bipartite graph with $|\mathcal{V}_1| = N$ and $|\mathcal{V}_2| = M$ that is complete, i.e. it has every possible edge between the two sets of vertices (the matrix $B$ in (1.3) is composed by all 1).

A *walk* of length $k$ in a graph $\mathcal{G}$ is an alternating sequence of vertices and edges (non necessarily distinct) $v_0\, e_1\, v_1 \ldots v_{k-1}\, e_k\, v_k$ starting and ending on a vertex such that $e_n = (v_{n-1}, v_n)$ with $n = 1, \ldots, k$. Of course in simple graphs the edge that connects vertex $v_{n-1}$ with $v_n$ is uniquely identified, so that one can write simply a sequence of vertices. A *trail* is a walk such that all of the edges are distinct. A *path* is a walk such that all of the vertices and edges are distinct. A graph is *connected* if there is a path connecting every pairs of vertices and it is *disconnected* if is not connected. The *distance* between two vertices is the length of the shortest path joining them (if there is no path joining them by convention the distance is infinite). We will also say that a vertex $j$ is the $k$-th neighbour of a vertex $i$ if it is at a distance $k$ from $i$. A walk is *closed* if the starting and ending vertex is the same. A closed trail is called *circuit*. A closed path is called *cycle* (or *loop*). A circuit that passes through all the edges of the graph is called *Eulerian*. If a graph has a Eulerian circuit it is called Eulerian. A cycle that passes through all the vertices is called *Hamiltonian*. If a graph has a Hamiltonian cycle the graph is called Hamiltonian. If the graph contains no cycles the graph is a called a *forest*; if in addition it is connected the graph is called a *tree*. A forest is a collection of trees.

A *subgraph* of a graph $\mathcal{G} = (\mathcal{V}, \mathcal{E})$ is a graph $\mathcal{G}' = (\mathcal{V}', \mathcal{E}')$ such that $\mathcal{V}' \subseteq \mathcal{V}$ and $\mathcal{E}' \subseteq \mathcal{E}$. A subgraph $\mathcal{G}'$ of a graph $\mathcal{G}$ is said to be *spanning* (or a *factor*) if $\mathcal{V}' = \mathcal{V}$. It is in general interesting to look at specific spanning subgraphs of a graph $\mathcal{G}$. A factor which is $k$-regular is called *$k$-factor*. The adjacency matrix $A$ of a $k$-factor on a simple graph has exactly $k$ entries 1 in each row and therefore in each column, i.e. has to satisfy the constraints

$$\begin{aligned} &\sum_{j=1}^{|\mathcal{V}|} A_{ij} = k\,, && i \in [|\mathcal{V}|] \\ &A_{ij} \in \{0,1\}\,, && A_{ij} \le G_{ij} \end{aligned} \tag{1.4}$$

where $G$ is the adjacency matrix of the whole graph $\mathcal{G}$. A 1-factor is also called a (perfect) *matching*. If the graph $G = \mathcal{K}_{N,N}$ then an 1-factor is usually called an *assignment*. A 2-factor is a *loop covering* of the graph. An Hamiltonian cycle can be seen as a 2-factor with only one cycle. If the spanning subgraph is a tree (forest) the subgraph is called a *spanning tree* (*forest*).

### 1.1.1 Random Graphs

It is often useful, not only on a purely mathematical level, but also from a physics point of view, to introduce graphs generated according to a certain probability distribution, the so-called *random graphs* [Bol98]. More precisely a random graph is a couple $\mathcal{G}_P = (G, P)$ where $G$ is a set of graphs and $P$ is a probability law defined over $G$. In general one is interested in studying typical properties of the ensemble and not in a particular realization. This can be done by averaging over the ensemble of graphs, i.e. given an observable $F$ to perform

$$\langle F \rangle \equiv \sum_{\mathcal{G} \in G} P[\mathcal{G}] F(\mathcal{G})\,. \tag{1.5}$$

Three popular models of random graphs with $N$ vertices are the following

- *Gilbert* ensemble $\mathcal{G}_{N,N_E}$: $G$ is the set of graphs of $N$ vertices with a fixed number of edges $N_E$. The probability law is uniform over all the set $G$, that is

$$P[\mathcal{G}] = \binom{\binom{N}{2}}{N_E}^{-1}\,. \tag{1.6}$$

- *Erdős-Rényi* (ER) ensemble $\mathcal{G}_{N,p}$: here every possible edge of the graph is included with a probability $p$ independent from every other edge. The probability to extract a graph with $N_E$ edges is binomial

$$P[\mathcal{G}] = \binom{\binom{N}{2}}{N_E} p^{N_E} (1-p)^{\binom{N}{2}-N_E} . \tag{1.7}$$

- *Random Regular Graphs* (RRG) $\mathcal{G}_{N,c}$: the graph is extracted with uniform measure from the set $G$ of all possible $c$-regular graphs.

We will now analyze some basic properties of these three ensembles. The basic difference between the Gilbert and the ER ensemble is that in the second one there are no correlation between the links. As a neat result the number of edges is itself a random variable with mean

$$\langle N_E \rangle = p \binom{N}{2} . \tag{1.8}$$

In the ER ensemble when $N$ goes to infinity, different (typical) properties of the graph can be derived according to different choices of the scaling of $p$ with $N$. For example when $p \simeq N^{-\alpha}$ with $\alpha > 1$, the graph is almost surely a forest, for $p \simeq \ln N/N$ becomes connected, whereas when $p \simeq (\ln N + \ln(\ln N))/N$ the typical graph becomes Hamiltonian [Bol98, HW06]. Here we will simply set $p = c/N$. The same applies to the Gilbert ensemble: in the following we will set $N_E = cN/2$; this will be useful because in the limit $N \to \infty$ the Gilbert and the ER ensemble share common properties. One important observable to look at is the *degree distribution* $p_k(N)$, that is the probability that, for a random vertex of the graph, the degree is equal to $k$

$$p_k(N) = \binom{N-1}{k} \left(\frac{c}{N}\right)^k \left(1 - \frac{c}{N}\right)^{N-1-k} \tag{1.9}$$

which still follows a binomial distribution. When $N$ tends to infinity, with $c$ finite, the degree distribution tends to a Poisson distribution

$$p_k = \frac{c^k}{k!} e^{-c} \tag{1.10}$$

with parameter $c$ which therefore can be identified with the mean degree or connectivity of the graph $\langle k \rangle = c$. Since the typical degree of the graph stays finite in the thermodynamic limit (and in the RRG case it is fixed by definition), these graphs are called *sparse* random graphs. The Gilbert and ER random graphs are also called *Poissonian*, because their degree distribution follows a Poisson distribution. Another quantity of interest is the degree distribution $\tilde{p}_k$ of a vertex which is an end of a randomly chosen edge of the random graph. This will be of course equal to the probability to extract a vertex with degree $k$ times the number of possible ways $k$ of arrange an edge on this vertex times a normalization

$$\tilde{p}_k = \frac{k p_k}{c} = \frac{c^{k-1}}{(k-1)!} e^{-c} . \tag{1.11}$$

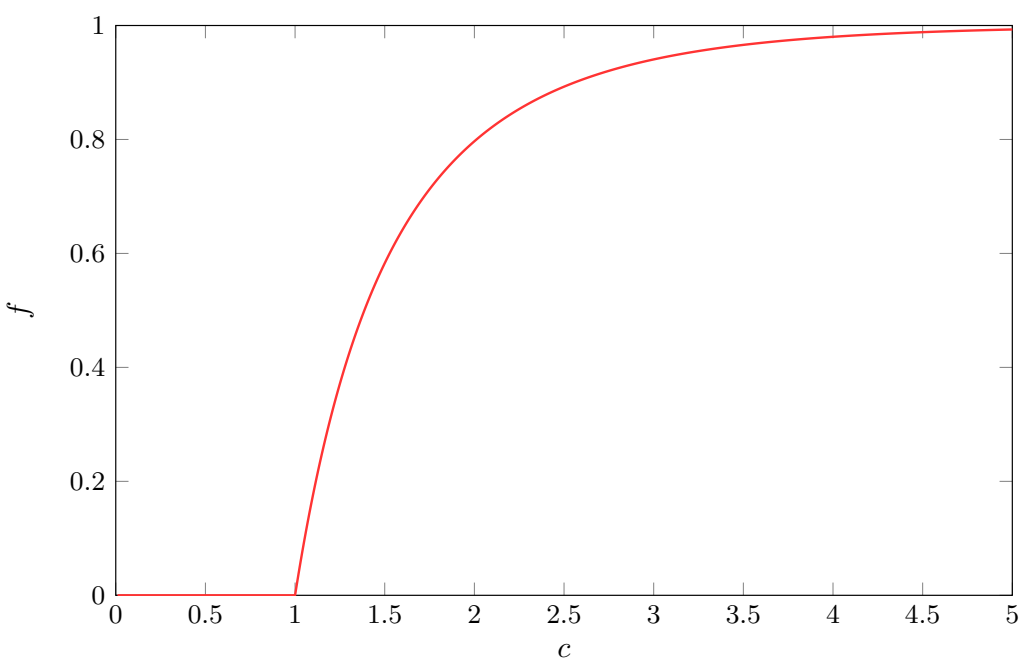

**Figure 1.2.** Fraction of vertices belonging to the giant component as a function of the mean degree, see equation (1.15).

The average degree of these vertices (also called *edge perspective degree*) is trivially $\sum_k k\,\tilde{p}_k = c+1$, i.e. the randomly selected edge plus one contribution coming from $c$ *excess edges*. Of course this result differs from the RRG case, for which the excess edge contribution is always $c-1$. Another important property of these three ensembles of random graphs is the fact that they are *locally tree-like*. To see this explicitly let us compute for $N \to \infty$ the probability of having a subgraph with $n$ vertices which is a tree. Since a tree with $n$ vertices has $n-1$ edges, this probability is proportional to

$$\frac{N!}{(N-n)!}\left(\frac{c}{N}\right)^{n-1} = Nc^{n-1} + O(1)\,, \tag{1.12}$$

i.e. it scales linearly with the size of the graph. If one instead wants to compute the probability to have a subgraph which is a loop of length $n$, one needs to add to a tree of $n$ vertices one link, that is to multiply the previous quantity by a factor $c/N$. This probability is therefore depressed by a factor $N$. If one adds other edges between the vertices of the loop this probability is even lower. This does not mean that loops are not present in principle, but they diverge with $N$. In fact, using the tree-like property, the (average) number of vertices within a distance $l$ from a center vertex grows as $c^l$; when this quantity becomes of order $N$ some loops can appear. Therefore the length of the loops scales logarithmically with $N$

$$l \sim \frac{\ln N}{\ln c}\,. \tag{1.13}$$

One last thing to cite is how the graph changes by varying the mean connectivity $c$. Intuitively one expects a phase transition tuning $c$: for $c \ll 0$ the typical graph is formed by an high number of isolated vertices, whereas as we increase $c$ the number of connected components decrease. It is interesting therefore to study the *size* of the largest component $\mathcal{M}(c)$ in the typical realization of a ER or Gilbert random graph. An important result of Erdős and Rényi states that, the size of the largest component behaves for $N \to \infty$ as

$$\mathcal{M}(c) = \begin{cases} \frac{\ln N}{a} + O(\ln\ln N) & \text{for } c < 1\,, \\ O(N^{2/3}) & \text{for } c = 1\,, \\ fN + O(\sqrt{N}) & \text{for } c > 1\,, \end{cases} \tag{1.14}$$

where $a \equiv c - 1 - \ln c$ and $f$ satisfies the following transcendental equation

$$1 - f = e^{-cf} . \tag{1.15}$$

For $c > 1$ a *giant component* appears, since it contains an extensive number $fN$ of vertices. For the analogies with percolation [SA14] this phase transition is also called *percolation* transition. It is easy to derive equation (1.15) using a simple argument. In order to evaluate the probability $1 - f$ that a vertex $i$ is not part of the giant component we need to impose that $i$ is not connected to no vertex $j$ which is not in the giant component. We have two possibilities: in the first case $i$ is connected to $j$ and $j$ does not belong to the giant component, or $i$ is not connected to $j$. In the first case the probability is $p(1 - f)$ whereas in the second is simply $1 - p$. Since there are $N - 1$ vertices $j$, the total probability is

$$1 - f = (1 - pf)^{N-1} , \tag{1.16}$$

which is equal to (1.15) in the limit $N \to \infty$. $f(c)$ is plotted in Fig. 1.2.

We refer to [MMZ01, EMH04] for the analysis of large deviations properties of sparse random graph and their connection with the Potts model of statistical physics.

## 1.2 Models on Graphs

In this section we want to study and develop general statistical techniques for models defined on graphs. To do this it is essential the notion of *factor graph*, which is a very useful formalism in order to keep things as general as possible, as we shall see in a while. A factor graph is simply a bipartite graph $G = (\mathcal{V}, \mathcal{F}, E)$ where $\mathcal{V}$ is the first set of vertices called *variable nodes* and $\mathcal{F}$ the second one, usually denominated *function nodes*. It is convention to use the variables $i, j, \ldots$ and $a, b, \ldots$ to identify variable and function nodes respectively. The factor graph formalism is very useful in statistical physics for the following reason: one can associate to variable nodes the degrees of freedom of the problem under consideration (e.g. spin variables), whereas the function nodes represent interactions between them. A one-body interaction (external field) is simply modeled by a function node $a$ with connectivity $|\partial a| = 1$ whereas a two-body interaction is characterized by having $|\partial a| = 2$. Multi-spins interactions are characterized by $\partial a > 2$. From now on, we shall not include anymore external fields in the set of function nodes $\mathcal{F}$, since they will be added by hand; we will also set $|\mathcal{V}| = N$ and $|\mathcal{F}| = M$.

In statistical physics one usually has a model with $N$ variables or spins $\underline{\sigma} \equiv \sigma_1, \ldots, \sigma_N$, which are taking values in some set $\chi$ (either discrete or continuous). One then has a joint distribution of these variables which we assume to take the general form [MM09]

$$\mu(\underline{\sigma}) \equiv \frac{1}{Z} \prod_{i=1}^{N} \psi_i(\sigma_i) \prod_{a=1}^{M} \psi_a(\underline{\sigma}_{\partial a}) . \tag{1.17}$$

where $\underline{\sigma}_{\partial a}$ is the set of all spins (variable nodes) which are neighbors of the function node $a$. In general, if $\mathcal{S}$ is a set of variable nodes, we will denote by $\underline{\sigma}_S \equiv \{\sigma_i\,;\, i \in \mathcal{S}\}$. As we have anticipated, in (1.17) we have isolated, for convenience, the one-body interactions $\psi_i(\sigma_i)$ from higher ones which are encoded in the function $\psi_a(\underline{\sigma}_{\partial a})$ with $|\partial a| \geq 2$. The functions $\psi_i$ and $\psi_a$ are non-negative by definition and assume real values; $\psi_i$ is usually called *bias*, because it is the equivalent of a magnetic field on site $i$, whereas $\psi_a$ is called *compatibility function*. $Z$ is instead a normalization (the partition function). A particular feature of the joint probability distribution (1.17), is that two variable nodes interact only through the variable nodes which are interposed between them. In mathematical terms, if we have three disjoint sets of variables $A$, $B$, $S \subseteq [N]$ such that there is no path going from $A$ to $B$ without passing through $S$ ($S$ *separates* $A$ and $B$), then the variables $\underline{\sigma}_A$ and $\underline{\sigma}_B$ are *conditionally independent*

$$\mu(\underline{\sigma}_{A\cup B}|\underline{\sigma}_S) = \mu(\underline{\sigma}_A|\underline{\sigma}_S)\mu(\underline{\sigma}_B|\underline{\sigma}_S)\,. \tag{1.18}$$

This is also called *global Markov property* for obvious reasons. Intuitively, when one conditions over the variables belonging to the set $S$ one is graphically canceling all the $\underline{\sigma}_S$, so that the resulting graph is disconnected and $\underline{\sigma}_A$ is independent from $\underline{\sigma}_B$. The great advantage of (1.17) is that one can determine the model by simply giving the form of the bias and of the compatibility functions. A very simple example is given by the $p-$spin Ising model in which every interaction $a \in \mathcal{F}$ has degree $|\partial a| = p$

$$\begin{aligned} \psi_i(\sigma_i) &= e^{\beta h_i \sigma_i} \\ \psi_a(\underline{\sigma}_{\partial a}) &= e^{\beta J_a \sigma_{i_1} \dots \sigma_{i_p}} \end{aligned} \tag{1.19a}$$

where $\beta$ is the inverse temperature, $h_i$ is the local magnetic field and $J_a$ is the magnitude of the $p$-body interaction. A simple picture of this factor graph representation is given in Fig. 1.3 for the $p = 2$ case on a two-dimensional lattice. One usually represents graphically variable nodes with circles, function nodes with black squares, and site-dependent magnetic fields with squares hatched inside. In general the representation (1.17) can be expressed in the usual Boltzmann-Gibbs way by defining an Hamiltonian which is expressed in terms of $\psi_i$ and $\psi_a$ as

$$H(\underline{\sigma}) = -\frac{1}{\beta}\left[\sum_a \ln \psi_a(\underline{\sigma}_{\partial a}) + \sum_i \ln \psi_i(\sigma_i)\right] \tag{1.20}$$

The principal target of statistical physics is to compute the partition function of the system (from which we can derive all other physical quantities of interest), or at least, marginals in an efficient way. For example the one-point marginal $\mu_i$ and the marginal corresponding to the $a$-th function node $\mu_a$

$$\mu_i(\sigma_i) = \sum_{\underline{\sigma}\setminus\sigma_i} \mu(\underline{\sigma})\,, \tag{1.21a}$$

$$\mu_a(\underline{\sigma}_{\partial a}) = \sum_{\underline{\sigma}\setminus\underline{\sigma}_{\partial a}} \mu(\underline{\sigma})\,, \tag{1.21b}$$

where we have used the symbol $\sum$ to denote the sum, or the integral over all values that the variable nodes can assume in $\chi$. It is evident that computing (1.21), for example in the Ising model case, requires $2^{N-1}$ steps, which exponential in $N$, this being usually a very large number. Therefore one is forced to resort to some type of approximation. Those types of approximation are the subject of the following subsections.

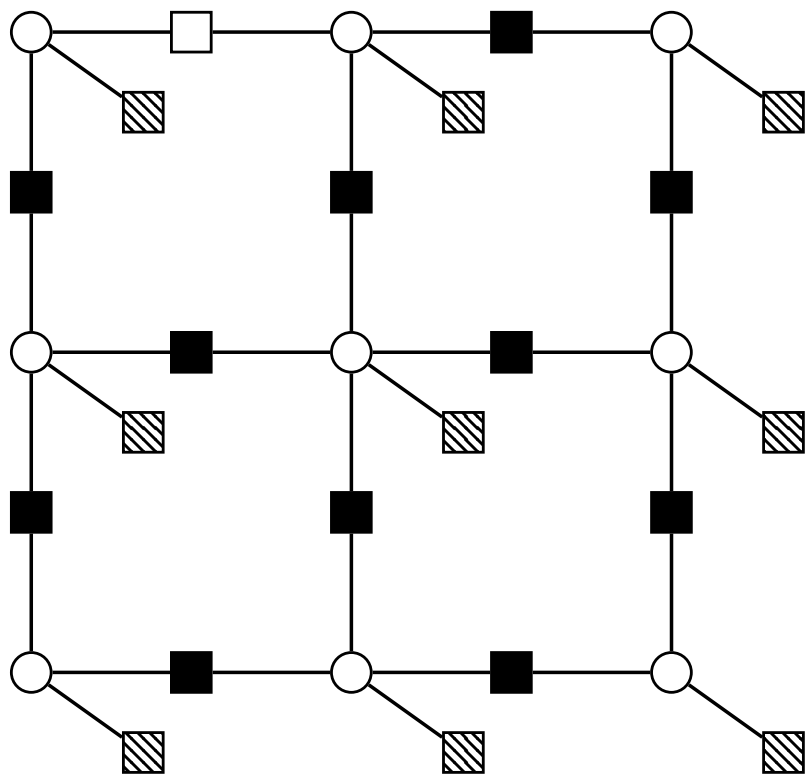

**Figure 1.3.** Factor Graph representation of the $2d$ Ising model with external magnetic fields (represented by a square hatched inside).

### 1.2.1 The "naive" mean field approximation

The first approximation one can have in mind is to write the total joint probability distribution $\mu$ as a product of *beliefs* $b_i$

$$\mu(\underline{\sigma}) \approx \prod_{i=1}^{N} b_i(\sigma_i)\,, \tag{1.22}$$

with $0 \leq b_i(\sigma_i) \leq 1$ for every $i = [N]$ and

$$\sum_{\sigma_i} b_i(\sigma_i) = 1\,. \tag{1.23}$$

What we are really doing here is neglecting interactions and approximate the system as if every spin lives in an *effective* local magnetic field parametrized by $b_i(\sigma_i)$. In addition we are implicitly saying that $\mu_i(\sigma_i) \approx b_i(\sigma_i)$, i.e. $b_i(\sigma_i)$ is our "belief" of what the true marginal is. This (very crude) approximation is called *naive mean field approximation* (MF). Obviously (1.22) is not the true form of the equilibrium probability distribution. But what we can do is to minimize the Gibbs free energy

$$G[\mu] \equiv \sum_{\underline{\sigma}} \mu(\underline{\sigma}) H(\underline{\sigma}) + \frac{1}{\beta} \sum_{\underline{\sigma}} \mu(\underline{\sigma}) \ln \mu(\underline{\sigma})\,, \tag{1.24}$$

subject to constraint (1.23) in the particular subspace of probability distribution specified by (1.22). This variational procedure provides the optimal choices for the beliefs $b_i(\sigma_i)$. Let us do a simple, well-known example: the Ising model on a generic graph $\mathcal{G}$

$$H(\underline{\sigma}) = -\sum_{(i,j)\in\mathcal{G}} J_{ij}\sigma_i\sigma_j - \sum_i h_i\sigma_i \tag{1.25}$$

Since in this case the spins have only two degrees of freedom $\sigma_i = \pm 1$ the only parametrization possible for the beliefs is

$$b_i(\sigma_i) = \frac{1 + m_i \sigma_i}{2}, \tag{1.26}$$

where $m_i$ are $N$ variational parameters that physically represent local magnetizations

$$m_i = \langle \sigma_i \rangle_{\mathrm{MF}} \equiv \sum_i \sigma_i \, b_i(\sigma_i). \tag{1.27}$$

Inserting (1.26) into (1.24) we get

$$\begin{aligned} G[\underline{m_i}] = &- \sum_{(i,j)\in\mathcal{G}} J_{ij} m_i m_j - \sum_i h_i m_i \\ &+ \frac{1}{\beta} \sum_i \left[ \frac{1-m_i}{2} \ln\left(\frac{1-m_i}{2}\right) + \frac{1+m_i}{2} \ln\left(\frac{1+m_i}{2}\right) \right]. \end{aligned} \tag{1.28}$$

Differentiating the Gibbs free energy with respect to $m_i$ we find the famous mean field equations for the local magnetizations

$$m_i = \tanh\left[ \beta \left( \sum_{j\in\partial i} J_{ij} m_j + h_i \right) \right]. \tag{1.29}$$

When the graph $\mathcal{G}$ is an hypercubic lattice in $d$ dimensions with all the interactions being uniform and ferromagnetic ($J_{ij} = J > 0$) and with $h_i = h$ (so that the magnetizations are uniform) we have

$$m = \tanh\left[\beta \left(2dJm + h\right)\right]. \tag{1.30}$$

When $h = 0$ this equation of state predicts a phase transition at the critical temperature $T_c = 2dJ$ and a non zero value of the total magnetization $m$ at $T < T_c$. This is clearly a wrong prediction in one dimension, where the Ising model has no phase transition. In $d = 2$ the critical temperature predicted by the mean field approximation is $T_c = 4J$, to be compared with Onsager's exact solution $T_c = 2J/\ln(1+\sqrt{2}) \simeq 2.269J$. It can be shown that the critical exponents predicted by the mean field approximation are correct only for $d > d_u$ where the upper critical dimension $d_u = 4$ in this case. Nonetheless we remind that this approximation is exact in the limit $d \to \infty$ i.e. when the spins are located on the vertices of a complete graph $\mathcal{K}_N$. The ferromagnetic fully-connected Ising model is called *Curie-Weiss* model; its Hamiltonian is

$$H(\underline{\sigma}) = -\frac{J}{2N} \sum_{i\neq j} \sigma_i \sigma_j - h \sum_i \sigma_i. \tag{1.31}$$

Using a Gaussian transformation (as we will often do), the partition function of this model can be written as

$$Z = \sqrt{\frac{\beta J N}{2\pi}} \int dm \, e^{-N\left[-\frac{\beta J}{2} m^2 + \ln(2\cosh(\beta J m + \beta h))\right]}, \tag{1.32}$$

so that using the saddle-point approximation we recover (1.29) by performing the derivative of the exponent.

### 1.2.2 The Bethe-Peierls approximation

One can do better than naive mean field since it completely neglects correlations. In first approximation, we can take into account correlations through variables that are involved in the same interaction, which we expect to be the ones more correlated between each other. The factorization

$$\mu(\underline{\sigma}) \approx \prod_{i=1}^{N} b_i(\sigma_i)^{1-|\partial i|} \prod_{a=1}^{M} b_a(\underline{\sigma}_{\partial a}) \,, \tag{1.33}$$

with the conditions $0 \leq b_i, b_a \leq 1$ and

$$\sum_{\underline{\sigma}_{\partial a \setminus i}} b_a(\underline{\sigma}_{\partial a}) = b_i(\sigma_i) \,, \tag{1.34a}$$

$$\sum_{\sigma_i} b_i(\sigma_i) = 1 \,, \tag{1.34b}$$

is called the *Bethe-Peierls approximation*. The factor $b_i(\sigma_i)^{-|\partial i|}$ was added to avoid over-counting of $\sigma_i$. The question is the same as before: how much the beliefs $b_i(\sigma_i)$ and $b_a(\sigma_{\partial a})$ are similar to the *true* marginals respectively $\mu_i(\sigma_i)$ and $\mu_a(\sigma_{\partial a})$? The general variational procedure is always the same: one defines a Lagrangian which is function of the beliefs and of the Lagrange multipliers that impose constraints on them. Then one takes derivatives of this Lagrangian and finds the beliefs that extremize it. We remind to [Lup17] for these details. To give a flavor of what's going on, let us take, as before, the (pairwise) Ising model case; the beliefs satisfying conditions (1.34) are

$$b_i(\sigma_i) = \frac{1 + m_i \sigma_i}{2} \,, \tag{1.35a}$$

$$b_{(ij)}(\sigma_i, \sigma_j) = \frac{1 + m_i \sigma_j + m_j \sigma_j + m_{ij} \sigma_i \sigma_j}{4} \,. \tag{1.35b}$$

Now we have $M$ additional variational parameters $m_{ij}$ with respect to naive mean field: we are extending the domain in the space of probability distributions where we want to find a minimum of the Gibbs free energy. On the hypercubic lattice, the Bethe-Peierls approximation predicts a phase transition at the critical temperature

$$T_c = 2J \left[ \ln \left( \frac{d}{d-1} \right) \right]^{-1} \,, \tag{1.36}$$

i.e. it correctly predicts no phase transition in $d = 1$ and it gives a critical temperature $T_c = 2J / \ln 2 \simeq 2.885J$ in $d = 2$ which is still higher than the exact result $T_c = 2J / \ln(1 + \sqrt{2}) \simeq 2.269J$ but better than the naive mean field estimate.

Let us now see when the Bethe-Peierls approximation is exact. Note that (1.33), in general, is even not normalized! Since we have considered correlations involving variables connected to the same function node $a$, it is clear that we are simply saying that two variables $i, j \in \partial a$ interact only through $a$ and with no one else, so that in general

$$\mu(\underline{\sigma}_{\partial a \setminus i} | \sigma_i) = \prod_{k \in \partial a \setminus i} \mu(\sigma_k | \sigma_i) \,, \tag{1.37}$$

i.e. they are conditionally independent only via $a$. This means that for every couple of variables $i$ and $j$ there is only one path joining them. With this reasoning we expect that, if the factor graph is a *tree*, the Bethe-Peierls is exact, i.e.

$$\mu(\underline{\sigma}) = \prod_{i=1}^{N} \mu_i(\sigma_i)^{1-|\partial i|} \prod_{a=1}^{M} \mu_a(\underline{\sigma}_{\partial a}) \,. \tag{1.38}$$

We can prove this result by induction on the number of interactions. For $M = 1$ equation (1.38) is trivially satisfied. Let us suppose now that (1.38) is valid on the tree $\mathcal{T}$ which is composed of $M-1$ function nodes and let us prove that this is valid also for the tree $\mathcal{T}'$ which has the same interactions of $\mathcal{T}$ plus one, that we will call $a$, which is attached to the vertex $i$. Thanks to the global Markov property (1.18), the joint probability distribution of all the variables $\underline{\sigma}'$ in the tree $\mathcal{T}'$ is

$$\mu_{\mathcal{T}'}(\underline{\sigma}') = \mu_{\mathcal{T}}(\underline{\sigma})\mu(\underline{\sigma}_{\partial a \setminus i}|\sigma_i) = \mu_{\mathcal{T}}(\underline{\sigma})\frac{\mu_a(\underline{\sigma}_{\partial a})}{\mu_i(\sigma_i)} \,, \tag{1.39}$$

which is in the same form of (1.38) because the factor $\mu_a(\underline{\sigma}_{\partial a})$ takes into account the new interaction $a$ whereas the factor $\mu_i(\sigma_i)$ takes into account the change of the degree of $i$ when $a$ is attached to the tree $\mathcal{T}$. Of course on tree graphical models things are easy; however the Bethe-Peierls approximation could work well whenever on the factor graph the correlations between distant variables decay fast enough. One important example is that of random (factor) graph with the property to be locally tree-like as happens for the Gilbert, ER and RRG ensemble, in which typical loops are of the size $\ln N$ as we have examined in subsection 1.1.1.

Finally we mention that having found the beliefs that minimize the Gibbs free energy, one can compute all the quantities of interests from the *Bethe free energy* which is written as

$$\begin{aligned}\beta F_{\mathrm{B}}(b) = & \sum_a \sum_{\underline{\sigma}_{\partial a}} b_a(\underline{\sigma}_{\partial a}) \ln \frac{b_a(\underline{\sigma}_{\partial a})}{\psi_a(\underline{\sigma}_{\partial a})} + \sum_i (1-|\partial i|) \sum_{\sigma_i} b_i(\sigma_i) \ln b_i(\sigma_i) \\ & - \sum_i \sum_{\sigma_i} b_i(\sigma_i) \ln \psi_i(\sigma_i)\end{aligned} \tag{1.40}$$

### 1.2.3 Belief Propagation

Suppose we have a tree graphical model. We can introduce two quantities $Z_{i\to a}(\sigma_i)$ and $\hat{Z}_{a\to i}(\sigma_i)$ defined on a certain edge $(i,a)$ of the factor graph called *cavity partition functions*. $Z_{i\to a}(\sigma_i)$ is the partition function for the subtree rooted at $i$ having removed interaction $a$ and conditioned over the value of variable $i$ to be $\sigma_i$. Analogously $\hat{Z}_{a\to i}(\sigma_i)$ is the partition function for the subtree rooted at $i$ having removed all the interactions $b \in \partial i \setminus a$ and conditioned over the value of variable $i$ to be $\sigma_i$. Those quantities satisfy the recursion relations

$$Z_{i\to a}(\sigma_i) = \psi_i(\sigma_i) \prod_{b \in \partial i \setminus a} \hat{Z}_{b\to i}(\sigma_i) \tag{1.41a}$$

$$\hat{Z}_{a\to i}(\sigma_i) = \sum_{\underline{\sigma}_{\partial a \setminus i}} \psi_a(\underline{\sigma}_{\partial a}) \prod_{j \in \partial a \setminus i} Z_{j\to a}(\sigma_j) \,. \tag{1.41b}$$

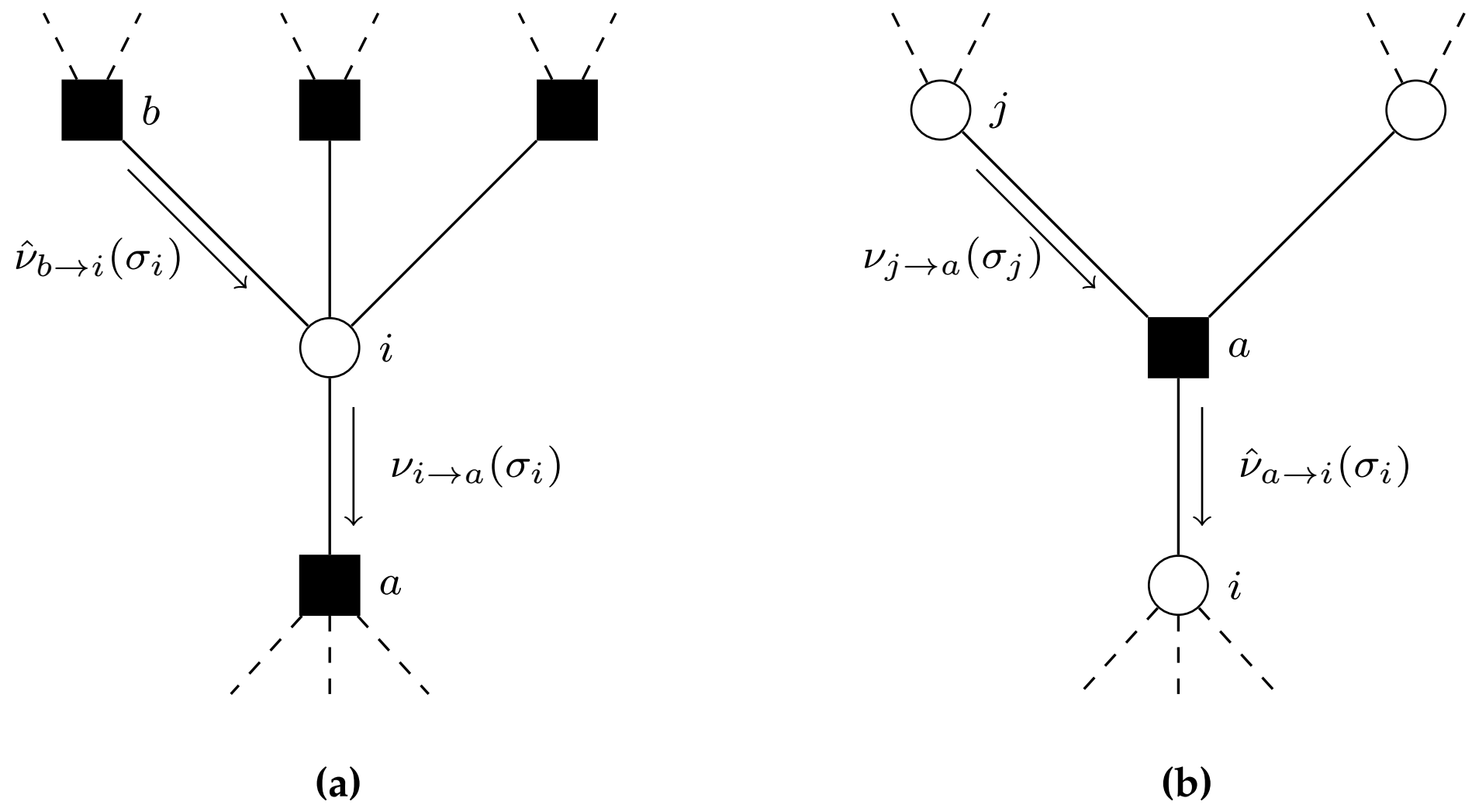

**Figure 1.4.** Parts of the graph involved in the computation of $\nu_{i\to a}(\sigma_i)$ (left panel) and $\hat{\nu}_{a\to i}(\sigma_i)$ (right panel). These quantities are written in the BP equations (1.44) in function respectively of $\hat{\nu}_{b\to i}(\sigma_i)$ and $\nu_{j\to a}(\sigma_j)$ (displayed also in the figures), with $b \in \partial i \backslash a$ and $j \in \partial a \backslash i$.

Since we are on a tree, the marginals (1.21) can be written easily as a product of cavity partition functions. Therefore we get

$$\mu_i(\sigma_i) = \frac{1}{Z}\psi_i(\sigma_i) \prod_{b\in\partial i} \hat{Z}_{b\to i}(\sigma_i) \tag{1.42a}$$

$$\mu_a(\underline{\sigma}_{\partial a}) = \frac{1}{Z}\psi_a(\underline{\sigma}_{\partial a}) \prod_{j\in\partial a} Z_{j\to a}(\sigma_j)\,, \tag{1.42b}$$

We can also introduce two quantities $\nu_{i\to a}(\sigma_i)$ and $\hat{\nu}_{a\to i}(\sigma_i)$ defined on a certain edge $(i,a)$ of the factor graph called *cavity marginals* or *messages* defined as

$$\nu_{i\to a}(\sigma_i) = \frac{Z_{i\to a}}{\sum_{\sigma_i} Z_{i\to a}} \tag{1.43a}$$

$$\hat{\nu}_{a\to i}(\sigma_i) = \frac{\hat{Z}_{a\to i}}{\sum_{\sigma_i} \hat{Z}_{a\to i}} \tag{1.43b}$$

More precisely, $\nu_{i\to a}(\sigma_i)$ is the marginal of variable $\sigma_i$ in a modified graphical model in which we have removed interaction $a$; $\hat{\nu}_{a\to i}(\sigma_i)$ is instead the marginal of $\sigma_i$ when all function nodes $b \in \partial i \backslash a$ are removed from the factor graph. One can write iterative equations for the cavity marginals analogous to 1.41 which are called *Belief-Propagation equations* (BP)

$$\nu_{i\to a}(\sigma_i) = \frac{1}{z_{i\to a}}\psi_i(\sigma_i) \prod_{b\in\partial i\backslash a} \hat{\nu}_{b\to i}(\sigma_i) \tag{1.44a}$$

$$\hat{\nu}_{a\to i}(\sigma_i) = \frac{1}{\hat{z}_{a\to i}} \sum_{\underline{\sigma}_{\partial a\backslash i}} \psi_a(\underline{\sigma}_{\partial a}) \prod_{j\in\partial a\backslash i} \nu_{j\to a}(\sigma_j) \tag{1.44b}$$

where $z_{i\to a}$ and $\hat{z}_{a\to i}$ are normalization constants. These equations can also be used as an *algorithm*, called *message passing*, since they can be solved iteratively starting from the leaves and iterating into the core of the tree; the use of equations (1.44) are preferred to (1.41) because they involve quantities that are normalized. The time needed scales is at most $Mkq^k$ where $k = \max_a |\partial a|$ and $q$ is the (maximum) number of degrees of freedom a variable can take. Nevertheless BP equations can be used over all types of graphs. One can prove, in addition, that the BP equations are the same self-consistent one derives with the variational approach, and therefore the BP fixed point are in biunivocal relation with stationary point of the Bethe free energy. In terms of messages equations (1.42) can be rewritten as

$$\mu_i(\sigma_i) = \frac{1}{z_i}\psi_i(\sigma_i)\prod_{b\in\partial i}\hat{\nu}_{b\to i}(\sigma_i) \tag{1.45a}$$

$$\mu_a(\underline{\sigma}_{\partial a}) = \frac{1}{z_a}\psi_a(\underline{\sigma}_{\partial a})\prod_{j\in\partial a}\nu_{j\to a}(\sigma_j)\,. \tag{1.45b}$$

In (1.45) we have also introduced new normalizations $z_i$ and $z_a$; they are given by

$$z_i = \sum_{\sigma_i}\psi_i(\sigma_i)\prod_{b\in\partial i}\hat{\nu}_{b\to i}(\sigma_i) \tag{1.46a}$$

$$z_a = \sum_{\underline{\sigma}_{\partial a}}\psi_a(\underline{\sigma}_{\partial a})\prod_{j\in\partial a}\nu_{j\to a}(\sigma_j)\,. \tag{1.46b}$$

Note that the one point marginal, using BP equations can be written simply as

$$\mu_i(\sigma_i) = \frac{1}{z_{ia}}\hat{\nu}_{b\to i}(\sigma_i)\,\nu_{i\to b}(\sigma_i)\,, \qquad \forall b\in\partial i \tag{1.47}$$

where

$$z_{ia} = \frac{z_i}{z_{i\to a}} = \sum_{\sigma_i}\nu_{i\to a}(\sigma_i)\hat{\nu}_{a\to i}(\sigma_i)\,. \tag{1.48}$$

Marginalizing (1.45b) one obtains an analogous expression for the same marginal and normalization but this time expressed as

$$z_{ia} = \frac{z_a}{\hat{z}_{a\to i}}\,. \tag{1.49}$$

These relations turn out to be useful in evaluating the Bethe free energy (1.40) in terms of local contributions written in terms of messages. The partition function in fact can be written as

$$Z = z_i\prod_{b\in\partial i}\sum_{\sigma_i}\hat{Z}_{b\to i}(\sigma_i) \tag{1.50}$$

as can be directly inspected by expressing the definition of $z_i$ (1.46a) in terms of cavity partition functions (1.43b). Plugging again the definitions (1.43) into $\hat{z}_{i\to a}$ and $\hat{z}_{a\to i}$ and using BP equations (1.41) we get the essential relations

$$z_{i\to a} = \frac{\sum_{\sigma_i} Z_{i\to a}(\sigma_i)}{\prod_{b\in\partial i\setminus a}\sum_{\sigma_i}\hat{Z}_{b\to i}(\sigma_i)}\,, \tag{1.51a}$$

$$\hat{z}_{a\to i} = \frac{\sum_{\sigma_i}\hat{Z}_{a\to i}(\sigma_i)}{\prod_{j\in\partial a\setminus i}\sum_{\sigma_j} Z_{j\to a}(\sigma_j)}\,. \tag{1.51b}$$

Starting from (1.50), and repetitively using (1.51) together with (1.48) and (1.49), the partition function can be written as

$$Z = z_i \prod_{b\in\partial i} \hat{z}_{b\to i} \prod_{j\in\partial b\setminus i} z_{j\to b} \prod_{c\in\partial j\setminus b} \hat{z}_{c\to j} \cdots = \frac{\prod_i z_i \prod_a z_a}{\prod_{(i,a)} z_{ia}} \tag{1.52}$$

so that the Bethe free energy is

$$-\beta F_{\mathrm{B}}(\nu,\hat{\nu}) = \sum_i \ln z_i + \sum_a \ln z_a - \sum_{(i,a)} \ln z_{ia}\,. \tag{1.53}$$

The terms comparing in the Bethe free energy (1.53) have a direct physical interpretation. Clearly $F_i = \ln z_i$ is a site term, that measures the free energy change when the site $i$ and all its edges are added; $F_a = \ln z_a$ is a local interaction term that gives the free energy change when the function node $a$ is added to the factor graph. Finally $F_{ia} = \ln z_{ia}$ is an edge term, which takes into account the fact that in adding vertex $i$ and $a$, the edge $(i,a)$ is counted twice.

## 1.3 Spin Glasses

Until now, apart for a short digression on random graphs, we have described "pure" models, with the classical example of the Ising model and its mean-field version, which are by now paradigmatic models describing the phenomenology of the ferromagnetic-paramagnetic phase transition.

However, life is not perfect. In real world, in fact, materials are always characterized by the presence of some kind of disorder. To make a few examples, disorder can be induced because of the presence of impurities or defects in the lattice or because of randomness in the position of the spins. When the disorder is relevant [Har74], it affects completely the low-temperature behavior of the material, producing interesting new phenomenons. The materials exhibiting this low-temperature phase were called *random magnets* or *spin-glasses*. At the microscopical level the spin-glass phase is characterized by configurations in which the spins are *frozen* in certain random directions in space. In order to produce such a phase, disorder must fulfill two requirements. Firstly the disorder must be *quenched*, i.e. it is fixed or, at least, it evolves on a time scale much larger than the typical one of the spins. Secondly it must produce *frustration*, that is it must generate competing interactions between the spins. The simplest example of frustration is depicted in Fig. (1.5), where we have a triangular plaquette with two interactions being ferromagnetic and one antiferromagnetic. In this case the ground state of the system becomes degenerate, since one cannot satisfy all the interaction by choosing a direction for the three Ising spins. When this happens the plaquette is said to be frustrated [Frö85]. A plaquette $\mathcal{P}$ of arbitrary length is frustrated if

$$\tau_{\mathcal{P}} = \prod_{(i,j)\in\partial\mathcal{P}} J_{ij} < 0\,, \tag{1.54}$$

i.e. the product of all interaction on the plaquette is negative.

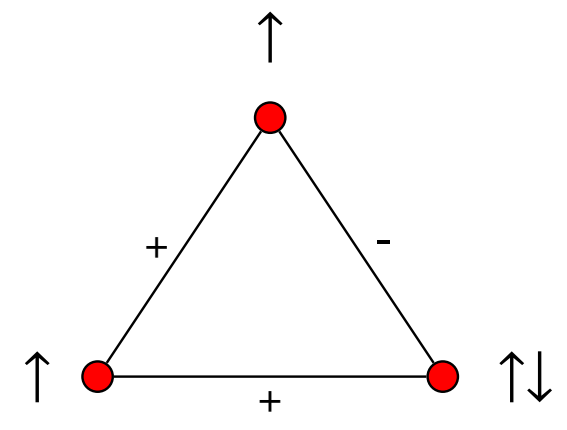

**Figure 1.5.** The simplest example of frustration.

It is important to stress that not any disorder induces frustration. When this happens, the disorder is said to be *irrelevant*. To make an example consider the Mattis model, which is simply an Ising model on the complete graph, the interaction being chosen as

$$J_{ij} = \xi_i \xi_j \,, \tag{1.55}$$

and $\xi_i = \pm 1$ with equal probability. Every plaquette $\mathcal{P}$ is not frustrated because

$$\prod_{(i,j)\in\partial\mathcal{P}} \xi_i \xi_j = 1 \,, \tag{1.56}$$

since the variable $\xi_i$ appears twice for every site $i$ in the plaquette. In general, for an Ising model with random two-body interactions, disorder is irrelevant if there exist a set of variables $\varepsilon_i = \pm 1$ such that with a *gauge transformation* of the spins and coupling

$$\begin{aligned} \sigma_i' &= \varepsilon_i \, \sigma_i \,, \\ J_{ij}' &= J_{ij} \, \varepsilon_i \, \varepsilon_j \,, \end{aligned} \tag{1.57}$$

frustration can be avoided from every plaquette. In the simple example of the Mattis model these variables are identified simply by $\varepsilon_i = \xi_i$.

Experimentally the first samples that were used to study the effects of quenched disorder and frustration were diluted magnetic alloys. These are materials where small quantity of impurities (usually iron or manganese) were inserted at random locations in a substrate consisting of a noble metal, such as gold, silver or copper. The magnetic moments of the $d$-shell electrons of the impurities polarize the $s$-shell conduction electrons of the substrate; this polarization can be positive or negative, depending to the distance from the impurity. The effective interaction between magnetic moments can be modeled by the so called *Ruderman - Kittel - Kasuya - Yosida* (RKKY) interaction

$$J_{\mathbf{xy}} = J_0 \, \frac{\cos\left(2\mathbf{k}_F \cdot \mathbf{r} + \phi_0\right)}{r^3} \,, \tag{1.58}$$

where $\mathbf{k}_F$ is the Fermi wave vector, $\mathbf{r}$ is the distance between the sites $\mathbf{x}$ and $\mathbf{y}$ and $J_0$, $\phi_0$ are constants depending on the material. Due to the random position of the impurities (which play the role of quenched disorder), it is clear that frustration effects are created by the random strength and sign of the interactions between magnetic moments.

Spin glass behaviors were observed in many materials in which the conditions of quenched disorder and frustration are fulfilled. The interactions need not even to be magnetic: spin-glass phenomenons have been observed also in ferroelectric-antiferroelectric mixtures, where the electric dipole plays an analogous role of the magnetic moment. This *universal* behavior of spin-glasses has encouraged people to construct simple statistical-mechanical models that maintain the essential physical features we want to study. These models, that we will introduce in the next subsection, have proved to be able to justify at the theoretical level many surprising effects spin-glasses exhibit, such as memory effects (e.g. *aging*), slowdown of the dynamics and chaos in temperature. All these phenomenons suggests that the spin-glass phase is characterized by the presence of many metastable states, separated by high energetic barriers proportional to the volume of the sample.

### 1.3.1 The Edwards Anderson model and its mean field version

In this subsection we will describe the models that were introduced to explain spin glass behavior. Let us start by placing $N$ spins on a generic graph $\mathcal{G} = (\mathcal{V}, \mathcal{E})$ with vertex set $\mathcal{V}$ and edge set $\mathcal{E}$. Then for every bond of the lattice $(i, j) \in \mathcal{E}$ we extract an interaction according to probability distribution $P(J_{ij})$ and we assign to this instance an energy of the Ising-type

$$H_J[\sigma] = - \sum_{(i,j)\in\mathcal{E}} J_{ij}\, \sigma_i\, \sigma_j - \sum_{i\in\mathcal{V}} h_i\, \sigma_i\,. \tag{1.59}$$

We have inserted an additional index $J$ to the Hamiltonian, in order to stress the dependence on the particular realization of the disorder. When the graph $\mathcal{G}$ is a $d$-dimensional hypercubic lattice, the model is called the *Edwards-Anderson model*. This was firstly introduced in [EA75] in 1975, as a simple generalization of Ising model to disordered systems. Typical probability distributions that are usually chosen for the couplings are for example the Gaussian one

$$P(J_{ij}) = \frac{1}{\sqrt{2\pi J^2}}\, e^{-\frac{(J_{ij}-J_0)^2}{2J^2}}\,, \tag{1.60}$$

or the bimodal

$$P(J_{ij}) = p\delta\left(J_{ij} - 1\right) + (1-p)\delta\left(J_{ij} + 1\right)\,. \tag{1.61}$$

We shall denote by a bar $\overline{\cdot}$ the average over the disordered couplings of a physical quantity $A_J$

$$\overline{A_J} \equiv \int \prod_{(i,j)\in\mathcal{E}} dJ_{ij}\, P(J_{ij})\, A_J\,. \tag{1.62}$$

The first problem to deal with is how to treat disorder. In principle every physical quantity $A_J$ will depend on the particular realization of the disorder. However common experience on macroscopic samples shows that, measuring the same observables on two different samples in the same external conditions, one must obtain the same result. If this is the case, the observable $A_J$ is said to be *self-averaging*. In mathematical terms, a self-averaging quantity can be identified, in the thermodynamic limit, by its average over the disorder,

$$A \equiv \lim_{N\to\infty} \overline{A_J}\,, \tag{1.63}$$

since its variance

$$\overline{A_J^2} - \overline{A_J}^2 = O\left(\frac{1}{N}\right) \tag{1.64}$$

vanishes in the thermodynamic limit. An example of a self-averaging quantity is the free energy density

$$f_J \equiv -\frac{1}{N\beta} \ln Z_J\,, \tag{1.65}$$

where $Z_J$ is the partition function of the system

$$Z_J = \sum_\sigma e^{-\beta H_J[\sigma]}\,. \tag{1.66}$$

Note also that $Z_J$ is not a self-averaging quantity. The second problem to take into account is how to average over the disorder the free energy density

$$f = -\frac{1}{N\beta} \overline{\ln Z_J}\,. \tag{1.67}$$

On first impact this job looks somewhat difficult to accomplish, since we have to average the logarithm of the partition function. One could be tempted to replace the previous average by

$$f_{\text{ann}} \equiv -\frac{1}{N\beta} \ln \overline{Z_J}\,, \tag{1.68}$$

which is tremendously easier. However the two average, of course, give different results. Expression (1.67) is called *quenched average* whereas (1.68) is called *annealed average*. Looking at the two definitions above, one has a clear indication of the physical difference between the two. In the quenched average (1.67) one first sums over spins degrees of freedom, then one takes the logarithm and then one averages the disorder out. Therefore from the point of view of the spins, disorder is fixed, i.e. quenched. Instead, for the annealed average (1.68), spins and disorder fluctuate together. The two terms, quenched and annealed, are borrowed from metallurgy. Quenching indicates a very rapid cooling of the sample. Indeed this technique is used to increase the hardness of metallic materials, since quenching can reduce their crystal grain size. Annealing instead suggests a slow cooling of the sample. Indeed the word "annealing" appears also in the name of a celebrated Monte-Carlo algorithm, the *simulated annealing* [KGV83], in which, in order to sample correctly phase space, one decreases slowly the temperature, in order to avoid the possibility of remaining stuck in local minima. The annealed approximation (1.68) of the true quenched average (1.67), can work reasonably well at high temperature, where thermal fluctuations destroy the effects coming from frustration. At low temperature, however, it fails completely. Nonetheless, for the convexity properties of the logarithm

$$f \geq f_{\text{ann}}\,, \tag{1.69}$$

so that the annealed approximation gives a lower bound to the true value, giving a quick idea of what is going on at high temperatures.

In order to deal with quenched averages, one has to introduce some kind of representation of the logarithm. This is what the *replica trick* does which is the identity

$$\overline{\ln Z_J} = \lim_{n\to 0} \frac{\overline{Z_J^n} - 1}{n} = \lim_{n\to 0} \frac{\ln \overline{Z_J^n}}{n} . \tag{1.70}$$

In the *replica method* one introduces an integer number $n$ of identical copies of the system and evaluates the average of the replicated partition function $\overline{Z_J^n}$, which has the same difficulty of computing an annealed average. The price to pay is that one must then perform the limit of vanishing number of replicas, trying to continue analytically the result obtained previously for integer $n$. In principle one can also use more exotic expression of the logarithm. One interesting integral representation (that will be also used extensively throughout this thesis) is

$$\ln Z_J = \int_0^{+\infty} \frac{dt}{t} \left[ e^{-t} - e^{-Z_J t} \right] . \tag{1.71}$$

Then one expands the exponential $e^{-Z_J t}$ and performs the average of every term of the series. Note that this expression does not involve no zero replicas limit, but instead, one needs an information coming from all the average of integer powers of the partition function.

The replica method was applied to the study of the Edwards-Anderson model, but its solution is still lacking and it is still a matter of debate if there is a spin glass phase in $d = 3$ with external field. The infinite range version of the Edwards-Anderson model is called *Sherrington-Kirkpatrick* model (SK), and it appeared only some months after the paper of Edwards and Anderson [SK75, KS78]. The Hamiltonian of the SK model is

$$H_J[\sigma] = -\sum_{i<j} J_{ij}\,\sigma_i\,\sigma_j - h\sum_{i=1}^{N} \sigma_i . \tag{1.72}$$

The application of the replica method to this model certified the success of the replica method. In the next chapter we will analyze in detail its complete solution found by Giorgio Parisi in a series of papers [Par79a, Par79b, Par80a, Par80b]. In the following we will describe briefly which is the phenomenology that is obtained in infinite dimensions.

### 1.3.2 Pure states

In our study of spin glasses in infinite dimensions we will see that, in the $N \to \infty$ limit, below a certain critical temperature $T_c$, the space of spin states is divided in many *valleys*, with infinite barriers separating them, i.e. we will have *ergodicity breaking*. Therefore, the system will explore only a small part of the all free energy landscape and the Gibbs measure splits into disconnected subcomponents $\langle\cdot\rangle_\alpha$

$$\langle\cdot\rangle = \sum_\alpha w_\alpha \langle\cdot\rangle_\alpha , \tag{1.73}$$

called *pure states*, since they cannot themselves be decomposed into other subcomponents [Par88]. In the previous equation $\alpha$ runs over all the valleys and $w_\alpha$ is the statistical weight of valley $\alpha$. Clearly it can be written as

$$w_\alpha = \frac{Z_\alpha}{Z}, \tag{1.74}$$

where

$$Z_\alpha = \sum_{\sigma \in \alpha} e^{-\beta H[\sigma]}, \tag{1.75}$$

is the partition function of all degrees of freedom belonging to valley $\alpha$. Pure states satisfy the so called *clustering property*, i.e. the connected correlation function vanishes in the limit of large distances

$$\left\langle \sigma_i \sigma_j \right\rangle_c \equiv \left\langle \sigma_i \sigma_j \right\rangle - \left\langle \sigma_i \right\rangle \left\langle \sigma_j \right\rangle \overset{|i-j| \to \infty}{\longrightarrow} 0. \tag{1.76}$$

The whole Gibbs measure, in general does not satisfy it.

To make a concrete example, consider the case of the Ising model under the Curie critical temperature. We have only two pure states: the one with positive magnetization $\langle \cdot \rangle_+$ and the one with negative magnetization $\langle \cdot \rangle_-$. In absence of an external magnetic field the probability of such components is equal, so that

$$\langle \cdot \rangle = \frac{1}{2} \langle \cdot \rangle_+ + \frac{1}{2} \langle \cdot \rangle_- . \tag{1.77}$$

One can select the state with positive (or negative) magnetization by simply applying a positive (or negative) magnetic field. In spin glasses things are much more complicated, since under the spin glass critical temperature there will be an *extensive* number of valleys into which spins could freeze and one does not know, in general, how to apply such an external field in order to select only one state. In addition, at variance with usual non disordered systems, symmetry breaking occurs at *all* temperature below $T_c$. This means that starting from a temperature $T < T_c$ and decreasing it of an infinitesimal amount $dT$, every valley $\alpha$, found at temperature $T$ splits into several valleys at temperature $T - dT$. Therefore in the spin glass phase the system is critical at all temperatures $T < T_c$ and the free energy is always a non-analytic function.

### 1.3.3 Overlaps and order parameters

As I have said previously, in the spin glass phase every spin is frozen along a certain (random) direction and the free energy landscape splits in many valleys separated by infinitely high barriers. The total magnetization of the system

$$m = \frac{1}{N} \sum_{i=1}^{N} \langle \sigma_i \rangle , \tag{1.78}$$

where $\langle \sigma_i \rangle$ is the thermal average of the spin $i$, does not distinguish between the paramagnetic phase and the spin-glass phase since it vanishes in both cases. A parameter which distinguish between these phases in a dynamical way is the

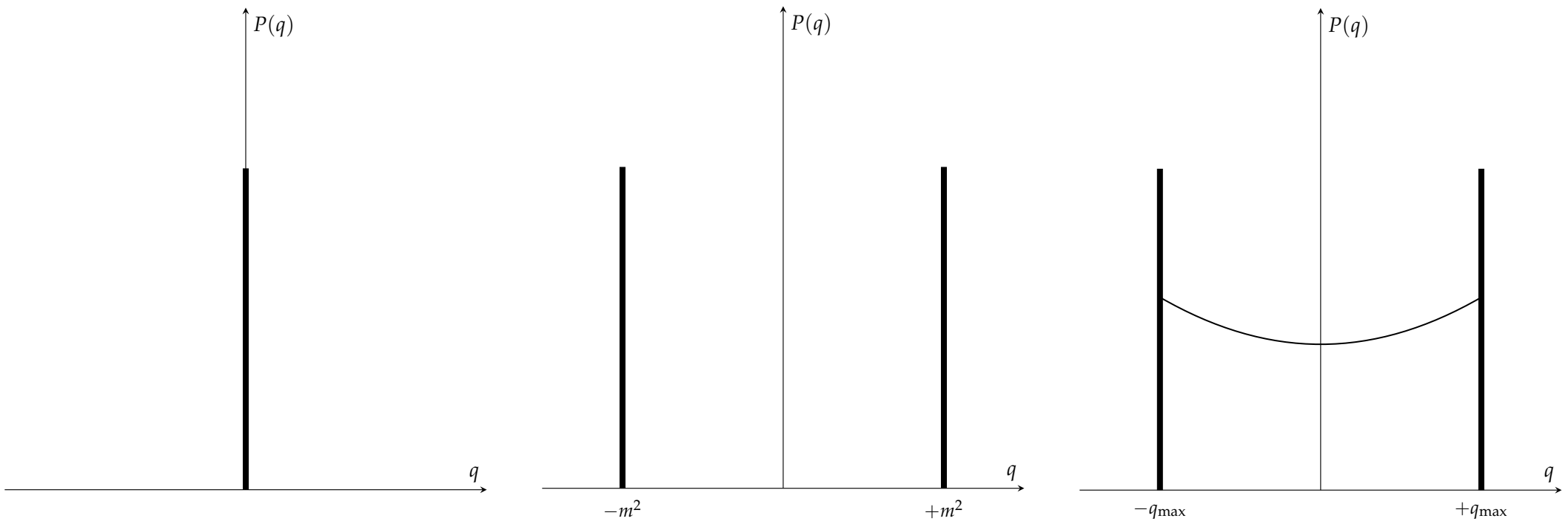

**(a)** Paramagnetic phase **(b)** Ferromagnetic phase **(c)** Spin glass phase

**Figure 1.6.** The probability distribution of the overlaps $P(q)$ in different phases.

*Edwards-Anderson* (EA) order parameter [EA75, FH93], defined in terms of time averages $\langle\cdot\rangle_t$ as

$$q_{\mathrm{EA}} \equiv \lim_{t\to\infty} \lim_{N\to\infty} \frac{1}{N} \sum_{i=1}^{N} \langle \sigma_i(t_0)\sigma_i(t_0+t)\rangle_t \,, \tag{1.79}$$

where $t_0$ is a time reference. Note how the order of the limits is of great importance. In fact performing the thermodynamic limit first, one does not allow the system, if there is ergodicity breaking, to explore all the free energy landscape, making $q_{\mathrm{EA}}$ non-vanishing. If the limits were inverted, the system has the possibility to escape from barriers, so that for symmetry arguments the final result would be always zero in both paramagnetic and spin glass phases. Note that this definition can depend on the sample, since we have performed no disorder average. However one can prove [MPV87] that in the $N \to \infty$ limit, $q_{\mathrm{EA}}$ is indeed a self-averaging quantity (so that one can substitute the sum over sites with the disorder average). In terms of ensemble averages $q_{\mathrm{EA}}$ reads

$$q_{\mathrm{EA}} = \frac{1}{N} \sum_{i=1}^{N} \sum_{\alpha} w_\alpha \, \langle\sigma_i\rangle_\alpha^2 = \sum_\alpha w_\alpha \, q_{\alpha\alpha} \,, \tag{1.80}$$

i.e. $q_{\mathrm{EA}}$ represents physically the average over the valleys of the square of the single valley magnetization. In the second equality of (1.80), the EA order parameter has been rewritten in terms of $q_{\alpha\alpha}$

$$q_{\alpha\alpha} = \frac{1}{N} \sum_{i=1}^{N} \langle\sigma_i\rangle_\alpha^2 \,, \tag{1.81}$$

which is called *self-overlap*. Indeed it is part of a much more general quantity called *overlap*

$$q_{\alpha\beta} \equiv \frac{1}{N} \sum_{i=1}^{N} \langle\sigma_i\rangle_\alpha \langle\sigma_i\rangle_\beta \,, \tag{1.82}$$

which measures the degree of similarity between two states[1] $\alpha$ and $\beta$. Indeed one can define the overlap $q_{\sigma\tau}$ among two configurations $\sigma$, $\tau$ as

$$q_{\sigma\tau} \equiv \frac{1}{N}\sum_{i=1}^{N}\sigma_i\tau_i\,, \tag{1.83}$$

and see that the overlap between two states $\alpha$ and $\beta$, can be written in terms of the overlap among configurations belonging to the states

$$q_{\alpha\beta} = \frac{1}{Z_\alpha Z_\beta}\sum_{\sigma\in\alpha}\sum_{\tau\in\beta} e^{-\beta H[\sigma]}e^{-\beta H[\tau]}q_{\sigma\tau}\,. \tag{1.84}$$

The overlap $q_{\alpha\beta}$ can assume values between $-1$ and $1$ depending on the degree of correlation of two states: if they are not correlated their overlap is zero; the maximum value of correlation is instead achieved when $\alpha = \beta$, i.e. by the self-overlap (1.81). The overlap $q_{\sigma\tau}$ induces a metric over the ensemble of states in the system through the *Hamming distance* given by

$$d_{\sigma\tau} \equiv \frac{1-q_{\sigma\tau}}{2}\,. \tag{1.85}$$

Having defined the overlap between different states, one can introduce another order parameter which takes into account "inter-valley" contributions (which are absent in the EA order parameter) and which measures, for a given sample, the mean square local equilibrium magnetization

$$q_J = \frac{1}{N}\sum_{i=1}^{N}\langle\sigma_i\rangle^2 = \sum_{\alpha\beta} w_\alpha w_\beta\, q_{\alpha\beta} \tag{1.86}$$

Differently from the EA order parameter, this quantity can be proved to be non self-averaging. These two quantities, $q_{\rm EA}$ and $q \equiv \overline{q_J}$, are of great importance, because they can be related to experimentally measurable quantities, as we will see in the next chapter. For the moment let us say that their difference

$$\Delta \equiv q_{\rm EA} - q\,, \tag{1.87}$$

is such that $\Delta = 0$ when there is only one pure state, and it is $\Delta > 0$ when there is ergodicity breaking.

Since in the spin glass phase there will be an enormous number of pure states it is convenient to introduce a distribution of their overlap

$$P(q) = \overline{P_J(q)} = \overline{\sum_{\alpha\beta} w_\alpha w_\beta\, \delta(q - q_{\alpha\beta})}\,. \tag{1.88}$$

In the paramagnetic phase, where we have only one pure state, the distribution $P(q)$ has only a delta function in the origin, see Fig. 1.6a. In the ferromagnetic phase, where we have two pure states, instead, the $P(q)$ develops a delta function in $m^2$ (given by the overlap $q_{++}$ and $q_{--}$), and another delta function in $-m^2$

[1] The words "states" and "valleys" have the same meaning.

(given by the mixed overlaps $q_{+-}$ and $q_{-+}$), as in Fig. 1.6b. In the spin glass phase, see Fig. 1.6c, between the two delta functions in $\pm q_{\max}$, the distribution of the overlaps develops a continuous part, due to the possibility of having many different overlaps among different states. In this manner the distribution of the overlaps $P(q)$ can be regarded as the true order parameter for spin glasses. The fact that the order parameter is a *function*, is necessary because one needs an infinite number of parameters to describe the spin glass phase, at variance to what happens in ordinary phase transitions. As we will see, $P(q)$ can be evaluated in mean field by means of the replica method.

## 1.4 Combinatorial Optimization

Beside condensed matter, spin glasses were also useful for many multidisciplinary applications in different areas of knowledge. One of the major of these applications include *optimization* [MM11]. Optimization is ubiquitous also in daily life problems. To make an example, suppose you are a courier and every day you have to deliver, starting from a storehouse, $N$ packages in $N$ different places of your city returning, at the end of the day, to the starting point. The head of your shipping company takes care not only of the satisfaction of the customers, but also is very much interested in minimizing the costs of expeditions, in order to maximize the profits. In order to satisfy your boss, you need to construct a *cost function* which you should minimize given the locations of your expeditions. This function can be constructed on the base of the relevance of the many variables involved; for example, in first approximation, you might think that the costs involved depend only on the total distance traveled and the satisfaction of your customers can be measured as the average time spent for an expedition. Minimizing the total distance traveled might be a very good idea in order to fulfill both the (frustrating) requests. However this might be a very crude approximation if the city in which you are living is, for example, Rome, where traffic jams are the most relevant variable.

Another example of application comes from machine learning. In the most simple case, you want to construct a device such that it is capable to distinguish if in a picture, given as input, there is a cat or a dog. One can use in this case a particular *neural network* called *perceptron*, which takes as input the $N$ pixels $\xi_i$ of a picture, and associates an output $+1$, if the picture is a dog, and $-1$ if it is a cat. The particular mapping from the input to the $\pm 1$ output is specified by $N$ parameters $J_i$ called *synapses*, which has to be chosen in an optimal way in order to realize the right mapping. So your strategy is the following: you collect pictures of different cats and dogs (the so-called *training set*) and a *teacher* trains your neural network giving him correct examples of which picture is of a cat and which is of a dog. During the training process one has to tune the synapses in such a way that all the examples we have proposed possess the correct label. Therefore in the training one must look for the configuration (or configurations) of synapses $J_i^*$ that minimize again a cost function, which in this case counts the number of errors the neural networks does on the training set.

These two are very simple examples, but optimization problems are really used everywhere, from the physical design of computers, to cryptography, from their use in economics and financial markets to robotics, automatized systems and artificial intelligence. Let us now define, from the mathematical point of view, what an optimization problem is. We define an optimization problem as a couple $(\mathcal{S}, E)$ composed by a space of configurations $\mathcal{S}$ and a function $E$ which associates a cost to each configuration. If the space of configuration contains a finite number of elements the optimization problem is said to be *combinatorial*. The final goal is to find the configuration that minimizes the cost function

$$x^* = \operatorname*{Argmin}_{x \in \mathcal{S}} E(x) \tag{1.89}$$

and the minimum cost itself

$$E(x^*) = \min_{x \in \mathcal{S}} E(x) \,. \tag{1.90}$$

For every optimization problem, one can also formulate a *decision* version in which it is asked if there is or there is not a configuration of cost less than a given value $C$. In the following we will always work with optimization problems defined on a weighted graph $\mathcal{G} = (\mathcal{V}, \mathcal{E})$ with weights $w_e \geq 0$, for every $e \in \mathcal{E}$. We define an *instance* of an optimization problem as a realization of the graph $\mathcal{G}$ and of the weights. In this thesis we are interested in configurations $\mathcal{S} = \{\mathcal{I}\}$ that are particular sets of *spanning subgraphs* $\mathcal{I} \subseteq \mathcal{G}$ of the graph $\mathcal{G}$. Of course we have to ensure that in $\mathcal{G}$ the set of spanning subgraph $\mathcal{S}$ we are looking at is non-empty. Given a spanning subgraph $\mathcal{I} \in \mathcal{S}$, we will associate to it a cost of the type

$$E(\mathcal{I}) = \sum_{e \in \mathcal{I}} w_e \,. \tag{1.91}$$

The optimization problem consists in finding the minimum spanning subgraph $\mathcal{I}^*$ and its cost. Depending on the class of spanning subgraph $\mathcal{S}$ we are interested in, the combinatorial optimization problem has a different name. We list here some examples of combinatorial optimization problems

- *Minimum Spanning Tree* (MST): here the set of spanning subgraph $\mathcal{S}$ is given by all the spanning trees; the weighted graph $\mathcal{G}$ must be connected.
- *Matching Problem*: here we look for all the 1-factors or perfect matching of the graph $\mathcal{G}$ which we denote by $\mathcal{M}_1$. If $\mathcal{G} = \mathcal{K}_{N,N}$ the problem is called *assignment problem*.
- *2-matching* or *2-factor Problem*: $\mathcal{S}$ is the set of all 2-factors that will be denoted by $\mathcal{M}_2$.
- *Traveling Salesman Problem* (TSP): $\mathcal{S}$ is the set of all the Hamiltonian cycles, which we will denote by $\mathcal{H}$.

In particular the last three problems will be discussed extensively throughout in this thesis. We say that an *algorithm* solves a combinatorial optimization problem if for every instance possible it gives the optimal configuration. The *size* of a combinatorial optimization problem is a quantity that identifies the amount of memory needed to store an instance; e.g. the number of vertices $|\mathcal{V}| = N$ of the graph $\mathcal{G}$. *Complexity theory*, which dates back to the Eulerian circuit problem solved by Euler in 1736, classifies combinatorial optimization problems according to their "level of hardness" (i.e. to the time it needs to be solved) and how this time changes varying the size of the problem. Intuitively a problem is "harder" if it needs more time to be solved, but this definition of course must not depend nor on the computer nor on the compiler used. The precise definition of "hardness" or *complexity* will require too much formalism and we remand to references [HW06, MM11] for the details. Here we will adopt an intuitive and simple point of view. Since the complexity of a problem of a fixed size will change even from instance to instance, the theory of complexity is based on the *worst possible instance*, that is on the instance that needs the most time to be solved. This is useful also because once a given algorithm has been proved to scale with the size in a certain way, then every other instance will be solved at most in the same time. The first complexity class we mention is the P or *polynomial* one where there are problems which can be solved by an algorithm in a time that scales polynomially with the size. Examples of combinatorial optimization problems in this class are the MST and the matching problem. The NP or *non-deterministic polynomial* class contains all the problems that can be solved in polynomial time by a non-deterministic algorithm. Basically this means that if someone gives you the optimal configuration, you can compute its cost polynomially. In is obvious that P$\subseteq$NP. The NP-*complete* class instead contains all the problems that are at least as hard as the NP ones. Examples are the TSP and the graph partitioning problem. Basically in the NP-complete class the best algorithm known scales exponentially with the size. From these definitions, it follows that if one NP-complete problem can be solved in polynomial time, then P=NP. Proving whether P=NP or P$\neq$NP is one of the millennium problems[2].

The interest in the physics community on combinatorial optimization problems came in particular from their *random* version, in which some parameters of the cost function itself are random variables. The simplest way to introduce randomness is to consider the weights $w_e$ independent and identically distributed random variables. In this case the problem under consideration is renamed with the prefix "*random*", e.g. *random matching problem* (RMP), *random 2-factor problem*, *random traveling salesman problem* (RTSP). Random-link matching problems along with the random-link traveling salesman problem have been the first class of optimization problems to be studied by statistical physics techniques [Orl85, MP86a], the link probability distribution $\rho(w)$ being controlled near the origin, by the exponent $r$

$$\rho_r(w) \sim w^r \,. \tag{1.92}$$

Indeed, as we shall see, the behavior of the distribution $\rho(w)$ near the origin is the only relevant information in the limit of large number of vertices $N$ [MP85]. Since there are no correlations between variables these models can be see as *mean-field* ones. In order to study their finite-dimensional counterparts, one has to introduce *correlations* between the weights. Consider when the graph $\mathcal{G}$ with $|\mathcal{V}| = N$ is embedded in $\mathbb{R}^d$, that is for each $i \in [N] = 1, 2, \dots, N$ we associate a point $x_i \in \mathbb{R}^d$, and for $e = (i, j)$ with $i, j \in [N]$ we introduce a cost which is a function of their

[2]see http://www.claymath.org/millennium-problems/p-vs-np.

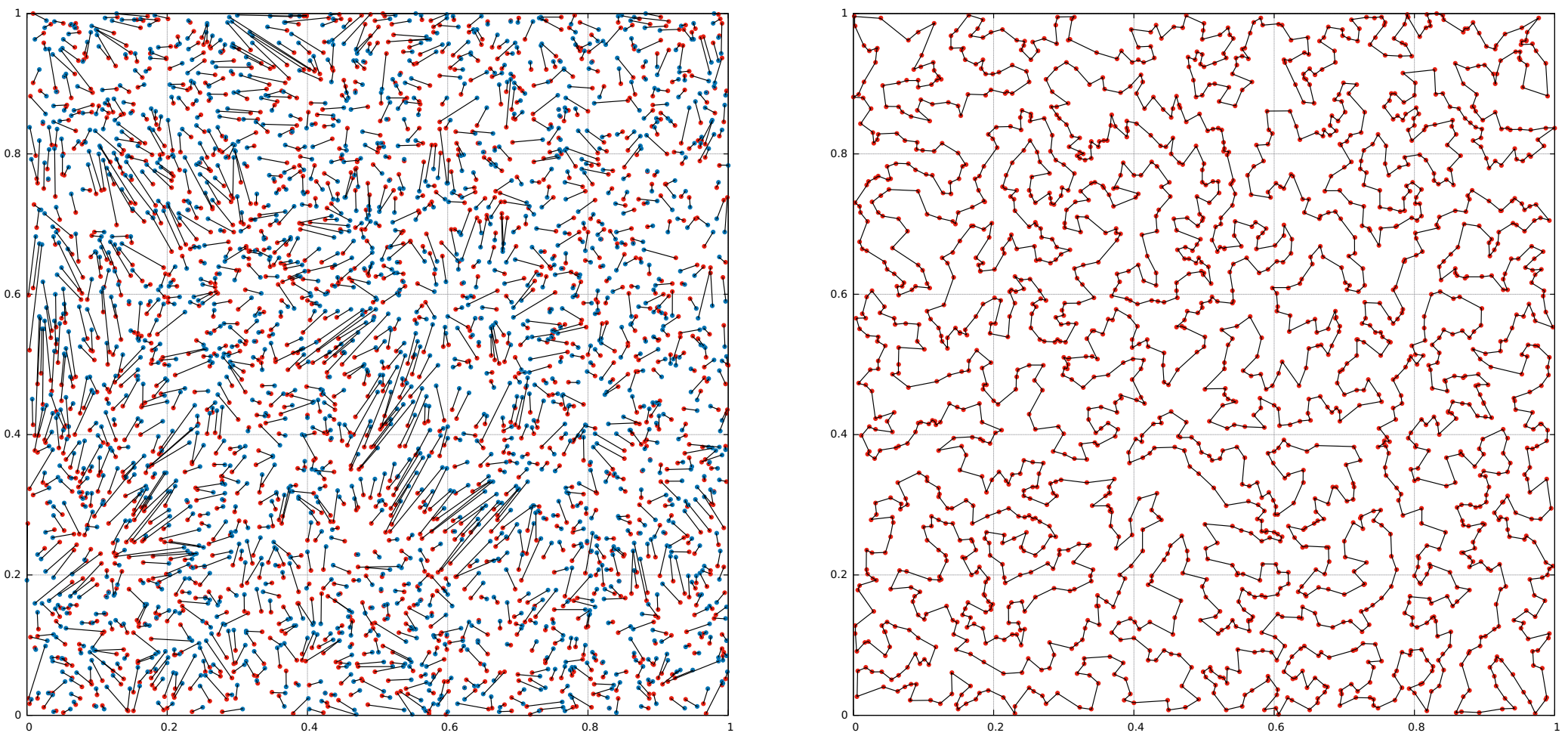

**(a)** Assignment with $N = 2000$. **(b)** TSP with $N = 2000$.

**Figure 1.7.** Optimal assignment (left panel) and TSP (right one) for two different random instances throwing points in the square and using $p = 1$ as introduced respectively in (1.94) and (1.93).

Euclidean distance

$$w_e = |x_i - x_j|^p \tag{1.93}$$

with $p \in \mathbb{R}$. When we deal with a bipartite graph, as for example the complete bipartite one $\mathcal{K}_{N,N}$, we shall distinguish with a different letter the two sets of vertices in $\mathbb{R}^d$ that is the red $\{r_i\}_{i\in[N]}$ and the blue $\{b_i\}_{i\in[N]}$ points. The edges connect red with blue points with a cost again of the type

$$w_e = |r_i - b_j|^p \,. \tag{1.94}$$

The *random Euclidean* versions of the previously mentioned problems are obtained by generating randomly and uniformly the positions of points in a space $\Omega_d$ of dimension $d$, which we will suppose to be, for definiteness, the $d$-dimensional hypercube $[0,1]^d$. As shown in [MP88] the random-link limit is achieved by sending both $d$ and $p$ to infinity with their ratio fixed

$$\frac{d}{p} = r + 1\,, \tag{1.95}$$

where $r$ is the same defined in (1.92).

In Fig. 1.7a we have depicted the optimal solution of an instance of a random Euclidean assignment problem on the complete bipartite graph $\mathcal{G} = \mathcal{K}_{N,N}$ and for $\Omega_d = [0,1]^d$ with $d = 2$, whereas in Fig. 1.7b we report the optimal solution of a random Euclidean TSP instance on the complete graph $\mathcal{G} = \mathcal{K}_N$ in the same 2-dimensional space. In both mean-field and Euclidean random versions we have introduced, one is interested in evaluating average properties of the solution and, in particular, the average optimal cost (aoc)

$$\overline{E} \equiv \overline{E(\mathcal{I}^*)}\,, \tag{1.96}$$

where we have denoted by a bar the average over all possible realizations of the disorder. Clearly, this is at variance with the point of view of computational complexity where one focuses on the worst case scenario. In general, the typical instance of a random combinatorial optimization problem (RCOP) can be very different from the worst case [Mer02]. However one can generate really hard instances tuning certain parameters of the model and observe abrupt changes of the typical computational complexity. One important example is the perceptron (with Ising-type synapses $W_i = \pm 1$) that was mentioned before, the tuning parameter being the *capacity* $\alpha = M/N$ defined as the ratio between the size of the training set and the number of synapses. When $\alpha$ is small, the problem is *underconstrained*, so that we expect that there are an exponential number of configurations for the synapses $J_i$ giving zero errors on the training set. On the other hand when $\alpha$ is large the problem is *overconstrained* and we expect the teacher to be the only solution. In both those cases any smart algorithm is able to find easily a solution. The real hard instances are selected when $\alpha$ is close to a critical value $\alpha_c$ which is nor to small nor too large. At this critical value the system undergoes a first order phase transition to perfect generalization (i.e. for every $\alpha > \alpha_c$ the only solution is the teacher) [GD89, Gyö90, STS90]. Another archetypal examples which show similar phenomenology are the random K-SAT problem [MZ97, Mon+99] and the previously mentioned TSP [GW96]. Away from these critical values of parameters typical instances are, instead, easy to solve.

This sudden change of behavior can be seen as phase transitions in physical systems [MMZ01] and, for this reason, can be studied with techniques developed in statistical mechanics. The general way of describing a random combinatorial optimization problem is to consider the cost function $E$ as the energy of a fictitious physical system at a certain temperature [KGV83, Sou86, FA86]. Finding the minimum of the cost function is perfectly equivalent to study the low temperature properties of this physical system. This approach has been tremendously fruitful: it turned out that, specially in mean field cases, the general theory of spin glasses and disordered systems could help not only to calculate those quantities at the analytical level using techniques like replica and cavity method [MPV87], but also to shed light on the design of new algorithms to find their solution. Celebrated is the result for the asymptotic value of the average optimal cost in the random assignment problem obtained by Mézard and Parisi [MP85] using the replica method. The same result was obtained later via the cavity method [MP86b, PR02].

The interplay between physics and optimization, is not unilateral. Many physics models have ground states that can be mapped into combinatorial optimization problems. An important and known example is the 2-dimensional spin glass whose ground state corresponds to find the minimum matching of the frustrated plaquettes [Frö85], therefore providing a polynomial algorithm able to find it. Other two important examples are the connections between the coloring problem with the antiferromagnetic Potts model and the maximal-flow problem (polynomial) with the random field Ising model (RFIM).

# Chapter 2

# The Sherrington-Kirkpatrick Model

In this chapter we will analyze in detail the Sherrington-Kirkpatrick model, i.e. the mean field model for spin glasses we have introduced in section 1.3. We will solve the model using replicas [MPV87, Dot00, Nis01] discussing the phase diagram of the model and the physical implications of the results.

For completeness, it is useful to cite here that a different point of view on spin glasses is given by the work of Thouless Anderson and Palmer [TAP77] which derived the so called known as TAP equations, i.e. self-consistent equations for the local magnetization of a given sample. From those equations they were able to predict how physical quantities behave for low temperatures and near the spin glass transition point well before the celebrated Parisi's solution using the replica method. TAP equations contain much more information than the ones we will derive with the replica method since their solutions (which are exponential in $N$) may correspond to metastable states [BM80]. The cavity method [MPV86], then, was developed not only to derive TAP equations from a different point of view, but also as an alternative to replicas. After its application on the SK model it was used not only to build a solid theory on spin glasses with finite connectivity, but also to construct new algorithms for optimization problems. The "cavity method" is the statistical physics name for "belief propagation" we have encountered in the previous chapter.

## 2.1 The Replicated partition function

As we have anticipated in section 1.3, in order to perform the average over the disorder of the logarithm of the partition function, one uses the replica trick (1.70), introducing $n$ non-interacting copies of the systems, performing the average over the disorder for $n$ integers, and then performing the analytical continuation for $n \to 0$. All these technical computations are at the base of the replica method. The replicated partition function for SK model is

$$Z^n = \sum_{\{\sigma_i^a\}} e^{\beta \sum_{i<j} J_{ij} \sum_a \sigma_i^a \sigma_j^a + \beta h \sum_i \sum_a \sigma_i^a} , \tag{2.1}$$

where we have dropped, for simplicity, the index dependence $J$ of the partition function. We extract the couplings according to a Gaussian probability distribution with mean and variance scaling with $N$ respectively as

$$\begin{aligned}\overline{J_{ij}} &= \frac{J_0}{N} \\ \overline{J_{ij}^2} - \overline{J_{ij}}^2 &= \frac{J^2}{N}\end{aligned} \tag{2.2}$$

in order to assure that that all is well defined in the thermodynamic limit. The average over the disorder is now simple to perform and we get

$$\overline{Z^n} = e^{\frac{\beta^2 J^2 N n}{4}} \sum_{\{\sigma_i^a\}} e^{\frac{\beta^2 J^2}{2N} \sum_{a<b} \left(\sum_i \sigma_i^a \sigma_i^b\right)^2 + \frac{\beta J_0}{2N} \sum_a \left(\sum_i \sigma_i^a\right)^2 + \beta h \sum_a \sum_i \sigma_i^a} . \tag{2.3}$$

In order to perform the sum over the spins independently from site to site we insert Gaussian integrals using the formula

$$e^{\frac{b^2}{4a}} = \sqrt{\frac{a}{\pi}} \int dx\, e^{-ax^2+bx} , \tag{2.4}$$

which is called *Hubbard-Stratonovich* transformation (HS). Since we have to quantities squared we have to introduce two types of integrals namely

$$e^{\frac{\beta^2 J^2}{2N}\left(\sum_i \sigma_i^a \sigma_i^b\right)^2} = \sqrt{\frac{N\beta^2 J^2}{2\pi}} \int dq_{ab}\, e^{-\frac{N\beta^2 J^2}{2} q_{ab}^2 + \beta^2 J^2 q_{ab} \sum_i \sigma_i^a \sigma_i^b} \tag{2.5a}$$

$$e^{\frac{\beta J_0}{2N}\left(\sum_i \sigma_i^a\right)^2} = \sqrt{\frac{N\beta J_0}{2\pi}} \int dm_a\, e^{-\frac{N\beta J_0}{2} m_a^2 + \beta J_0 m_a \sum_i \sigma_i^a} \tag{2.5b}$$

Dropping for simplicity the irrelevant constants in front of the integrals we obtain

$$\overline{Z^n} = \int \prod_{a<b} dq_{ab} \int \prod_a dm_a\, e^{-NS[q,m]} , \tag{2.6}$$

where we have defined an action $S\,[q,m]$

$$S[q,m] \equiv \frac{\beta^2 J^2}{2} \sum_{a<b} q_{ab}^2 + \frac{\beta J_0}{2} \sum_a m_a^2 - \frac{\beta^2 J^2 n}{4} - \ln \mathcal{Z}\,[q,m] , \tag{2.7}$$

and a one site partition function

$$\mathcal{Z}[q,m] \equiv \sum_\sigma e^{-\beta \mathcal{H}[\sigma]} = \sum_{\{\sigma^a\}} e^{\beta^2 J^2 \sum_{a<b} q_{ab} \sigma^a \sigma^b + \beta \sum_a (J_0 m_a + h)\sigma^a} . \tag{2.8}$$

The neat result is that we have decoupled the sites, but at the price to insert an interaction between replicas given by the the new Hamiltonian

$$\mathcal{H}[\sigma] = -\beta J^2 \sum_{a<b} q_{ab}\, \sigma^a \sigma^b - \sum_a \left(J_0 m_a + h\right) \sigma^a \tag{2.9}$$

Note that this result is valid, at the leading order in $N$, for every disorder distribution, since higher moments need to scale with an higher power in $N$ in the denominator. This is valid in general on fully connected topologies thanks to the central limit theorem. This result is not valid, instead, in *dilute* systems (i.e. models defined on graph with finite connectivity in the thermodynamic limit), where one needs all sets of multi-overlaps $q_{a_1 \dots a_p}$ with $p \geq 1$ to describe them. To give an example, on might look at [Mot87, GDD90] (and references therein) where it was studied using replicas, spin glasses on tree graphs and on RRG topologies. We will show some other examples of this in Part II of this thesis, where we will analyze random-link combinatorial optimization problems such as the RMP or the RTSP.

### 2.1.1 Saddle point equations

The average of the replicated partition function (2.6) is now in the right form to apply the saddle-point method. However, in order to do that, we have to exchange the order of the limits $N \to \infty$ and $n \to 0$. In fact, if we want to do things correctly, we need to perform the $n \to 0$ limit first and only *then* the $N \to \infty$ one. It has been proved rigorously by Hemmen and Palmer [HP79] that this change of limit can be done in the SK model. The averaged free energy density is therefore given by

$$\beta \overline{f} = \lim_{n \to 0} \min_{q,m} \frac{S[q,m]}{n} . \tag{2.10}$$

The saddle-point equations are

$$m_a = \langle \sigma^a \rangle_{\mathcal{Z}} \tag{2.11a}$$

$$q_{ab} = \langle \sigma^a \sigma^b \rangle_{\mathcal{Z}} , \tag{2.11b}$$

where $\langle \cdot \rangle_{\mathcal{Z}}$ is the average with respect to the weight $e^{-\beta \mathcal{H}}$. The variables that one has introduced for technical convenience now acquire physical meaning. In fact one can prove, repeating the replica computation, that equations (2.11a), (2.11b) can be rewritten as

$$m_a = \overline{\langle \sigma_i^a \rangle} \tag{2.12a}$$

$$q_{ab} = \overline{\langle \sigma_i^a \sigma_i^b \rangle} , \tag{2.12b}$$

where now the average $\langle \cdot \rangle$ is with respect to the original replicated Hamiltonian. Physically $m_a$ can be identified with the usual ferromagnetic order parameter and $q_{ab}$ with the spin glass order parameter.

## 2.2 RS Ansatz

The saddle-point equations (2.11a), (2.11b) are in general too difficult to solve. It is reasonable to search solutions in a subspace of the whole saddle-points giving an *ansatz*, or an explicit dependence of the saddle-points on replica indices. On physical grounds we expect that this dependence should not affect the physics of the system, because replicas have been introduced artificially. It seems natural to

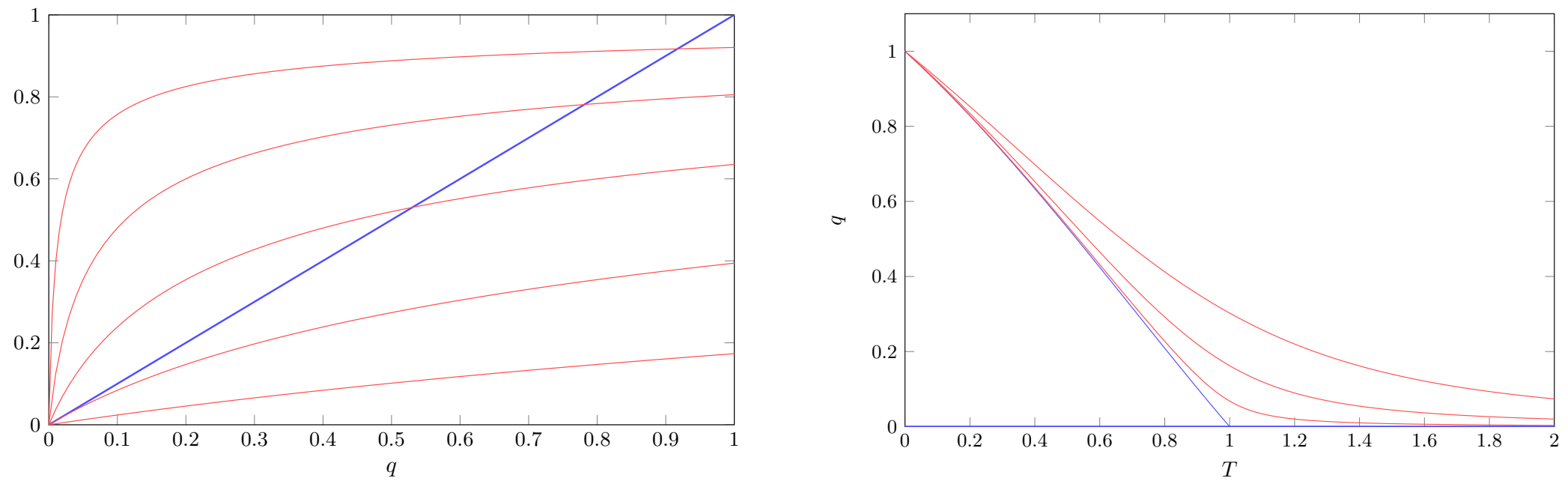

**(a)** Red lines are the plot of the right hand side of (2.17b) with $T = 0.1, 0.25, 0.5, 1, 2$ (from top to bottom).

**(b)** The blue lines are the solutions of (2.17b) for $h = 0$. The red curve are the solutions for $h = 0.5, 0.25, 0.1$ (from top to bottom).

**Figure 2.1.** In (a) is represented the graphical solution of equation (2.17b) for $J_0 = h = 0$ and $J = 1$. For $T = J = 1$ the right hand side of (2.17b) is tangent to the straight line. In (b) the overlap is plotted as a function of temperature for $J_0 = 0$, $J = 1$ and for several values of the external magnetic field.

assume that

$$m_a = m \tag{2.13a}$$

$$q_{ab} = q(1 - \delta_{ab}) \tag{2.13b}$$

which is called the *replica symmetric ansatz* (RS). The tricky part to compute is the one-site partition function. The sum over spins can be performed using an HS transformation

$$\begin{aligned} \ln \mathcal{Z} &= -\frac{n\beta^2 J^2}{2} q + \ln \sum_{\{\sigma^a\}} e^{\frac{\beta^2 J^2}{2} q (\sum_a \sigma^a)^2 + \beta (J_0 m + h) \sum_a \sigma^a} \\ &\simeq -\frac{n\beta^2 J^2}{2} q + n \int Dz \, \ln 2 \cosh \left[\beta \left(J\sqrt{q}\, z + J_0 m + h\right)\right] \end{aligned} \tag{2.14}$$

where $Dz$ is the Gaussian measure [1]

$$Dz \equiv \frac{e^{-z^2/2}}{\sqrt{2\pi}} \,. \tag{2.15}$$

In this way the averaged free energy density (2.10) becomes

$$\beta \overline{f}_{\mathrm{RS}} = -\frac{\beta^2 J^2}{4}(1-q)^2 + \frac{\beta J_0}{2} m^2 - \int Dz \, \ln 2 \cosh \left[\beta \left(J\sqrt{q}\, z + J_0 m + h\right)\right] . \tag{2.16}$$

Note that, in the limit $n \to 0$, the coefficient of the $q^2$ terms has changed sign. This is due to the fact that the number of replica pairs changes sign at $n = 1$. This implies that one has to *maximize* with respect to $q$ the free energy density. The coefficient of $m^2$, instead, is not affected by this limit, so that one continues

[1] This notation will be used throughout this thesis.

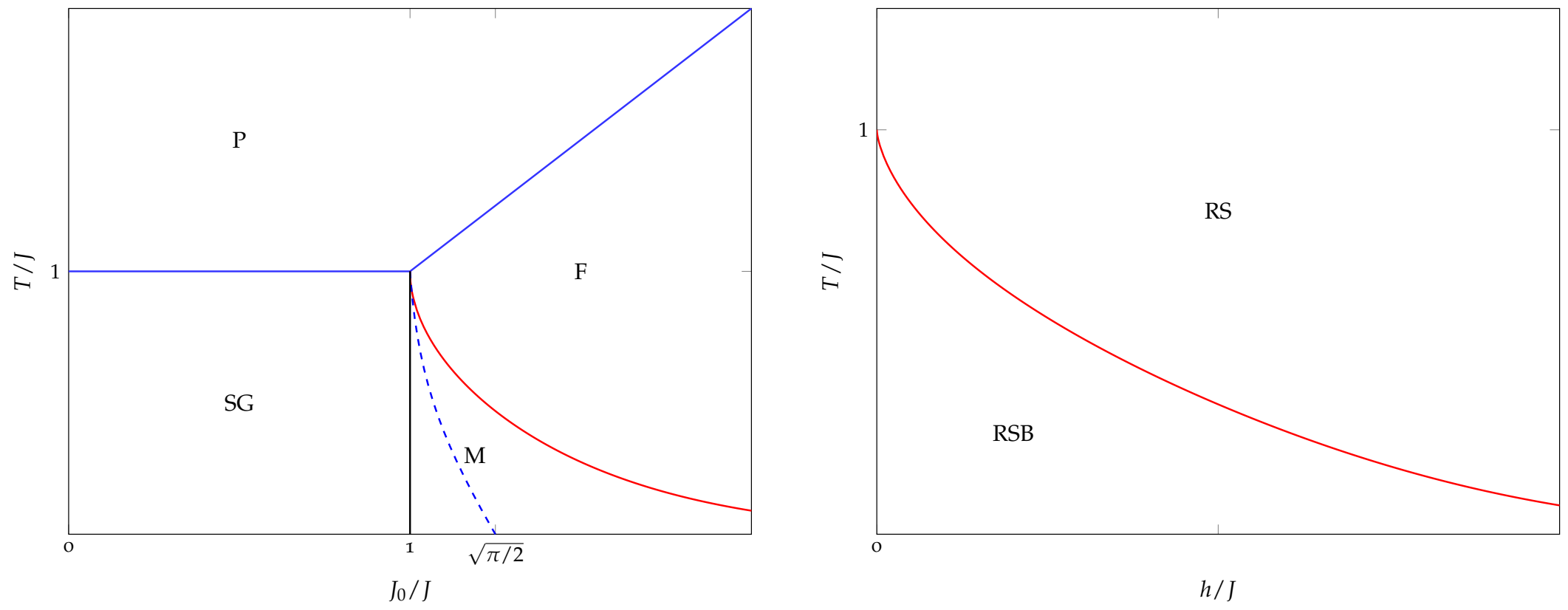

**(a)** $T/J$ vs $J_0/J$ plane. Blue lines are the phase boundaries obtained using the RS ansatz. The dashed one actually disappears accounting RSB effects and it is substituted by the black vertical one.

**(b)** The dAT line in the $T/J$ vs $h/J$ plane. Asymptotic behaviors for small and large $h$ are given in equations (2.38) and (2.39) respectively.

**Figure 2.2.** Phase Diagram and the dAT line (shown in red).

to minimize the free energy with respect to this parameter. The saddle-point equations now become

$$m = \int Dz \tanh\left[\beta\left(J\sqrt{q}\,z + J_0 m + h\right)\right] \tag{2.17a}$$

$$q = \int Dz \tanh^2\left[\beta\left(J\sqrt{q}\,z + J_0 m + h\right)\right] \tag{2.17b}$$

### 2.2.1 Phase Diagram

By studying the RS saddle-point equations (2.17a), (2.17b) one can obtain phase boundaries predicted by the RS ansatz. Let us start by analyzing the case in which there is no external magnetic field $h = 0$.

If the mean of the disorder distribution is $J_0 = 0$, for symmetry arguments the magnetization vanishes and there is no ferromagnetic phase. In addition, $q = 0$ is always a solution. In order to find the temperature below which there can be a solution with $q > 0$ (spin glass phase), we expand for small $q$ the free energy density, obtaining

$$\beta \overline{f}_{\mathrm{RS}} = -\frac{\beta^2 J^2}{4} - \ln 2 - \frac{\beta^2 J^2}{4}\left(1 - \beta^2 J^2\right) q^2 + O\left(q^3\right) . \tag{2.18}$$

Imposing that the coefficient of $q^2$ vanishes at the critical point, we find that

$$T_{\mathrm{SG}} = J . \tag{2.19}$$

We plot in Fig. 2.1a a graphical solution of the saddle-point equation (2.1a) with $J = 1$. In (2.1b) we plot the overlap as a function of temperature, not only for $h = 0$, but also for non-vanishing magnetic field, and still $J_0 = 0$. For $h = 0$ and $T \to T_{\mathrm{SG}}$ the non trivial solution for the overlap vanishes linearly as $q \simeq 1 - T/T_{\mathrm{SG}}$.

If $J_0 > 0$ it is possible to have a ferromagnetic solution $m > 0$. Expanding the saddle-point equations (2.17a), (2.17b) we find

$$m = \beta J_0 m + O(q) \tag{2.20a}$$
$$q = \beta^2 J^2 q + \beta^2 J_0^2 m^2 , \tag{2.20b}$$

from which we find that the boundary between the ferromagnetic phase and the paramagnetic phase is

$$T_c = J_0 . \tag{2.21}$$

The boundary between the paramagnetic phase and the spin glass phase is still given by (2.19), until temperature (2.21) is reached. These two phase boundaries are represented with a blue full line in Fig. (2.2a). The last boundary, between the spin glass and the ferromagnetic phase, can only be obtained numerically, and it is represented as a dashed blue line in the same figure. However this last phase boundary is wrong, because the RS ansatz fails in the region below the line plotted in red. Only taking into account replica symmetry breaking (RSB) we will see that the true solution is given by the vertical black line $J_0 = J$ [Tou80]; between the spin glass and the ferromagnetic phase, there will be a mixed phase region, characterized by ferromagnetic order, but with RSB.

### 2.2.2 The negative entropy problem

To check our results, let us evaluate the entropy at zero temperature, for $J_0 = h = 0$. The free energy density is in this case

$$\beta \overline{f}_{\text{RS}} = -\frac{\beta^2 J^2}{4}(1-q)^2 - \int Dz \, \ln 2 \cosh \left[\beta J \sqrt{q}\, z\right] . \tag{2.22}$$

For small temperature, the saddle-point equation (2.17b) gives

$$q = 1 - \sqrt{\frac{2}{\pi}} \frac{T}{J} + O\left(T^2\right) . \tag{2.23}$$

Inserting this expression into (2.22) we have

$$\overline{f}_{\text{RS}} = -\sqrt{\frac{2}{\pi}} J + \frac{T}{2\pi} + O\left(T^2\right) , \tag{2.24}$$

from which we derive that the ground-state entropy is

$$\lim_{T \to 0} \overline{S}_{\text{RS}}(T) = -\frac{1}{2\pi} , \tag{2.25}$$

which is clearly unphysical. From this result we understand that something must be wrong with the RS ansatz (2.13).

### 2.2.3 Stability Analysis

In this subsection we will perform the stability analysis of the RS saddle-point, finding the line separating the regions where the RS ansatz is correct from the one where it is wrong. This line is called the *de Almeida-Thouless* line (dAT), from the name of the authors that originally performed this calculation [AT78]. We will resume it, for simplicity, in the simple case $J_0 = 0$, where there is no $m_a$ order parameter and then we will describe what happens in the general case. To check the stability of the RS saddle-point $q$ we need to compute the Hessian matrix. Performing the derivatives we get

$$G_{(ab)(cd)} \equiv \frac{1}{\beta^2 J^2} \frac{\partial^2 S}{\partial q_{ab} \partial q_{cd}}\bigg|_{\text{RS}} = \delta_{(ab)(cd)} - \beta^2 J^2 \left( \langle \sigma^a \sigma^b \sigma^c \sigma^d \rangle_{\mathcal{Z}} - \langle \sigma^a \sigma^b \rangle_{\mathcal{Z}} \langle \sigma^c \sigma^d \rangle_{\mathcal{Z}} \right) \tag{2.26}$$

The RS saddle-point will be stable if, every of the $\frac{1}{2}n(n-1)$ eigenvalues of this matrix is non-negative. There are three different entries of the matrix $G_{(ab)(cd)}$

$$\begin{aligned} P &\equiv G_{(ab)(ab)} = 1 - \beta^2 J^2 \left( 1 - \langle \sigma^a \sigma^b \rangle_{\mathcal{Z}}^2 \right) , \\ Q &\equiv G_{(ab)(ac)} = -\beta^2 J^2 \left( \langle \sigma^b \sigma^c \rangle_{\mathcal{Z}} - \langle \sigma^a \sigma^b \rangle_{\mathcal{Z}}^2 \right) , \qquad b \neq c , \\ R &\equiv G_{(ab)(cd)} = -\beta^2 J^2 \left( \langle \sigma^a \sigma^b \sigma^c \sigma^d \rangle_{\mathcal{Z}} - \langle \sigma^a \sigma^b \rangle_{\mathcal{Z}}^2 \right) , \qquad (a,b) \neq (c,d) . \end{aligned} \tag{2.27}$$

$P$, $Q$ and $R$ appear in one line of the matrix $G_{(ab)(cd)}$ respectively 1, $2(n-2)$ and $\frac{1}{2}(n-2)(n-3)$ times. In these expressions we identify with $q = \langle \sigma^a \sigma^b \rangle_{\mathcal{Z}}$ and with

$$r \equiv \langle \sigma^a \sigma^b \sigma^c \sigma^d \rangle_{\mathcal{Z}} = \int Dz \tanh^4 \left[ \beta \left( J\sqrt{q}\, z + h \right) \right] . \tag{2.28}$$

The eigenvalues of the matrix $G_{(ab)(cd)}$ can be studied for generic $n$ and the eigenvalue equation is

$$\sum_{c<d} G_{(ab)(cd)} \eta_{cd} = \lambda \eta_{ab} . \tag{2.29}$$

One can classify the eigenvectors in three different ways (each one forming a "*sector*") depending on the symmetry it has in replica space.

- *Longitudinal* sector: since the lines of $G_{(ab)(cd)}$ have the same sum, surely there has to be the eigenvector that is completely symmetric under permutations

  $$\eta^{(1)}_{ab} = \eta , \quad a \neq b . \tag{2.30}$$

  The corresponding eigenvalue (which is non-degenerate) is

  $$\lambda_1(n) = P + 2(n-2)Q + \frac{1}{2}(n-2)(n-3)R . \tag{2.31}$$

- *Anomalous* sector: the second type of eigenvectors are those that are completely symmetric except for one of the replica indices (for example 1)

  $$\begin{aligned} \eta^{(2)}_{ab} &= \eta^{(2)}_1 , \quad a = 1 \text{ or } b = 1 \\ \eta^{(2)}_{ab} &= \eta^{(2)}_2 , \quad a, b \neq 1 . \end{aligned} \tag{2.32}$$

Imposing orthogonality of $\eta_{ab}^{(2)}$ with $\eta_{ab}^{(1)}$ we find the corresponding eigenvalue to be

$$\lambda_2(n) = P + (n-4)Q - (n-3)R\,, \tag{2.33}$$

which is $(n-1)$ times degenerate.

- *Replicon* sector: here there are the eigenvectors that are completely symmetric except for two of the replica indices (for example 1 and 2)

$$\begin{aligned} \eta_{12}^{(3)} &= \eta_1^{(3)}\,, \\ \eta_{1a}^{(3)} &= \eta_{2a}^{(3)} = \eta_2^{(3)}\,, \qquad a \neq 1,2 \\ \eta_{ab}^{(3)} &= \eta_3^{(3)}\,, \qquad\qquad a,b \neq 1,2\,. \end{aligned} \tag{2.34}$$

Imposing orthogonality of $\eta_{ab}^{(3)}$ with both $\eta_{ab}^{(1)}$ and $\eta_{ab}^{(2)}$ we find the corresponding eigenvalue to be

$$\lambda_3 = P - 2Q + R\,, \tag{2.35}$$

which is $\frac{1}{2}n(n-3)$ times degenerate.

We have found the whole possible eigenvalues, since summing the dimensions of the three subspaces we get $\frac{1}{2}n(n-1)$. Next we can perform the zero $n$ limit and we have

$$\begin{aligned} \lambda_1 = \lambda_2 &= P - 4Q + 3R \\ \lambda_3 &= P - 2Q + R\,, \end{aligned} \tag{2.36}$$

The simplest case where one can check stability is in the paramagnetic phase, were both $q = r = 0$. It directly follows that the only non vanishing element is $P$, the diagonal one, and therefore $\lambda_1 = \lambda_2 = \lambda_3 = P = 1 - \beta^2 J^2 > 0$, i.e. $T > J$, which is exactly the boundary of the paramagnetic phase we have derived before. We conclude that, for $J_0 = 0$, the RS ansatz is stable in the paramagnetic phase. Extending the following stability study to $J_0 > 0$ one finds that the whole paramagnetic phase is correctly described by RS. In the ferromagnetic and spin-glass phase, instead, one finds that $\lambda_1 = \lambda_2 > 0$ for every value of $T$, $h$ and $J_0$. Problems arise examining the third eigenvalue $\lambda_3$ which is called the *replicon* [BM78]. It turns out that there is a region in the $(T, h, J_0)$ space such that the condition $\lambda_3 > 0$ is violated. For $J_0 = 0$, the dAT line of instability $\lambda_3 = 0$, can be expressed as

$$\int Dz\, \mathrm{sech}^4\left[\beta\left(J\sqrt{q}\,z + h\right)\right] = \frac{T^2}{J^2}\,. \tag{2.37}$$

We plot this line in the $(T/J, h/J)$ plane in Fig. 2.2b. Several properties of this line are deducible. In the limit of small external magnetic field we have

$$\frac{T_{\mathrm{SG}} - T_{\mathrm{AT}}(h)}{T_{\mathrm{SG}}} \simeq \left(\frac{3}{4}\right)^{1/3} \left(\frac{h}{J}\right)^{2/3}\,, \tag{2.38}$$

whereas in the large $h$ limit we have

$$T_{\mathrm{AT}}(h) \simeq \frac{4J}{2\sqrt{2\pi}} e^{-\frac{h^2}{2J^2}}\,. \tag{2.39}$$

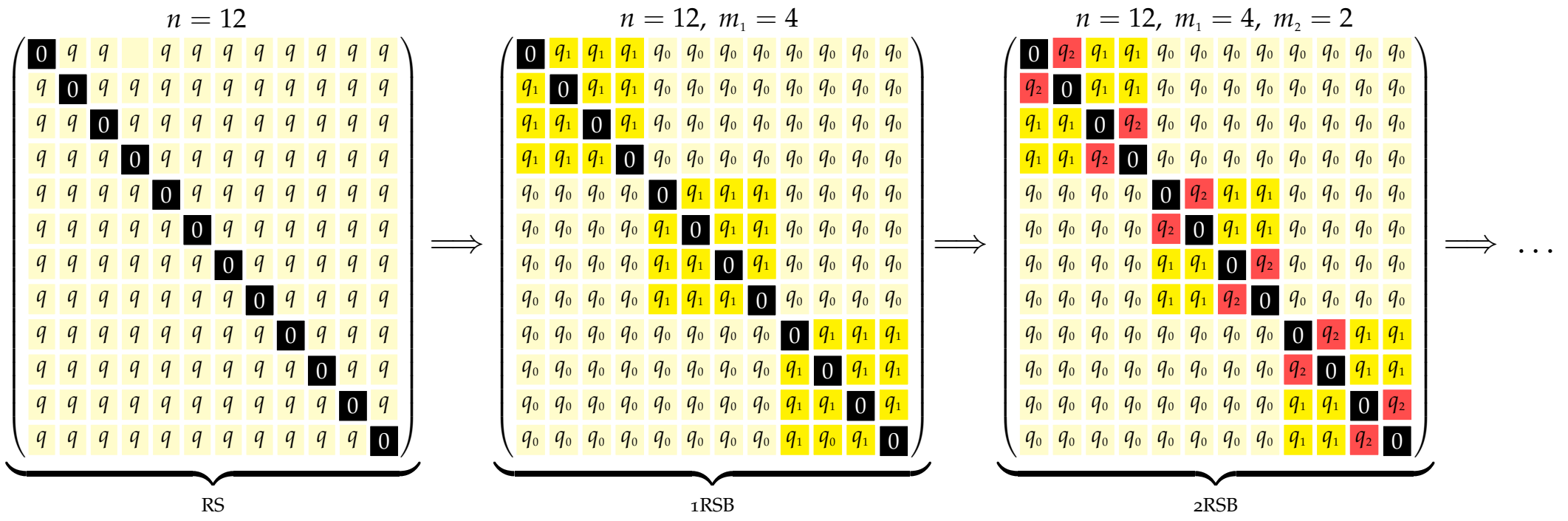

**Figure 2.3.** Parisi's scheme of RSB

This means that, no matter how strong the magnetic field is, there is always a non-vanishing temperature $T_{AT}(h)$ under which there is RSB. This is a feature of fully connected models. Defining instead the problem on a finite connectivity system, such as on a Bethe lattice, one finds that for $T = 0$ there is a critical value of the magnetic field $h_c < \infty$ such that for $h > h_c$ RSB does not occur. This suggests that going in finite dimension the effects of RSB level off. A long-standing question is if in finite dimension (the EA model) the dAT line completely disappears or survives. Even after the great efforts in numerical simulations, with the construction of specifically designed computers (Janus collaboration), a definite answer to this question has not been found yet; at the present state there are two antagonist theories for spin glasses in finite dimension: the Parisi's RSB picture (that we will describe in the next section) and the *droplet* picture developed by McMillan [McM84], Bray and Moore [BM87] and Fisher and Huse [FH86, FH87, FH88].

In the $(J_0/J, T/J)$ plane, it suffices to say that the equation of the stability line is modified by simply making the substitution $h \to J_0 m$; its plot is reported, for $J_0/J \geq 1$ in Fig. 2.2a in red. For $J_0/J \leq 1$ the dAT line simply coincides with the paramagnetic spin glass phase boundary. One last thing to notice is that no RSB occurs in $m_a$, since the expression of the replicon eigenvalue does not change when $J_0 \neq 0$.

## 2.3 Breaking the Replica Symmetry

After the failure of RS ansatz people started to search a new parametrization of the overlap matrix $q_{ab}$. Unfortunately there is no standard procedure to find the correct one, and the number of possible parametrizations are infinite. The first proposal has been made by Bray and Moore [BM78], which subdivided the replicas in two groups of size $m_1$ and $n - m_1$ giving 3 different values of the overlaps (2 for the diagonal blocks and 1 for the off-diagonal one). No solution different from the RS one has been found in this way. The first step in the right direction was done by Blandin [DDG06], who proposed to break replica symmetry by grouping replicas into different blocks of size $m_1 = 2$, assigning two different values $q_1$ and $q_0$ for the diagonal and off-diagonal blocks respectively. The parametrization of Blandin was generalized by Parisi [Par79b], which used the block size $m_1$ as a variational parameter. In other terms Parisi proposed an ansatz of this form

$$q_{ab} = \begin{cases} q_1, & \text{if } I\left(\frac{a}{m_1}\right) = I\left(\frac{b}{m_1}\right), \\ q_0, & \text{if } I\left(\frac{a}{m_1}\right) \neq I\left(\frac{b}{m_1}\right), \\ 0, & \text{if } a = b\,, \end{cases} \tag{2.40}$$

where $1 \leq m_1 \leq n$ and $I(x)$ is the integer valued function which is equal to the smallest integer larger than or equal to $x$. This parametrization is known as *one-step* replica symmetry breaking (1RSB), see Fig. 2.3. Denoting with $S_n$ the permutation group of $n$ elements, it is evident that the overlap matrix $q_{ab}$ is left invariant under the $S_n$ subgroup $(S_{m_1})^{\otimes n/m_1} \otimes S_{m_1}$, where $(S_{m_1})^{\otimes n/m_1}$ is the direct product of $n/m_1$ groups $S_{m_1}$. Note that, before doing the $n \to 0$ limit, $n/m_1$ must be an integer. This ansatz was found to have the same issues of the RS result (namely the negative entropy problem and a negative eigenvalue in the replicon sector), but comparison with numerical simulations showed that there where several improvements compared to the RS ansatz. This suggested to improve the 1RSB results by breaking replica symmetry with multiple steps [Par79a], approximating the true solution better and better at each step [Par80a]. For example, the *two-step* replica symmetry breaking (2RSB) consists in dividing every group of $m_1$ replicas into $m_1/m_2$ blocks of $m_2$ replicas, see Fig. 2.3. The generalization to $k$-steps ($k$RSB) is immediate. We introduce a set of integers $n = m_0 \geq m_1 \geq \cdots \geq m_k \geq m_{k+1} = 1$ such that $m_i/m_{i+1}$, $i = 0, \ldots k$ are integers. The overlap matrix is then parametrized as follows

$$q_{ab} = \begin{cases} q_i, & \text{if } I\left(\frac{a}{m_i}\right) = I\left(\frac{b}{m_i}\right) \text{ and } I\left(\frac{a}{m_{i+1}}\right) \neq I\left(\frac{b}{m_{i+1}}\right), \\ 0, & \text{if } a = b\,, \end{cases} \tag{2.41}$$

where $i = 0, \ldots, k$. It was found that, in order to obtain the right solution, one must send the number of breaking $k$ to infinity. In this limit one obtains the so called *full*-replica symmetry breaking (fRSB) solution. As we will see, in this limit the overlap becomes a function of the interval $[0, 1]$ [Par80b, Par83] and this particular parametrization will have consequences of the structure of the states of the spin glass phase. The Parisi solution has been proved rigorously to give the correct result by the work of Guerra and Toninelli [GT02, Gue03] and Talagrand [Tal06] who also proved the uniqueness of the analytical continuation from an integer to a real number of replicas when one performs the $n \to 0$ limit.

### 2.3.1 1RSB

We start by examining the 1RSB ansatz given by (2.40). We remind that there is no need to break replica symmetry for the magnetizations parameters $m_a = m$.

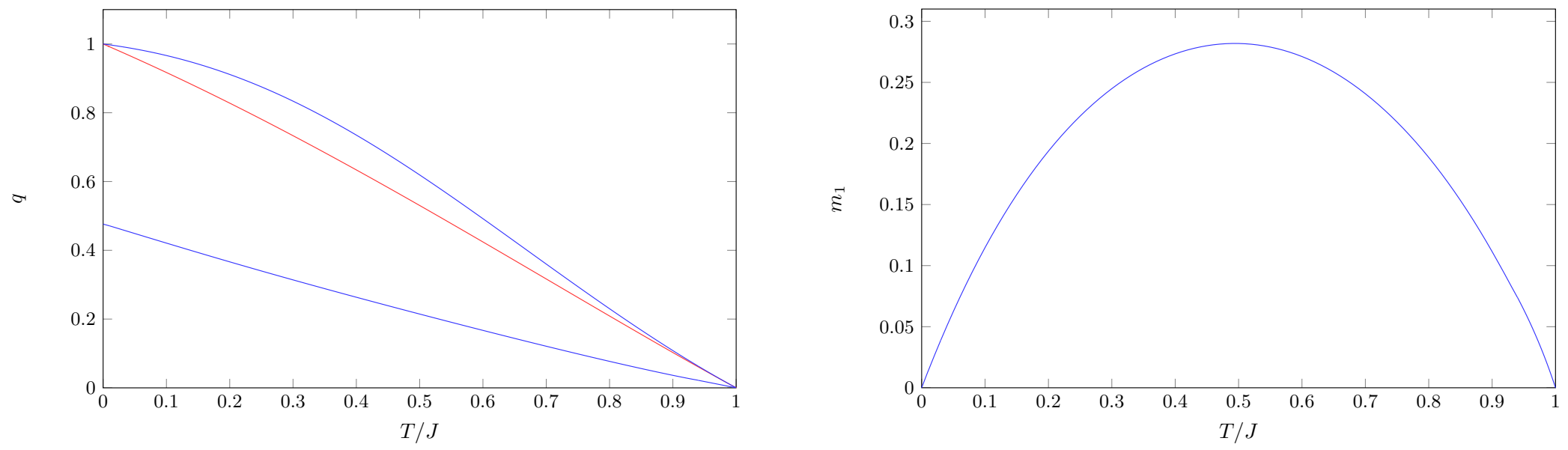

**(a)** In blue $q_1$ and $q_0$ (from top to bottom), in red the overlap $q$ in the RS case.

**(b)** $m_1$

**Figure 2.4.** Plot of $q_0$ and $q_1$(left panel) and of $m_1$ (right panel) vs $T/J$ obtained solving numerically the 1RSB saddle-point equations. In the left panel it is also shown in red the overlap $q$ obtained using RS.

The Hamiltonian (2.9) is therefore

$$-\beta\mathcal{H} = \frac{\beta^2 J^2}{2}\left[q_0\left(\sum_{a=1}^{n}\sigma^a\right)^2 + (q_1 - q_0)\sum_{i=1}^{n/m_1}\left(\sum_{a_i=1}^{m_1}\sigma^{a_i}\right)^2 - nq_1\right] + \beta(J_0 m + h)\sum_{a=1}^{n}\sigma^a . \tag{2.42}$$

As usual, for every squared term we introduce an Hubbard-Stratonovich transformation in order to linearize it

$$\begin{aligned} e^{-\beta\mathcal{H}[\sigma]} &= e^{-\frac{n\beta^2 J^2}{2}q_1}\int Du \int \prod_{i=1}^{n/m_1} Dv_i\, e^{\beta\sum_{i=1}^{n/m_1}\left[J\sqrt{q_0}\,u + J_0 m + h + J\sqrt{q_1 - q_0}\,v_i\right]\sum_{a_i=1}^{m_1}\sigma^{a_i}} \\ &= e^{-\frac{n\beta^2 J^2}{2}q_1}\int Du \left[\int Dv\, e^{\beta\left[J\sqrt{q_0}\,u + J_0 m + h + J\sqrt{q_1 - q_0}\,v\right]\sum_{a=1}^{m_1}\sigma^a}\right]^{n/m_1} \end{aligned} \tag{2.43}$$

Introducing the quantity

$$\Xi \equiv \beta\left[J\sqrt{q_0}u + J\sqrt{q_1 - q_0}\,v + J_0 m + h\right] , \tag{2.44}$$

and summing over all the spins we get

$$\begin{aligned} \ln\mathcal{Z} &= -\frac{n\beta^2 J^2}{2}q_1 + \ln\int Du\left[\int Dv\,(2\cosh\Xi)^{m_1}\right]^{n/m_1} \\ &\simeq -\frac{n\beta^2 J^2}{2}q_1 + \frac{n}{m_1}\int Du \ln\int Dv\,(2\cosh\Xi)^{m_1} . \end{aligned} \tag{2.45}$$

The quadratic term in the free energy is

$$\begin{aligned} \lim_{n\to 0}\frac{1}{n}\sum_{a<b} q_{ab}^2 &= \lim_{n\to 0}\frac{1}{2n}\left[n^2 q_0^2 + \frac{n}{m_1}m_1^2(q_1^2 - q_0^2) - nq_1^2\right] \\ &= \frac{1}{2}\left[(m_1 - 1)q_1^2 - m_1 q_0^2\right] . \end{aligned} \tag{2.46}$$

Note that in the limit $n \to 0$ the parameter $m_1$ that initially was an integer satisfying $1 \leq m_1 \leq n$ now becomes a continuous parameter defined on the interval $0 \leq m_1 \leq 1$. The same holds for the other variational parameters $q_0$, $q_1$ and $m$. The 1RSB free energy is therefore

$$\begin{aligned}\beta \overline{f}_{1\text{RSB}} = & \frac{\beta^2 J^2}{4}\left[(m_1-1)q_1^2 - m_1 q_0^2 + 2q_1 - 1\right] + \frac{\beta J_0}{2}m^2 \\ & - \frac{1}{m_1}\int Du \ln \int Dv \,(2\cosh \Xi)^{m_1} \,.\end{aligned} \tag{2.47}$$

The next step is to perform derivatives with respect to $m$, $q_0$, $q_1$ and $m_1$. We report here the first three since they are more useful for physical interpretation

$$m = \int Du \,\frac{\int Dv \,(\cosh \Xi)^{m_1} \tanh \Xi}{\int Dv \,(\cosh \Xi)^{m_1}}\,, \tag{2.48a}$$

$$q_0 = \int Du \left[\frac{\int Dv \,(\cosh \Xi)^{m_1} \tanh \Xi}{\int Dv \,(\cosh \Xi)^{m_1}}\right]^2, \tag{2.48b}$$

$$q_1 = \int Du \,\frac{\int Dv \,(\cosh \Xi)^{m_1} \tanh^2 \Xi}{\int Dv \,(\cosh \Xi)^{m_1}}\,. \tag{2.48c}$$

One can interpret these equations as follows. In the first one, the integrand of $Du$ represents the magnetization within a block of the 1RSB matrix, as can be verified using the saddle-point equation (2.11a) in the RSB framework. Then this value is averaged over all blocks using the usual Gaussian weight. Using the saddle-point equation for $q_{ab}$ (2.11b) one can verify that the equation for $q_0$ is obtained assuming that the spins $\sigma^a$ and $\sigma^b$ belong to different blocks in the 1RSB overlap matrix. Indeed the argument of $Du$ in (2.48b) is the product of the magnetizations of two different blocks. In the same way $q_1$ can be obtained assuming that $\sigma^a$ and $\sigma^b$ belong to the same block as can be readily seen in equation (2.48c). From the Schwarz inequality it follows also that

$$q_0 \leq q_1\,. \tag{2.49}$$

Let us now examine what 1RSB predicts when $J_0 = h = 0$. In this case $m$ vanishes for every temperature confirming that there cannot be a ferromagnetic phase, as happens in the RS ansatz. The parameter $q_1$ is zero (and therefore also $q_0$) when $T \geq T_{\text{SG}}$ and correspondingly also $m_1 = 0$. In the paramagnetic phase 1RSB reproduces exactly the RS results. For $T < T_{\text{SG}}$ the parameter $q_1$ can instead be positive, and correspondingly $m_1$ decreases with temperature. Coherently with the stability analysis the spin glass transition predicted by 1RSB is the same of RS. In Fig. 2.4 we plot the behavior of the 1RSB parameters as function of $T/J$ whereas in Fig. 2.5 we plot the energy and entropy vs temperature obtained using the RS and 1RSB ansatz.

The results found by 1RSB suffers, however, of the same problems of RS, namely the negative entropy problem and replicon instability. For $J_0 = h = T = 0$ the entropy increases from $-1/2\pi \approx -0.159$ of the RS to $S_{1\text{RSB}}(0) \approx -0.01$ of 1RSB, as can be seen in Fig. 2.5b. Also the replicon, even if still negative, decreases in absolute value. This suggests that even though 1RSB is still wrong for

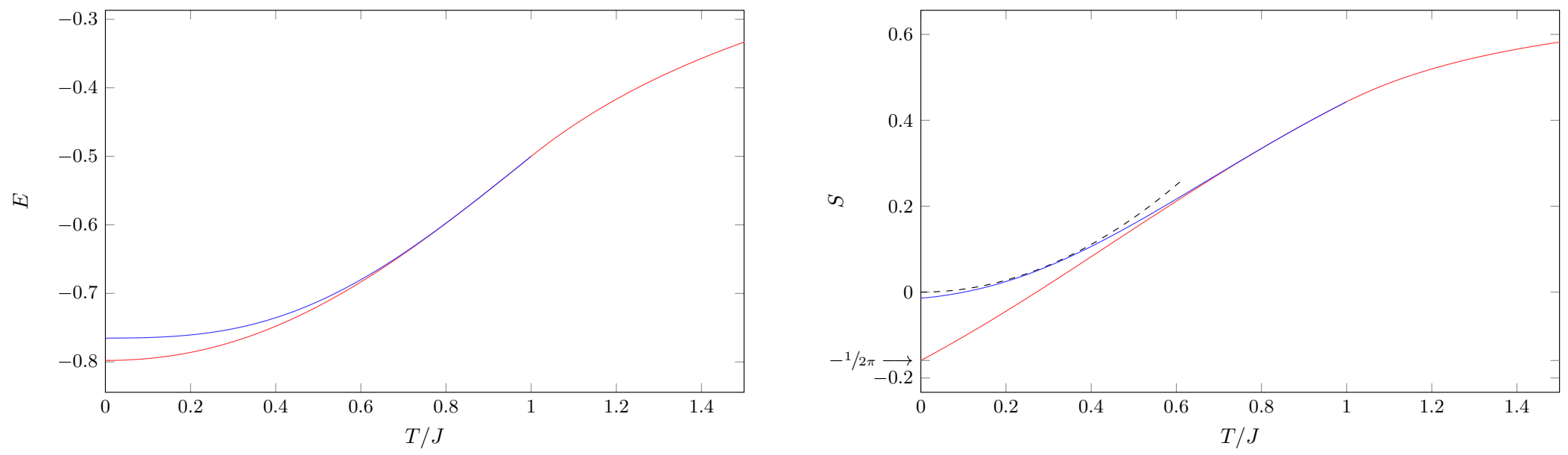

**(a)** Energy

**(b)** Entropy

**Figure 2.5.** Energy (left panel) and entropy (right panel) vs $T/J$ for $J_0 = h = 0$ obtained numerically using the RS ansatz (red line) and 1RSB ansatz (blue line). In the entropy plot the black and dashed line is the prediction $S \simeq T^2 \ln 2$ of [TAP77] using TAP equations for $T \to 0$.

the SK model, this is a better approximation to the solution. Therefore one can try to iterate this procedure via multiple breaking of replica symmetry.

### 2.3.2 fRSB

We generalize the computation of the previous subsection to generic $k$ using the parametrization of the overlap matrix given in (2.41) and then we send $k \to \infty$. The Hamiltonian (2.9) is

$$\begin{aligned} -\beta\mathcal{H} = & \frac{\beta^2 J^2}{2}\left[q_0\left(\sum_{a=1}^{n}\sigma^a\right)^2 + \sum_{j=1}^{k}(q_j - q_{j-1})\sum_{i=1}^{n/m_j}\left(\sum_{a_i=1}^{m_j}\sigma^{a_i}\right)^2 - nq_k\right] \\ & + \beta(J_0 m + h)\sum_{a=1}^{n}\sigma^a \,. \end{aligned} \tag{2.50}$$

Introducing several Hubbard-Stratonovich transformation we get

$$\begin{aligned} e^{-\beta\mathcal{H}[\sigma]} = & e^{-\frac{n\beta^2 J^2}{2}q_k}\int Dz_0 \prod_{j=1}^{k}\int\prod_{i=1}^{n/m_j} Dz_j^i\, e^{\beta\left[J\sqrt{q_0}\,z_0 + J_0 m + h\right]\sum_{a=1}^{n}\sigma^a} \\ & e^{\beta J\sum_{j=1}^{k}\sqrt{q_j - q_{j-1}}\sum_{i=1}^{n/m_j} z_j^i \sum_{a=1}^{m_j}\sigma^{a_i}} \,. \end{aligned} \tag{2.51}$$

Next we use the general decomposition of the sums $\sum_{a=1}^{m_h} \bullet = \sum_{i=1}^{m_h/m_{h+1}} \sum_{a_i=1}^{m_{h+1}} \bullet$ and $\sum_{i=1}^{m_h/m_l} \bullet = \sum_{j=1}^{m_h/m_{h+1}} \sum_{i_j=1}^{m_{h+1}/m_l} \bullet$ for every $h = 0, \dots, k$ and $l = h+1, \dots, k$ to regroup the integrals. At the first step (i.e. $h = 0$) we obtain

$$\begin{aligned} e^{-\beta\mathcal{H}[\sigma]} = & e^{-\frac{n\beta^2 J^2}{2}q_k}\int Dz_0 \left[\int Dz_1 \prod_{j=2}^{k}\int\prod_{i=1}^{m_1/m_j} Dz_j^i\, e^{\beta\left[J\sqrt{q_0}\,z_0 + J\sqrt{q_1 - q_0}\,z_1 + J_0 m + h\right]\sum_{a=1}^{m_1}\sigma^a}\right. \\ & \left. e^{\beta J\sum_{j=2}^{k}\sqrt{q_j - q_{j-1}}\sum_{i=1}^{m_1/m_j} z_j^i \sum_{a=1}^{m_j}\sigma^{a_i}}\right]^{n/m_1} \,. \end{aligned} \tag{2.52}$$

Iterating this procedure other $k-1$ times we obtain

$$e^{-\beta\mathcal{H}[\sigma]} = e^{-\frac{n\beta^2 J^2}{2}q_k}\int Dz_0\left[\int Dz_1\left[\int Dz_2\ldots\left[\int Dz_k\, e^{\Xi\sum_{a=1}^{m_k}\sigma^a}\right]^{m_{k-1}/m_k}\ldots\right]^{m_1/m_2}\right]^{n/m_1} \tag{2.53}$$

where we have redefined the quantity

$$\Xi \equiv \beta\left[J\sqrt{q_0}\,z_0 + J\sum_{j=1}^{k}\sqrt{q_j-q_{j-1}}\,z_j + J_0 m + h\right]. \tag{2.54}$$

Performing the sum over the spins and expanding for $n$ small we obtain

$$\ln\mathcal{Z} = -\frac{n\beta^2J^2}{2}q_k + \ln\int Dz_0\, L^{n/m_1}(z_0) \simeq -\frac{n\beta^2J^2}{2}q_k + \frac{n}{m_1}\int Dz_0 \ln L(z_0)\,, \tag{2.55}$$

where we have defined $L(z_0)$ to be

$$L(z_0) \equiv \int Dz_1\left[\int Dz_2\ldots\left[\int Dz_k\,\left(2\cosh\Xi\right)^{m_k}\right]^{m_{k-1}/m_k}\ldots\right]^{m_1/m_2}. \tag{2.56}$$

The quadratic term in the free energy instead reads

$$\begin{aligned}\lim_{n\to 0}\frac{1}{n}\sum_{a<b}q_{ab}^2 &= \lim_{n\to 0}\frac{1}{2n}\left[n^2q_0^2 + n\sum_{j=1}^{k}m_j(q_j^2-q_{j-1}^2) - nq_k^2\right] \\ &= \frac{1}{2}\sum_{j=0}^{k}(m_j-m_{j+1})\,q_j^2\,,\end{aligned} \tag{2.57}$$

having identified $m_0 = n$ and $m_{k+1} = 1$. In general for every function $f$ of the overlaps we have

$$\lim_{n\to 0}\frac{1}{n}\sum_{a<b}f(q_{ab}) = \frac{1}{2}\sum_{j=0}^{k}(m_j-m_{j+1})\,f(q_j)\,. \tag{2.58}$$

The free energy of the kRSB ansatz is therefore

$$\beta\overline{f}_{\text{kRSB}} = \frac{\beta^2J^2}{4}\left[\sum_{j=0}^{k}(m_j-m_{j+1})\,q_j^2 + 2q_k - 1\right] + \frac{\beta J_0}{2}m^2 - \frac{1}{m_1}\int Dz_0\ln L(z_0) \tag{2.59}$$

which correctly reduces to (2.47) for $k=1$. The parameters $m_i$ are now continuous, assuming values in the interval $[0,1]$ and they are ordered as follows $0 = m_0 \leq m_1 \leq \cdots \leq m_k \leq m_{k+1} = 1$. For finite $k$ the overlap parameters satisfy $q_i \leq q_{i+1}$. The next step is to derive the saddle-point equations for all the parameters and look to fixed point solutions. This is a really difficult task. However for $k$ sufficiently low, one can solve numerically these equations and one finds that the situations gets better and better increasing the value of $k$. For $k = 3$, for example, the zero-temperature entropy reduces in modulus to $S_{3\text{RSB}}(0) = -0.001$. One therefore looks for a stable solution in the $k \to \infty$ limit. In the infinite $k$ limit one has an infinite set of parameters $q_i$: the overlap becomes a continuous function $q(x)$ defined as the limit of the step-wise function

$$q(x) = q_i \qquad 0 \le m_i \le x \le m_{i+1} \le 1 \tag{2.60}$$

for $i = 0, \ldots, k$. The function $q(x)$ has the property $dq/dx \ge 0$. In the infinite $k$ limit one can replace $m_j - m_{j+1} \to -dx$ so that the quadratic term in the free energy (2.59) is

$$\lim_{k\to\infty} \sum_{j=0}^{k} (m_j - m_{j+1})\, q_j^2 = -\int_0^1 dx\, q^2(x)\,. \tag{2.61}$$

More work is required to find the limit in the last term of (2.59). The easiest derivation is that of Parisi [Par80a] and Duplantier [Dup81] which is based on the following identity

$$\int Dz\, f\,(uz + v) = \int Dz\, e^{uz\frac{d}{dv}} f(v) = e^{\frac{u^2}{2}\frac{d^2}{dv^2}} f(v)\,, \tag{2.62}$$

based on the fact that the Gaussian is the Green function of the Heat equation. The $L(z)$ term has exactly the same linear dependence on $z$. Therefore one can write a recursive relation

$$g(n,h) \equiv \int Dz_0\, L^{n/m_1}(z_0) = e^{\frac{J^2}{2} q_0 \frac{\partial^2}{\partial h^2}} \left[g(m_1,h)\right]^{n/m_1}\,, \tag{2.63}$$

in which the generic term is

$$g(m_i,h) = e^{\frac{J^2}{2}(q_i - q_{i-1})\frac{\partial^2}{\partial h^2}} \left[g(m_{i+1},h)\right]^{m_i/m_{i+1}}\,, \qquad i = 0, \ldots, k\,, \tag{2.64}$$

with $q_{-1} = 0$ and initial condition

$$g(m_{k+1},h) \equiv 2\cosh\left[\beta\,(J_0 m + h)\right]\,. \tag{2.65}$$

Sending $n \to 0$ and $k \to \infty$ equation (2.64) becomes a differential equation since $m_i - m_{i+1} \to dx$

$$g(x+dx,h) = e^{-\frac{J^2}{2} dq(x) \frac{\partial^2}{\partial h^2}} \left[g(x,h)\right]^{1+d\ln x}\,, \tag{2.66}$$

i.e.

$$\frac{\partial g}{\partial x} = -\frac{J^2}{2}\frac{dq}{dx}\frac{\partial^2 g}{\partial h^2} + \frac{g\ln g}{x}\,, \tag{2.67}$$

with boundary condition given by $g(1,h) = 2\cosh\left[\beta\,(J_0 m + h)\right]$. Taking the limit $n \to 0$ and $k \to \infty$ in (2.63) we have

$$\frac{1}{n}\ln g(n,h) \simeq \lim_{x\to 0} e^{\frac{J^2}{2} q(0) \frac{\partial^2}{\partial h^2}} \frac{\ln g(x,h)}{x}. \tag{2.68}$$

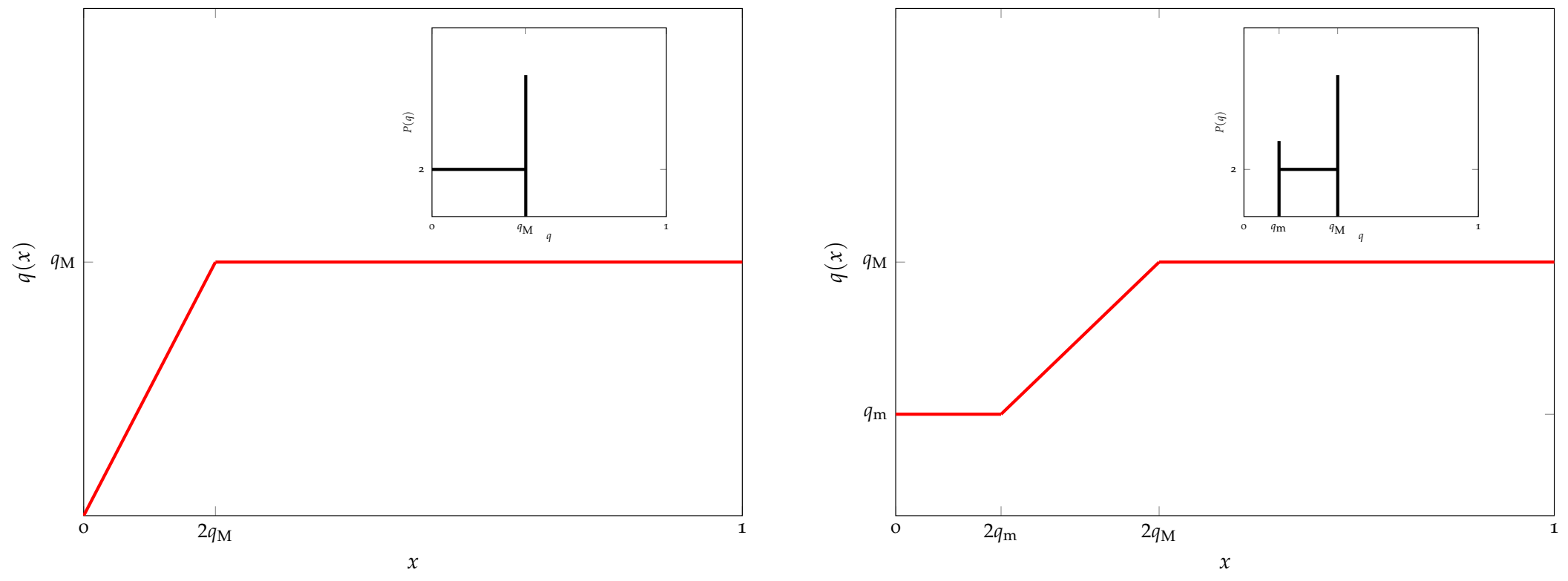

**(a)** $h = 0$: in this case $q_{\mathrm{m}}$ vanishes. The value of $q_{\mathrm{M}}$ is $\tau$.

**(b)** $0 < h < h_{\mathrm{dAT}}(T)$: here $q_{\mathrm{m}} \simeq h^{2/3}$ and $q_{\mathrm{M}} = \tau$. When $h \to h_{\mathrm{dAT}}$, $q_{\mathrm{m}} \to q_{\mathrm{M}}$ so that the two delta function fuse together and RS becomes exact.

**Figure 2.6.** Behavior of the overlap $q(x)$ and its distribution $P(q)$ (shown in the inset) for $\tau \ll 1$.

Defining

$$f(x,h) \equiv \frac{\ln g(x,h)}{x}, \tag{2.69}$$

the previous equation can rewritten as

$$\lim_{n\to 0} \frac{1}{n} \ln g(n,h) = \int Du\, f\left(0, J\sqrt{q(0)}\, u + h\right). \tag{2.70}$$

The fRSB free energy is therefore

$$\beta \overline{f}_{\mathrm{fRSB}} = -\frac{\beta^2 J^2}{4}\left[1 + \int_0^1 dx\, q^2(x) - 2q(1)\right] + \frac{\beta J_0}{2} m^2 - \int Du\, f\left(0, J\sqrt{q(0)}\, u + h\right), \tag{2.71}$$

where $f$ satisfies the *Parisi equation* [Par80a], which is a non-linear antiparabolic partial differential equation

$$\frac{\partial f}{\partial x} = -\frac{J^2}{2}\frac{dq}{dx}\left[\frac{\partial^2 f}{\partial h^2} + x\left(\frac{\partial f}{\partial h}\right)^2\right], \tag{2.72}$$

with boundary condition

$$f(1,h) = \ln 2 \cosh\left[\beta\left(J_0 m + h\right)\right]. \tag{2.73}$$

Note that for $q(x) = q$ constant we recover as it should the RS result. The stability of the fRSB has been checked by De Dominicis and Kondor [DDK83] who showed that the Parisi's solution is *marginally stable* since the replicon goes to 0 for $k \to \infty$. In this sense the SK model is *critical* for every temperature below the spin glass one. One can easily find the behavior of the overlap when we are near the spin-glass critical temperature. In this case $q(x)$ is expected to be small so one can expand the free energy (2.10) [BM78]. One gets the following general behavior for $\tau \equiv (T_{SG} - T)/T_{SG} \to 0$ and small fields

$$q(x) = \begin{cases} q_m\,, & 0 \le x \le x_m \\ \frac{x}{2}\,, & x_m \le x \le x_M \\ q_M\,, & x_M \le x \le 1 \end{cases} \tag{2.74}$$

with $x_m = 2q_m$, $x_M = 2q_M$ and

$$\begin{aligned} q_m &\simeq h^{2/3} \\ q_M &\simeq \tau \end{aligned} \tag{2.75}$$

i.e. $q_M$ is qualitatively independent on the external magnetic field. When $h \to h_{dAT}$, $q_m \to q_M$, so that the RS solution becomes exact, as it should. In addition $q_m \to q_M$ agrees with equation (2.38) derived in the context of the stability analysis. We plot in Fig. 2.6a these qualitative behavior. In turns out that qualitatively the previous picture continue to be true also at lower temperature, but with a generic function of $x$ between $x_m$ and $x_M$. Within this framework one can evaluate all the physical quantities of interest, for example the energy for $J_0 = h = 0$

$$E = -\frac{\beta J^2}{2}\left(1 + \frac{2}{n}\sum_{a<b} q_{ab}^2\right) = -\frac{\beta J^2}{2}\left(1 - \int_0^1 dx\, q^2(x)\right), \tag{2.76}$$

and the magnetic susceptibility

$$\chi = \beta\left(1 + \frac{2}{n}\sum_{a<b} q_{ab}\right) = \beta\left(1 - \int_0^1 dx\, q(x)\right) \tag{2.77}$$

Particularly interesting is the susceptibility because it allows to derive the vertical phase boundary derived by [Tou80] between the ferromagnetic and the spin glass phase of Fig. (2.2a). For $h = 0$ and $\tau \ll 1$, $\chi$ is constant

$$\chi = \beta(1-\tau) + O(\tau^2) = \frac{1}{J}\,. \tag{2.78}$$

It turns out that this result is valid for every $T \le T_{SG}$. One can see [Nis01] that the quadratic term of the expansion of the free energy in powers of $m$ is proportional to $(\chi^{-1} - J_0)$, signaling a phase transition when $J = J_0$. The plot of the susceptibility as a function of temperature is shown in Fig. 2.7a.

## 2.4 Replicas and Physics

The breaking of the replica symmetry may seem strange at first sight, because replicas have been introduced by our own hand with a subtle mathematical trick (1.70). It is important therefore to understand the physical meaning of replica symmetry breaking evaluating analytically some experimental measurable quantities [MPV87]. On the other hand it is also important how one can give an experimental evidence of RSB [Par02]. In the first subsection we will answer to the first question, showing how the probability distribution of the overlaps over states defined in (1.88) is equal to the corresponding one between two replicas [Par83]. In the next subsection we will answer to the second question.

### 2.4.1 Overlap Distribution

In this subsection we will see how to evaluate the distribution of the overlaps, defined in (1.88), using the replica method. Let us consider

$$q_J^{(1)} = \frac{1}{N} \sum_i \langle \sigma_i \rangle^2 \,. \tag{2.79}$$

Now we can use the decomposition in pure states (1.73) in order to have

$$q_J^{(1)} = \frac{1}{N} \sum_i \sum_{\alpha\beta} w_\alpha w_\beta \, \langle \sigma_i \rangle_\alpha \, \langle \sigma_i \rangle_\beta = \sum_{\alpha\beta} w_\alpha w_\beta \, q_{\alpha\beta} = \int dq \, P_J(q) \, q \,, \tag{2.80}$$

i.e. $q_J^{(1)}$ can be written as the first moment of the overlap distribution. Then we take the following quantity

$$q_J^{(2)} = \frac{1}{N^2} \sum_{i_1 i_2} \langle \sigma_{i_1} \sigma_{i_2} \rangle^2 \,, \tag{2.81}$$

and using the same technique we get

$$q_J^{(2)} = \frac{1}{N^2} \sum_{i_1 i_2} \sum_{\alpha\beta} w_\alpha w_\beta \, \langle \sigma_{i_1} \sigma_{i_2} \rangle_\alpha \, \langle \sigma_{i_1} \sigma_{i_2} \rangle_\beta \simeq \sum_{\alpha\beta} w_\alpha w_\beta \, q_{\alpha\beta}^2 = \int dq \, P_J(q) \, q^2 \tag{2.82}$$

where, in the second equality we have used the clustering property of pure states (1.76), that is valid in the large $N$ limit. In general we can perform a similar computation for $q_J^{(k)}$

$$q_J^{(k)} \equiv \frac{1}{N^k} \sum_{i_1 \dots i_k} \langle \sigma_{i_1} \dots \sigma_{i_k} \rangle^2 \simeq \int dq \, P_J(q) \, q^k \,, \tag{2.83}$$

which is related to the $k$-th moment of the overlap distribution. Performing the sample average we have

$$q^{(k)} \equiv \overline{q_J^{(k)}} = \int dq \, P(q) \, q^k \,. \tag{2.84}$$

The important point is that the above multi-point correlation function can be evaluated using also the replica approach. In the case $k = 1$ we have just performed the computation in subsection 2.1.1. Fixing two replicas $a \neq b$, we get

$$q^{(1)} = \lim_{n \to 0} \overline{\sum_{\{\sigma_i^c\}} \left( \frac{1}{N} \sum_i \sigma_i^a \sigma_i^b \right) e^{-\beta \sum_c H[\sigma^c]}} = \lim_{n \to 0} \overline{\langle \sigma_i^a \sigma_i^b \rangle} = \lim_{n \to 0} q_{ab} \,, \tag{2.85}$$

where $q_{ab}$ is the saddle-point overlap matrix. In a similar way we can derive an analogous result for $k$ generic

$$q^{(k)} = \lim_{n \to 0} \overline{\langle \sigma_{i_1}^a \dots \sigma_{i_k}^a \sigma_{i_1}^b \dots \sigma_{i_k}^b \rangle} = \lim_{n \to 0} q_{ab}^k \,. \tag{2.86}$$

This result is certainly true when we have replica symmetry. However, when we have replica symmetry breaking, we have many saddle-point solutions, connected by permutation transformations of rows and columns of $q_{ab}$. We have, therefore, to take into account all saddle-points solutions, i.e. to sum over all rows and columns of $q_{ab}$

$$q^{(k)} = \lim_{n\to 0} \frac{1}{n(n-1)} \sum_{a\neq b} q_{ab}^k . \tag{2.87}$$

Comparing equation (2.84) with (2.87) we obtain the replica representation of the overlap distribution

$$P(q) = \lim_{n\to 0} \frac{1}{n(n-1)} \sum_{a\neq b} \delta\left(q - q_{ab}\right) . \tag{2.88}$$

This equation is very important because it tells us that the fraction of elements in the saddle-point matrix $q_{ab}$ equal to $q$ is related to the probability that two pure states have overlap $q$. The EA order parameter can be written in terms of the replica overlap as

$$q_{\mathrm{EA}} = \max_{a,b} q_{ab} = \max_{x} q(x) . \tag{2.89}$$

We can therefore use the various ansatz used in the replica approach to acquire informations over the states. In the RS ansatz, for example, the distribution of overlaps is composed by only a delta function whereas in the 1RSB we will have two

$$P(q) = (1-m_1)\delta(q-q_1) + m_1\delta(q-q_0) . \tag{2.90}$$

In the fRSB we will instead have, using (2.58)

$$P(q) = -\lim_{k\to\infty} \sum_{j=0}^{k} (m_j - m_{j+1})\, \delta(q-q_j) = \int_0^1 dx\, \delta(q-q(x)) , \tag{2.91}$$

so that, if the function $q(x)$ is monotonous we can write the previous equation as

$$P(q) = \frac{dx(q)}{dq} . \tag{2.92}$$

$x(q)$, which is the cumulative of $P(q)$, gives the probability of finding two pure states with an overlap smaller or equal to $q$. In the fRSB ansatz $q(x)$ becomes a function, so that $P(q)$ develops a continuous part $p(q)$ between $q_{\mathrm{m}}$ and $q_{\mathrm{M}}$. In particular, when $\tau \ll 1$, we can use the explicit solution derived in equation (2.74) we have

$$P(q) = x_{\mathrm{m}}\delta(q-q_{\mathrm{m}}) + (1-x_{\mathrm{M}})\delta(q-q_{\mathrm{M}}) + p(q) , \tag{2.93}$$

where $p(q) = 2$. Plots of $P(q)$ near the spin glass critical temperature are depicted in the inset of Fig. 2.6a for $h = 0$ and of Fig. (2.6b) for $0 < h < h_{\mathrm{AT}}$.

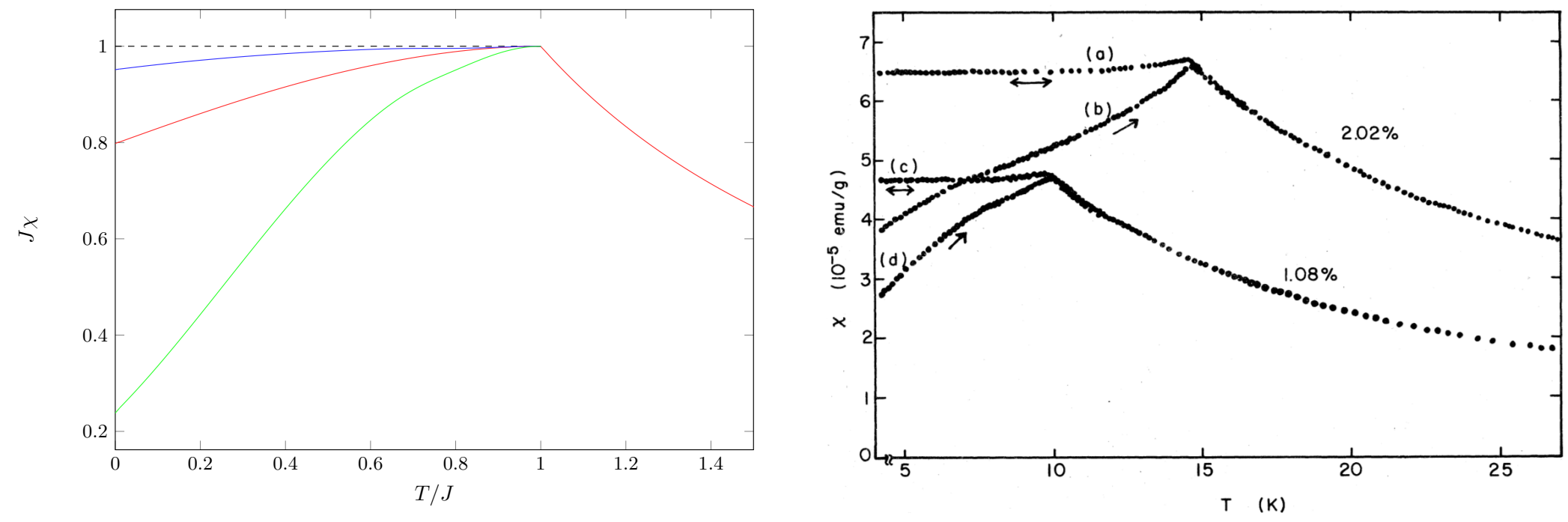

**(a)** The red and blue lines are respectively the RS and 1RSB results for the field-cooled susceptibility. The dashed black line is the fRSB limit. The green line is $\chi_{\text{ZFC}}$ in the 1RSB approximation.

**(b)** Lines (a) and (c) are the field-cooled susceptibilities, whereas (b) and (d) are the zero field-cooled susceptibilities. Reprinted from [NKH79].

**Figure 2.7.** Plot of the susceptibility as a function of temperature for the SK model (left panel) and for a sample of CuMn at different concentration of Mn impurities (right panel). The similarities between theoretical and experimental results are impressive.

### 2.4.2 Susceptibilities

What one measures experimentally are *susceptibilities* [BY86]. Defining the local magnetization $m_i = \langle \sigma_i \rangle$, the single site susceptibility is defined as

$$\chi_{ii} \equiv \frac{\partial m_i}{\partial h_i} = \beta \left(1 - m_i^2\right) , \tag{2.94}$$

where we have used the fluctuation-dissipation theorem in the second equality. Averaging over all sites and over the disorder we obtain the *equilibrium* susceptibility

$$\chi = \frac{1}{N} \sum_{i=1}^{N} \overline{\chi_{ii}} = \beta \left(1 - \overline{q_J}\right) = \beta \left(1 - \int dx\, P(q)\, q\right) , \tag{2.95}$$

where $q_J$ has been defined in (1.86). Actually this quantity is the same of (2.77). This susceptibility can be measured cooling a sample in a small magnetic field and it is also called for this reason *field cooled* susceptibility. Thanks to the external magnetic field, the system can jump over barriers and change state, allowing to better align with the field itself. Cooling the sample applying an even smaller field forces the system to remain in the same state. In this manner one measures experimentally the so called *zero field-cooled* susceptibility, which is related to the EA order parameter via

$$\chi_{\text{ZFC}} \equiv \beta \left(1 - q_{\text{EA}}\right) = \beta \left(1 - \max_x q(x)\right) . \tag{2.96}$$

The difference in these two susceptibilities under the critical temperature is responsible of the phenomenon of RSB. The comparison between analytical and experimental data for these susceptibilities is highlighted in Fig. 2.7. For other experimental results see [DJN99].

### 2.4.3 Ultrametricity

It is clear that the hierarchical way through which one construct the Parisi RSB scheme must reflect in some way in the structure of of the pure states. The distribution function $P(q)$ is insufficient to characterize and describe such non-trivial structure, because only one overlap between two replicas (or states) is involved. Therefore it is needed to inspect the properties of the joint probability distribution of more than one overlaps in order to study correlations between states [Méz+84]. The easiest case is the case of three overlaps $q_1$, $q_2$ and $q_3$

$$P(q_1, q_2, q_3) = \overline{\sum_{\alpha\beta\gamma} w_\alpha w_\beta w_\gamma \delta(q_1 - q_{\alpha\beta})\delta(q_2 - q_{\alpha\gamma})\delta(q_3 - q_{\beta\gamma})}\,, \tag{2.97}$$

which can be written in terms of replicas as

$$P(q_1, q_2, q_3) = \lim_{n\to 0} \frac{1}{n(n-1)(n-2)} \sum_{\substack{a\neq b\neq c \\ a\neq c}} \delta(q_1 - q_{ab})\delta(q_2 - q_{ac})\delta(q_3 - q_{bc})\,. \tag{2.98}$$

Using the fRSB ansatz, one can see that this probability is zero unless almost two overlaps are equal and in the case there is one that is different from the other two, this must be the smallest one. This means that the spin glass states are such that all the triangles are all equilateral or isosceles, with the different side which must be the smaller one. A space with this property is said to be *ultrametric* [RTV86, Méz+84, MV85]. In such a space, the usual triangular inequality $d_{ab} \leq d_{ac} + d_{bc}$ is replaced by

$$d_{ab} \leq \max(d_{ac}, d_{bc})\,, \tag{2.99}$$

which, for the overlaps reads

$$q_{ab} \geq \min(q_{ac}, q_{bc})\,, \tag{2.100}$$

# Part II

# Mean-Field

# Chapter 3

# Finite-size corrections in random matching problems

## 3.1 Introduction and main results

In this chapter we will focus on the study of the finite-size corrections of *random matching* problems in the mean field case, i.e. when the weights $w_e \geq 0$ defined on every edge $e$ of the graph $\mathcal{G}$ are chosen to be independent and identically distributed random variables. We remind that the we have denoted by $\mathcal{M}_1$ the set of all perfect matchings of the graph $\mathcal{G}$. Here we discuss in particular the case of the *random assignment problem* (RAP), in which the graph $\mathcal{G}$ is chosen to be the complete bipartite one $\mathcal{K}_{N,N}$. In this case a perfect matching $\pi \in \mathcal{M}_1$ is a permutation in the symmetric group $\mathcal{S}_N$ and can be represented by a square matrix with entries $\pi_{ij} \in \{0, 1\}$ for all $i \in [N]$ and $j \in [N]$ such that

$$\pi_{ij} = \begin{cases} 1 & \text{for } e = (i,j) \in \pi \\ 0 & \text{otherwise,} \end{cases} \tag{3.1}$$

with the constraints

$$\sum_{i=1}^{N} \pi_{ij} = \sum_{i=1}^{N} \pi_{ji} = 1 \quad \forall j \in [N]. \tag{3.2}$$

The matching cost associated with $\pi$ can be written as

$$E(\pi) = \sum_{i=1}^{N} \sum_{j=1}^{N} \pi_{ij} w_{ij}. \tag{3.3}$$

Only at the end of the chapter we present analogous computations for the complete graph case $\mathcal{G} = \mathcal{K}_{2N}$; this variation will be called simply *random matching problem* (RMP). From the point of view of computational complexity, matching problems are *simple* problems, being in the P complexity class, as Kuhn [Kuh55] proved with his celebrated *Hungarian algorithm* for the assignment problem. Very fast algorithms are nowadays available both to find perfect matchings and to solve the matching problem on a generic graph [Edm65, EK72, MV80, LP09].

Both the RMP and the RAP have been solved by Mézard and Parisi [MP85] by means of the replica trick. The random assignment problem and the random matching problem have also been generalized to the Euclidean case, in which the weights $w$ are functions of the distances between points associated with the vertices of the graph and the points are assumed to be randomly generated on a certain Euclidean domain [Sho85, MP88, Yuk98, LPS17]. Due to the underlying Euclidean structure, dimensionality plays an important role in the scaling of the optimal cost of random Euclidean matching problems [MP88, Car+14], and correlation functions can be introduced and calculated [BCS14, CS15b]. Euclidean matching problems proved to be deeply connected with Gaussian stochastic processes [BCS14, CS14] and with the theory of optimal transport [CS15a]. In the latter context, Ambrosio, Stra, and Trevisan [AST18] rigorously derived the asymptotic behavior of the average optimal cost for the two-dimensional random Euclidean assignment problem, previously obtained in Ref. [Car+14] using a proper scaling ansatz. For a recent review on random Euclidean matching problems, see Ref. [Sic17].

Remarkably enough, after the seminal works of Kirkpatrick, Gelatt, and Vecchi [KGV83], Orland [Orl85], and Mézard and Parisi, the application of statistical physics techniques to random optimization problems proved to be extremely successful in the study of the typical properties of the solutions, such as the large $N$ behavior of the average optimal cost

$$\overline{E} \equiv \overline{E(\pi^*)} = \overline{\min_{\pi \in \mathcal{M}_1} \sum_{i=1}^{N} \sum_{j=1}^{N} \pi_{ij} w_{ij}}, \tag{3.4}$$

but also in the development of algorithms to solve a given instance of the problem [Bap+13]. In formulating a combinatorial problem as a model in statistical mechanics, an artificial inverse temperature $\beta$ is introduced to define a Boltzmann weight $\exp(-\beta E)$ for each configuration. Of course, configurations of minimal energy are the only ones to contribute in the limit of infinite $\beta$. For example, in the assignment problem, the corresponding partition function for each instance is

$$Z[w] = \sum_{\pi} \left[ \prod_{j=1}^{N} \delta \left( 1 - \sum_{i=1}^{N} \pi_{ij} \right) \delta \left( 1 - \sum_{i=1}^{N} \pi_{ji} \right) \right] e^{-\beta E(\pi)}, \tag{3.5}$$

where the "energy" $E(\pi)$ is given by (3.3). Thermodynamic information is obtained from the average total free energy

$$\overline{F} \equiv -\frac{\overline{\ln Z}}{\beta}, \tag{3.6}$$

$$\overline{E} = \frac{\partial}{\partial \beta} \beta \overline{F}. \tag{3.7}$$

In the following we will assume the weights $w_{ij}$ to be independent and identically distributed random variables with probability distribution density $\rho_r(w)$ such that, in the neighborhood of the origin, $\rho_r$ can be written as

$$\rho_r(w) = w^r \sum_{k=0}^{\infty} \eta_k(r) w^k, \quad r > -1, \quad \eta_0(r) \neq 0. \tag{3.8}$$

In the previous expression, $\eta_k(r)$ are coefficients (possibly dependent on $r$) of the Maclaurin series expansion of the function $\rho_r(w)w^{-r}$, which is supposed to be analytic in the neighborhood of the origin. The constraint $r > -1$ is required to guarantee the integrability of the distribution near the origin. By the general analysis performed in Refs. [Orl85, MP85], which we will resume in Section 3.2, the average cost, in the asymptotic regime of an infinite number $N$ of couples of matched points, will depend on the power $r$ that appears in Eq. (3.8) only, aside from a trivial overall rescaling related to $\eta_0$. More precisely, if $\overline{E}_r$ is the average optimal cost obtained using the law $\rho_r$, then

$$\hat{E}_r = \lim_{N\to\infty} \frac{1}{N^{\frac{r}{r+1}}}\overline{E}_r = \frac{r+1}{[\eta_0\Gamma(r+1)]^{\frac{1}{r+1}}} J_r^{(r+1)} \tag{3.9a}$$

where

$$J_r^{(\alpha)} \equiv \int_{-\infty}^{+\infty} \hat{G}_r(-u)\,\mathcal{D}_u^{\alpha}\,\hat{G}_r(u)du \tag{3.9b}$$

(we will later specify the meaning of the fractional order derivative $\mathcal{D}_u^{\alpha}$). The function $\hat{G}_r(y)$ is the solution of the integral equation

$$\hat{G}_r(l) = \int_{-l}^{+\infty} \frac{(l+y)^r}{\Gamma(r+1)} e^{-\hat{G}_r(y)}\,dy \tag{3.9c}$$

and it is analytically known for $r = 0$ and, after a proper rescaling of its variable, in the $r \to \infty$ limit.

Our main results concern the finite-size corrections to the average optimal costs, and they will be presented in Section 3.3, extending the classical achievements in Refs. [MP87, PR02]. In particular, we obtain the expansion (the sum is absent for $r < 0$)

$$\hat{E}_r(N) = \hat{E}_r + \sum_{k=1}^{\lfloor r\rfloor+1} \Delta\hat{F}_r^{(k)} + \Delta\hat{F}_r^T + \Delta\hat{F}_r^F + o\left(\frac{1}{N}\right), \tag{3.10a}$$

where $\lfloor r\rfloor$ is the integer part of $r$, and the corrections have the structure

$$\Delta\hat{F}_r^{(k)} = \frac{\Delta\phi_r^{(k)}}{N^{\frac{k}{r+1}}}, \quad r \geq 0, \quad 1 \leq k \leq \lfloor r\rfloor + 1, \tag{3.10b}$$

$$\Delta\hat{F}_r^T = -\frac{1}{N}\frac{\Gamma(2r+2)J_r^{(0)}}{(r+1)\eta_0^{\frac{1}{r+1}}[\Gamma(r+1)]^{\frac{2r+3}{r+1}}} \tag{3.10c}$$

$$\Delta\hat{F}_r^F = -\frac{1}{N}\frac{1}{2\,[\eta_0\Gamma(r+1)]^{\frac{1}{r+1}}}\frac{1}{J_r^{(r+3)}}, \tag{3.10d}$$

$\Delta\phi_r^{(k)}$ being independent of $N$. In particular, for $r > 0$, we have that, provided $\eta_1 \neq 0$, the first finite-size correction is given by

$$\Delta\hat{F}_r^{(1)} = -\frac{\eta_1}{N^{\frac{1}{r+1}}}\frac{r+1}{\eta_0[\eta_0\Gamma(r+1)]^{\frac{2}{r+1}}} J_r^{(r)}. \tag{3.10e}$$

In our discussion, we will consider in particular two probability distribution densities, namely, the Gamma distribution

$$\rho_r^\Gamma(w) \equiv \frac{w^r e^{-w}\theta(w)}{\Gamma(r+1)}, \tag{3.11}$$

defined on $\mathbb{R}^+$, and the power-law distribution

$$\rho_r^\mathrm{P}(w) \equiv (r+1)w^r\theta(w)\theta(1-w), \tag{3.12}$$

defined on the compact interval $[0,1]$. In the previous expressions we have denoted by $\theta(w)$ the Heaviside theta function on the real line. Observe that, for the distribution $\rho_r^\Gamma$, we have

$$\eta_k^\Gamma(r) = \frac{1}{\Gamma(r+1)}\frac{(-1)^k}{k!}, \quad k \geq 0, \tag{3.13}$$

whereas in the case of $\rho_r^\mathrm{P}$,

$$\eta_k^\mathrm{P}(r) = (r+1)\,\delta_{k,0}, \quad k \geq 0. \tag{3.14}$$

The case $r = 0$ has already been considered by Mézard and Parisi [MP87] and subsequently revised and corrected by Parisi and Ratiéville [PR02]. In the case analyzed in their works, the contributions $\Delta\hat{F}_0^{(1)}$, $\Delta\hat{F}_0^T$, and $\Delta\hat{F}_0^F$ are of the same order. This is not true anymore for a generic distribution with $r \neq 0$. As anticipated, a relevant consequence of our evaluation is that, if $\eta_1 \neq 0$, for $r > 0$ the most important correction comes from $\Delta\hat{F}_r^{(1)}$ and scales as $N^{-\frac{1}{r+1}}$. It follows that, in order to extrapolate to the limit of an infinite number of points, the best choice for the law for random links (in the sense of the one that provides results closer to the asymptotic regime) is the pure power law $\rho_r^\mathrm{P}$, where only analytic corrections in inverse power of $N$ are present. Such a remark is even more pertinent in the limit when $r \to \infty$ at a fixed number of points, where the corrections $\Delta\hat{F}_r^{(k)}$ become of the same order of the leading term. Indeed, the two limits $r \to \infty$ and $N \to \infty$ commute only if the law $\rho_r^\mathrm{P}$ is considered.

The rest of the chapter is organized as follows. In Section 3.2 we review, in full generality, the calculation of the replicated partition function of the random assignment problem. In Section 3.3 we evaluate the finite-size corrections, discussing the different contributions and proving Eqs. (3.10). In Section 3.4 we evaluate the relevant $r \to \infty$ case, pointing out the non-commutativity of this limit with the thermodynamic limit. In Section 3.5 we provide the numerical values of the necessary integrals and we compare our prediction with numerical simulations for different values of $r$. In Section 3.6 we simply sketch the computation for the RMP case.

## 3.2 The replicated action

In the present section we perform a survey of the classical replica computation for the RAP, following the seminal works of Mézard and Parisi [MP85, MP87] (for a slightly different approach see also Ref. [Orl85]), but we do not adopt their choice to replace $\beta$ with $\beta/2$. As anticipated previously, the computation of the average of $\ln Z$ goes through the replica trick [EA75]

$$\overline{\ln Z} = \lim_{n\to 0} \frac{\overline{Z^n} - 1}{n}. \tag{3.15}$$

In other words, in order to compute $\overline{\ln Z}$ we introduce $n$ noninteracting replicas of the initial system, denoted by the index $a \in [n]$. For each $i \in [N]$, $2n$ replicated fields $\{\lambda_i^a, \mu_i^a\}_{a=1,\dots,n}$ appear to impose the constraints in Eq. (3.2), using the relation

$$\int_0^{2\pi} e^{ik\lambda} d\lambda = 2\pi\delta_{k\,0}. \tag{3.16}$$

We obtain

$$Z^n[w] = \left[\prod_{a=1}^{n}\prod_{i=1}^{N}\int_0^{2\pi}\frac{d\lambda_i^a}{2\pi}\int_0^{2\pi}\frac{d\mu_i^a}{2\pi}e^{i(\lambda_i^a+\mu_i^a)}\right]\prod_{i,j=1}^{N}\prod_{a=1}^{n}\left[1+e^{-i(\lambda_j^a+\mu_i^a)-\beta w_{ij}}\right]. \tag{3.17}$$

Let $\mathcal{P}([n])$ be the set of subsets of the set $[n]$ and for each subset $\alpha \in \mathcal{P}([n])$ let $|\alpha|$ be its cardinality. Then

$$\begin{aligned}\prod_{a=1}^{n}\left[1+e^{-i(\lambda_j^a+\mu_i^a)-\beta w_{ij}}\right] &= \sum_{\alpha\in\mathcal{P}([n])} e^{-\beta|\alpha|w_{ij}-i\sum_{a\in\alpha}(\lambda_j^a+\mu_i^a)}\\ &= 1+\sum_{p=1}^{n}e^{-\beta\, p\, w_{ij}}\sum_{\substack{\alpha\in\mathcal{P}([n])\\|\alpha|=p}} e^{-i\sum_{a\in\alpha}(\lambda_j^a+\mu_i^a)},\end{aligned} \tag{3.18}$$

where we have extracted the contribution from the empty set in the sum, which is 1, and we have partitioned the contribution from each subset of replicas in terms of their cardinality. This expression is suitable for the average on the costs. From the law $\rho_r$ we want to extract the leading term for large $\beta$ of the contribution of each subset $\alpha \in \mathcal{P}([n])$ with $|\alpha| = p$. In particular, we define

$$g_\alpha \equiv g_p \equiv \int_0^{+\infty} \rho_r(w)e^{-\beta p w}dw. \tag{3.19}$$

Due to the fact that short links only participate in the optimal configuration, approximating $\rho_r(w) \sim \eta_0 w^r$, we obtain that the minimal cost for each matched vertex is of the order $N^{-\frac{1}{r+1}}$ so the total energy $E$ and the free energy should scale as $N^{\frac{r}{r+1}}$, that is, the limits

$$\lim_{N\to\infty}\frac{1}{N^{\frac{r}{r+1}}}\overline{F} = \hat{F}, \tag{3.20}$$

$$\lim_{N\to\infty}\frac{1}{N^{\frac{r}{r+1}}}\overline{E} = \hat{E} \tag{3.21}$$

are finite. This regime can be obtained by considering in the thermodynamic limit

$$\beta = \hat{\beta}N^{\frac{1}{r+1}}, \tag{3.22}$$

where $\hat{\beta}$ is kept fixed. As a consequence we set

$$\hat{g}_p \equiv N g_p = N \int_0^{+\infty} \rho_r(w) e^{-p\hat{\beta} N^{\frac{1}{r+1}} w} dw = \sum_{k=0}^{+\infty} \frac{1}{N^{\frac{k}{r+1}}} \frac{\eta_k \Gamma(k+r+1)}{(\hat{\beta}p)^{k+r+1}}. \tag{3.23}$$

The replicated partition function can be written therefore as

$$\begin{aligned} \overline{Z^n} &= \left[ \prod_{a=1}^{n} \prod_{i=1}^{N} \int_0^{2\pi} \frac{d\lambda_i^a}{2\pi} \int_0^{2\pi} \frac{d\mu_i^a}{2\pi} e^{i(\lambda_i^a + \mu_i^a)} \right] \prod_{i,j=1}^{N} \left( 1 + \frac{T_{ij}}{N} \right) \\ &= \left[ \prod_{a=1}^{n} \prod_{i=1}^{N} \int_0^{2\pi} \frac{d\lambda_i^a}{2\pi} \int_0^{2\pi} \frac{d\mu_i^a}{2\pi} e^{i(\lambda_i^a + \mu_i^a)} \right] e^{\frac{1}{N} \sum_{i,j=1}^{N} \left( T_{ij} - \frac{T_{ij}^2}{2N} \right) + o\left(\frac{1}{N^2}\right)}, \end{aligned} \tag{3.24}$$

with

$$T_{ij} \equiv \sum_{\alpha \in \mathcal{P}([n])}' \hat{g}_\alpha \, e^{-i \sum_{a\in\alpha} (\lambda_j^a + \mu_i^a)} \tag{3.25}$$

where in the sum $\sum'$ on subsets the empty set is excluded. If we introduce, for each subset $\alpha \in \mathcal{P}([n])$, the quantities

$$\frac{x_\alpha + i y_\alpha}{\sqrt{2}} := \sum_{k=1}^{N} e^{-i \sum_{a\in\alpha} \lambda_k^a} \tag{3.26a}$$

$$\frac{x_\alpha - i y_\alpha}{\sqrt{2}} := \sum_{k=1}^{N} e^{-i \sum_{a\in\alpha} \mu_k^a} \tag{3.26b}$$

we can write

$$\sum_{i=1}^{N} \sum_{j=1}^{N} T_{ij} = \sum_{\alpha \in \mathcal{P}([n])}' \hat{g}_\alpha \frac{x_\alpha^2 + y_\alpha^2}{2}, \tag{3.27a}$$

$$\sum_{i=1}^{N} \sum_{j=1}^{N} T_{ij}^2 = \sum_{\alpha,\beta \in \mathcal{P}([n])}' \hat{g}_\alpha \hat{g}_\beta \frac{x_{\alpha\cup\beta}^2 + y_{\alpha\cup\beta}^2}{2}. \tag{3.27b}$$

As observed by Mézard and Parisi [MP87] and Parisi and Ratiéville [PR02], in Eq. (3.27b) we can constrain the sum on the right-hand side to the couples $\alpha, \beta \in \mathcal{P}([n])$ such that $\alpha \cap \beta = \emptyset$. Indeed, let us consider $\alpha, \beta \in \mathcal{P}([n])$ and $\alpha \cap \beta \neq \emptyset$. Then, defining $\alpha \triangle \beta \equiv (\alpha \cup \beta) \setminus (\alpha \cap \beta)$, we have that

$$\frac{x_{\alpha\cup\beta}^2 + y_{\alpha\cup\beta}^2}{2} = \sum_{l,k} \exp\left[ -2i \sum_{a\in\alpha\cap\beta} (\lambda_l^a + \mu_k^a) - i \sum_{b\in\alpha\triangle\beta} \left( \lambda_l^b + \mu_k^b \right) \right]. \tag{3.27c}$$

Due to Eq. (3.16) and to the presence of the coefficients $\exp\left(-2i\lambda_l^a - 2i\mu_k^a\right)$, the contribution of the term above will eventually be suppressed because of the integration over the Lagrange multipliers in the partition function. We can therefore simplify our calculation by substituting immediately

$$\sum_{i=1}^{N} \sum_{j=1}^{N} T_{ij}^2 = \sum_{\substack{\alpha,\beta \in \mathcal{P}([n]) \\ \alpha\cap\beta=\emptyset}}' \hat{g}_\alpha \hat{g}_\beta \frac{x_{\alpha\cup\beta}^2 + y_{\alpha\cup\beta}^2}{2}. \tag{3.27d}$$

We perform now a Hubbard–Stratonovich transformation, neglecting $o(N^{-2})$ terms in the exponent in Eq. (3.24), obtaining

$$e^{\frac{1}{N}\sum_{i,j=1}^{N}\left(T_{ij}-\frac{T_{ij}^2}{2N}\right)} = \left[\prod_{\alpha\in\mathcal{P}([n])}{}' \iint \frac{N dX_\alpha dY_\alpha}{2\pi\hat{g}_\alpha} e^{x_\alpha X_\alpha + y_\alpha Y_\alpha}\right] \times \\ \times \exp\left[-N\sum_{\alpha\in\mathcal{P}([n])}{}' \frac{X_\alpha^2+Y_\alpha^2}{2\hat{g}_\alpha} - \sum_{\substack{\alpha,\beta\in\mathcal{P}([n])\\ \alpha\cap\beta=\emptyset}}{}' \hat{g}_\alpha\hat{g}_\beta \frac{X_{\alpha\cup\beta}^2+Y_{\alpha\cup\beta}^2}{4\hat{g}_{\alpha\cup\beta}^2}\right], \tag{3.28}$$

up to higher order terms in the exponent. Now, let us observe that

$$x_\alpha X_\alpha + y_\alpha Y_\alpha = \left(\sum_{i=1}^{N} e^{-i\sum_{a\in\alpha}\lambda_i^a}\right)\frac{X_\alpha - iY_\alpha}{\sqrt{2}} + \left(\sum_{i=1}^{N} e^{-i\sum_{a\in\alpha}\mu_i^a}\right)\frac{X_\alpha + iY_\alpha}{\sqrt{2}}. \tag{3.29}$$

Introducing the function of $v_\alpha$,

$$z[v_\alpha] \equiv \left[\prod_{a=1}^{n}\int_0^{2\pi}\frac{d\lambda^a}{2\pi}e^{i\lambda^a}\right]\exp\left[v_\alpha e^{-i\sum_{b\in\alpha}\lambda^b}\right], \tag{3.30}$$

and the order parameters

$$Q_\alpha \equiv \frac{X_\alpha + iY_\alpha}{\sqrt{2}}, \tag{3.31}$$

we can write

$$\overline{Z^n} = \left[\prod_{\alpha\in\mathcal{P}([n])}{}' \frac{N}{2\pi\,\hat{g}_\alpha}\iint dQ_\alpha dQ_\alpha^*\right] e^{-NS[Q]-N\Delta S^T[Q]}, \tag{3.32a}$$

with

$$S[Q] = \sum_{\alpha\in\mathcal{P}([n])}{}' \left(\frac{|Q_\alpha|^2}{\hat{g}_\alpha} - \ln z\,[Q_\alpha] - \ln z\,[Q_\alpha^*]\right), \tag{3.32b}$$

$$\Delta S^T[Q] = \sum_{\substack{\alpha,\beta\in\mathcal{P}([n])\\ \alpha\cap\beta=\emptyset}}{}' \hat{g}_\alpha\hat{g}_\beta\frac{|Q_{\alpha\cup\beta}|^2}{2N\hat{g}_{\alpha\cup\beta}^2}, \tag{3.32c}$$

a form that is suitable to be evaluated, in the asymptotic limit for large $N$, by means of the saddle-point method. It is immediately clear that $\Delta S^T$ contains a contribution to the action that is $O(N^{-1})$ and therefore it can be neglected in the evaluation of the leading contribution. It follows that the stationarity equations are of the form

$$\frac{Q_\alpha^*}{\hat{g}_\alpha} = \frac{d\ln z[Q_\alpha]}{dQ_\alpha}, \tag{3.33a}$$

$$\frac{Q_\alpha}{\hat{g}_\alpha} = \frac{d\ln z[Q_\alpha^*]}{dQ_\alpha^*}. \tag{3.33b}$$

The application of the saddle-point method gives

$$\overline{Z^n} \simeq e^{-NS[Q^{\text{sp}}]-N\Delta S^T[Q^{\text{sp}}]-\frac{1}{2}\ln\det\mathbf{\Omega}[Q^{\text{sp}}]}, \tag{3.34}$$

where $\mathbf{\Omega}$ is the Hessian matrix of $S[Q]$ and $Q^{\text{sp}}$ is the saddle-point solution. As we will show below, the contribution $\ln\det\mathbf{\Omega}[Q^{\text{sp}}]$ provides finite-size corrections to the leading contribution of the same order of the corrections in $N\Delta S^T[Q^{\text{sp}}]$. Notice also how in describing the RAP a whole set of multi-overlaps $Q_\alpha$ and $Q^*_\alpha$ is needed, differently to what happens in fully connected systems as in the SK model analyzed in Chap. 2, where only the first two moments $m_a$ and $q_{ab}$ are needed. The RAP is a first example of *diluite* or *sparse* system: even if the problem is defined on a fully connected graph, the interactions contributing to the solution are only a tiny fraction of the total number of edges. Therefore, the effective graph obtained by cutting all the "most weighted" edges, is locally tree-like and this is the reason why belief propagation, which is exact on tree graphs, gives correct result also for the RAP, and in general, for the random-link models we analyze in this and subsequent chapters of Part II.

### 3.2.1 Replica symmetric ansatz and limit of vanishing number of replicas

To proceed with our calculation, we adopt, as usual in the literature, a RS ansatz for the solution of the saddle-point equations. A RS solution is of the form

$$Q_\alpha = Q^*_\alpha = q_{|\alpha|}. \tag{3.35}$$

In particular this implies that $Y_\alpha = 0$. In order to analytically continue to $n \to 0$ the value at the saddle-point of $S$ in Eq. (3.32b), let us first remark that under the assumption in Eq. (3.35)

$$\sum_{\alpha\in\mathcal{P}([n])}{}' \frac{|Q_\alpha|^2}{\hat{g}_\alpha} = \sum_{k=1}^{n}\binom{n}{k}\frac{q_k^2}{\hat{g}_k} = n\sum_{k=1}^{\infty}\frac{(-1)^{k-1}}{k}\frac{q_k^2}{\hat{g}_k} + o(n). \tag{3.36}$$

Moreover, as shown in Appendix A.1,

$$\sum_{\alpha\in\mathcal{P}([n])}{}' \ln z[Q_\alpha] = n\int_{-\infty}^{+\infty} dl\left[e^{-e^l} - e^{-G(l)}\right], \tag{3.37}$$

where

$$G(l) \equiv \sum_{k=1}^{\infty}(-1)^{k-1} q_k \frac{e^{lk}}{k!}. \tag{3.38}$$

In conclusion, under the RS ansatz in Eq. (3.35), the functional to be minimized is

$$\hat{\beta}\hat{F} = \sum_{k=1}^{\infty}\frac{(-1)^{k-1}}{k}\frac{q_k^2}{\hat{g}_k} - 2\int_{-\infty}^{+\infty} dl\left[e^{-e^l} - e^{-G(l)}\right]. \tag{3.39}$$

A variation with respect to $q_k$ gives the saddle-point equation

$$\frac{1}{k}\frac{q_k}{\hat{g}_k} = \int_{-\infty}^{+\infty} dy\, e^{-G(y)}\frac{e^{yk}}{k!} \tag{3.40}$$

which is to say

$$G(l) = \sum_{k=1}^{\infty} (-1)^{k-1} q_k \frac{e^{lk}}{k!} = \int_{-\infty}^{+\infty} dy\, e^{-G(y)} \sum_{k=1}^{\infty} (-1)^{k-1} k \hat{g}_k \frac{e^{(y+l)\,k}}{(k!)^2}\,. \tag{3.41}$$

This implies that

$$\sum_{k=1}^{\infty} \frac{(-1)^{k-1}}{k} \frac{q_k^2}{\hat{g}_k} = \sum_{k=1}^{\infty} (-1)^{k-1} q_k \int_{-\infty}^{+\infty} dy\, e^{-G(y)} \frac{e^{yk}}{k!} = \int_{-\infty}^{+\infty} dy\, G(y) e^{-G(y)}\,. \tag{3.42}$$

These formulas are for a general law $\rho_r$. Observe also that the expression of $\hat{g}_p$ is not specified. For finite $r$ and $N \to \infty$, Eq. (3.23) simplifies as

$$\lim_{N\to\infty} \hat{g}_p = \frac{\eta_0 \Gamma(r+1)}{(\hat{\beta} p)^{r+1}}. \tag{3.43}$$

We will restrict the analysis to the case in which Eq. (3.43) holds. Then Eq. (3.41) becomes

$$G_r(l) = \frac{\eta_0 \Gamma(r+1)}{\hat{\beta}^{r+1}} \int_{-\infty}^{+\infty} dy\, B_r(l+y) e^{-G_r(y)}, \tag{3.44}$$

with

$$B_r(x) \equiv \sum_{k=1}^{\infty} (-1)^{k-1} \frac{e^{xk}}{k^r (k!)^2}. \tag{3.45}$$

In Eq. (3.44), and in the following, we introduce the subindex $r$ to stress the dependence of $G$ and of the thermodynamical functionals on $r$. The average cost is therefore

$$\hat{E}_r = \frac{\partial}{\partial \hat{\beta}} \hat{\beta} \hat{F}_r = \frac{r+1}{\hat{\beta}} \int_{-\infty}^{+\infty} dy\, G_r(y) e^{-G_r(y)}\,. \tag{3.46}$$

Using the fact that (see Appendix A.2)

$$\lim_{\delta\to\infty} \frac{1}{\delta^r} B_r(\delta x) = \frac{x^r \theta(x)}{\Gamma(r+1)}, \tag{3.47}$$

if we introduce

$$\hat{G}_r(l) \equiv G_r\left( \frac{\hat{\beta}}{[\eta_0 \Gamma(r+1)]^{\frac{1}{r+1}}}\, l \right) \tag{3.48}$$

in the limit $\hat{\beta} \to +\infty$, the function $\hat{G}_r$ satisfies Eq. (3.9c) and the value of $\hat{E}_r$ is the one reported in Eq. (3.9a). In particular, at fixed $r$, if we consider the two laws $\rho_r^{\mathrm{P}}$ and $\rho_r^{\Gamma}$, the ratio between the corresponding average optimal costs is given by

$$\lambda_r \equiv \frac{\hat{E}_r^{\mathrm{P}}}{\hat{E}_r^{\Gamma}} = \left( \frac{\eta_0^{\Gamma}}{\eta_0^{\mathrm{P}}} \right)^{\frac{1}{r+1}} = [\Gamma(r+2)]^{-\frac{1}{r+1}}. \tag{3.49}$$

In the case $r = 0$, we have the classical result by Mézard and Parisi [MP85]

$$\hat{G}_0(l) = \ln(1+e^l), \tag{3.50a}$$

$$\hat{E}_0 = \frac{1}{\eta_0(0)} \int_{-\infty}^{+\infty} \frac{\ln(1+e^y)}{1+e^y} dy = \frac{1}{\eta_0(0)} \frac{\pi^2}{6}, \tag{3.50b}$$

a result that was later obtained with a cavity approach by Aldous [Ald01]. For the evaluation of the integral, see Appendix A.4.

## 3.3 Finite size corrections

The evaluation of the first-order corrections for a finite number of points has been considered in Refs. [MP87, PR02] in the $r = 0$ case. For this particular choice and assuming a distribution law $\rho_0^\Gamma$, a much stronger conjecture was proposed by Parisi [Par98] and later proved by Linusson and Wästlund [LW04] and Nair, Prabhakar, and Sharma [NPS05], that is, for every $N$,

$$\hat{E}_0^\Gamma(N) = H_{N,2} \equiv \sum_{k=1}^{N} \frac{1}{k^2}. \tag{3.51}$$

For large $N$, Parisi's formula implies

$$\hat{E}_0^\Gamma(N) = \frac{\pi^2}{6} - \frac{1}{N} + o\left(\frac{1}{N}\right). \tag{3.52}$$

Using instead the law $\rho_0^{\mathrm{P}}$ (uniform distribution on the interval) we have [MP87, PR02]

$$\hat{E}_0^{\mathrm{P}}(N) = \frac{\pi^2}{6} - \frac{1+2\zeta(3)}{N} + o\left(\frac{1}{N}\right) \tag{3.53}$$

from which we see that corrections for both laws are analytic, with the same inverse power of $N$, but different coefficients.

In their study of the finite-size corrections, the authors of Ref. [PR02] show that, in their particular case, there are two kind of finite-size corrections. The first one comes from the application of the saddle-point method, giving a series of corrections in the inverse powers of $N$. This contribution is the sum of two terms. The first term in this expansion corresponds to the contribution of the $\Delta S^T$ term given in Eq. (3.32c) appearing in the exponent in Eq. (3.34). The second term is related to the fluctuations, also appearing in Eq. (3.34), involving the Hessian of $S$. The second kind of corrections, instead, is due to the particular form of the law $\rho_r(w)$ for the random links and in particular to the series expansion in Eq. (3.23). This contribution can be seen at the level of the action $S$ in Eq. (3.32b), being

$$\frac{|Q_\alpha|^2}{\hat{g}_\alpha} \approx |Q_\alpha|^2 \frac{(\hat{\beta}|\alpha|)^{r+1}}{\eta_0 \Gamma(r+1)} - |Q_\alpha|^2 \frac{(r+1)}{N^{\frac{1}{r+1}}} \frac{\eta_1}{\eta_0^2} \frac{(\hat{\beta}|\alpha|)^r}{\Gamma(r+1)} + O\left(N^{-\frac{2}{r+1}}\right). \tag{3.54}$$

In full generality, the expansion of $1/\hat{g}_\alpha$ generates a sum over terms each one of order $N^{-\frac{k}{r+1}}$ with $k \geq 1$. All these corrections are $o(N^{-1})$ for $r \in (-1, 0)$, whereas the corrections obtained from the contributions with $1 \leq k \leq r+1$ are of the same order as the analytic term, or greater, for $r \geq 0$. In particular, if $\eta_1 \neq 0$, for $r > 0$ the $k = 1$ term provides the leading correction, scaling as $N^{-\frac{1}{r+1}}$. It is also evident that all these corrections are absent if $\eta_k = 0$ for $k \geq 1$, as it happens in the case of the $\rho_r^{\mathrm{P}}$ law.

### 3.3.1 Correction due to $\eta_1$

Let us consider the $r \geq 0$ case and let us restrict ourselves to the $k = 1$ term, of order $N^{-\frac{1}{r+1}}$ in Eq. (3.10). Its contribution to the total free energy is given by

$$\hat{\beta}\Delta\hat{F}_r^{(1)} = -\frac{r+1}{N^{\frac{1}{r+1}}}\frac{\eta_1}{\eta_0^2}\sum_{p=1}^{\infty}\frac{(-1)^{p-1}}{p}\frac{(\hat{\beta}p)^r}{\Gamma(r+1)}q_p^2, \tag{3.55}$$

where we already made a RS assumption and considered the $n \to 0$ limit. Imposing the saddle-point relation in Eq. (3.40) and using the limit in Eq. (3.43), we obtain

$$\begin{aligned} N^{\frac{1}{r+1}}\Delta\hat{F}_r^{(1)} &= -\frac{\eta_1(r+1)}{\hat{\beta}^2\eta_0}\int_{-\infty}^{+\infty} dy\, e^{-G_r(y)}\sum_{p=1}^{\infty}\frac{(-1)^{p-1}q_p e^{py}}{p\, p!} \\ &= -\frac{\eta_1(r+1)}{\hat{\beta}^2\eta_0}\int_{-\infty}^{+\infty} dy\, e^{-G_r(y)}\int_{-\infty}^{y} du\, G_r(u) \\ &= -\frac{\eta_1(r+1)}{\eta_0\left[\eta_0\Gamma(r+1)\right]^{\frac{2}{r+1}}}\int_{-\infty}^{+\infty} dy\, e^{-\hat{G}_r(y)}\int_{-\infty}^{y} du\, \hat{G}_r(u). \end{aligned} \tag{3.56}$$

To put the expression above in the form presented in Eq. (3.10e), observe that

$$\begin{aligned} \int_{-\infty}^{+\infty} dy\, e^{-\hat{G}_r(y)}\int_{-\infty}^{y} du\, \hat{G}_r(u) &= \int_{-\infty}^{+\infty} du\, \hat{G}_r(-u)\int_{-\infty}^{+\infty} dy\, e^{-\hat{G}_r(y)}\,\theta(y+u) \\ &= \int_{-\infty}^{+\infty}\hat{G}_r(-u)\,\mathcal{D}_u^r\,\hat{G}_r(u)du \equiv J_r^{(r)}, \end{aligned} \tag{3.57}$$

a structure that can be more useful for numerical evaluation, at least for $r$ integer. In this equation we have used Eq. (A.12) and we have introduced the Riemann–Liouville fractional derivative

$$\begin{aligned} \mathcal{D}_t^{\alpha} f(t) &\equiv \frac{d^{\lfloor\alpha\rfloor+1}}{dt^{\lfloor\alpha\rfloor+1}}\int_{-\infty}^{t} d\tau\,\frac{(t-\tau)^{\lfloor\alpha\rfloor-\alpha}}{\Gamma(\lfloor\alpha\rfloor-\alpha+1)}f(\tau), \\ \alpha \geq 0, &\quad f \in L_p(\Omega)\ \forall p \in \left[1, \frac{1}{\lfloor\alpha\rfloor-\alpha+1}\right), \end{aligned} \tag{3.58}$$

where $\Omega \equiv (-\infty, t)$ is the domain of integration (see Refs. [SKM93, Pod99] for further details).

### 3.3.2 Correction due to the saddle-point approximation

Let us now consider the corrections due to the saddle-point approximation. The first contribution is expressed by $\Delta S^T$, given in Eq. (3.32c). In the RS hypothesis, we have that

$$\sum_{\substack{\alpha,\beta\in\mathcal{P}([n])\\ \alpha\cap\beta=\emptyset}}{}' \hat{g}_\alpha\hat{g}_\beta\frac{|Q_{\alpha\cup\beta}|^2}{2N\hat{g}_{\alpha\cup\beta}^2} = \frac{1}{2N}\sum_{s=1}^{\infty}\sum_{t=1}^{\infty}\binom{n}{s,t,n-s-t}\frac{\hat{g}_s\hat{g}_t}{\hat{g}_{s+t}^2}q_{s+t}^2. \tag{3.59}$$

We can write the corresponding correction to the free energy as

$$\Delta \hat{F}_r^T = \frac{1}{2\hat{\beta}N} \sum_{s=1}^{\infty}\sum_{t=1}^{\infty}(-1)^{s+t-1}\frac{(s+t-1)!}{s!\,t!}\frac{\hat{g}_s\hat{g}_t}{\hat{g}_{s+t}^2}q_{s+t}^2. \tag{3.60}$$

In Appendix A.5 we show that the previous quantity can be written as

$$\begin{aligned}\Delta \hat{F}_r^T &= -\frac{\Gamma(2r+2)}{N\eta_0^{\frac{1}{r+1}}\Gamma^{2+\frac{1}{r+1}}(r+1)}\frac{1}{r+1}\int_{-\infty}^{+\infty} du\, \hat{G}_r(-u)\hat{G}_r(u) \\ &= -\frac{\Gamma(2r+2)}{N\eta_0^{\frac{1}{r+1}}\Gamma^{2+\frac{1}{r+1}}(r+1)}\frac{J_r^{(0)}}{r+1}.\end{aligned} \tag{3.61}$$

Another type of finite-size correction comes from the fluctuations around the saddle-point [PR02, Section B.3], related to the Hessian matrix $\mathbf{\Omega}$ appearing in Eq. (3.34). The evaluation of the contribution of the Hessian matrix is not trivial and it has been discussed by Mézard and Parisi [MP87] and later by Parisi and Ratiéville [PR02]. They proved that the whole contribution comes from a volume factor due to a non trivial metric $\hat{\mathbf{\Omega}}$ obtained from $\mathbf{\Omega}$ imposing the RS assumption. The volume factor arises because of the invariance of the action

$$Q_\alpha \to Q_\alpha\, e^{i\sum_{a\in\alpha}\theta_\alpha}, \tag{3.62}$$

which makes the RS saddle point(3.35) $n$ times degenerate. The metric $\hat{\mathbf{\Omega}}$ is such that

$$\ln\sqrt{\det\mathbf{\Omega}} = \ln\sqrt{\det\hat{\mathbf{\Omega}}}. \tag{3.63}$$

The $n\times n$ matrix $\hat{\mathbf{\Omega}}$ can be written as

$$\hat{\mathbf{\Omega}} = na_1\mathbf{\Pi} + (a_0 - a_1)\mathbb{I}_n, \tag{3.64}$$

where $\mathbb{I}_n$ is the $n\times n$ identity matrix and we have introduced the quantities

$$a_0 \equiv \sum_{p=1}^{\infty}\binom{n-1}{p-1}\frac{q_p^2}{\hat{g}_p}, \tag{3.65a}$$

$$a_1 \equiv \sum_{p=2}^{\infty}\binom{n-2}{p-2}\frac{q_p^2}{\hat{g}_p}, \tag{3.65b}$$

and $\mathbf{\Pi}$ is a projection matrix on the constant vector defined as

$$\mathbf{\Pi} \equiv \frac{\mathbb{J}_n}{n}, \tag{3.66}$$

where $\mathbb{J}_n$ is the $n\times n$ matrix with all entries equal to 1. The matrix $\mathbf{\Pi}$ has one eigenvalue equal to 1 and $n-1$ eigenvalues equal to 0. It follows that, because the two matrices $\mathbf{\Pi}$ and $\mathbb{I}_n$ obviously commute, $\hat{\mathbf{\Omega}}$ has one eigenvalue equal to $a_0+(n-1)a_1$ and $n-1$ eigenvalues equal to $a_0-a_1$. Its determinant is therefore simply given by

$$\det\hat{\mathbf{\Omega}} = (a_0-a_1)^{n-1}[a_0+(n-1)a_1]. \tag{3.67}$$

In the limit of $n \to 0$ we easily get

$$\begin{aligned} a_0 &= \sum_{p=1}^{\infty} (-1)^{p-1} \frac{q_p^2}{\hat{g}_p} = \sum_{p=1}^{\infty} (-1)^{p-1} p q_p \int_{-\infty}^{+\infty} dy\, e^{-G_r(y)} \frac{e^{py}}{p!} \\ &= \int_{-\infty}^{+\infty} dy\, e^{-G_r(y)} \frac{dG_r(y)}{dy} = \int_{-\infty}^{+\infty} e^{-\hat{G}_r(y)} \frac{d\hat{G}_r(y)}{dy} dy \\ &= e^{-\hat{G}_r(-\infty)} - e^{-\hat{G}_r(+\infty)} = 1 \end{aligned} \tag{3.68}$$

for all values of $r$. Similarly,

$$a_1 = -\sum_{p=2}^{\infty} (-1)^{p-1} (p-1) \frac{q_p^2}{\hat{g}_p} = -\sum_{p=1}^{\infty} (-1)^{p-1} (p-1) \frac{q_p^2}{\hat{g}_p} \tag{3.69}$$

so

$$\begin{aligned} a_0 - a_1 &= \sum_{p=1}^{\infty} (-1)^{p-1} p \frac{q_p^2}{\hat{g}_p} = \sum_{p=1}^{\infty} (-1)^{p-1} p^2 q_p \int_{-\infty}^{+\infty} dy\, e^{-G_r(y)} \frac{e^{py}}{p!} \\ &= \int_{-\infty}^{+\infty} dy\, e^{-G_r(y)} \frac{d^2}{dy^2} G_r(y) = \frac{[\eta_0 \Gamma(r+1)]^{\frac{1}{r+1}}}{\hat{\beta}} \int_{-\infty}^{+\infty} dy\, e^{-\hat{G}_r(y)} \frac{d^2}{dy^2} \hat{G}_r(y). \end{aligned} \tag{3.70}$$

Therefore,

$$\sqrt{\det \hat{\Omega}} = 1 + \frac{n}{2} \left[ \frac{a_1}{a_0 - a_1} + \ln(a_0 - a_1) \right] + o(n). \tag{3.71}$$

In conclusion, integrating by parts and using Eq. (A.14), we obtain

$$\Delta \hat{F}_r^F = -\lim_{\hat{\beta} \to \infty} \frac{1}{nN\hat{\beta}} \ln \sqrt{\det \hat{\Omega}} = -\frac{1}{2N [\eta_0 \Gamma(r+1)]^{\frac{1}{r+1}}} \frac{1}{J_r^{(r+3)}}. \tag{3.72}$$

### 3.3.3 Application: the $r = 0$ case

The results obtained in the $r = 0$ case, analyzed by Parisi and Ratiéville [PR02], can be easily recovered. From the general expression in Eq. (3.10), by setting $r = 0$, we get

$$\begin{aligned} \Delta \hat{F}_0 \equiv \Delta \hat{F}_0^{(1)} + \Delta \hat{F}_0^T + \Delta \hat{F}_0^F &= -\frac{1}{\eta_0(0) N} \left[ \left( 1 + \frac{\eta_1(0)}{\eta_0(0)} \right) \frac{J_0^{(0)}}{\eta_0(0)} + \frac{1}{2 J_0^{(3)}} \right] \\ &= -\frac{1}{\eta_0(0) N} \left[ \left( 1 + \frac{\eta_1(0)}{\eta_0(0)} \right) \frac{2\zeta(3)}{\eta_0(0)} + 1 \right], \end{aligned} \tag{3.73}$$

where we have used the results discussed in the Appendix A.4 for the two integrals involved in the expression above. Eqs. (3.52) and (3.53) are obtained using Eqs. (3.13) and (3.14), respectively.

## 3.4 The limiting case $r \to +\infty$

In this section we concentrate on the limiting case in which $r \to +\infty$. We can easily verify that, in the weak sense,

$$\lim_{r\to+\infty} \rho_r^{\mathrm{P}}(w) = \delta(w-1) \tag{3.74}$$

so all the weights become equal to unity. We expect therefore that

$$\lim_{r\to+\infty} \hat{E}_r^{\mathrm{P}}(N) = 1, \tag{3.75}$$

independently of $N$. The average cost obtained using $\rho_r^\Gamma$ instead diverges and it is therefore more interesting to consider the modified law

$$\rho_r^\gamma(w) \equiv \frac{(r+1)^{r+1}}{\Gamma(r+1)} w^r e^{-(r+1)w}\theta(w) \xrightarrow{r\to\infty} \delta(w-1). \tag{3.76}$$

According to our general discussion, we have that

$$\eta_k^\gamma(r) = \frac{(r+1)^{k+r+1}}{\Gamma(r+1)} \frac{(-1)^k}{k!}, \quad k \geq 0, \tag{3.77}$$

implying that, independently of $N$,

$$\hat{E}_r^\gamma(N) = \frac{1}{r+1} \hat{E}_r^\Gamma(N) \tag{3.78}$$

and therefore

$$\hat{E}_r^\gamma = \frac{\Gamma(r+2)^{\frac{1}{r+1}}}{r+1} \hat{E}_r^{\mathrm{P}} \tag{3.79}$$

in the limit of infinite $N$. In particular,

$$\lim_{r\to+\infty} \hat{E}_r^\gamma = \frac{1}{e}. \tag{3.80}$$

It follows that, even though the two laws $\rho^{\mathrm{P}}$ and $\rho^\gamma$ both converge to the same limiting distribution, according to our formulas, the corresponding average costs are *not* the same. This is due to the fact that the two limits $N \to +\infty$ and $r \to +\infty$ do not commute for the law $\rho^\gamma$, because of the presence of $O(N^{-\frac{k}{r+1}})$ corrections that give a leading contribution if the $r \to \infty$ limit is taken first.

To look into more details in the $r \to +\infty$ limit, we find it convenient, when looking at the saddle-point solution, to perform a change of variables, following the approach in Refs. [HBM98, PR01], that is, writing

$$\mathcal{G}_r(x) \equiv \hat{G}_r \left[ \Gamma^{\frac{1}{r+1}}(r+2) \left( \frac{1}{2} + \frac{x}{r+1} \right) \right] \tag{3.81}$$

then Eq. (3.9c) becomes

$$\mathcal{G}_r(x) = \int_{-x-r-1}^{+\infty} dt \left( 1 + \frac{x+t}{r+1} \right)^r e^{-\mathcal{G}_r(t)}, \tag{3.82}$$

so, in the $r \to +\infty$ limit

$$\mathcal{G}_\infty(x) = e^x \int_{-\infty}^{+\infty} dt\, e^{t-\mathcal{G}_\infty(t)}\,. \tag{3.83}$$

If we set $\mathcal{G}_\infty(x) = ae^x$, with

$$a = \int_{-\infty}^{+\infty} dt\, e^{t-\mathcal{G}_\infty(t)} \tag{3.84}$$

we recover

$$a = \int_{-\infty}^{+\infty} dt\, e^{t-ae^t} = \int_0^{+\infty} dz\, e^{-az} = \frac{1}{a} \Rightarrow \mathcal{G}_\infty(x) = e^x. \tag{3.85}$$

From Eq. (3.9a) with the change of variable in Eq. (3.81), we get

$$\hat{E}_r = \left(\frac{r+1}{\eta_0(r)}\right)^{\frac{1}{r+1}} \int_{-\infty}^{+\infty} dx\, \mathcal{G}_r(x) e^{-\mathcal{G}_r(x)}, \tag{3.86}$$

so

$$\hat{E}_\infty = \lim_{r\to+\infty} \left(\frac{r+1}{\eta_0(r)}\right)^{\frac{1}{r+1}} = \lim_{r\to+\infty} \eta_0(r)^{-\frac{1}{r+1}} = \begin{cases} 1 & \text{for } \rho_r^{\mathrm{P}}, \\ \frac{1}{e} & \text{for } \rho_r^{\gamma}, \end{cases} \tag{3.87}$$

in agreement with the previous results. Let us now evaluate the integrals appearing in the finite-size corrections in Eq. (3.10). Let us first start with the $\Delta F_r^T$ and the $\Delta F_r^F$ corrections. From the definition, for large $r$,

$$\begin{aligned} J_r^{(0)} &= \int_{-\infty}^{+\infty} \hat{G}_r(y)\hat{G}_r(-y) dy \\ &= \int_{-\infty}^{+\infty} dy \int_{-y}^{\infty} dt_1 \frac{(t_1+y)^r}{\Gamma(r+1)} e^{-\hat{G}_r(t_1)} \int_y^{\infty} dt_2 \frac{(t_2+y)^r}{\Gamma(r+1)} e^{-\hat{G}_r(t_2)} \\ &= \Gamma^{\frac{1}{r+1}}(r+2) \int_{-\infty}^{+\infty} dt_1 \int_{-\infty}^{+\infty} dt_2\, K_r(t_1,t_2)\, e^{-\mathcal{G}_r(t_1)-\mathcal{G}_r(t_2)} \end{aligned} \tag{3.88}$$

where

$$\begin{aligned} K_r(t_1,t_2) &\equiv \int_{-x_1-r-1}^{x_2} \frac{(t+t_1+r+1)^r (t_2-t)^r}{(r+1)^{2r+1}} dt \\ &= \left(1+\frac{x_1+x_2}{r+1}\right)^{2r+1} \frac{\Gamma^2(r+1)}{\Gamma(2r+2)} \end{aligned} \tag{3.89}$$

and therefore for large $r$,

$$\begin{aligned} J_r^{(0)} &\simeq \Gamma^{\frac{1}{r+1}}(r+2) \frac{\Gamma^2(r+1)}{\Gamma(2r+2)} \left(\int_{-\infty}^{+\infty} e^{x-e^x} dx\right)^2 \\ &= \Gamma^{\frac{1}{r+1}}(r+2) \frac{\Gamma^2(r+1)}{\Gamma(2r+2)} \simeq \frac{\Gamma^{2+\frac{1}{r+1}}(r+1)}{\Gamma(2r+2)}, \end{aligned} \tag{3.90}$$

so we get

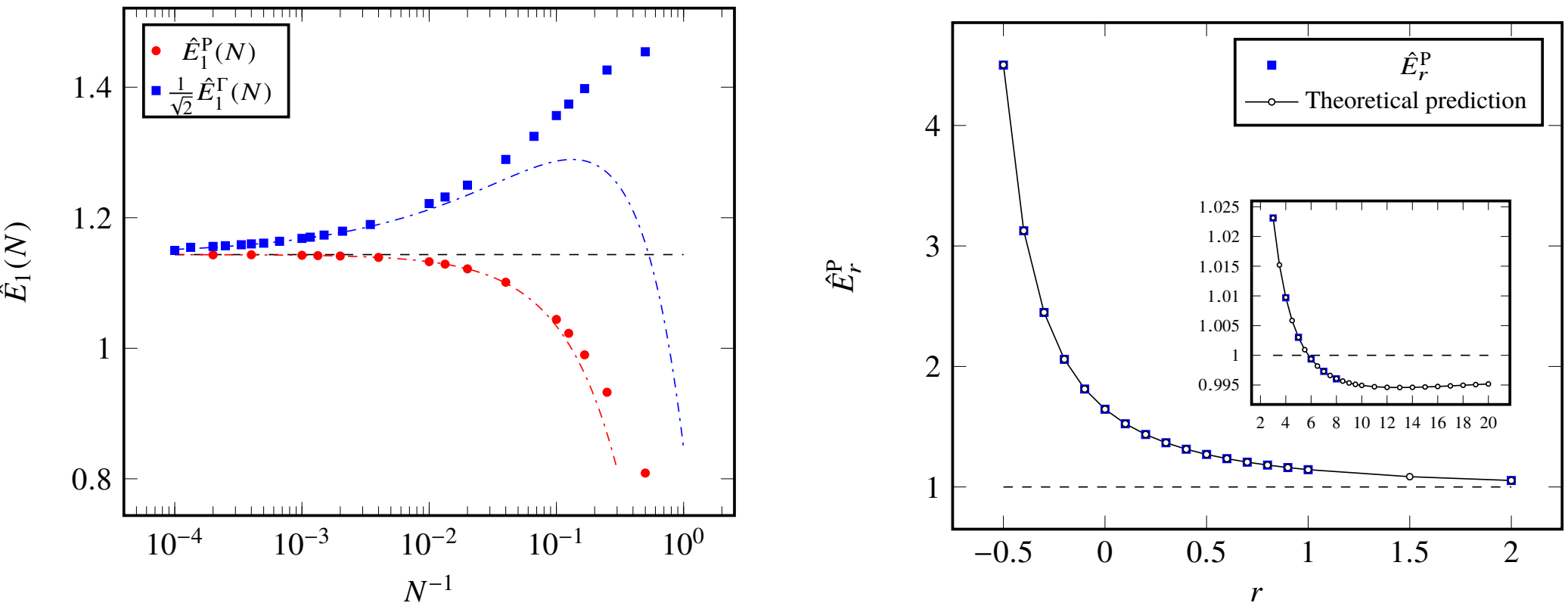

**(a)** Numerical results for $\hat{E}_1^{\mathrm{P}}(N)$ and $\hat{E}_1^{\Gamma}(N)$ for several values of $N$. Note that finite-size corrections have a different sign for $N \to +\infty$. We have represented also the theoretical predictions for both cases obtained including the finite-size corrections up to $O(N^{-1})$.

**(b)** Theoretical prediction of $\hat{E}_r^{\mathrm{P}}$ for several values of $r$ (solid line), compared with our numerical results. The dashed line is the large-$r$ asymptotic estimate, equal to 1. Error bars do not appear because they are smaller than the marks in the plot. The values for $\lambda_r \hat{E}_r^{\Gamma}$ almost coincide with the values of $\hat{E}_r^{\mathrm{P}}$ (see Table 3.2) and are not represented.

**Figure 3.1**

$$\Delta \hat{F}_r^T \simeq -\frac{1}{N\eta_0^{\frac{1}{r+1}}(r)r} \tag{3.91}$$

a contribution that vanishes as $r^{-1}$ both for the law $\rho_r^{\mathrm{P}}$ and for the law $\rho_r^{\gamma}$ (indeed we know that, in this case, all corrections must vanish when $r \to \infty$ at fixed $N$). In particular, if we consider the law $\rho_r^{\Gamma}$, we have, in the $r \to \infty$ limit

$$\Delta \hat{F}_\infty^T = -\frac{1}{eN}. \tag{3.92}$$

Similarly, for large $r$, we have that

$$\frac{1}{J_r^{(r+3)}} = \frac{\Gamma(r+2)^{\frac{1}{r+1}}}{r+1}\left[\int_{-\infty}^{+\infty} dx\, e^{-\mathcal{G}_r(x)} \frac{d^2}{dx^2}\mathcal{G}_r(x)\right]^{-1} = \frac{1}{e} \tag{3.93}$$

and therefore

$$\Delta \hat{F}_r^F \simeq -\frac{1}{2N\eta_0^{\frac{1}{r+1}}(r)r} = \frac{\Delta \hat{F}_r^T}{2}. \tag{3.94}$$

Instead, if we consider $\Delta \hat{F}_r^{(1)}$, we have that

$$\begin{aligned} J_r^{(r)} &= \left[\frac{\Gamma(r+2)^{\frac{1}{r+1}}}{r+1}\right]^2 \int_{-\infty}^{+\infty} du\, e^{-\mathcal{G}_r(u)} \int_{-\infty}^{u} dv\, \mathcal{G}_r(v) \\ &\xrightarrow{r\to\infty} \frac{1}{e^2}\int_{-\infty}^{+\infty} du\, e^{-e^u}\int_{-\infty}^{u} dv\, e^v = \frac{1}{e^2}, \end{aligned} \tag{3.95}$$

finally obtaining

$$\Delta \hat{F}_r^{(1)} \simeq -\frac{\eta_1(r)}{N^{\frac{1}{r+1}} \eta_0^{1+\frac{2}{r+1}}(r) r} \tag{3.96}$$

so that, considering the law $\rho_r^\gamma$, if we send $r \to \infty$ *before* taking the limit $N \to \infty$, $\Delta \hat{F}_{+\infty}^{(1)} \sim O(1)$ and we get a new contribution to the average optimal cost

$$\hat{E} = \hat{E}_\infty + \sum_{k=1}^{\infty} \Delta \hat{F}_\infty^{(k)} = \frac{1}{e} + \frac{1}{e^2} + \dots \tag{3.97}$$

a series where we miss the contributions of order $N^{\frac{k}{r+1}}$ for $k \geq 2$, and that we know will sum to 1.

## 3.5 Numerical results

In this section we discuss some numerical results. First, we present a numerical study of our theoretical predictions obtained in the previous sections. Second, we compare with numerical simulations, in which the random assignment problem is solved using an exact algorithm.

The evaluation of all quantities in Eq. (3.10) depends on the solution of Eq. (3.9c). We solved numerically this equation for general $r$ by a simple iterative procedure. In particular, for $r > 0$ we generated a grid of $2K - 1$ equispaced points in an interval $[-y_{\max}, y_{\max}]$ and we used a discretized version of the saddle-point equation in Eq. (3.9c) in the form

$$\hat{G}_r^{[s+1]}(y_i) = \frac{y_{\max}}{K} \sum_{k=2K-i}^{2K} \frac{(y_i + y_k)^r e^{-\hat{G}_r^{[s]}(y_k)}}{\Gamma(r+1)}, \tag{3.98}$$

with $y_i = \frac{i-K}{K} y_{\max}, , i = 0, 1, \dots, 2K$. We imposed as the initial function $\hat{G}_r^{[0]}$ of the iterative procedure

$$\hat{G}_r^{[0]}(y_i) \equiv \hat{G}_0(y_i) = \ln(1 + e^{y_i}) \,. \tag{3.99}$$

We observed that the quantity

$$\Delta G_r^{[s]} = \sum_{i=0}^{2K} \left| \hat{G}_r^{[s]}(y_i) - \hat{G}_r^{[s-1]}(y_i) \right| \tag{3.100}$$

decays exponentially with $s$ and therefore convergence is very fast. For our computation, we used typically 30 iterations.

For $r < 0$ the term $(l + y)^r$ in the saddle-point equation is divergent in $y = -l$ and Eq. (3.98) cannot be adopted. We have therefore rewritten the saddle-point equation using an integration by parts, obtaining

$$\hat{G}_r(l) = \int_{-l}^{+\infty} dy \, \frac{(l + y)^{r+1} \, e^{-\hat{G}_r(y)}}{\Gamma(r+2)} \frac{d\hat{G}_r(y)}{dy}. \tag{3.101}$$

| $r$ | $\eta_0^{\frac{1}{r+1}}\hat{E}_r$ | $(N\eta_0^2)^{\frac{1}{r+1}}\frac{\eta_0\Delta\hat{F}_r^{(1)}}{\eta_1}$ | $N\eta_0^{\frac{1}{r+1}}\Delta\hat{F}_r^T$ | $N\eta_0^{\frac{1}{r+1}}\Delta\hat{F}_r^F$ |
|---|---|---|---|---|
| -0.5 | 1.125775489 | -2.777285153 | -3.917446075 | -1.192663973 |
| -0.4 | 1.334614017 | -2.952484269 | -3.665262242 | -1.250475151 |
| -0.3 | 1.471169704 | -2.921791666 | -3.324960744 | -1.222990786 |
| -0.2 | 1.558280634 | -2.784084499 | -2.984917100 | -1.157857158 |
| -0.1 | 1.612502443 | -2.600804197 | -2.675513663 | -1.079610016 |
| 0 | 1.644934067 | -2.404113806 | -2.404113806 | -1 |
| 0.1 | 1.662818967 | -2.215821874 | -2.168528577 | -0.924257491 |
| 0.2 | 1.671039856 | -2.038915744 | -1.966713438 | -0.854501434 |
| 0.3 | 1.672729262 | -1.877696614 | -1.792481703 | -0.791231720 |
| 0.4 | 1.670005231 | -1.732453452 | -1.641566768 | -0.734262435 |
| 0.5 | 1.664311154 | -1.602337915 | -1.510248399 | -0.683113178 |
| 0.6 | 1.656639222 | -1.486024319 | -1.395391897 | -0.637204338 |
| 0.7 | 1.647677145 | -1.382051819 | -1.294397704 | -0.595951390 |
| 0.8 | 1.637905005 | -1.288993419 | -1.205124002 | -0.558807473 |
| 0.9 | 1.627659755 | -1.205532353 | -1.125808312 | -0.525279810 |
| 1 | 1.617178636 | -1.130489992 | -1.054997763 | -0.494933215 |
| 2 | 1.519733739 | -0.670341811 | -0.626403698 | -0.303146650 |
| 3 | 1.446919560 | -0.461144035 | -0.431759755 | -0.211631545 |
| 4 | 1.393163419 | -0.346056113 | -0.324185048 | -0.159938240 |
| 5 | 1.352087648 | -0.274505368 | -0.257174804 | -0.127356338 |
| 6 | 1.319651066 | -0.226200326 | -0.211931870 | -0.105200594 |
| 7 | 1.293333076 | -0.191617643 | -0.179566694 | -0.089276830 |
| 8 | 1.271505390 | -0.165752490 | -0.155385461 | -0.077340947 |
| 9 | 1.253073980 | -0.145742887 | -0.136697943 | -0.068095120 |
| 10 | 1.237277174 | -0.129842072 | -0.121861122 | -0.060741591 |

**Table 3.1.** Numerical values of the rescaled corrections appearing in Eqs. (3.10) for different values of $r$.

After discretizing the previous equation, we used the same algorithm described for the $r \geq 0$ case (for a discussion on the uniqueness of the solution of Eq. (3.9c), see Ref. [Sal15]). In Table 3.1 we present our numerical results for the quantities

$$\eta_0^{\frac{1}{r+1}}\hat{E}_r, \quad (N\eta_0^2)^{\frac{1}{r+1}}\frac{\eta_0\Delta\hat{F}_r^{(1)}}{\eta_1}, \quad N\eta_0^{\frac{1}{r+1}}\Delta\hat{F}_r^T, \quad N\eta_0^{\frac{1}{r+1}}\Delta\hat{F}_r^F$$

for different values of $r$. Observe that the quantities appearing in the expansion in Eqs. (3.10) can be calculated using these values for any $\rho_r$ at given $r$, in addition to simple prefactors depending on the chosen distribution $\rho_r$.

In order to test our analysis for correction terms, we performed a direct sampling on a set of instances. Previous simulations have been reported, for example, in Refs. [MP85, Bru+91, HBM98, LO93]. In our setting, each realization of the matching problem has been solved by a `C++` implementation of the Jonker-Volgenant algorithm [JV87].

We first evaluated the asymptotic average optimal costs $\hat{E}_r^{\mathrm{P}}$ and $\hat{E}_r^{\Gamma}$, obtained,

| $r$ | $\hat{E}_r^{\mathrm{P}}$ | $\lambda_r \hat{E}_r^{\Gamma}$ | $\hat{E}_r^{\mathrm{P}}$ [HBM98] | $\hat{E}_r^{\mathrm{P}}$ [LO93] | Th. prediction |
|---|---|---|---|---|---|
| -0.5 | 4.5011(3) | 4.504(1) | – | – | 4.503101957 |
| -0.4 | 3.12611(5) | 3.1268(2) | – | – | 3.126825159 |
| -0.3 | 2.4484(1) | 2.4488(3) | – | – | 2.448788557 |
| -0.2 | 2.0593(5) | 2.0593(3) | – | – | 2.059601452 |
| -0.1 | 1.8127(3) | 1.8126(2) | – | – | 1.812767212 |
| 0 | 1.64500(5) | 1.6449(2) | 1.645(1) | 1.6450(1) | 1.644934067 |
| 0.1 | 1.5245(2) | 1.5253(9) | – | – | 1.524808331 |
| 0.2 | 1.4356(2) | 1.4357(5) | – | – | 1.435497487 |
| 0.3 | 1.3670(1) | 1.3670(4) | – | – | 1.367026464 |
| 0.4 | 1.31323(6) | 1.3132(3) | – | – | 1.3132296 |
| 0.5 | 1.27007(8) | 1.2697(4) | – | – | 1.270107121 |
| 0.6 | 1.2350(1) | 1.2348(3) | – | – | 1.234960167 |
| 0.7 | 1.20585(6) | 1.2062(6) | – | – | 1.205907312 |
| 0.8 | 1.18143(3) | 1.1812(7) | – | – | 1.181600461 |
| 0.9 | 1.16099(8) | 1.1605(6) | – | – | 1.161050751 |
| 1 | 1.14344(7) | 1.1433(4) | 1.143(2) | – | 1.14351798 |
| 2 | 1.05371(1) | 1.054(1) | 1.054(1) | 1.054(1) | 1.053724521 |
| 3 | 1.02311(1) | 1.0288(9) | 1.0232(1) | 1.0236(2) | 1.023126632 |
| 4 | 1.009690(4) | 1.010(3) | 1.0098(1) | – | 1.009736514 |
| 5 | 1.00303(2) | 1.005(3) | 1.00306(8) | 1.0026(8) | 1.003027802 |

**Table 3.2.** Numerical results for the average optimal cost for different values of $r$ and theoretical predictions. The value $\hat{E}_r^{\mathrm{P}}$ from Ref. [HBM98], due to a different convention adopted in that paper, is obtained as

$$\hat{E}_r^{\mathrm{P}} = \left( \frac{2\pi^{\frac{r+1}{2}} \Gamma(r+1)}{\Gamma\left(\frac{r+1}{2}\right) \Gamma(r+2)} \right)^{\frac{1}{r+1}} \beta_{\mathrm{num}}(r+1))$$

from Table II therein. The data for $\hat{E}_r^{\mathrm{P}}$ from Ref. [LO93] have been obtained via a linear fit, using a fitting function in the form of Eq. (3.102).

for different values of $r$, using the laws $\rho_r^{\mathrm{P}}$ and $\rho_r^{\Gamma}$, respectively. In the case of the law $\rho_r^{\mathrm{P}}$, the asymptotic estimate for $\hat{E}_r^{\mathrm{P}}$ has been obtained using the fitting function

$$f^{\mathrm{P}}(N) = \alpha_r^{\mathrm{P}} + \frac{\beta_r^{\mathrm{P}}}{N}, \tag{3.102}$$

with $\alpha_r^{\mathrm{P}}$ and $\beta_r^{\mathrm{P}}$ fitting parameters to be determined, $\alpha_r^{\mathrm{P}}$ corresponding to the value of the average optimal cost in the $N \to \infty$ limit. For a given value of $r$, we averaged over $I_N$ instances for each value of $N$ accordingly with the table below.

| $N$ | 500 | 750 | 1000 | 2500 | 5000 |
|---|---|---|---|---|---|
| $I_N$ | 100000 | 75000 | 50000 | 20000 | 10000 |

Similarly, the asymptotic average optimal cost $\hat{E}_r^{\Gamma}$ has been obtained using a fitting function in the form

$$f^{\Gamma}(N) = \begin{cases} \alpha_r^{\Gamma} + \beta_r^{\Gamma} N^{-1} + \gamma_r^{\Gamma} N^{-\frac{1}{r+1}} & \text{for } -\frac{1}{2} \leq r < 1 \\ \alpha_r^{\Gamma} + \gamma_r^{\Gamma} N^{-\frac{1}{r+1}} + \delta_r^{\Gamma} N^{-\frac{2}{r+1}} & \text{for } r \geq 1. \end{cases} \tag{3.103}$$

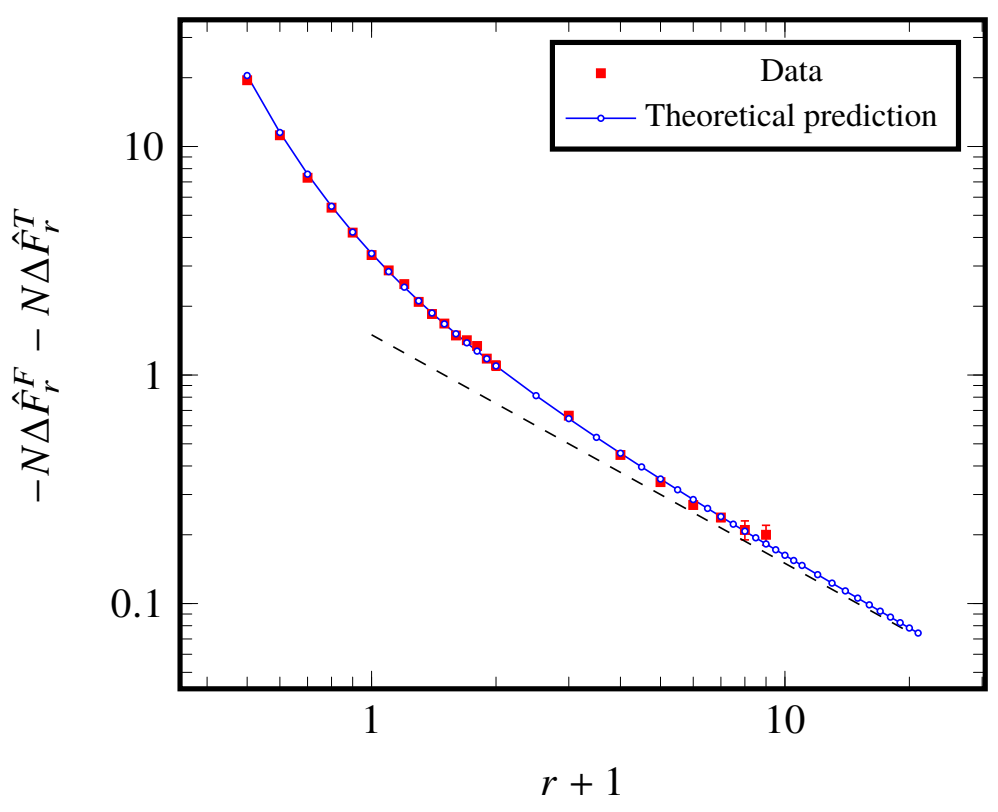

**(a)** Numerical estimates of $\Delta\hat{F}^T_r + \Delta\hat{F}^F_r$ for several values of $r$ (red squares) and theoretical prediction (blue line) obtained using the law $\rho^{\mathrm{P}}_r$. The dashed line is the large-$r$ asymptotic estimate.

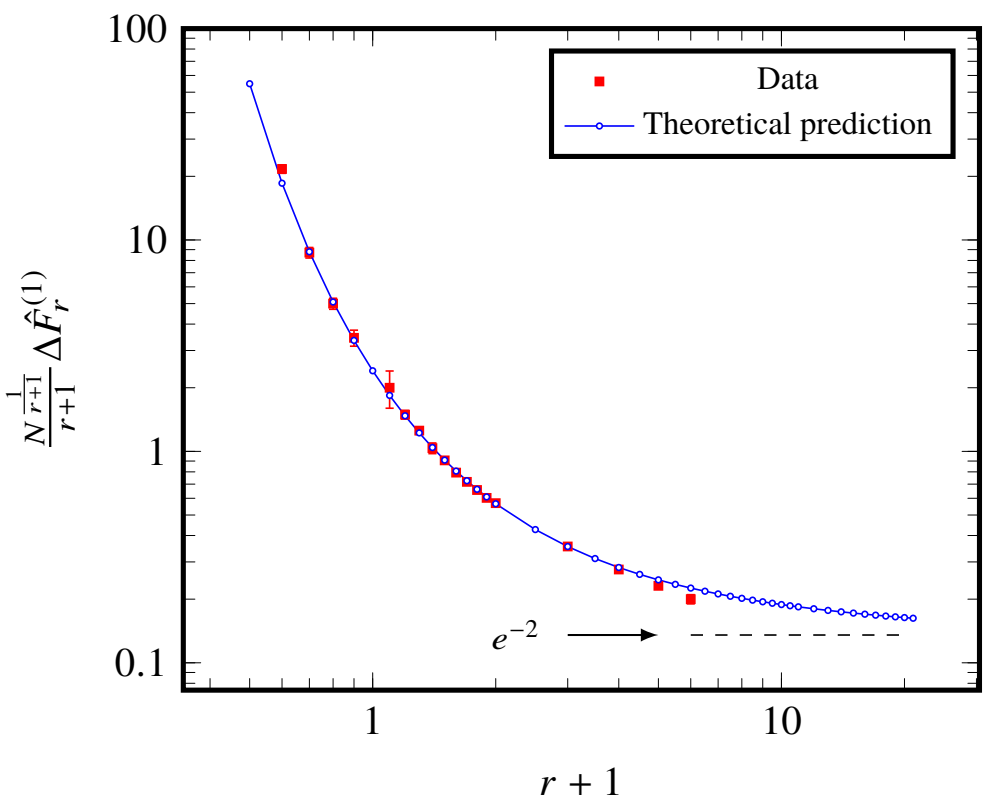

**(b)** Numerical estimates of $\Delta\hat{F}^{(1)}_r$ for several values of $r$ (red squares) and theoretical prediction (blue line with circles) obtained using the law $\rho^{\Gamma}_r$. The dashed line is the large-$r$ asymptotic estimate.

**Figure 3.2.** Plot of the finite-size corrections $\Delta\hat{F}^T_r + \Delta\hat{F}^F_r$ for the law $\rho^{\mathrm{P}}_r$ (left panel) and for $\Delta\hat{F}^{(1)}_r$ in the $\rho^{\Gamma}_r$ case (right panel). Observe that in this last plot a discrepancy between the theoretical prediction and the numerical results appears for $r \geq 1$: we interpret this fact as a consequence of the similar scaling of $\Delta\hat{F}^{(1)}_r$ and $\Delta\hat{F}^{(2)}_r$ for $r \gg 1$, which makes the numerical evaluation of the single contribution $\Delta\hat{F}^{(1)}$ difficult.

We adopted therefore a three-parameter fitting function, constructed according to Eq. (3.10) including the finite-size correction up to $o(N^{-1})$ for $r \geq 0$ and up to $O(N^{-2})$ for $\frac{1}{2} \leq r < 2$. As in the case before, the asymptotic estimation for $\hat{E}^{\Gamma}_r$ is given by $\alpha^{\Gamma}_r$. Our data were obtained extrapolating the $N \to \infty$ limit from the average optimal cost for different values of $N$. The investigated sizes and the number of iterations were the same adopted for the evaluation of $\hat{E}^{\mathrm{P}}_r$. To better exemplify the main differences in the finite-size scaling between the $\rho^{\mathrm{P}}_r$ case and the $\rho^{\Gamma}_r$ case, we have presented the numerical results for $r = 1$ in Fig. 3.1a. In the picture, it is clear that the asymptotic value $\hat{E}^{\mathrm{P}}_1 = \frac{1}{\sqrt{2}}\hat{E}^{\Gamma}_1$ is the same in the two cases, as expected from Eq. (3.49), but the finite-size corrections are different both in sign and in their scaling properties.

In Table 3.2 we compare the results of our numerical simulations with the ones in the literature (when available) for both $\hat{E}^{\mathrm{P}}_r(N)$ and $\lambda_r\hat{E}^{\Gamma}_r(N)$, $\lambda_r$ being defined in Eq. (3.49). In Fig. 3.1b we plot our theoretical predictions and the numerical results that are presented in Table 3.2.

Let us now consider the finite-size corrections. In the case of the $\rho^{\mathrm{P}}_r$ law, the $O(N^{-1})$ corrections are given by $\Delta\hat{F}^T_r + \Delta\hat{F}^F_r$ and no nonanalytic corrections to the leading term appear. We obtain the finite-size corrections from the data used for Table 3.2, using Eq. (3.102) but fixing $\alpha^{\mathrm{P}}_r$ to the average optimal cost $\hat{E}^{\mathrm{P}}_r$ given by the theoretical prediction in Table 3.1 and therefore with one free parameter only, namely, $\beta^{\mathrm{P}}_r$. In Fig. 3.2a we compare our predictions for $\Delta\hat{F}^T_r + \Delta\hat{F}^F_r$, deduced by the values in Table 3.1, with the results of our numerical simulations for different values of $r$.

In the case of the $\rho_r^\Gamma$ law with $r > 0$, the first correction to the average optimal cost is given by $\Delta\hat{F}_r^{(1)}$, whereas $\Delta\hat{F}_r^{(1)}$ is $o(N^{-1})$ for $r < 0$. Again, this correction can be obtained by a fit of the same data used to extrapolate the average optimal cost, fixing the fitting parameter $\alpha_r^\Gamma$ in Eq. (3.103) to the theoretical prediction, and performing a two parameters fit in which the quantity $\gamma_r^\Gamma$ appearing in Eq. (3.103) corresponds to $\Delta\hat{F}_r^{(1)}$. In Fig. 3.2b we compare our prediction for $\Delta\hat{F}_r^{(1)}$, given in Table 3.1, with the results of our fit procedure for $\gamma_r^\Gamma$ for $-\frac{1}{2} < r \leq 5$. Observe that the numerical evaluation of the single contribution $\Delta\hat{F}_r^{(1)}$ is not possible for $r = 0$. In this case, the result of our fit for the $O(N^{-1})$ correction was $\beta_r^\Gamma + \gamma_r^\Gamma = -0.97(4)$, to be compared with the theoretical prediction $N(\Delta\hat{F}_0^{(1)} + \Delta\hat{F}_0^F + \Delta\hat{F}_0^T) = -0.998354732\ldots$.

## 3.6 The random matching problem

Similar conclusion can be deduced for the random matching problem case where the underlying graph is the complete one with $2N$ points $\mathcal{K}_{2N}$. The partition function is

$$Z[w] = \sum_{\pi} \left[ \prod_{j=1}^{2N} \delta\left(1 - \sum_{i=1}^{2N} \pi_{ij}\right)\right] e^{-\beta E(\pi)} . \tag{3.104}$$

where $\pi$ can assume as before 0 if there is no link in the matching and 1 otherwise. The energy $E(\pi)$ is defined now as

$$E(\pi) = \sum_{i<j}^{2N} \pi_{ij} w_{ij} . \tag{3.105}$$

Using the replica trick the replicated partition function can be written in the following way [PR02]

$$\overline{Z^n} = \left[ \prod_{\alpha \in \mathcal{P}([n])}{}' \sqrt{\frac{N}{2\pi\, \hat{g}_\alpha}} \int dQ_\alpha \right] e^{-NS[Q] - N\Delta S^T[Q]}, \tag{3.106a}$$

with

$$S[Q] = \sum_{\alpha \in \mathcal{P}([n])}{}' \left( \frac{Q_\alpha^2}{2\hat{g}_\alpha} - 2 \ln z\,[Q_\alpha] \right), \tag{3.106b}$$

$$\Delta S^T[Q] = \sum_{\substack{\alpha,\beta \in \mathcal{P}([n]) \\ \alpha \cap \beta = \emptyset}}{}' \hat{g}_\alpha \hat{g}_\beta \frac{|Q_{\alpha\cup\beta}|^2}{4N\hat{g}_{\alpha\cup\beta}^2}, \tag{3.106c}$$

The computation of the average optimal cost is basically the same as before. In the replica symmetric ansatz, one introduces a generating function of the overlaps $G$ as in (3.38) which, in the low temperature limit must rescaled this time by

$$\hat{G}_r(l) \equiv G_r\left( \frac{\hat{\beta}}{[2\eta_0 \Gamma(r+1)]^{\frac{1}{r+1}}}\, l \right) \tag{3.107}$$

in order to satisfy the same saddle point equation (3.9c) of the RAP. The average optimal cost of the RMP in the thermodynamic limit is

$$\hat{E}_r = \lim_{N\to\infty} \frac{1}{N^{\frac{r}{r+1}}} \overline{E}_r = \frac{r+1}{[2\eta_0\Gamma(r+1)]^{\frac{1}{r+1}}} J_r^{(r+1)}\,, \tag{3.108}$$

i.e. $2^{1/(r+1)}$ times smaller than the corresponding one of the bipartite case [MP85]. In particular, for $r=0$, using the solution (3.50a) we obtain

$$\hat{E}_0 = \frac{1}{\eta_0(0)} \frac{\pi^2}{12}\,. \tag{3.109}$$

Analogously the finite-size corrections $\Delta\hat{F}_r^T$ and $\Delta\hat{F}_r^{(1)}$ can be evaluated in the same way and one gets

$$\Delta\hat{F}_r^T = -\frac{1}{2^{1+\frac{1}{r+1}}N} \frac{\Gamma(2r+2) J_r^{(0)}}{(r+1)\eta_0^{\frac{1}{r+1}} [\Gamma(r+1)]^{\frac{2r+3}{r+1}}}\,, \tag{3.110a}$$

$$\Delta\hat{F}_r^{(1)} = -\frac{\eta_1}{N^{\frac{1}{r+1}}} \frac{r+1}{2\eta_0[\eta_0\Gamma(r+1)]^{\frac{2}{r+1}}} J_r^{(r)}\,, \tag{3.110b}$$

i.e. they are $2^{1+\frac{1}{r+1}}$ and $2^{\frac{2}{r+1}}$ smaller than the respective quantities in the RMP reported in equations (3.10c) and (3.10e). This determinant correction is much more involved and was examined in detail by [PR02] in the case $r=0$. However, their formulas can be easily extended to the case $r\neq 0$. The final result is

$$\Delta\hat{F}_r^F = \frac{1}{N} \frac{1}{[2\eta_0\Gamma(r+1)]^{\frac{1}{r+1}}} \sum_{p=0}^{\infty} \frac{I_{2p+1}}{2p+1} \tag{3.111}$$

where

$$I_p \equiv \int_0^{\infty} dt\, \mathrm{Tr} H_t^p \tag{3.112}$$

and $H_t$ is the integral operator

$$H_t(x,y) \equiv e^{-\frac{\hat{G}_r(x+t)+\hat{G}_r(y+t)}{2}} \frac{(x+y)^r}{\Gamma(r+1)} \theta(x+y)\,. \tag{3.113}$$

This formula is reminiscent in form of the finite-size corrections of spin glass models on locally tree-like random graphs [Fer+13, Luc+14]; the finite-size corrections can be interpreted as a sum over non-self-intersecting loops of the graph each loop contributing with a combinatorial factor and a free energy contribution due to the addition of the loop to an infinite tree. This form of the corrections appear also in fluctuations around instanton expansions for the determination of the large order behavior of a quantum field theory [BP78, MPR17]. Here only odd-length loops appear. $I_p$ can be written as

$$I_p = \int_0^{\infty} dt \int_{-\infty}^{+\infty} \prod_{k=1}^{p} dx_k \frac{e^{-\sum_{k=1}^{p}\hat{G}_r(x_k+t)}}{\Gamma^p(r+1)} (x_1+x_2)^r\theta(x_1+x_2)\dots(x_p+x_1)^r\theta(x_p+x_1)\,. \tag{3.114}$$

This expression will be examined and simplified in the general $r$ case in Appendix A.6; here we limit ourselves to the limiting $r \to \infty$ case, where some exact results can be achieved. Thanks to the solution for $r \to \infty$ given in (3.81) and (3.85) we perform the change of variables $x_k = \Gamma^{\frac{1}{r+1}}(r+2)\left(\frac{1}{2} + \frac{x'_k}{r+1}\right)$ for every $k = 1, \ldots, p$ and $t = \Gamma^{\frac{1}{r+1}}(r+2)\frac{t'}{r+1}$ in expression (3.114), getting

$$\begin{aligned} I_p = & \frac{\Gamma^{\frac{p(r+1)+1}{r+1}}(r+2)}{(r+1)^{p+1}} \int_0^\infty dt \int_{-\infty}^{+\infty} dx_1 \ldots dx_p \frac{e^{-\sum_{k=1}^p \mathcal{G}_r(x_k+t)}}{\Gamma^p(r+1)} \\ & \times \theta\left[\Gamma^{\frac{1}{r+1}}(r+2)\left(1+\frac{x_1+x_2}{r+1}\right)\right] \ldots \theta\left[\Gamma^{\frac{1}{r+1}}(r+2)\left(1+\frac{x_p+x_1}{r+1}\right)\right] \\ & \times \left(1+\frac{x_1+x_2}{r+1}\right)^r \ldots \left(1+\frac{x_p+x_1}{r+1}\right)^r \end{aligned} \tag{3.115}$$

In the infinite $r$ limit the argument of the theta functions is always positive so that

$$\begin{aligned} I_p &= \frac{\Gamma^{\frac{1}{r+1}}(r+2)}{r+1} \int_0^\infty dt \int_{-\infty}^{+\infty} dx_1 \ldots dx_p e^{-e^{x_1+t}} \ldots e^{-e^{x_p+t}} e^{2x_1} \ldots e^{2x_p} \\ &= \frac{\Gamma^{\frac{1}{r+1}}(r+2)}{r+1} \int_0^\infty dt \left[\int_{-\infty}^{+\infty} dx\, e^{-e^{x+t}} e^{2x}\right]^p . \end{aligned} \tag{3.116}$$

Setting $a \equiv e^t > 0$ we have that the integral in square brackets is

$$\int_{-\infty}^{+\infty} dx e^{-ae^x} e^{2x} = \int_0^{+\infty} dy\, e^{-ay} y = -\frac{d}{da}\int_0^{+\infty} dy\, e^{-ay} = \frac{1}{a^2} \tag{3.117}$$

We finally have

$$I_p = \frac{\Gamma^{\frac{1}{r+1}}(r+2)}{r+1} \int_0^{+\infty} dt\, e^{-2pt} = \frac{\Gamma^{\frac{1}{r+1}}(r+2)}{2(r+1)p} . \tag{3.118}$$

Confronting this equation with (3.112) we also get $\mathrm{Tr} H_t^p \propto e^{-2pt}$ a form that was verified in [LPS17] in the case $r = 0$ for large values of $p$, where the integral is dominated by the region of small $t$. Plugging this result into (3.111) we obtain

$$\Delta \hat{F}_r^F = \frac{1}{2(r+1)N}\left[\frac{r+1}{2\eta_0}\right]^{\frac{1}{r+1}} \sum_{p=0}^\infty \frac{1}{(2p+1)^2} = \frac{\pi^2}{16(r+1)N}\left[\frac{r+1}{2\eta_0}\right]^{\frac{1}{r+1}} \tag{3.119}$$

# Chapter 4

# The random fractional matching problem

## 4.1 Introduction

In the present chapter, we will apply the replica formalism to a different type of matching problem, namely the random fractional matching problem (RFMP), the linear relaxation of the matching problem. We start considering a weighted complete graph $\mathcal{K}_{2N}$. The weights $\mathbf{w} \equiv \{w_e\}_e$ associated to the edges $\mathcal{E} \equiv \{e\}_e$ are non-negative independent random variables, identically distributed according to a probability density $\varrho(w)$, exactly as in the RMP case. In the RFMP we search for the set of quantities $\mathbf{m} \equiv \{m_e\}_e$ that minimize the cost

$$E[\mathbf{m}, \mathbf{w}] \equiv \sum_{e \in \mathcal{E}} m_e w_e, \tag{4.1a}$$

with the additional constraints

$$m_e \in [0, 1] \quad \forall e \in \mathcal{E}, \quad \sum_{e \to v} m_e = 1 \quad \forall v \in \mathcal{V}. \tag{4.1b}$$

In the previous expression, the sum $\sum_{e \to v}$ runs over all edges having $v$ as an endpoint. It is easy to show for general graphs that the problem has semi-integer solutions, i.e. optimal configurations $\mathcal{M}$ with $m_e \in \{0, 1/2, 1\}$. $\mathcal{M}$ contains only (odd) cycles and edges that do not share their endpoints (see Fig. 4.1, center). It is expected that the aoc of the RFMP is less than or equal to the aoc of the RMP obtained with the same weight probability density $\varrho$, due to the fact that all matching configurations feasible for the RMP, are also feasible for the RFMP.

If defined on bipartite graphs, the relaxed problem above has integer solutions only, due to the absence of odd cycles, and therefore it is equivalent to the standard RAP. Noticeably, the BP algorithm, deeply related to the cavity method, is able to recover the optimal solution of an assignment problem when the solution is unique, and its computational complexity is on par with that one of the best alternative solvers [BSS08]. The statistical physics of linear or convex relaxations of discrete optimization problems has been actively investigated in recent years [TH16, JMRT16]. In an interesting variation of the RFMP, that we will call "loopy

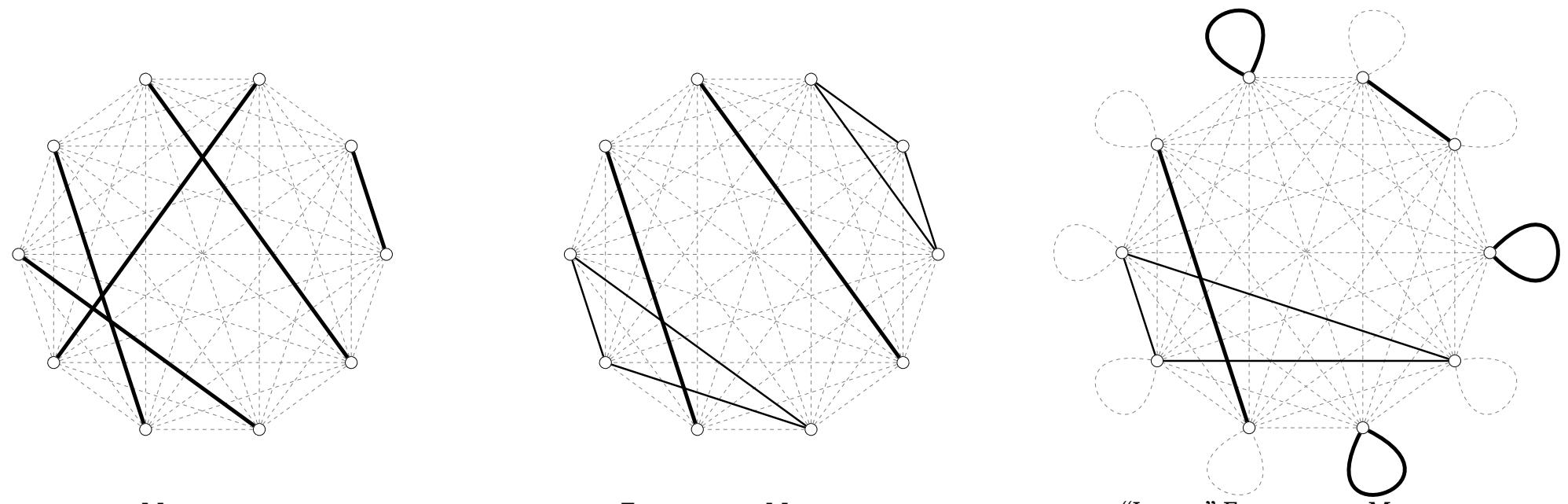

**Figure 4.1.** *On the left*, complete graph $\mathcal{K}_{10}$ with a usual matching on it: the matching cost in this case is simply the sum of the weights $w_e$ of the matching edges $e \in \mathcal{M}$ (thick edges). *In the center*, the same graph with a fractional matching on it: in this case, cycles are allowed and the contribution to the matching cost of an edge of weight $w_e$ belonging to a cycle is $w_e/2$ (thin edges), $w_e$ otherwise (thick edges). *On the right*, the graph $\hat{\mathcal{K}}_{10}$ obtained allowing loops on each vertex of $\mathcal{K}_{10}$, with a fractional matching on it. From the left to the right, progressively more matching configurations are allowed. In particular, any feasible configuration for the usual matching problem is feasible for the fractional matching problem; moreover, any feasible configuration for the fractional matching problem is feasible for the loopy fractional matching problem.

RFMP", an additional non-negative weight $w_v$ is associated to each vertex $v \in \mathcal{V}$ of the graph. Each weight $w_v$ is a random variable extracted independently from all other weights in the problem with the same distribution of the edge weights $\varrho(w)$. The loopy RFMP corresponds therefore to a RFMP defined on a graph $\hat{\mathcal{K}}_{2N}$ obtained allowing self-loops on $\mathcal{K}_{2N}$ (see Fig. 4.1, right). The cost is defined as

$$E[\mathbf{m},\mathbf{w}] \equiv \sum_{e\in\mathcal{E}} m_e w_e + 2\sum_{v\in\mathcal{V}} m_v w_v, \tag{4.2a}$$

with the constraints

$$m_e \in [0,1] \quad \forall e \in \mathcal{E}, \quad m_v \in [0,1] \quad \forall v \in \mathcal{V}, \quad \sum_{e\to v} m_e + 2m_v = 1 \quad \forall v \in \mathcal{V}. \tag{4.2b}$$

Wästlund proved that, in the loopy RFMP, in the optimal configuration $m_e \in \{0, 1/2, 1\}$ $\forall e$ and $m_v \in \{0, 1/2\}$ $\forall v$ [Wäs10]. Remarkably, he also obtained the expression for the aoc for any $N$ assuming $\varrho(w) = e^{-w}\theta(w)$, namely

$$\overline{\min_{\mathbf{m}} E[\mathbf{m},\mathbf{w}]} = \sum_{n=1}^{2N} \frac{(-1)^{n-1}}{n^2} = \frac{\pi^2}{12} - \frac{1}{8N^2} + o\left(\frac{1}{N^2}\right). \tag{4.2c}$$

Again, since the loopy RFMP is a relaxed version of the RFMP, that in turn is a relaxed version of the RMP, it follows from Eq. (4.2c) that all three RCOPs have the same asymptotic aoc, equal to $\pi^2/12$. The presence of cycles in the RFMP and cycles and loops in the loopy RFMP does not affect, therefore, the value of the aoc, but the finite-size corrections only.

In the following, we will study, using the replica approach, both the RFMP and the loopy RFMP, and their finite-size corrections. We shall consider weight probability densities $\varrho(w)$ with non-negative support and such that $\varrho(w) = 1 - \mu w + o(w)$ for $w \to 0^+$. In particular, if the weights are exponentially distributed we have $\mu = 1$, whereas $\mu = 0$ if they are uniformly distributed. We will show, in particular, that the RMP saddle-point solution naturally appears as asymptotic solution of the problem, alongside with another saddle-point solution of higher cost corresponding to the solution to the RTSP. Moreover, we will evaluate the finite-size corrections to the aoc on the matching saddle-point, obtaining as the main result of this chapter the closed formula

$$E_\varepsilon(\mu) = \frac{\zeta(2)}{2} + \frac{1}{2N}\left[(\mu-1)\zeta(3) + \frac{1-\varepsilon}{4}\zeta(2)\right] + o\left(\frac{1}{N}\right). \tag{4.3}$$

Here $\zeta(z)$ is the Riemann zeta function and, since $\zeta(2) = \pi^2/6$, we obtain the result of Eq. (3.109) at the leading order (in the cases we are considering here $\eta_0(0)$ is always 1). The parameter $\varepsilon$ takes value $+1$ if self-loops are allowed in the model, $-1$ otherwise, i.e., when $m_{ii}$ is not present. The analytic predictions from the replica calculation are supported by the numerical simulations in the last section of the chapter.

## 4.2 Replica calculation

In the spirit of the seminal works of Orland [Orl85] and Mézard and Parisi [MP85], let us first write down the partition function for the RFMP, both in its usual and in its loopy version. As anticipated, we will consider a random-link formulation of the problem on the complete graph with $2N$ vertices $\mathcal{V} = \{i\}_{i=1,\dots,2N}$, where the weights $\{w_{ij}\}_{ij}$ are non-negative independent random variables identically distributed with distribution $\varrho(w)$. As anticipated, in the following, we will consider a particular class of probability densities, i.e., we will assume that $\varrho(w) = 1 - \mu w + o(w)$ for $w \to 0^+$. The uniform distribution on the unit interval, $\varrho(w) = \theta(w)\theta(1-w)$, and the exponential distribution on the positive real axis, $\varrho(w) = e^{-w}\theta(w)$, belong to this class, with $\mu = 0$ and $\mu = 1$ respectively. Due to the fact that all distributions in this class have the same limit for $w \to 0^+$, i.e., $\lim_{w\to 0^+} \varrho(w) = 1$, we expect that the asymptotic aoc will be the same for all of them, as it happens in the RMP. Indeed, in the general framework of the analysis of the RMP performed in the previous chapter, it is easily seen that, whereas the asymptotic cost only depends on the behavior of the first term in the Maclaurin expansion of $\varrho(w) \simeq \varrho_0 w^r$, the $O(1/N)$ finite-size corrections depend on the expansion up to, at least, the second term, and the power $r$ also affects the scaling of the corrections themselves.

Let us start observing that, in the RFMP, the occupation number $m_{ij} = m_{ji}$ of the edge $(i,j)$ between the node $i$ and the node $j$ can assume the values

$$m_{ij} \in \{0,1,2\}, \quad \text{with the constraint } \sum_{j=1}^{2N} m_{ij} + \varepsilon m_{ii} = 2, \quad 1 \le i \le 2N, \tag{4.4}$$

where $\varepsilon = +1$ if loops are allowed, $\varepsilon = -1$ otherwise. The parameter $\varepsilon$, therefore, allows us to switch between the two variations of the model described in the

Introduction. The cost of a given matching configuration is

$$E_\varepsilon[\mathbf{m},\mathbf{w}] = \frac{1}{2}\sum_{i\leq j}\left(1+\varepsilon\delta_{ij}\right) m_{ij}w_{ij}. \tag{4.5}$$

For calculation convenience we consider, for each edge $e$, $m_e \in \{0,1,2\}$ and not $m_e \in \{0, 1/2, 1\}$ as in the Introduction. This fact does not affect the results apart from the necessary rescaling of the cost that indeed we have introduced in Eq. (4.5). The partition function for a given instance of the problem can be written as

$$Z_\varepsilon(\beta) \equiv \sum_{m_{ij}\in\{0,1,2\}} \left[\prod_{i=1}^{2N} \mathbb{I}\left(\sum_{j=1}^{2N} m_{ij} + \varepsilon m_{ii} = 2\right)\right] e^{-2\beta N E_\varepsilon[\mathbf{m},\mathbf{w}]}, \tag{4.6}$$

where $\mathbb{I}(\bullet)$ is an indicator function that is equal to one when the condition in the brackets is satisfied, and zero otherwise. The aoc is recovered as

$$E_\varepsilon(\mu) \equiv \overline{\min_{\mathbf{m}} E_\varepsilon[\mathbf{m},\mathbf{w}]} = -\lim_{\beta\to+\infty}\frac{1}{2N}\frac{\partial \overline{\ln Z(\beta)}}{\partial\beta}. \tag{4.7}$$

Note that we have made explicit the dependence of the aoc on the value of $\mu = -\left.\partial_w \varrho\right|_{w=0}$. From the results in Ref. [Wäs10], we know that $E_\varepsilon(\mu) = O(1)$, i.e., the aoc is not extensive, and the cost density scales as $O\left(1/N\right)$: this is indeed expected, due to the fact that the shortest link amongst $N$ scales as $1/N$. We have therefore rescaled $\beta$ in the exponent of the partition function accordingly, in such a way that a finite thermodynamical limit at fixed $\beta$ can be obtained, and an extensive functional $2NE_\varepsilon(\mu) = O(N)$ appears in the exponent in the low-temperature regime.

To average over the disorder, we use the following integral representation of the Kronecker delta,

$$\delta_{a,0} = \int_0^{2\pi}\frac{d\lambda}{2\pi}\,e^{i\lambda a}, \tag{4.8}$$

and we apply, as usual, the replica trick. The average replicated partition function for the fractional matching problem can be written as

$$\overline{Z^n_\varepsilon} = \left[\prod_{a=1}^{n}\prod_{i=1}^{2N}\int_0^{2\pi}\frac{e^{-2i\lambda_i^a}d\lambda_i^a}{2\pi}\right]\prod_{i<j}\left(1+\frac{T_{ij}}{N}\right)\prod_{i=1}^{2N}\left(1+\frac{\varepsilon+1}{2}\frac{R_i}{N}\right). \tag{4.9}$$

In Eq. (4.9) we have introduced the quantities

$$\begin{aligned}1+\frac{T_{ij}}{N} &\equiv \overline{\prod_{a=1}^{n}\left[1+e^{i\lambda_i^a+i\lambda_j^a-\beta N w_{ij}}+e^{2i\lambda_i^a+2i\lambda_j^a-2\beta N w_{ij}}\right]}\\ &= 1+\frac{1}{N}\sum_{\substack{\alpha\cap\beta=\emptyset\\ \alpha\cup\beta\neq\emptyset}}\hat{g}_{|\alpha|+2|\beta|}\,e^{i\sum_{a\in\alpha}\left(\lambda_i^a+\lambda_j^a\right)+2i\sum_{b\in\beta}\left(\lambda_i^b+\lambda_j^b\right)},\end{aligned} \tag{4.10a}$$

and, in the presence of loops, the on-site contribution

$$1+\frac{R_i}{N} \equiv \overline{\prod_{a=1}^{n}\left[1+e^{2i\lambda_i^a-2\beta N w_{ii}}\right]} = 1+\frac{1}{N}\sum_{\alpha\neq\emptyset}\hat{g}_{2|\alpha|}\,e^{2i\sum_{a\in\alpha}\lambda_i^a}. \tag{4.10b}$$

We remind that, in Eqs. (4.10) the sums run over the elements of $\mathcal{P}([n])$, set of subsets of $[n] \equiv \{1, \dots, n\}$, and we have denoted the cardinality of $\alpha \in \mathcal{P}([n])$ by $|\alpha|$. We have also introduced the quantity

$$\begin{aligned} \hat{g}_p &\equiv N \int_0^{+\infty} e^{-\beta p N w} \varrho(w) dw \\ &= g_p \left[1 - \mu \frac{g_p}{N} + o\left(\frac{1}{N}\right)\right], \quad \text{where } g_p \equiv \frac{1}{\beta p}. \end{aligned} \tag{4.11}$$

Using the previous equation, and in order to evaluate the aoc and its first finite-size correction, we use the fact that

$$\begin{aligned} 2R_i = 2\sum_{\alpha \neq \emptyset} \hat{g}_{2|\alpha|}\, e^{2i\sum_{a\in\alpha}\lambda_i^a} &= \sum_{\alpha \neq \emptyset} \left[g_{|\alpha|} - \frac{2g_{2|\alpha|}^2}{N} + o\left(\frac{1}{N}\right)\right] e^{2i\sum_{a\in\alpha}\lambda_i^a} \\ &\sim T_{ii} + O\left(\frac{1}{N}\right), \end{aligned} \tag{4.12}$$

since $2g_{2p} = g_p$. In the last step we have used the fact that

$$T_{ii} = \sum_{\substack{\alpha\cap\beta=\emptyset \\ \alpha\cup\beta\neq\emptyset}} \hat{g}_{|\alpha|+2|\beta|}\, e^{2i\sum_{a\in\alpha}\lambda_i^a + 4i\sum_{b\in\beta}\lambda_i^b} \tag{4.13}$$

gives zero contribution, unless $\beta = \emptyset$, due to the overall constraint imposed by the integration on $\{\lambda_i^a\}$. Neglecting $O(1/N)$ terms in the exponent (i.e., $O(1/N^2)$ to the cost), we can write the partition function as

$$\overline{Z_\varepsilon^n} = \left[\prod_{a=1}^{n}\prod_{i=1}^{2N}\int_0^{2\pi} \frac{e^{-2i\lambda_i^a} d\lambda_i^a}{2\pi}\right] e^{\frac{1}{2N}\sum_{i,j}\left(T_{ij} - \frac{T_{ij}^2}{2N}\right) + \frac{\varepsilon-1}{4N}\sum_{i=1}^{2N} T_{ii} + O\left(\frac{1}{N}\right)}. \tag{4.14}$$

We introduce now the placeholders

$$q_{\alpha,\beta} \equiv \mathbb{I}(\alpha\cap\beta=\emptyset)\sum_{i=1}^{2N} e^{i\sum_{a\in\alpha}\lambda_i^a + 2i\sum_{b\in\beta}\lambda_i^b}, \tag{4.15}$$

that allow us to write

$$\sum_{i,j} T_{ij} + \frac{\varepsilon-1}{2}\sum_i T_{ii} = \sum_{\substack{\alpha\cap\beta=\emptyset \\ \alpha\cup\beta\neq\emptyset}} \hat{g}_{|\alpha|+2|\beta|}\, q_{\alpha,\beta}^2 + \frac{\varepsilon-1}{2}\sum_{\alpha\neq\emptyset} \hat{g}_{|\alpha|}\, q_{\emptyset,\alpha}, \tag{4.16a}$$

$$\sum_{i,j} T_{ij}^2 = {\sum_{\alpha,\beta|\hat{\alpha},\hat{\beta}}}' \hat{g}_{|\alpha|+2|\beta|}\, \hat{g}_{|\hat{\alpha}|+2|\hat{\beta}|}\, q^2_{\alpha\triangle\hat{\alpha},\beta\cup\hat{\beta}\cup(\alpha\cap\hat{\alpha})}. \tag{4.16b}$$

In the previous equations $\alpha\triangle\beta \equiv (\alpha\setminus\beta)\cup(\beta\setminus\alpha)$ is the symmetric difference of the sets $\alpha$ and $\beta$, we have denoted by

$${\sum_{\alpha,\beta|\hat{\alpha},\hat{\beta}}}' = \sum_{\substack{\alpha\cap\beta=\emptyset \\ \alpha\cup\beta\neq\emptyset}}\sum_{\substack{\hat{\alpha}\cap\hat{\beta}=\emptyset \\ \hat{\alpha}\cup\hat{\beta}\neq\emptyset}} \mathbb{I}(\beta\cap\hat{\beta}=\emptyset)\,\mathbb{I}((\alpha\cup\hat{\alpha})\cap(\beta\cup\hat{\beta})=\emptyset). \tag{4.17}$$

Denoting by

$$\varphi_{\alpha,\beta} \equiv \sum_{\sigma,\rho|\hat{\sigma},\hat{\rho}}{}' \hat{g}_{|\rho|+2|\sigma|}\, \hat{g}_{|\hat{\rho}|+2|\hat{\sigma}|}\, \mathbb{I}\left(\rho \triangle \hat{\rho} = \alpha\right) \mathbb{I}\left(\sigma \cup \hat{\sigma} \cup (\rho \cap \hat{\rho}) = \beta\right), \tag{4.18}$$

and using a Hubbard–Stratonovich transformation in the form

$$\begin{aligned} &\exp\left(\frac{\hat{g}_{|\alpha|+2|\beta|} - {}^{1}/_{2N}\,\varphi_{\alpha,\beta}}{2N} q_{\alpha,\beta}^2\right) \\ &= \sqrt{\frac{N}{2\pi\left(\hat{g}_{|\alpha|+2|\beta|} - {}^{\varphi_{\alpha,\beta}}/_{2N}\right)}} \int_{-\infty}^{+\infty} dQ_{\alpha,\beta} \exp\left(-\frac{NQ_{\alpha,\beta}^2}{2\hat{g}_{|\alpha|+2|\beta|} - {}^{\varphi_{\alpha,\beta}}/_{N}} + Q_{\alpha,\beta}\, q_{\alpha,\beta}\right) \\ &= \sqrt{\frac{N}{2\pi\hat{g}_{|\alpha|+2|\beta|}}} \int_{-\infty}^{+\infty} dQ_{\alpha,\beta} \exp\left[-\frac{NQ_{\alpha,\beta}^2}{2\hat{g}_{|\alpha|+2|\beta|}} - \frac{\varphi_{\alpha,\beta} Q_{\alpha,\beta}^2}{4\hat{g}_{|\alpha|+2|\beta|}^2} + Q_{\alpha,\beta} q_{\alpha,\beta} + O\left(\frac{1}{N}\right)\right] \end{aligned} \tag{4.19}$$

we can finally introduce the order parameters $Q_{\alpha,\beta}$ as follows

$$\begin{aligned} &\left[\prod_{a=1}^{n}\prod_{i=1}^{2N}\int_0^{2\pi}\frac{e^{-2i\lambda_i^a}d\lambda_i^a}{2\pi}\right] e^{\frac{1}{2N}\sum_{i,j}\left(T_{ij}-\frac{T_{ij}^2}{2N}\right)+\frac{\varepsilon-1}{4N}\sum_{i=1}^{2N}T_{ii}} \simeq \left[\prod_{\substack{\alpha\cap\beta=\emptyset\\ \alpha\cup\beta\neq\emptyset}}\int dQ_{\alpha,\beta}\sqrt{\frac{N}{2\pi\hat{g}_{|\alpha|+2|\beta|}}}\right] \\ &\times \exp\left[-N\sum_{\substack{\alpha\cap\beta=\emptyset\\ \alpha\cup\beta\neq\emptyset}}\frac{Q_{\alpha,\beta}^2}{2\hat{g}_{|\alpha|+2|\beta|}} + 2N\ln z[\mathbf{Q}] - \sum_{\alpha,\beta|\hat{\alpha},\hat{\beta}}{}'\frac{g_{|\alpha|+2|\beta|}\,g_{|\hat{\alpha}|+2|\hat{\beta}|}}{4g_{|\alpha|+|\hat{\alpha}|+2|\beta|+2|\hat{\beta}|}^2} Q_{\alpha\triangle\hat{\alpha},\beta\cup\hat{\beta}\cup(\alpha\cap\hat{\alpha})}^2\right]. \end{aligned} \tag{4.20}$$

The expression of $\ln z[\mathbf{Q}]$, in which we have exponentiated the integration on $\{\lambda_i^a\}$, is given in Eq. (B.1) in B.1. Eq. (4.20) generalizes the equivalent expression for the partition function obtained for the RMP in Refs. [MP85, MP87, PR02], that is recovered imposing that $Q_{\alpha,\beta} \equiv Q_\beta \delta_{|\alpha|,0}$.

Observe that in Eq. (4.20) $\hat{g}_p$ appears, a quantity defined in Eq. (4.11). We have that $\lim_{N\to+\infty}\hat{g}_p = g_p$. However, expanding for large $N$ the quantities $\hat{g}_p$ in Eq. (4.20), new $1/N$ finite-size corrections to the cost will appear[1]. In B.1 we show that the replicated action can be finally written as

$$\overline{Z^n} \simeq \left[\prod_{\substack{\alpha\cap\beta=\emptyset\\ \alpha\cup\beta\neq\emptyset}}\int dQ_{\alpha,\beta}\sqrt{\frac{N}{2\pi\hat{g}_{|\alpha|+2|\beta|}}}\right] e^{-N\mathcal{S}[\mathbf{Q}]} \tag{4.21a}$$

$$\mathcal{S}[\mathbf{Q}] \equiv S[\mathbf{Q}] + \Delta S^{\mathrm{T}}[\mathbf{Q}] + \Delta S^{\varrho}[\mathbf{Q}] + o\left(1/N\right).$$

[1] As we have seen in the previous chapter, in the case of a generic distribution $\varrho(w) = w^r[\varrho_0 + \varrho_1 w + o(w)]$ these corrections will scale as $N^{-\frac{k}{r+1}}$ with $k \in \mathbb{N}$ and they will be *dominant* respect to all other corrections, that are $O(1/N)$, for all values $k$ such that $k < r+1$, and of the same order for $k = r+1$ if $r+1$ is a natural number. Here only the $k = 1$ term appears, because $r = 0$, without any anomalous contribution.

The three contributions appearing in the previous expression are

$$S[\mathbf{Q}] \equiv \sum_{\substack{\alpha\cap\beta=\emptyset\\ \alpha\cup\beta\neq\emptyset}} \frac{Q^2_{\alpha,\beta}}{2g_{|\alpha|+2|\beta|}} - 2\ln z_0[\mathbf{Q}], \tag{4.21b}$$

$$\Delta S^{\mathrm{T}}[\mathbf{Q}] \equiv \frac{1}{N}\sum_{\alpha,\beta|\hat\alpha,\hat\beta}{}' \frac{g_{|\alpha|+2|\beta|}g_{|\hat\alpha|+2|\hat\beta|}}{4g^2_{|\alpha\triangle\hat\alpha|+2|\beta|+2|\hat\beta|+2|\alpha\cap\hat\alpha|}} Q^2_{\alpha\triangle\hat\alpha,\beta\cup\hat\beta\cup(\alpha\cap\hat\alpha)} - \frac{\varepsilon-1}{2N}\sum_{\alpha\neq\emptyset} g_{|\alpha|}\frac{\partial \ln z_0[\mathbf{Q}]}{\partial Q_{0,\alpha}}, \tag{4.21c}$$

$$\Delta S^{\varrho}[\mathbf{Q}] \equiv \frac{\mu}{2\beta N}\sum_{\substack{\alpha\cap\beta=\emptyset\\ \alpha\cup\beta\neq\emptyset}} Q^2_{\alpha,\beta}, \tag{4.21d}$$

where

$$z_0[\mathbf{Q}] \equiv \lim_{N\to\infty} z[\mathbf{Q}] = \left[\prod_{a=1}^{n}\int_0^{2\pi}\frac{e^{-2i\lambda^a}d\lambda^a}{2\pi}\right]\exp\left[\sum_{\substack{\alpha\cap\beta=\emptyset\\ \alpha\cup\beta\neq\emptyset}} Q_{\alpha,\beta}\, e^{i\sum_{a\in\alpha}\lambda^a+2i\sum_{b\in\beta}\lambda^b}\right] \tag{4.21e}$$

is the (leading) one-site partition function. It follows that $S$ is the leading term in $\mathcal{S}$. The $\Delta S^{\mathrm{T}}$ term contains the finite-size correction due to the re-exponentiation and to the one-site partition function $z$. Finally, the $\Delta S^{\varrho}$ term contains an additional contribution due to the finite-size corrections to $g_p$ appearing in Eq. (4.11). Observe, once again, that this contribution is absent in the case of flat distribution.

The integral in Eq. (4.21a) can now be evaluated using the saddle-point method. The saddle-point equation for $Q_{\alpha,\beta}$ is

$$\frac{Q_{\alpha,\beta}}{g_{|\alpha|+2|\beta|}} = 2\frac{\partial \ln z_0[\mathbf{Q}]}{\partial Q_{\alpha,\beta}} = 2\left\langle e^{i\sum_{a\in\alpha}\lambda^a+2i\sum_{b\in\beta}\lambda^b}\right\rangle_{z_0}, \tag{4.22}$$

where $\langle\bullet\rangle_{z_0}$ is the average performed respect to the one-site partition function $z_0$. Denoting by $\mathbf{Q}^{\mathrm{sp}}$ the solution of Eq. (4.22), the replicated partition function becomes

$$\overline{Z^n} \simeq e^{-N\mathcal{S}[\mathbf{Q}^{\mathrm{sp}}]-\frac{\ln\det\mathbf{\Omega}[\mathbf{Q}^{\mathrm{sp}}]}{2}}, \tag{4.23}$$

where $\mathbf{\Omega}$ is the Hessian matrix of $S[\mathbf{Q}]$ evaluated on the saddle-point $\mathbf{Q}^{\mathrm{sp}}$, solution of Eq. (4.22), i.e.,

$$\Omega_{\alpha\beta,\hat\alpha\hat\beta}[\mathbf{Q}^{\mathrm{sp}}] \equiv \sqrt{g_{|\alpha|+2|\beta|}g_{|\hat\alpha|+2|\hat\beta|}}\left.\frac{\partial^2 S[\mathbf{Q}]}{\partial Q_{\alpha,\beta}\partial Q_{\hat\alpha,\hat\beta}}\right|_{\mathbf{Q}=\mathbf{Q}^{\mathrm{sp}}} \tag{4.24}$$

The additional term $-1/2\ln\det\mathbf{\Omega}[\mathbf{Q}^{\mathrm{sp}}]$ provides, in general, a nontrivial finite-size correction to the leading term [MP87, PR02].

### 4.2.1 Replica symmetric ansatz and matching saddle-point

To proceed further, let us assume that a replica-symmetric ansatz holds, i.e., we search for a solution of the saddle-point equation in the form

$$Q_{\alpha,\beta} \equiv Q_{|\alpha|,|\beta|}. \tag{4.25}$$

This is a common and successful assumption in the study of random combinatorial optimization problems [MP85, MP86a, MP88] that greatly simplifies the calculation. The stability of a replica symmetric solution is however not obvious and, in general, a replica symmetry breaking might occur. Here this assumption will be justified *a posteriori*, on the basis of the agreement between the analytical predictions and the numerical computation. Using the replica symmetric hypothesis, the leading term in Eq. (4.21b) becomes

$$\begin{aligned} S[\mathbf{Q}] = & \sum_{p+q\geq 1} \binom{n}{p\ q} \frac{Q^2_{p,q}}{2g_{p+2q}} - 2\ln z_0[\mathbf{Q}] \\ & \xrightarrow{n\to 0} n \sum_{p+q\geq 1} \frac{(-1)^{p+q-1}}{p+q} \binom{p+q}{q} \frac{Q^2_{p,q}}{2g_{p+2q}} - 2n \lim_{n\to 0} \frac{\ln z_0[\mathbf{Q}]}{n}. \end{aligned} \tag{4.26}$$

In the previous expression, and in the following, we will adopt the notation

$$\binom{a}{b_1 \cdots b_s} \equiv \frac{\Gamma(a+1)}{\Gamma\left(a+1-\sum_{i=1}^s b_i\right)\prod_{i=1}^s \Gamma(b_i+1)}. \tag{4.27}$$

Even under the replica symmetric hypothesis, the evaluation of $\lim_{n\to 0} \frac{1}{n} \ln z_0[\mathbf{Q}]$ remains nontrivial. However, in B.1 we show that a special replica symmetric saddle-point solution exists, namely

$$Q^{\text{sp}}_{p,q} = \delta_{p,0} Q_{0,q} \equiv \delta_{p,0} Q_q, \tag{4.28}$$

corresponding to the replica-symmetric saddle-point solution of the RMP. This fact is not surprising: as anticipated in the Introduction, the aoc of the RFMP coincides with the aoc of the RMP in the $N \to +\infty$ limit, and, indeed, the evaluation of $z_0[\mathbf{Q}]$ on the matching saddle-point can be performed exactly, and coincides with the one of the RMP [MP85, MP87, PR02]

$$\ln z_0[\mathbf{Q}^{\text{sp}}] = n \int_{-\infty}^{+\infty} \left(e^{-e^x} - e^{-G(x)}\right) dx, \tag{4.29a}$$

$$G(x) \equiv \sum_{k=1}^{\infty} \frac{(-1)^{k-1}}{k!} Q_k e^{xk}. \tag{4.29b}$$

The saddle-point equations become

$$Q_{p,q} = \delta_{p,0} Q_q = \frac{\delta_{p,0}}{\beta} \int_{-\infty}^{+\infty} \frac{e^{qy-G(y)}}{q!} dy, \tag{4.30}$$

implying the self-consistent equation for $G$ given by

$$G(x) = \frac{1}{\beta} \int_{-\infty}^{+\infty} B(x+y) e^{-G(y)} dy, \tag{4.31}$$

$$B(x) \equiv \sum_{k=1}^{\infty} (-1)^{k-1} \frac{e^{kx}}{\Gamma^2(k+1)}. \tag{4.32}$$

As expected, the leading contribution corresponds therefore to the RMP free-energy,

$$\begin{aligned}\frac{1}{n}S[\mathbf{Q}^{\text{sp}}] &= \beta\sum_{k=1}^{\infty}(-1)^{k-1}Q_k^2 - 2\int_{-\infty}^{+\infty}dy\left[e^{-y}-e^{-G(y)}\right]\\ &= \int_{-\infty}^{+\infty}dy\,G(y)e^{-G(y)} - 2\int_{-\infty}^{+\infty}dy\left[e^{-y}-e^{-G(y)}\right].\end{aligned} \tag{4.33}$$

In the $\beta \to +\infty$ limit we can introduce

$$\hat{G}(x) \equiv G(\beta x). \tag{4.34}$$

and we can use, as shown in Appendix A.2 the fact that

$$\lim_{\beta\to+\infty} B(\beta x) = \theta(x). \tag{4.35}$$

We can solve for $\hat{G}$ as

$$\hat{G}(x) = \int_{-x}^{+\infty} e^{-\hat{G}(y)}dy \Longrightarrow \hat{G}(x) = \ln(1+e^x). \tag{4.36}$$

We finally obtain that, in both the considered formulations, the aoc of the RFMP is equal to the aoc of the RMP, as anticipated in the Introduction

$$\lim_{N\to+\infty} E_\varepsilon(\mu) = \frac{1}{2}\int_{-\infty}^{+\infty} dl\,\hat{G}(l)e^{-\hat{G}(l)} = \frac{\pi^2}{12}. \tag{4.37}$$

### 4.2.2 Finite-size corrections

In Eq. (4.23) three contributions to the finite-size corrections appear, namely $\Delta S^{\varrho}$, $\Delta S^{\mathrm{T}}$, and $1/2 \ln\det\mathbf{\Omega}$. The first contribution can be evaluated straightforwardly as

$$\begin{aligned}\Delta S^{\varrho}[\mathbf{Q}^{\text{sp}}] &= \frac{\mu}{2N}\sum_{\beta\neq\emptyset}Q_{|\beta|}^2 = \frac{\mu}{2N}\sum_{q=1}^{\infty}\binom{n}{q}Q_q^2 = \frac{n\mu}{2N}\sum_{q=1}^{\infty}\frac{(-1)^{q-1}}{q}Q_q^2 + o(n)\\ &= \frac{n\mu}{2N\beta}\int_{-\infty}^{+\infty}dy\int_0^{+\infty}dx\,G(y-x)e^{-G(y)} + o(n).\end{aligned} \tag{4.38}$$

In the $\beta \to +\infty$ limit we obtain

$$\lim_{\beta\to+\infty}\lim_{n\to 0}\frac{\Delta S^{\varrho}[\mathbf{Q}^{\text{sp}}]}{n\beta} = \frac{\mu}{2N}\int_{-\infty}^{+\infty}dy\int_0^{+\infty}dx\,\hat{G}(y-x)e^{-\hat{G}(y)} = \frac{\mu}{N}\zeta(3). \tag{4.39}$$

The last integration is performed explicitly in Appendix A.4. The $\Delta S^{\mathrm{T}}$ contribution depends on $\varepsilon$, i.e., on the presence or not of self-loops. In B.2 we show that

$$\lim_{\beta\to+\infty}\lim_{n\to 0}\frac{\Delta S^{\mathrm{T}}[\mathbf{Q}^{\text{sp}}]}{n\beta} = -\frac{\varepsilon-1}{4N}\zeta(2) - \frac{1}{N}\zeta(3). \tag{4.40}$$

Finally, the fluctuation contribution

$$\frac{\Delta E^{\mathbf{\Omega}}}{N} \equiv \lim_{\beta\to+\infty}\lim_{n\to 0}\frac{\ln\det\mathbf{\Omega}[\mathbf{Q}^{\text{sp}}]}{4n\beta}. \tag{4.41}$$

The expression of this fluctuation term is more involved than the corresponding one for the RMP, that has been studied in Refs. [MP87, PR02]. As it happens in that case, an exact evaluation through the replica formalism is quite complicated (we give some details in B.3). However, observe that $\Delta E^{\mathbf{\Omega}}$ depends on neither $\varepsilon$ nor $\mu$. We can therefore avoid the complex, direct evaluation and extract its value from Wästlund's formula in Eq. (4.2c) for the aoc in the loopy RFMP with exponentially distributed weights, comparing our result in this specific case with Wästlund's one. In particular, Eq. (4.2c) predicts no $1/N$ corrections in the loopy RFMP, implying the simple result

$$\Delta E^{\mathbf{\Omega}} = 0. \tag{4.42}$$

Moreover, in the spirit of the analysis in Ref. [AT78], the fact that $\Delta E^{\mathbf{\Omega}}$ is a finite and well-defined quantity also suggests that the Hessian $\mathbf{\Omega}$ remains positive definite within the replica symmetric ansatz for $\beta \to +\infty$, and, therefore, that the replica symmetric solution remains stable.

Collecting all contributions, we can finally write down a general expression for the aoc for the RFMP, and its finite-size corrections, as

$$E_\varepsilon(\mu) = \frac{\zeta(2)}{2} + \frac{1}{2N}\left[(\mu-1)\zeta(3) + \frac{1-\varepsilon}{4}\zeta(2)\right] + o\left(\frac{1}{N}\right). \tag{4.43}$$

## 4.3 Numerical results

The analytic results in Eq. (4.43) have been verified numerically using the `LEMON` graph library [DJK11]. For each one of the considered models, the results have been obtained averaging over at least $3 \cdot 10^6$ instances for each value of $N$. In Fig. 4.2 we show our results, both for the case of uniform weight distribution ($\mu = 0$) and the case of exponential weight distribution ($\mu = 1$). In both cases it is evident that the aoc of the RMP is greater than the corresponding aoc of the RFMP, and similarly $E_{-1}(\mu) \geq E_{+1}(\mu)$, as expected. The asymptotic formula, Eq. (4.43), has been verified for all cases. We performed moreover a parametric fit of our data using the fitting function

$$E = \frac{\pi^2}{12} + \frac{a}{2N} + \frac{b}{4N^2} + \frac{c}{8N^3} \tag{4.44a}$$

to verify our predictions for the RFMP in two ways, i.e., either assumiming all parameters free, or fixing $a$ equal to our analytical prediction (when available) to improve our evaluation of $b$ and $c$. We also performed, for comparison, a similar fit for the RMP, using the fitting function

$$E = \frac{\pi^2}{12} + \frac{a}{2N} + \frac{\hat{c}}{(2N)^{3/2}} + \frac{b}{4N^2}, \tag{4.44b}$$

to take into account the anomalous scaling of the corrections in this case [LPS17]. The results are summarized in Table 4.1. It is remarkable that the $N^{-3/2}$ correction introduced in Eq. (4.44a) for the RMP cost, that is numerically present in agreement with the results in Ref. [LPS17], is absent in all considered variants of the

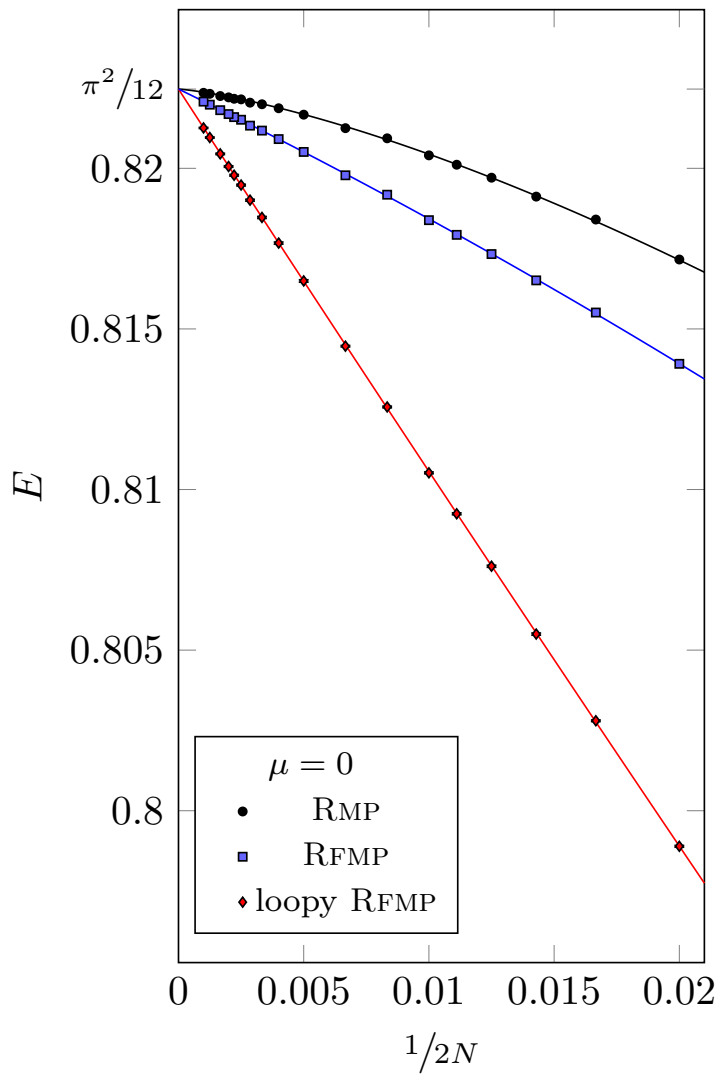

**(a)** $\mu = 0$ case. The smooth lines correspond to the fits obtained using the asymptotic theoretical predictions for both the asymptotic aoc and the $1/N$ corrections in the RFMP and the loopy RFMP in Eq. (4.44a) (see last column of Table 4.1). The black line corresponds to the fit obtained using Eq. (4.44b) for the RMP data (see Table 4.1).

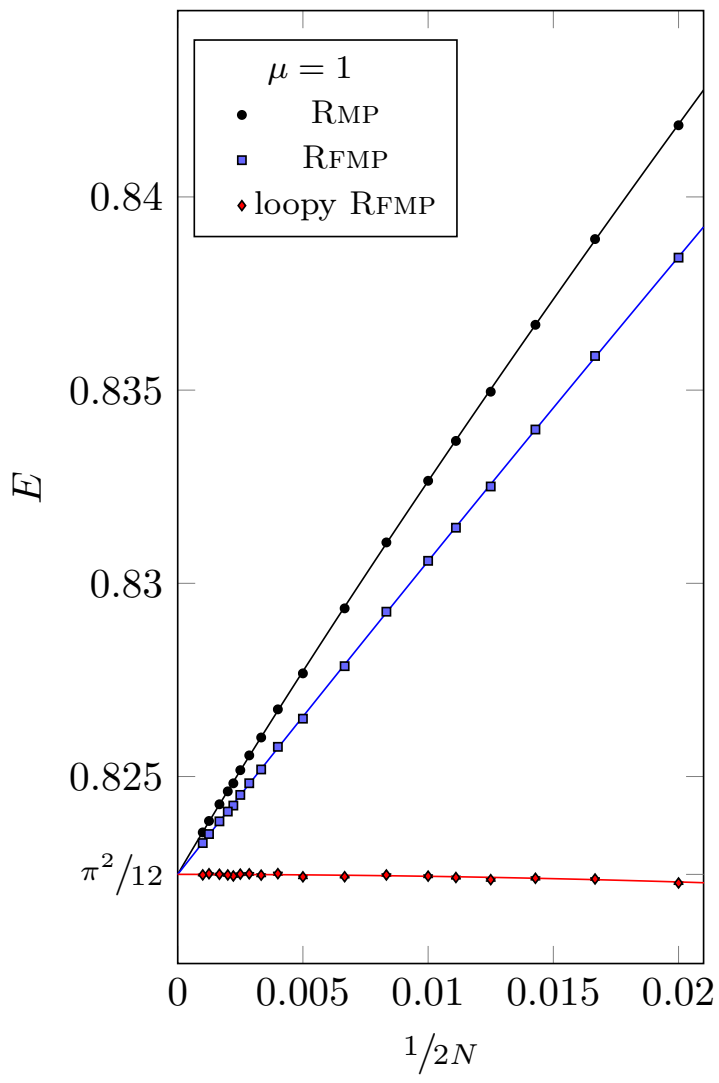

**(b)** $\mu = 1$ case. The blue line corresponds to the fit obtained using the asymptotic theoretical prediction for the RFMP in Eq. (4.44a) (see last column of Table 4.1). The red line is Wästlund's formula in Eq. (4.2c) for the loopy RFMP, that is exact for all values of $N$. Finally, the black line corresponds to the fit obtained using Eq. (4.44b) for the RMP data (see Table 4.1).

**Figure 4.2.** aoc for the RMP, the RFMP and the loopy RFMP in the case of both uniform (left panel) and exponential distribution (right panel) for the weights. Error bars are represented but smaller than the markers.

RFMP, as we numerically verified and as analytically predicted in Eq. (4.2c) for the loopy RFMP with exponential weight distribution.

Eq. (4.43) allows us to make predictions about differences of aoc for different types of models, due to the fact that we have isolated the different contributions depending on the presence of loops, or on the chosen distribution $\varrho$. For example, we expect that

$$\delta E_\ell := E_{-1}(\mu) - E_{+1}(\mu) = \frac{\zeta(2)}{4N} + o\left(\frac{1}{N}\right). \tag{4.45a}$$

Similarly, we have that

$$\delta E_\varrho := E_\varepsilon(1) - E_\varepsilon(0) = \frac{\zeta(3)}{2N} + o\left(\frac{1}{N}\right). \tag{4.45b}$$

Both the relations above have been verified numerically. Our results are shown in Fig. 4.3 and they are in agreeement with Eqs. (4.45).

| | Problem | Theoretical $a$ | $a, b, c/\hat{c}$ free $a$ | $b$ | $c$ or $\hat{c}$ | $b, c$ free $b$ | $c$ |
|---|---|---|---|---|---|---|---|
| $\mu = 0$ | Rmp | – | $-0.049(4)$ | $0.03(9)$ | $-1.54(4)$ | – | – |
| | Rfmp $\varepsilon = +1$ | $-1.2020569\dots$ | $-1.204(1)$ | $1.26(5)$ | $-1.0(4)$ | $1.19(2)$ | $-0.5(2)$ |
| | Rfmp $\varepsilon = -1$ | $-0.3795898\dots$ | $-0.381(1)$ | $-2.43(5)$ | $4.7(4)$ | $-2.50(2)$ | $5.2(2)$ |
| $\mu = 1$ | Rmp | – | $1.131(3)$ | $-0.03(8)$ | $-1.13(3)$ | – | – |
| | Rfmp $\varepsilon = +1$ | $0$ | $-0.001(1)$ | $-0.48(5)$ | $0.5(5)$ | $-0.53(2)$ | $0.9(2)$ |
| | Rfmp $\varepsilon = -1$ | $0.8224670\dots$ | $0.821(1)$ | $-1.15(6)$ | $0.5(5)$ | $-1.19(3)$ | $0.9(3)$ |

**Table 4.1.** Results of a fitting procedure of the aoc obtained numerically compared with the theoretical predictions. Wästlund's formula predicts $a = 0$ and $b = -c = -1/2$ for the aoc of the RFMP with $\mu = 1$ and $\epsilon = +1$.

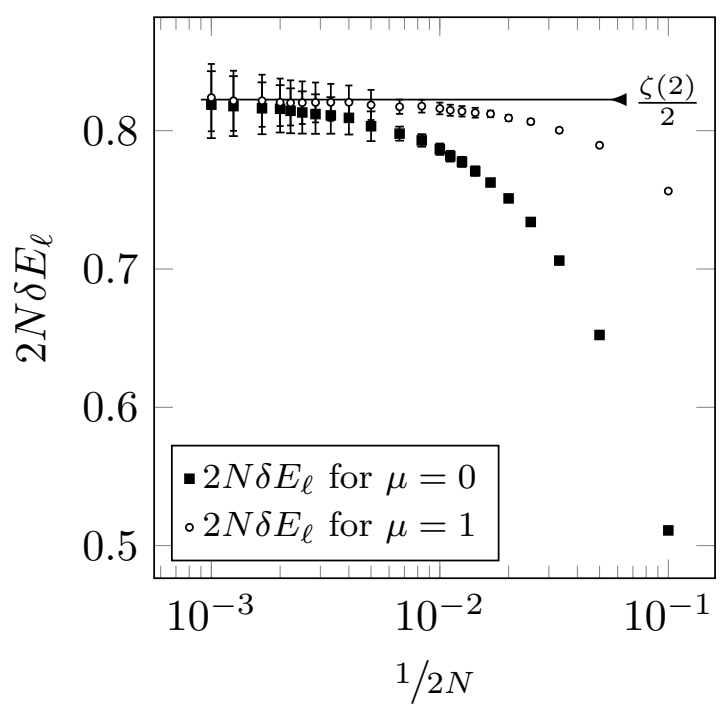

**(a)** Difference $\delta E_\ell$ between the aoc obtained with loops and the aoc obtained without loops, both for $\mu = 1$ (exponentially distributed weights) and for $\mu = 0$ (uniformly distributed weights). The smooth line is the predicted asymptotic behavior given Eq. (4.45a).

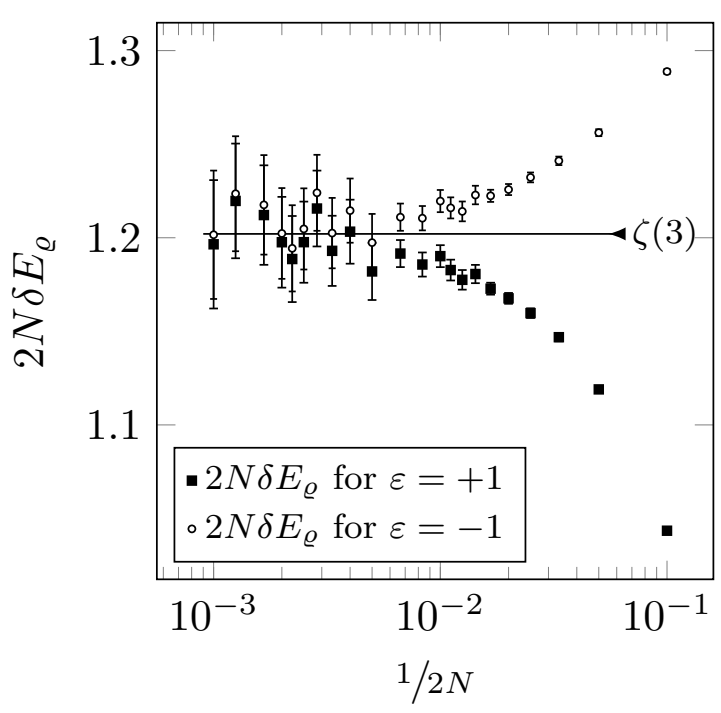

**(b)** Difference $\delta E_\varrho$ between the aoc obtained with exponentially distributed weights and the aoc obtained with uniformly distributed weights, both for $\epsilon = 1$ (with loops) and $\varepsilon = -1$ (without loops). The smooth line is the predicted asymptotic behavior given Eq. (4.45b).

**Figure 4.3.** Plots of the differences $\delta E_\ell$ and $\delta E_\varrho$ as functions of $N$. In both plots error bars are represented but smaller than the markers.

## 4.4 Conclusions

The RFMP on the complete graph $\mathcal{K}_{2N}$ generalizes the more famous RMP, allowing cycles in the optimal solution. Here we have studied, using the replica formalism, its aoc and we have compared it with the aoc of the RMP. We have considered the case of random weights on the graph edges distributed according to $\varrho(w) = 1 - \mu w + o(w)$, and we have evaluated the asymptotic expression of the aoc on the matching saddle-point, obtaining for the RFMP the same asymptotic aoc of the RMP. Remarkably enough, another saddle-point solution naturally appears in the calculation, corresponding to the RTSP solution, whose average optimal cost is however higher.

We have also explicitly obtained the finite-size corrections to the asymptotic aoc in two variations of the model, the standard RFMP and the loopy RFMP, in which each node can be matched to itself (loop). For the latter model, in particular, an explicit formula for the aoc had been found by Wästlund for any value of $N$ in the case of exponentially distributed weights [Wäs10]. The two models have different, non-trivial finite-size corrections.

By means of our approach, we have been able to separate the different contributions in the finite-size corrections coming from different details of the problem, namely a first one due to the possibility of having loops, a second one due to the chosen weight distribution and a third one that is independent from the aforementioned characteristics. The first and the second contributions have been explicitly derived, whereas the third one has been obtained comparing Wästlund's result with ours in the corresponding specific case. We have been able to write down an explicit expression for the aoc for all the considered cases up to $o(1/N)$ corrections, given in Eq. (4.3). We finally verified our results in all the considered cases comparing them with the values for the aoc obtained numerically. Going beyond the $1/N$ corrections, numerical results also suggest that the anomalous scaling $1/N^{3/2}$ in higher order corrections in the RMP, discussed in Ref. [LPS17] and verified in the present work, does not appear in the RFMP.

Despite the fact that an explicit expression for the aoc has been found for the RFMP in all the considered formulations of it, a number of open problems still remains. For example, a complete replica calculation of the contribution $\Delta E^{\mathbf{\Omega}}$ deriving from $\ln\det\mathbf{\Omega}$ is still missing. We have been able to estimate this contribution relying on Wästlund's results. However, an explicit replica calculation is still interesting for a series of reasons. One of them is that, as shown in B.3, $\Delta E^{\mathbf{\Omega}}$ appears to be the sum of two terms, the first one identical to a corresponding one appearing in the RMP that is known to be nonzero, the second one that we expect, from Wästlund's formula, to be exactly opposite to the former one, in such a way that $\Delta E^{\mathbf{\Omega}} = 0$. The presence of this cancellation suggests that the contribution of cycles plays an important role in the anomalous scaling of higher corrections in the RMP, and in its absence in the RFMP. Remembering also that an explicit formula for the $1/N$ correction in the RMP is still missing, further investigations in this direction are in order, to shed some light on the problem of finite-size corrections both in the RFMP and in the RMP even beyond the $O(1/N)$ terms.

# Chapter 5

# Replica-cavity connection

In this chapter we analyze the key connection between cavity and replica approaches for both the RMP and RTSP. While in the RMP both the replica and cavity method have permitted to derive average properties of the solution as long as finite-size corrections, instead in the case of the RTSP one has some unresolved technical complications using replicas [MP86a] when dealing with the limit of low-temperatures. This technical problem does not allow to write down direct expressions for the average optimal cost and finite-size corrections as can be done easily for the RMP. On the other hand the average optimal cost of the RTSP can be easily derived in the thermodynamic limit, by means of the cavity method [MP86b, KM89]; the result when $r = 0$ (flat or exponential distribution of the weights) is

$$\hat{E}_0 = \frac{1}{2} \int dy\, \hat{G}(y)\, (1 + \hat{G}(y))\, e^{-\hat{G}(y)} \tag{5.1}$$

with $\hat{G}(y)$ satisfying

$$\hat{G}(x) = \int_{-x}^{+\infty} dy\ (1 + \hat{G}(y))\, e^{-\hat{G}(y)}\,. \tag{5.2}$$

Notice the similarity with the corresponding equations of the RMP of Chap. 3. The expression (5.1) was proved to be analogous to the rigorous result of Wästlund [Wäs10] by [PW17].

In order to bypass those mathematical difficulties of replicas for the RTSP, it is in principle useful to derive a relation that connects quantities belonging to those two different words: replica and cavity. Here we prove that this relation is

$$q_k = \hat{g}_k\, \overline{m^k}\,, \tag{5.3}$$

where $q_k$ is the RS replica multi-overlap, whereas $\overline{m^k}$ are the moments of the cavity magnetization distribution function. This equation was already mentioned in [MP86b], but never proved, and it is true for both the RMP and the RTSP. We will prove (5.3) by using BP equations as introduced in Chap. 1. This approach is of course equivalent to standard cavity arguments which were obtained both for the RMP [PR01, MP86b] and RTSP [MP86b, KM89] using a representation of the partition function borrowed from the theory of polymers [Gen72]. For simplicity we limit ourselves to the case $r = 0$, but this result is valid also in the generic $r$ case. The rest of the chapter is organized as follows: in Section 5.1 we will start by analyzing the simpler RMP. In Section 5.2 we will prove (5.3) for the RTSP, using the BP equations of the random-link 2-factor problem, which is known to converge, for large number of points, to the RTSP [Fri02].

## 5.1 Matching warm-up

Given a generic graph $\mathcal{G} = (\mathcal{V}', \mathcal{E}')$ with weights $w_e$ for every $e \in \mathcal{E}$ and cardinality of the vertex set $|\mathcal{V}'| = N$, we can associate a factor graph representation $F(\mathcal{G}) = (\mathcal{V}, \mathcal{F}, \mathcal{E})$ in order to describe the matching problem using BP equations. This factor graph representation is depicted in Fig. 5.1 and can be obtained as follows. To every edge of the graph $\mathcal{G}$ one associates a variable node $e \in \mathcal{V}$ on which it lives an occupation number $n_e = \{0, 1\}$; to every vertex of the graph $\mathcal{G}$, instead, it corresponds a function node $a \in \mathcal{F}$ which checks that only one of the variable nodes $\underline{n}_{\partial a}$ incident on $a$ assumes the value 1. In addition, on every variable node $e$ an external field is present due to the presence of a random weight $w_e$ on every edge of the original graph. With those considerations in mind, one identifies the compatibility functions defined in (1.17) with

$$\psi_e(n_e) = e^{-\beta n_e w_e} \tag{5.4a}$$

$$\psi_a(\underline{n}_{\partial a}) = \mathbb{I}\left[\sum_{e \in \partial a} n_e = 1\right] \tag{5.4b}$$

and therefore, the probability measure of the matching problem is

$$\mu(\underline{n}) = \frac{1}{Z} \prod_{a \in \mathcal{F}} \mathbb{I}\left[\sum_{e \in \partial a} n_e = 1\right] \prod_{e \in \mathcal{V}} e^{-\beta n_e w_e} \tag{5.5}$$

Denoting by $a$ and $b$ the two factor nodes incident on the variable node $e$, the BP equations (1.44) are rewritten as

$$\nu_{e \to a}(n_e) \propto \hat{\nu}_{b \to e}(n_e)\, e^{-\beta n_e w_e}\,, \tag{5.6a}$$

$$\hat{\nu}_{a \to e}(n_e) \propto \sum_{\underline{n}_{\partial a \setminus e}} \mathbb{I}\left[\sum_{e \in \partial a} n_e = 1\right] \prod_{k \in \partial a \setminus e} \nu_{k \to a}(n_k)\,. \tag{5.6b}$$

Next we compute the ratio

$$\frac{\hat{\nu}_{a \to e}(0)}{\hat{\nu}_{a \to e}(1)} = \sum_{k \in \partial a \setminus e} \frac{\nu_{k \to a}(1)}{\nu_{k \to a}(0)} = \sum_{k \in \partial a \setminus e} \frac{\hat{\nu}_{c \to k}(1)}{\hat{\nu}_{c \to k}(0)} e^{-\beta w_k} \tag{5.7}$$

where, for every fixed $k$, we have denoted by $c$ the other factor node incident on $k$ together with $a$. The *cavity field* is defined as

$$h_{a \to e} \equiv -\frac{1}{\hat{\beta}} \ln\left(\frac{\hat{\nu}_{a \to e}(0)}{\hat{\nu}_{a \to e}(1)}\right) \tag{5.8}$$

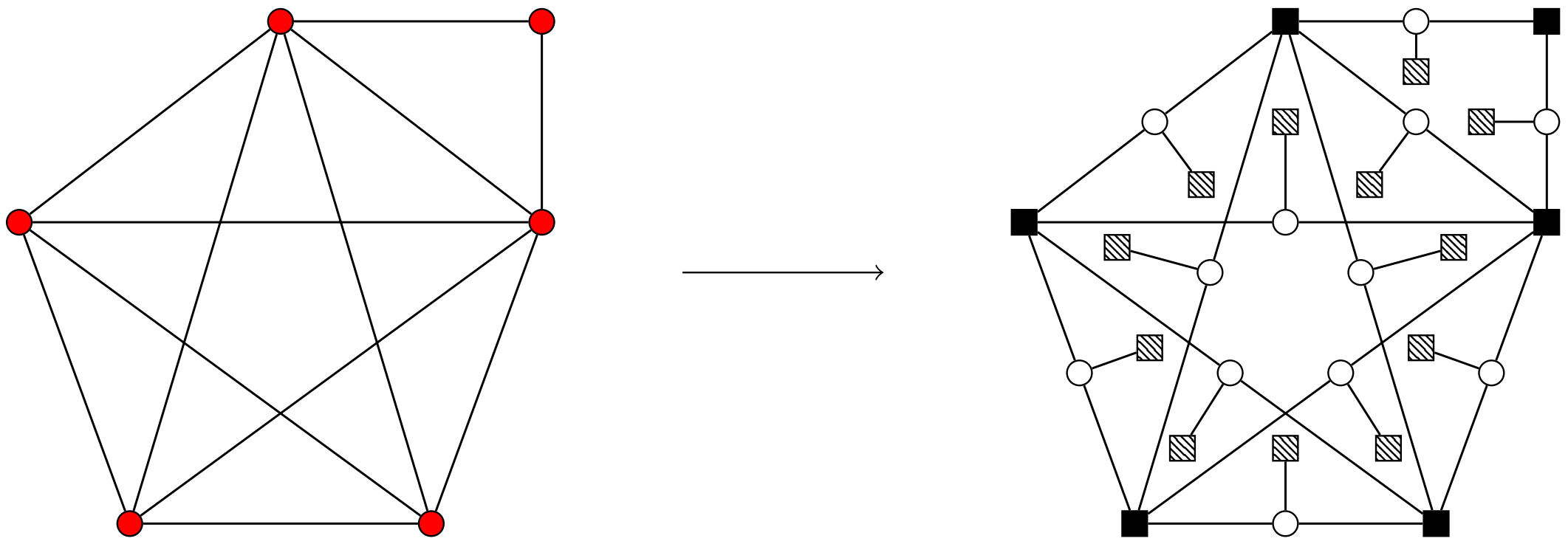

**Figure 5.1.** Factor graph representation of the matching and 2-factor problems for a given graph $\mathcal{G}$ with $N = 6$ vertices. Note how the vertices of the graph becomes interactions (factor nodes), whereas the edges of the graph become the variable nodes. For every variable nodes an external field is present due to the (random) weight of the edge and it is represented as a squared hatched inside.

and it satisfies the following self-consistent relation

$$h_{a\to e} = -\frac{1}{\hat{\beta}} \ln \sum_{k\in\partial a\setminus e} e^{-\hat{\beta}(N w_k - h_{c\to k})}\,, \tag{5.9}$$

where we have rescaled the temperature with $N$ as in (3.22). The previous equation is the analogous of cavity magnetization equations derived by Mézard and Parisi [MP86b]; those magnetizations are related to the cavity field by the simple change of variable $m = e^{\hat{\beta} h}$. In the low temperature limit $\hat{\beta} \to \infty$, the resulting cavity field is the minimum of the incoming ones

$$h_{a\to e} = \min_{k\in\partial a\setminus e} (N w_k - h_{c\to k}) \tag{5.10}$$

From now on we will focus on the problem defined on the fully connected topology $\mathcal{K}_N$ with $N$ even. In this case, easing the notation for convenience, equation (5.9) is rewritten as

$$h_e = -\frac{1}{\hat{\beta}} \ln \sum_{k=1}^{N-1} e^{-\hat{\beta}(N w_k - h_k)} \tag{5.11}$$

We can write a distributional version of last equation by introducing the probability density of the cavity fields $P(h)$

$$\begin{aligned} P(h) &= \int \prod_{k=1}^{N-1} dh_k\, dw_k\, P(h_k)\, \rho(w_k)\, \delta\left(h + \frac{1}{\hat{\beta}} \ln \sum_{k=1}^{N-1} e^{-\hat{\beta}(N w_k - h_k)}\right) \\ &= \int \frac{du\, d\hat{u}}{2\pi} e^{-iu\hat{u}} \delta\left(h + \frac{1}{\hat{\beta}} \ln u\right) \left[\int d\varphi\, dw\, P(\varphi)\, \rho(w)\, e^{i\hat{u} e^{-\hat{\beta}(Nw-\varphi)}}\right]^{N-1} . \end{aligned} \tag{5.12}$$

Expanding in series the exponential and performing the average over the disorder distribution $\rho(w)$ as in (3.23), we obtain

$$P(h) = \int \frac{du d\hat{u}}{2\pi} e^{-iu\hat{u}} \delta\left(h + \frac{\ln u}{\hat{\beta}}\right) \left[1 + \frac{1}{N} \sum_{k=1}^{\infty} \frac{(i\hat{u})^k}{k!} \hat{g}_k \int d\varphi\, P(\varphi)\, e^{\hat{\beta} k \varphi}\right]^{N-1} \tag{5.13}$$

and in the thermodynamic limit

$$P(h) \simeq \int \frac{du d\hat{u}}{2\pi} e^{-iu\hat{u}} \delta\left(h + \frac{\ln u}{\hat{\beta}}\right) e^{\sum_{k=1}^{\infty} \frac{(i\hat{u})^k}{k!} \hat{g}_k \int d\varphi\, P(\varphi)\, e^{\hat{\beta} k \varphi}} . \tag{5.14}$$

Introducing the integral representation of the delta function we have

$$\begin{aligned} P(h) &= \int \frac{du d\hat{u}}{2\pi} \frac{d\hat{h}}{2\pi} e^{-iu\hat{u} - ih\hat{h}} u^{-\frac{i\hat{h}}{\hat{\beta}}} \exp\left[\sum_{k=1}^{\infty} \frac{(i\hat{u})^k}{k!} \hat{g}_k \int d\varphi\, P(\varphi)\, e^{\hat{\beta} k \varphi}\right] \\ &= \int \frac{du d\hat{u}}{2\pi} \frac{d\hat{h}}{2\pi} \int_0^{\infty} dt\, e^{-iu\hat{u} - ih\hat{h} - tu} \frac{t^{\frac{i\hat{h}}{\hat{\beta}} - 1}}{\Gamma\left(\frac{i\hat{h}}{\hat{\beta}}\right)} \exp\left[\sum_{k=1}^{\infty} \frac{(i\hat{u})^k}{k!} \hat{g}_k \int d\varphi\, P(\varphi)\, e^{\hat{\beta} k \varphi}\right] \\ &= \int \frac{d\hat{h}}{2\pi} \int_0^{\infty} dt\, e^{-ih\hat{h}} \frac{t^{\frac{i\hat{h}}{\hat{\beta}} - 1}}{\Gamma\left(\frac{i\hat{h}}{\hat{\beta}}\right)} \exp\left[\sum_{k=1}^{\infty} \frac{(-t)^k}{k!} \hat{g}_k \int d\varphi\, P(\varphi)\, e^{\hat{\beta} k \varphi}\right] . \end{aligned} \tag{5.15}$$

Next we define

$$q_k \equiv \hat{g}_k \int d\varphi\, P(\varphi) e^{\hat{\beta} k \varphi} , \tag{5.16a}$$

$$\hat{G}(y) \equiv \sum_{k=1}^{\infty} (-1)^{k-1} q_k \frac{e^{\hat{\beta} y k}}{k!} \tag{5.16b}$$

and we prove a posteriori that those quantities are the equivalent respectively of equations (3.35) and (3.38) i.e. the RS replica overlap and its generating function. Using those definitions the probability distribution of the cavity field is written as

$$P(h) = \hat{\beta} \int \frac{d\hat{h}}{2\pi} dy \frac{e^{i\hat{h}(y-h)}}{\Gamma\left(\frac{i\hat{h}}{\hat{\beta}}\right)} e^{-\hat{G}(y)} \tag{5.17}$$

so that $q_k$ is, by definition

$$q_k = \hat{g}_k \int dh\, P(h)\, e^{\hat{\beta} k h} = \hat{\beta} \hat{g}_k \int dy \frac{e^{\hat{\beta} y k}}{\Gamma(k)} e^{-\hat{G}(y)} , \tag{5.18}$$

that is, we have correctly recovered replica formula (3.40). Therefore the RS replica overlaps $q_k$ are related to the moments of the cavity magnetizations [MP86b] via (5.3). In the low temperature limit the probability distribution can be also computed directly from (5.17) by using the fact that $\Gamma(x) \simeq 1/x$ when $x \to 0$

$$P(h) = \int dy \left[\frac{d}{dy} \int \frac{d\hat{h}}{2\pi} e^{i\hat{h}(y-h)}\right] e^{-\hat{G}(y)} = \hat{G}'(h)\, e^{-\hat{G}(h)} , \tag{5.19}$$

which coincides with the result in Ref. [PR01].

## 5.2 Traveling Salesman Problem

Let us now turn to the TSP. For a generic graph $\mathcal{G} = (\mathcal{V}', \mathcal{E}')$, with $|\mathcal{V}'| = N$, its factor graph representation $F(\mathcal{G}) = (\mathcal{V}, \mathcal{F}, \mathcal{E})$, is different from that of the matching problem for because one has to impose not only that on each vertices must depart two edges, but also that every configuration has only one cycle. The first condition guarantees only to find a valid 2-factor configuration. To enforce the unique cycle condition, let us consider the set of edges $\delta(\mathcal{S}) \subset \mathcal{E}'$ with $\mathcal{S} \subset \mathcal{V}'$ defined as follows

$$\delta(\mathcal{S}) \equiv \{e = (ij) \in \mathcal{V}';\ i \in \mathcal{S},\ j \in \mathcal{V}' \backslash \mathcal{S}\} \tag{5.20}$$

i.e. the set of edges that have one end in $\mathcal{S}$ and the other in $\mathcal{V}' \backslash \mathcal{S}$. It is now clear that one must have at least 2 edges departing from each $\mathcal{S}$ i.e.

$$\sum_{e \in \delta(\mathcal{S})} n_e \geq 2, \qquad \forall \mathcal{S} \subset \mathcal{V}',\ \mathcal{S} \neq \emptyset. \tag{5.21}$$

Because of those additional constraints (which are $2^N - 2$), the factor graph representation becomes much more involved. In the following we will not care about those additional constraints, because one can prove, by means both of replica arguments [MPV86] or by using rigorous results [Fri02] that the average optimal cost of the RTSP and of the random-link 2-factor problem is the same in the thermodynamic limit. For the 2-factor problem the factor graph representation is the same of the matching problem; the only difference is the form constraint in (5.4b), which is substituted by

$$\psi_a(\underline{n}_{\partial a}) = \mathbb{I}\left[\sum_{e \in \partial a} n_e = 2\right]. \tag{5.22}$$

The BP equations (1.44) for the 2-factor problem are

$$\nu_{e \to a}(n_e) \propto \hat{\nu}_{b \to e}(n_e)\, e^{-\beta n_e w_e} \tag{5.23a}$$

$$\hat{\nu}_{a \to e}(n_e) \propto \sum_{\underline{n}_{\partial a \backslash e}} \mathbb{I}\left[\sum_{e \in \partial a} n_e = 2\right] \prod_{k \in \partial a \backslash e} \nu_{k \to a}(n_k) \tag{5.23b}$$

Using the same notation of the previous section, the ratio of messages becomes

$$\frac{\hat{\nu}_{a \to e}(0)}{\hat{\nu}_{a \to e}(1)} = \frac{\sum_{\substack{k,l \in \partial a \backslash e \\ k \neq l}} \frac{\nu_{k \to a}(1)}{\nu_{k \to a}(0)} \frac{\nu_{l \to a}(1)}{\nu_{l \to a}(0)}}{\sum_{k \in \partial a \backslash e} \frac{\nu_{k \to a}(1)}{\nu_{k \to a}(0)}} = \frac{\sum_{\substack{k,l \in \partial a \backslash e \\ k \neq l}} \frac{\hat{\nu}_{c \to k}(1)}{\hat{\nu}_{c \to k}(0)} \frac{\hat{\nu}_{c \to l}(1)}{\nu_{c \to l}(0)} e^{-\beta w_k} e^{-\beta w_l}}{\sum_{k \in \partial a \backslash e} \frac{\hat{\nu}_{c \to k}(1)}{\hat{\nu}_{c \to k}(0)} e^{-\beta w_k}} \tag{5.24}$$

The cavity field (5.8) is therefore

$$h_{a \to e} = -\frac{1}{\hat{\beta}} \ln \left[\frac{\sum_{\substack{k,l \in \partial a \backslash e \\ k \neq l}} e^{-\hat{\beta}(N w_k - h_{c \to k})} e^{-\hat{\beta}(N w_l - h_{c \to l})}}{\sum_{k \in \partial a \backslash e} e^{-\hat{\beta}(N w_k - h_{c \to k})}}\right] \tag{5.25}$$

The previous equation is again the analogous of cavity magnetization equations derived by Mézard and Parisi [MP86b] and in the low-temperature limit as in [KM89] we have

$$h_{a\to e} = \mathrm{mmin}_{k\in\partial a\setminus e}\left(Nw_k - h_{c\to k}\right) . \tag{5.26}$$

where we have denoted by $\mathrm{mmin}_k(x_k)$ the second smallest element of the quantities $x_k$. Focusing on the fully connected topology we can proceed by writing a distributional equation for the probability density of cavity fields as we have done in the previous section

$$\begin{aligned} P(h) &= \int \prod_{k=1}^{N-1} dh_k\, dw_k\, P(h_k)\,\rho(w_k)\,\delta\left(h + \frac{1}{\hat\beta}\ln\frac{\sum_{k<l} e^{-\hat\beta(Nw_k-h_k)}e^{-\hat\beta(Nw_l-h_l)}}{\sum_k e^{-\hat\beta(Nw_k-h_k)}}\right) \\ &= \int \frac{du\, d\hat u}{2\pi}\frac{dv\, d\hat v}{2\pi} e^{-iu\hat u - iv\hat v}\delta\left(h + \frac{1}{\hat\beta}\ln\frac{v}{u}\right)\int\prod_{k=1}^{N-1} dh_k\, dw_k\, P(h_k)\,\rho(w_k) \\ &\times \exp\left[i\hat v \sum_{k<l} e^{-\hat\beta(Nw_k-h_k)}e^{-\hat\beta(Nw_l-h_l)} + i\hat u \sum_k e^{-\hat\beta(Nw_k-h_k)}\right] . \end{aligned} \tag{5.27}$$

Using an Hubbard-Stratonovich transformation

$$\exp\left[\frac{i\hat v}{2}\left(\sum_k e^{-\hat\beta(Nw_k-h_k)}\right)^2\right] = \int Dz\, \exp\left[\sqrt{i\hat v}\, z \sum_k e^{-\hat\beta(Nw_k-h_k)}\right] \tag{5.28}$$

we have

$$\begin{aligned} P(h) &= \int \frac{du\, d\hat u}{2\pi}\frac{dv\, d\hat v}{2\pi} Dz\, e^{-iu\hat u - iv\hat v}\delta\left(h + \frac{1}{\hat\beta}\ln\frac{v}{u}\right) \\ &\left[\int d\varphi\, dw\, P(\varphi)\,\rho(w)\exp\left[-\frac{1}{2}\left(\sqrt{i\hat v}\, e^{-\hat\beta(Nw-h)}\right)^2 + \sqrt{i\hat v}\, e^{\hat\beta(w-h)}\left(z + \frac{i\hat u}{\sqrt{i\hat v}}\right)\right]\right]^{N-1} . \end{aligned} \tag{5.29}$$

In the previous equation we notice the presence of the generating function of *Hermite polynomials*

$$e^{-\frac{t^2}{2}+xt} = \sum_{k=0}^{\infty} \mathrm{He}_k(x)\frac{t^k}{k!} , \tag{5.30}$$

so that we can easily perform the average over the disorder distribution $\rho(w)$, getting

$$\begin{aligned} P(h) &= \int \frac{du\, d\hat u}{2\pi}\frac{dv\, d\hat v}{2\pi} Dz\, e^{-iu\hat u - iv\hat v}\delta\left(h + \frac{1}{\hat\beta}\ln\frac{v}{u}\right) \\ &\times \left[1 + \frac{1}{N}\sum_{k=1}^{\infty}\frac{\left(\sqrt{i\hat v}\right)^k}{k!}\hat g_k\, \mathrm{He}_k\left(z + \frac{i\hat u}{\sqrt{i\hat v}}\right)\int d\varphi\, P(\varphi) e^{\hat\beta h k}\right]^{N-1} . \end{aligned} \tag{5.31}$$

In the limit of large number of points we obtain

$$P(h) = \int \frac{du\, d\hat{u}}{2\pi} \frac{dv\, d\hat{v}}{2\pi} Dz\, e^{-iu\hat{u}-iv\hat{v}} \delta\left(h + \frac{1}{\hat{\beta}} \ln \frac{v}{u}\right) e^{\sum_{k=1}^{\infty} \frac{(\sqrt{i\hat{v}})^k}{k!} \mathrm{He}_k\left(z+\frac{i\hat{u}}{\sqrt{i\hat{v}}}\right) q_k}, \tag{5.32}$$

where it was used the definition (5.16a) also in this case. We now show that $q_k$ corresponds to the RS multi-overlap of Ref. [MP86a]. Introducing the integral representation of the delta function and performing the $v$ and $\hat{v}$ integrals we have

$$P(h) = \int \frac{du\, d\hat{u}}{2\pi} \frac{d\hat{h}}{2\pi} \int_0^\infty dt\, e^{-iu\hat{u}-ih\hat{h}} \frac{u^{\frac{i\hat{h}}{\hat{\beta}}} t^{\frac{i\hat{h}}{\hat{\beta}}-1}}{\Gamma\left(\frac{i\hat{h}}{\hat{\beta}}\right)} \int Dz\, e^{\sum_{k=1}^{\infty} \frac{(i\sqrt{t})^k}{k!} \mathrm{He}_k\left(z+\frac{\hat{u}}{\sqrt{t}}\right) q_k}, \tag{5.33}$$

so that (5.16a) is by definition

$$\begin{aligned} q_p &= \hat{g}_p \int \frac{du\, d\hat{u}}{2\pi} \int_0^\infty dt\, e^{-iu\hat{u}} \frac{u^p t^{p-1}}{\Gamma(p)} \int Dz\, e^{\sum_{k=1}^{\infty} \frac{(i\sqrt{t})^k}{k!} \mathrm{He}_k\left(z+\frac{\hat{u}}{\sqrt{t}}\right) q_k} \\ &= \hat{g}_p \frac{d^p}{d\lambda^p} \int \frac{du\, d\hat{u}}{2\pi} \int_0^\infty dt\, e^{-iu(\hat{u}+i\lambda)} \frac{t^{p-1}}{\Gamma(p)} \int Dz\, e^{\sum_{k=1}^{\infty} \frac{(i\sqrt{t})^k}{k!} \mathrm{He}_k\left(z+\frac{\hat{u}}{\sqrt{t}}\right) q_k} \Bigg|_{\lambda=0} \\ &= \hat{g}_p \frac{d^p}{d\lambda^p} \int_0^\infty dt\, \frac{t^{p-1}}{\Gamma(p)} \int Dz\, e^{\sum_{k=1}^{\infty} \frac{(i\sqrt{t})^k}{k!} \mathrm{He}_k\left(z-\frac{i\lambda}{\sqrt{t}}\right) q_k} . \Bigg|_{\lambda=0} \end{aligned} \tag{5.34}$$

Next using the fact that for every function $f$ the following relation is valid

$$\frac{d^p}{d\lambda^p} f\left(\frac{\lambda}{a} + z\right)\Bigg|_{\lambda=0} = \frac{1}{a^p} \frac{d^p}{dz^p} f(z), \tag{5.35}$$

we have

$$q_p = (-1)^p \hat{g}_p \int_0^\infty \frac{dt}{t} \frac{(i\sqrt{t})^p}{\Gamma(p)} \int Dz\, \frac{d^p}{dz^p} \exp\left[\sum_{k=1}^{\infty} \frac{q_k}{k!} (i\sqrt{t})^k \mathrm{He}_k(z)\right]. \tag{5.36}$$

Using $p$ integrations by parts and the representation of the Hermite polynomials

$$\mathrm{He}_p(z) = (-1)^p e^{z^2/2} \frac{d^p}{dz^p} e^{-z^2/2}, \tag{5.37}$$

we finally get

$$q_p = (-1)^p \hat{g}_p \int_0^\infty \frac{dt}{t} \frac{(i\sqrt{t})^p}{\Gamma(p)} \int Dz\, \mathrm{He}_p(z) \exp\left[\sum_{k=1}^{\infty} \frac{q_k}{k!} (i\sqrt{t})^k \mathrm{He}_k(z)\right], \tag{5.38}$$

which is exactly the saddle point equation obtained using the replica method for the RTSP as in Ref. [MP86a, equation 3.7]. Therefore also for the RTSP equation (5.3) is valid.

Note that (5.38) is difficult to analyze, once introduced the generating function (3.38), in the low temperature limit. The distributional equation of the cavity fields at zero temperature, instead, is very easy to study and gives

$$P(h) = G'(h) G(h) e^{-G(h)}, \tag{5.39}$$

with $G(h)$ satisfying the saddle point equation (5.2).

# Part III

# Finite Dimension

# Chapter 6

# Euclidean TSP in one dimension

## 6.1 Introduction

The traveling salesman problem (TSP) is one of the most studied combinatorial optimization problems, because of the simplicity in its statement and the difficulty in its solution. Given $N$ cities and $N(N-1)/2$ values that represent the cost paid for traveling between all pairs of them, the TSP consists in finding the tour that visits all the cities and finally comes back to the starting point with the least total cost to be paid for the journey. The TSP is the archetypal problem in combinatorial optimization [Law+85]. Its first formalization can be probably traced back to the Austrian mathematician Karl Menger, in the 1930s [Men32], but it is yet extensively investigated. As it belongs to the class of NP-complete problems, see Karp and Steele in [Law+85], the study of the TSP could shed light on the famous P vs NP problem. Many problems in various fields of science (computer science, operational research, genetics, engineering, electronics and so on) and in everyday life (lacing shoes, Google maps queries, food deliveries and so on) can be mapped on a TSP or a variation of it, see for example Ref. [Rei94, Chap. 3] for a non-exhaustive list. Interestingly, the complexity of the TSP seems to remain high even if we try to modify the problem. For example, the Euclidean TSP, where the costs to travel from cities are the Euclidean distances between them, remains NP-complete [Pap77]. The bipartite TSP, where the cities are divided in two sub-sets and the tour has to alternate between them, is NP-complete too, as its Euclidean counterpart. Here we are interested in random Euclidean versions of the TSP as they were introduced in Sect. 1.4. We will focus on two types of graph, the complete $\mathcal{K}_N$ and the complete bipartite one $\mathcal{K}_{N,N}$.

Previous investigations [Car+14, CS15a, CS15b] suggested that the Euclidean matching problem is simpler to deal with in its bipartite version, at least in two dimensions. This idea encouraged us to consider first the bipartite TSP, starting from the one dimensional case, whose analysis has often enabled progress in the study of higher-dimensional cases. This will indeed, be again the case: the same problem in two dimension will be analyzed and solved in Chapter 8. On the complete graph $\mathcal{K}_N$, instead, the results found in one dimension, which we will present here, cannot be extended in higher dimensions as in the bipartite counterpart.

This chapter is organized as follows: in Sect. 6.2 we shall introduce a representation in terms of permutations of the two models, which we will be useful in order to find their solutions. In the complete bipartite case, in particular, one needs a couple of permutations to identify the configurations. In this way we also establish a very general connection between the bipartite TSP and a much simpler model, which is in the P complexity class, the assignment problem. Always using our representation, in Sect. 6.3 we can provide the explicit solution of the bipartite case for every instance of the disorder (that is, for every position of the points) in the one dimensional case when the cost is a convex and increasing function of the Euclidean distance between the cities. In particular this result will be valid in the $p > 1$ case for the cost function given in (1.94). In the same section, we exploit our explicit solution to compute the average optimal cost for an arbitrary number of points, when they are chosen with uniform distribution in the unit interval, and we present a comparison with the results of numerical simulations. In subsection 6.3.4 we discuss also the behavior of the cost in the thermodynamic limit of an infinite number of points. Here the results can be extended to more general distribution laws for the points.

In section 6.4 we shall analyze the complete graph case. When in the weights (1.93) we set $p > 1$ the solution directly follows from the complete bipartite results. In addition, we will solve completely the problem also when $0 < p < 1$, for every number of points $N$. In the $p < 0$ case, instead we will find that the solution (and the number of possible solutions for a given instance) depends on the value of $N$. Finally we compute the average optimal costs in the various cases and we compare them with numerical simulations.

## 6.2 Representations in terms of permutations

### Bipartite case

Here we shall consider the complete bipartite graph $\mathcal{K}_{N,N}$. Let $\mathcal{S}_N$ be the group of permutation of $N$ elements. For each $\sigma, \pi \in \mathcal{S}_N$, the sequence for $i \in [N]$

$$\begin{aligned} e_{2i-1} &= (r_{\sigma(i)}, b_{\pi(i)}) \\ e_{2i} &= (b_{\pi(i)}, r_{\sigma(i+1)}) \end{aligned} \tag{6.1}$$

where $\sigma(N+1)$ must be identified with $\sigma(1)$, defines a Hamiltonian cycle. More properly, it defines a Hamiltonian cycle $h \in \mathcal{H}$ with starting vertex $r_1 = r_{\sigma(1)}$ with a particular orientation, that is

$$h[(\sigma, \pi)] \equiv (r_1 b_{\pi(1)} r_{\sigma(2)} b_{\pi(2)} \cdots r_{\sigma(N)} b_{\pi(N)}) = (r_1 C), \tag{6.2}$$

where $C$ is an open walk which visit once all the blue points and all the red points with the exception of $r_1$. Let $C^{-1}$ be the open walk in opposite direction. This defines a new, dual, couple of permutations which generate the same Hamiltonian cycle

$$h[(\sigma, \pi)^\star] \equiv (C^{-1} r_1) = (r_1 C^{-1}) = h[(\sigma, \pi)], \tag{6.3}$$

since the cycle $(r_1 C^{-1})$ is the same as $(r_1 C)$ (traveled in the opposite direction). By definition

$$h[(\sigma,\pi)^\star] = (r_1 b_{\pi(N)} r_{\sigma(N)} b_{\pi(N-1)} r_{\sigma(N-1)} \cdots b_{\pi(2)} r_{\sigma(2)} b_{\pi(1)})\,. \tag{6.4}$$

Let us introduce the cyclic permutation $\tau \in \mathcal{S}_N$, which performs a left rotation, and the inversion $I \in \mathcal{S}_N$. That is $\tau(i) = i+1$ for $i \in [N-1]$ with $\tau(N) = 1$ and $I(i) = N+1-i$. In the following we shall denote a permutation by using the second raw in the usual two-raw notation, that is, for example $\tau = (2, 3, \cdots, N, 1)$ and $I = (N, N-1, \dots, 1)$. Then

$$h[(\sigma,\pi)^\star] = h[(\sigma \circ \tau \circ I, \pi \circ I)]\,. \tag{6.5}$$

There are $N!\,(N-1)!/2$ Hamiltonian cycles for $\mathcal{K}_{N,N}$. Indeed the couples of permutations are $(N!)^2$ but we have to divide them by $2N$ because of the $N$ different starting points and the two directions in which the cycle can be traveled.

### Complete case

Here we shall consider the complete graph $\mathcal{K}_N$ with $N$ vertices. This graph has $\frac{(N-1)!}{2}$ Hamiltonian cycles. Indeed, each permutation $\pi$ in the symmetric group of $N$ elements, $\pi \in \mathcal{S}_N$, defines a Hamiltonian cycle on $\mathcal{K}_N$. The sequence of points $(\pi(1), \pi(2), \dots, \pi(N), \pi(1))$ defines a closed walk with starting point $\pi(1)$, but the same walk is achieved by choosing any other vertex as starting point and the walk in the opposite order, that is, $(\pi(1), \pi(N), \dots, \pi(2), \pi(1))$ corresponds to the same Hamiltonian cycle. As the cardinality of $\mathcal{S}_N$ is $N!$ we get that the number of Hamiltonian cycles in $\mathcal{K}_N$ is $\frac{N!}{2 \cdot N}$.

There is another way to associate permutations to Hamiltonian cycles. Let $\pi^k := \pi \circ \pi^{k-1}$ for integer $k$ and $\pi^0$ be the identity function. Of course $\pi^N = \pi^0$ $\pi \in \mathcal{S}_N$. A permutation $\pi \in \mathcal{S}_N$ is said to be a $k$-cycle if it formed by a unique cycle of length $k$ and $N-k$ fixed points. There are $\frac{1}{k}\frac{N!}{(N-k)!}$ $k$-cycles in $\mathcal{S}_N$. Let us consider now the orbit of the point $j$ under the action of $\pi$, that is the sequence of points $(\pi^0, \pi, \pi^2, \dots, \pi^N)(j)$, with $j \in [N]$. This sequence defines a Hamiltonian cycle if and only if the permutation $\pi$ is an $N$-cycle. If $\pi$ is an $N$-cycle also $\pi^{-1}$ is an $N$-cycle. It provides the same closed walk in the opposite direction. As the cardinality of the $N$-cycles in $\mathcal{S}_N$ is $(N-1)!$ we get, once more, that the number of Hamiltonian cycles in $\mathcal{K}_N$ is $\frac{(N-1)!}{2}$.

## 6.3 Euclidean Bipartite TSP

### 6.3.1 Comparison with the assignment problem

From (6.1) and weights of the form (1.94), we get an expression for the total cost

$$E[h[(\sigma,\pi)]] = \sum_{i \in [N]} \left[ |r_{\sigma(i)} - b_{\pi(i)}|^p + |r_{\sigma \circ \tau(i)} - b_{\pi(i)}|^p \right]\,. \tag{6.6}$$

Now we can re-shuffle the sums and we get

$$\begin{aligned} E[h[(\sigma,\pi)]] &= \sum_{i\in[N]} |r_i - b_{\pi\circ\sigma^{-1}(i)}|^p + \sum_{i\in[N]} |r_i - b_{\pi\circ\tau^{-1}\circ\sigma^{-1}(i)}|^p \\ &= E[m(\pi\circ\sigma^{-1})] + E[m(\pi\circ\tau^{-1}\circ\sigma^{-1})] \end{aligned} \tag{6.7}$$

where $E[m(\lambda)]$ is the total cost of the assignment $m$ in $\mathcal{K}_{N,N}$ associated to the permutation $\lambda \in \mathcal{S}_N$

$$E[m(\lambda)] = \sum_{i\in[N]} |r_i - b_{\lambda(i)}|^p \,. \tag{6.8}$$

The duality transformation (6.5), that is

$$\sigma \to \sigma\circ\tau\circ I \tag{6.9}$$

$$\pi \to \pi\circ I\,, \tag{6.10}$$

interchanges the two matchings because

$$\mu_1 := \pi\circ\sigma^{-1} \to \pi\circ I\circ I\circ\tau^{-1}\circ\sigma^{-1} = \pi\circ\tau^{-1}\circ\sigma^{-1} \tag{6.11a}$$

$$\mu_2 := \pi\circ\tau^{-1}\circ\sigma^{-1} \to \pi\circ I\circ\tau^{-1}\circ I\circ\tau^{-1}\circ\sigma^{-1} = \pi\circ\sigma^{-1} \tag{6.11b}$$

where we used

$$I\circ\tau^{-1}\circ I = \tau\,. \tag{6.12}$$

The two matchings corresponding to the two permutations $\mu_1$ and $\mu_2$ have no edges in common and therefore each vertex will appear twice in the union of their edges. Remark also that

$$\mu_2 = \mu_1\circ\sigma\circ\tau^{-1}\circ\sigma^{-1} \tag{6.13}$$

which means that $\mu_1$ and $\mu_2$ are related by a permutation which has to be, as it is $\tau^{-1}$, a unique cycle of length $N$. It follows that, if $h^*$ is the optimal Hamiltonian cycle and $m^*$ is the optimal assignment,

$$E[h^*] \;\geq\; 2\,E[m^*]\,. \tag{6.14}$$

In the case of the Euclidean assignment the scaling of the average optimal cost is known in every dimensions and for every $p > 1$ [Car+14]:

$$\overline{E[\mu^*]} \sim \begin{cases} N^{1-\frac{p}{2}} & d=1\,; \\ N^{1-\frac{p}{2}}(\log N)^{\frac{p}{2}} & d=2\,; \\ N^{1-\frac{p}{d}} & d>2\,. \end{cases} \tag{6.15}$$

The scaling shows an anomalous behavior at lower dimension differently from what occurs for the matching problem on the complete graph $\mathcal{K}_N$ where in any dimension the scaling with the number of points is always $N^{1-\frac{p}{d}}$. Indeed, also for the monopartite Euclidean TSP (that is on $\mathcal{K}_N$) in [BHH59] it has been shown that for $p = 1$, in a finite region, with probability 1, the total cost scales according to $N^{1-\frac{p}{d}}$ in any dimension.

### 6.3.2 Solution in $d = 1$ for all instances

Here we shall concentrate on the one-dimensional case, where both red and blue points are chosen uniformly in the unit interval $[0, 1]$. In our analysis we shall make use of the results for the Euclidean assignment problem in one dimension of [BCS14] which have been obtained when in (1.94) is set $p > 1$. In this work it is showed that sorting both red and blue points in increasing order, the optimal assignment is defined by the identity permutation $\mathbb{1} = (1, 2, \dots, N)$. From now on, we will assume $p > 1$ and that both red and blue points are ordered, i.e. $r_1 \le \cdots \le r_N$ and $b_1 \le \cdots \le b_N$. Let

$$\tilde{\sigma}(i) = \begin{cases} 2i - 1 & i \le (N+1)/2 \\ 2N - 2i + 2 & i > (N+1)/2 \end{cases} \tag{6.16}$$

and

$$\tilde{\pi}(i) = \tilde{\sigma} \circ I(i) = \tilde{\sigma}(N + 1 - i) = \begin{cases} 2i & i < (N+1)/2 \\ 2N - 2i + 1 & i \ge (N+1)/2 \end{cases} \tag{6.17}$$

the couple $(\tilde{\sigma}, \tilde{\pi})$ will define a Hamiltonian cycle $\tilde{h} \in \mathcal{H}$. More precisely, according to the correspondence given in (6.1), it contains the edges for even $N$,

$$\tilde{e}_{2i-1} = \begin{cases} (r_{2i-1}, b_{2i}) & i \le N/2 \\ (r_{2N-2i+2}, b_{2N-2i+1}) & i > N/2 \end{cases} \tag{6.18a}$$

$$\tilde{e}_{2i} = \begin{cases} (b_{2i}, r_{2i+1}) & i < N/2 \\ (b_N, r_N) & i = N/2 \\ (b_{2N-2i+1}, r_{2N-2i}) & N/2 < i < N \\ (b_1, r_1) & i = N \end{cases} \tag{6.18b}$$

while for $N$ odd

$$\tilde{e}_{2i-1} = \begin{cases} (r_{2i-1}, b_{2i}) & i < (N-1)/2 \\ (r_N, b_N) & i = (N-1)/2 \\ (r_{2N-2i+2}, b_{2N-2i+1}) & i > (N-1)/2 \end{cases} \tag{6.19a}$$

$$\tilde{e}_{2i} = \begin{cases} (b_{2i}, r_{2i+1}) & i < (N-1)/2 \\ (b_{2N-2i+1}, r_{2N-2i}) & (N-1)/2 < i < N \\ (b_1, r_1) & i = N\,. \end{cases} \tag{6.19b}$$

The main ingredient of our analysis is the following

**Proposition 6.3.1.** *For a convex and increasing cost function the optimal Hamiltonian cycle is provided by $\tilde{h}$.*

This cycle is the analogous of the *criss-cross* solution introduced by Halton [Hal95] (see Fig. 6.1). In his work, Halton studied the optimal way to lace a shoe. This problem can be seen as a peculiar instance of a 2-dimensional bipartite Euclidean TSP with the parameter which tunes the cost $p = 1$. One year later, Misiurewicz [Mis96] generalized Halton's result giving the least restrictive requests on the 2-dimensional TSP instance to have the criss-cross cycle as solution. Other generalizations of these works have been investigated in more recent papers [Pol02, GT17]. We will show that the same criss-cross cycle has the lowest cost for the Euclidean bipartite TSP in one dimension, provided that $p > 1$. To do this, we will prove in a novel way the optimality of the criss-cross solution, suggesting two moves that lower the energy of a tour and showing that the only Hamiltonian cycle that cannot be modified by these moves is $\tilde{h}$.

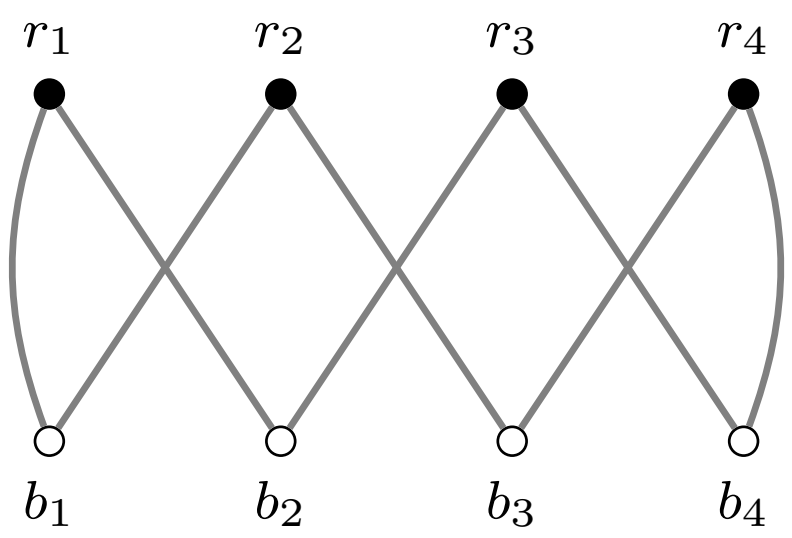

**Figure 6.1.** The optimal Hamiltonian cycle $\tilde{h}$ for $N = 4$ blue and red points chosen in the unit interval and sorted in increasing order.

We shall make use of the following moves in the ensemble of Hamiltonian cycles. Given $i, j \in [N]$ with $j > i$ we can partition each cycle as

$$h[(\sigma,\pi)] = (C_1 r_{\sigma(i)} b_{\pi(i)} C_2 b_{\pi(j)} r_{\sigma(j+1)} C_3), \tag{6.20}$$

where the $C_i$ are open paths in the cycle, and we can define the operator $R_{ij}$ that exchanges two blue points $b_{\pi(i)}$ and $b_{\pi(j)}$ and reverses the path between them as

$$\begin{aligned} h[R_{ij}(\sigma,\pi)] &\equiv (C_1 r_{\sigma(i)} [b_{\pi(i)} C_2 b_{\pi(j)}]^{-1} r_{\sigma(j+1)} C_3) \\ &= (C_1 r_{\sigma(i)} b_{\pi(j)} C_2^{-1} b_{\pi(i)} r_{\sigma(j+1)} C_3)\,. \end{aligned} \tag{6.21}$$

Analogously by writing

$$h[(\sigma,\pi)] = (C_1 b_{\pi(i-1)} r_{\sigma(i)} C_2 r_{\sigma(j)} b_{\pi(j)} C_3) \tag{6.22}$$

we can define the corresponding operator $S_{ij}$ that exchanges two red points $r_{\sigma(i)}$ and $r_{\sigma(j)}$ and reverses the path between them

$$\begin{aligned} h[S_{ij}(\sigma,\pi)] &\equiv (C_1 b_{\pi(i-1)} [r_{\sigma(i)} C_2 r_{\sigma(j)}]^{-1} b_{\pi(j)} C_3) \\ &= (C_1 b_{\pi(i-1)} r_{\sigma(j)} C_2^{-1} r_{\sigma(i)} b_{\pi(j)} C_3)\,. \end{aligned} \tag{6.23}$$

Two couples of points $(r_{\sigma(k)}, r_{\sigma(l)})$ and $(b_{\pi(j)}, b_{\pi(i)})$ have the same orientation if $(r_{\sigma(k)} - r_{\sigma(l)})(b_{\pi(j)} - b_{\pi(i)}) > 0$. Remark that as we have ordered both set of points this means also that $(\sigma(k), \sigma(l))$ and $(\pi(j), \pi(i))$ have the same orientation.

Then

**Lemma 1.** *Let $E[(\sigma,\pi)]$ be the cost defined in (6.6). Then $E[R_{ij}(\sigma,\pi)] - E[(\sigma,\pi)] > 0$ if the couples $(r_{\sigma(j+1)}, r_{\sigma(i)})$ and $(b_{\pi(j)}, b_{\pi(i)})$ have the same orientation and $E[S_{ij}(\sigma,\pi)] - E[(\sigma,\pi)] > 0$ if the couples $(r_{\sigma(j)}, r_{\sigma(i)})$ and $(b_{\pi(j)}, b_{\pi(i-1)})$ have the same orientation.*

*Proof.*

$$E[R_{ij}(\sigma,\pi)] - E[(\sigma,\pi)] = w_{(r_{\sigma(i)},b_{\pi(j)})} + w_{(b_{\pi(i)},r_{\sigma(j+1)})} - w_{(r_{\sigma(i)},b_{\pi(i)})} - w_{(b_{\pi(j)},r_{\sigma(j+1)})} \tag{6.24}$$

and this is the difference between two matchings which is positive if the couples $(r_{\sigma(j+1)}, r_{\sigma(i)})$ and $(b_{\pi(j)}, b_{\pi(i)})$ have the same orientation (as shown in [McC99, BCS14] for a weight which is an increasing convex function of the Euclidean distance). The remaining part of the proof is analogous. ■

**Lemma 2.** *The only couples of permutations $(\sigma,\pi)$ with $\sigma(1) = 1$ such that both $(\sigma(j+1),\sigma(i))$ have the same orientation as $(\pi(j),\pi(i))$ and $(\pi(j),\pi(i-1))$ and $(\sigma(j),\sigma(i))$, for each $i,j \in [N]$ are $(\tilde{\sigma},\tilde{\pi})$ and its dual $(\tilde{\sigma},\tilde{\pi})^\star$.*

*Proof.* We have to start our Hamiltonian cycle from $r_{\sigma(1)} = r_1$. Next we look at $\pi(N)$, if we assume now that $\pi(N) > 1$, there will be a $j$ such that our cycle would have the form $(r_1 C_1 r_{\sigma(j)} b_1 C_2 b_{\pi(N)})$, if we assume $j > 1$ then $(1,\sigma(j))$ and $(\pi(N),1)$ have opposite orientation, so that necessarily $\pi(N) = 1$. In the case $j = 1$ our Hamiltonian cycle is of the form $(r_1 b_1 C)$, that is $(b_1 C r_1)$, and this is exactly of the other form if we exchange red and blue points. We assume that it is of the form $(r_1 C b_1)$; the other form would give, at the end of the proof, $(\tilde{\sigma},\tilde{\pi})^\star$. Now we shall proceed by induction. Assume that our Hamiltonian cycle is of the form $(r_1 b_2 r_3 \cdots x_k C y_k \cdots b_3 r_2 b_1)$ with $k < N$, where $x_k$ and $y_k$ are, respectively, a red point and a blue point when $k$ is odd and vice versa when $k$ is even. Then $y_{k+1}$ and $x_{k+1}$ must be in the walk $C$. If $y_{k+1}$ it is not the point on the right of $x_k$ the cycle has the form $(r_1 b_2 r_3 \cdots x_k y_s C_1 y_{k+1} x_l \cdots y_k \cdots b_3 r_2 b_1)$ but then $(x_l, x_k)$ and $(y_{k+1}, y_s)$ have opposite orientation, which is impossible, so that $s = k+1$, that is the point on the right of $x_k$. Where is $x_{k+1}$? If it is not the point on the left of $y_k$ the cycle has the form $(r_1 b_2 r_3 \cdots x_k y_{k+1} \cdots y_l x_{k+1} C_1 x_s \cdots y_k \cdots b_3 r_2 b_1)$, but then $(x_s, x_{k+1})$ and $(y_k, y_l)$ have opposite orientation, which is impossible, so that $s = k+1$, that is the point on the left of $y_k$. We have now shown that the cycle has the form $(r_1 b_2 r_3 \cdots y_{k+1} C x_{k+1} \cdots b_3 r_2 b_1)$ and can proceed until $C$ is empty. ■

The case with $N = 3$ points is explicitly investigated in Appendix C.1.

Now that we have understood what is the optimal Hamiltonian cycle, we can look in more details at what are the two matchings which enter in the decomposition we used in (6.7). As $\tilde{\pi} = \tilde{\sigma} \circ I$ we have that

$$I = \tilde{\sigma}^{-1} \circ \tilde{\pi} = \tilde{\pi}^{-1} \circ \tilde{\sigma}. \tag{6.25}$$

As a consequence both permutations associated to the matchings appearing in (6.7) for the optimal Hamiltonian cycle are involutions:

$$\tilde{\mu}_1 \equiv \tilde{\pi} \circ \tilde{\sigma}^{-1} = \tilde{\sigma} \circ I \circ \tilde{\sigma}^{-1} = \tilde{\sigma} \circ \tilde{\pi}^{-1} = \left[\tilde{\pi} \circ \tilde{\sigma}^{-1}\right]^{-1} \tag{6.26a}$$

$$\tilde{\mu}_2 \equiv \tilde{\pi} \circ \tau^{-1} \circ \tilde{\sigma}^{-1} = \tilde{\sigma} \circ I \circ \tau^{-1} \circ I \circ \tilde{\pi}^{-1} = \left[\tilde{\pi} \circ \tau^{-1} \circ \tilde{\sigma}^{-1}\right]^{-1}, \tag{6.26b}$$

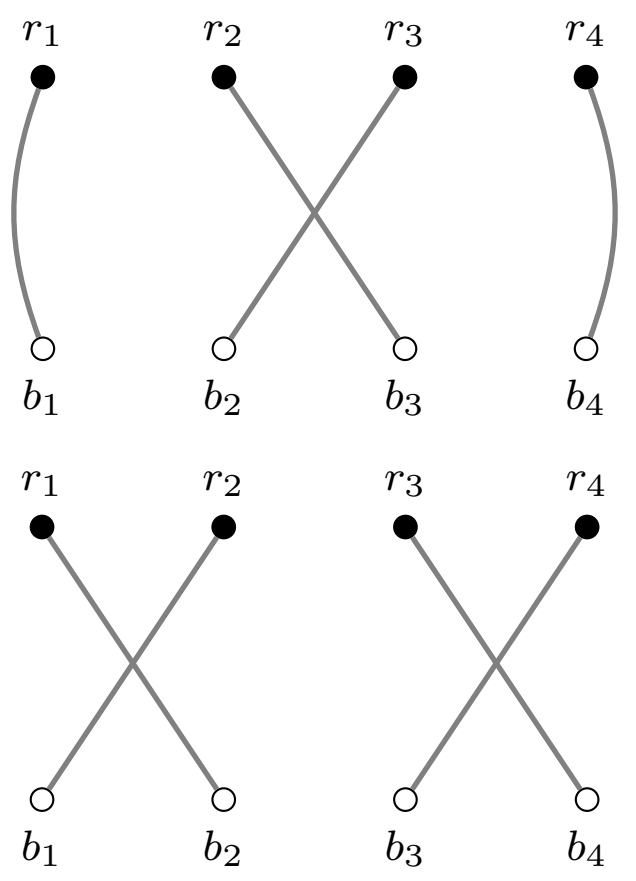

**(a)** $N = 4$.

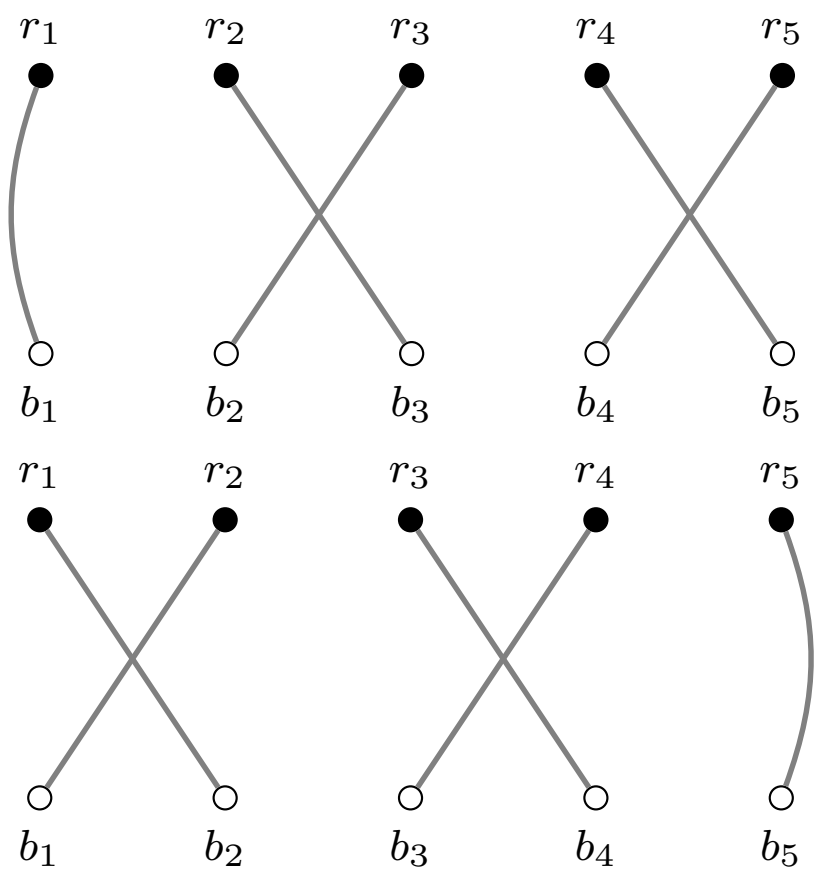

**(b)** $N = 5$.

**Figure 6.2.** Decomposition of the optimal Hamiltonian cycle $\tilde{h}$ in two disjoint matchings $\tilde{\mu}_1$ and $\tilde{\mu}_2$ for $N = 4$ (left panel) and $N = 5$ (right panel).

where we used (6.12). This implies that those two permutations have at most cycles of period two, a fact which reflects a symmetry by exchange of red and blue points. When $N$ is odd it happens that

$$I \circ \tilde{\sigma} \circ I = \tilde{\sigma} \circ \tau^{-\frac{N-1}{2}}, \tag{6.27}$$

so that

$$I \circ \tilde{\pi} \circ I = I \circ \tilde{\sigma} \circ I \circ I = \tilde{\sigma} \circ \tau^{-\frac{N-1}{2}} \circ I = \tilde{\pi} \circ I \circ \tau^{-\frac{N-1}{2}} \circ I = \tilde{\pi} \circ \tau^{\frac{N-1}{2}}. \tag{6.28}$$

It follows that the two permutations in (6.26a) and (6.26b) are conjugate by $I$

$$I \circ \tilde{\pi} \circ \tau^{-1} \circ \tilde{\sigma}^{-1} \circ I = \tilde{\pi} \circ \tau^{\frac{N-1}{2}} \circ \tau \circ \tau^{\frac{N-1}{2}} \circ \tilde{\sigma}^{-1} = \tilde{\pi} \circ \tilde{\sigma}^{-1} \tag{6.29}$$

so that, in this case, they have exactly the same numbers of cycles of order 2. Indeed we have

$$\tilde{\mu}_1 = (2, 1, 4, 3, 6, \dots, N-1, N-2, N) \tag{6.30a}$$
$$\tilde{\mu}_2 = (1, 3, 2, 5, 4, \dots N, N-1) \tag{6.30b}$$

and they have $\frac{N-1}{2}$ cycles of order 2 and 1 fixed point. See Fig. 6.2b for the case $N = 5$.

In the case of even $N$ the two permutations have not the same number of cycles of order 2, indeed one has no fixed point and the other has two of them. More explicitly

$$\tilde{\mu}_1 = (2, 1, 4, 3, 6, \dots, N, N-1) \tag{6.31a}$$
$$\tilde{\mu}_2 = (1, 3, 2, 5, 4, \dots N-1, N-2, N) \tag{6.31b}$$

See Fig. 6.2a for the case $N = 4$.

### 6.3.3 Evaluation of the cost

Here we will evaluate the cost of the optimal Hamiltonian cycle $\tilde{h}$ for $\mathcal{K}_{N,N}$,

$$E_{N,N}(\tilde{h}) = |r_1 - b_1|^p + |r_N - b_N|^p + \sum_{i=1}^{N-1} \left[|b_{i+1} - r_i|^p + |r_{i+1} - b_i|^p\right]. \quad (6.32)$$

Assume that both red and blue points are chosen according to the law $\rho$ and let

$$\Phi_\rho(x) := \int_0^x ds\, \rho(s) \quad (6.33)$$

be its *cumulative*. The probability that, chosen $N$ points at random, the $k$-th is in $[x, x+dx]$ is

$$P_k(x)dx = k \binom{N}{k} \Phi_\rho^{k-1}(x) \left[1 - \Phi_\rho(x)\right]^{N-k} \rho(x)dx \quad (6.34)$$

Given two sequences of $N$ points, the probability for the difference $\phi_k$ in the position between the $(k+1)$-th and the $k$-th points is

$$\begin{aligned} &\Pr{}_\rho\left[\phi_k \in d\phi\right] = d\phi_k \int dx\, dy\, P_k(x) P_{k+1}(y) \delta(\phi_k - y + x) \\ &= k(k+1)\binom{N}{k}\binom{N}{k+1} d\phi_k \int dx\, dy\, \rho(x)\, \rho(y) \delta(\phi_k - y + x) \\ &\times \Phi_\rho(y)\, \left[1 - \Phi_\rho(x)\right] \left[\Phi_\rho(x)\Phi_\rho(y)\right]^{k-1} \left[(1 - \Phi_\rho(x))\, (1 - \Phi_\rho(y))\right]^{N-k-1}. \end{aligned} \quad (6.35)$$

Let us now focus on the simple case in which the law $\rho$ is flat, where $\Phi_\rho(x) = x$ and (6.34) reduces to

$$P_k(x) = \frac{\Gamma(N+1)}{\Gamma(k)\,\Gamma(N-k+1)}\, x^{k-1}(1-x)^{N-k}. \quad (6.36)$$

We we also limit ourself to the simple case $p = 2$. We remind to Appendix E.2 for the evaluation of the generic $p$ case, which needs the use of the *Selberg integrals*. We obtain

$$\overline{|r_1 - b_1|^2} = \overline{|r_N - b_N|^2} = \frac{2N}{(N+1)^2(N+2)} \quad (6.37)$$

whereas

$$\begin{aligned} \overline{|b_{k+1} - r_k|^2} = \overline{|r_{k+1} - b_k|^2} &= \int_0^1 dx\, dy\, P_k(x)\, P_{k+1}(y)\, (x-y)^2 \\ &= \frac{2(k+1)(N-k+1)}{(N+1)^2(N+2)} \end{aligned} \quad (6.38)$$

so that

$$\sum_{k=1}^{N-1} \frac{2(k+1)(N-k+1)}{(N+1)^2(N+2)} = \frac{1}{3}\frac{(N+6)(N-1)}{(N+1)(N+2)}. \quad (6.39)$$

In conclusion, the average cost for the flat distribution and $p = 2$ is exactly

$$\overline{E^{(2)}_{N,N}} = \frac{2}{3}\frac{N^2 + 4N - 3}{(N+1)^2}. \quad (6.40)$$

If we recall that for the assignment the average optimal total cost is exactly $\frac{1}{3}\frac{N}{N+1}$, the difference between the average optimal total cost of the bipartite TSP and twice the assignment is

$$\frac{2}{3}\left[\frac{N^2+4N-3}{(N+1)^2}-\frac{N}{N+1}\right]=\frac{1}{3}\frac{N-1}{(N+1)^2}\geq 0 \tag{6.41}$$

and vanishes for infinitely large $N$. Remark that the limiting value is reached from above for the TSP and from below for the assignment. We plot in Fig. 6.3 the numerical results of the average optimal cost for different number of points.

It is also interesting to look at the contribution from the two different matchings in which we have subdivided the optimal Hamiltonian cycle. In the case of $N$ odd we have for one of them the average cost

$$\frac{2N}{(N+1)^2(N+2)}+2\sum_{k=1}^{\frac{N-1}{2}}\frac{4k(N-2k+2)}{(N+1)^2(N+2)}=\frac{1}{3}\frac{N^2+4N-3}{(N+1)^2} \tag{6.42}$$

and also for the other

$$\frac{2N}{(N+1)^2(N+2)}+2\sum_{k=1}^{\frac{N-1}{2}}\frac{2(2k+1)(N-2k+1)}{(N+1)^2(N+2)}=\frac{1}{3}\frac{N^2+4N-3}{(N+1)^2}\,. \tag{6.43}$$

In the case of $N$ even we have for the matching with two fixed points the average cost

$$\frac{4N}{(N+1)^2(N+2)}+2\sum_{k=1}^{\frac{N-2}{2}}\frac{2(2k+1)(N-2k+1)}{(N+1)^2(N+2)}=\frac{1}{3}\frac{N^2+4N-6}{(N+1)^2}\,, \tag{6.44}$$

while for the other with no fixed points

$$2\sum_{k=1}^{\frac{N-2}{2}}\frac{4k(N-2k+2)}{(N+1)^2(N+2)}=\frac{1}{3}\frac{N^2+4N}{(N+1)^2}, \tag{6.45}$$

which then has a cost higher at the order $N^{-2}$.

### 6.3.4 Asymptotic analysis for the optimal average cost

Motivated by the preceding discussion, one can try to perform a more refined analysis in the thermodynamic limit. In the asymptotic regime of large $N$, in fact, only the term with a sum on $i$ in (6.32) will contribute, and each of the two terms will provide an equal optimal matching contribution. Proceeding as in the case of the assignment [BCS14, CS14], one can show that the random variables $\phi_k$ defined above Eq. (6.35) converge (in a weak sense specified by Donsker's theorem) to $\phi(s)$, which is a difference of two Brownian bridge processes [CDS17]. One can write the re-scaled average optimal cost as

$$\overline{E_p}\equiv\lim_{N\to\infty}N^{\frac{p}{2}-1}\,\overline{E^{(p)}_{N,N}} \tag{6.46}$$

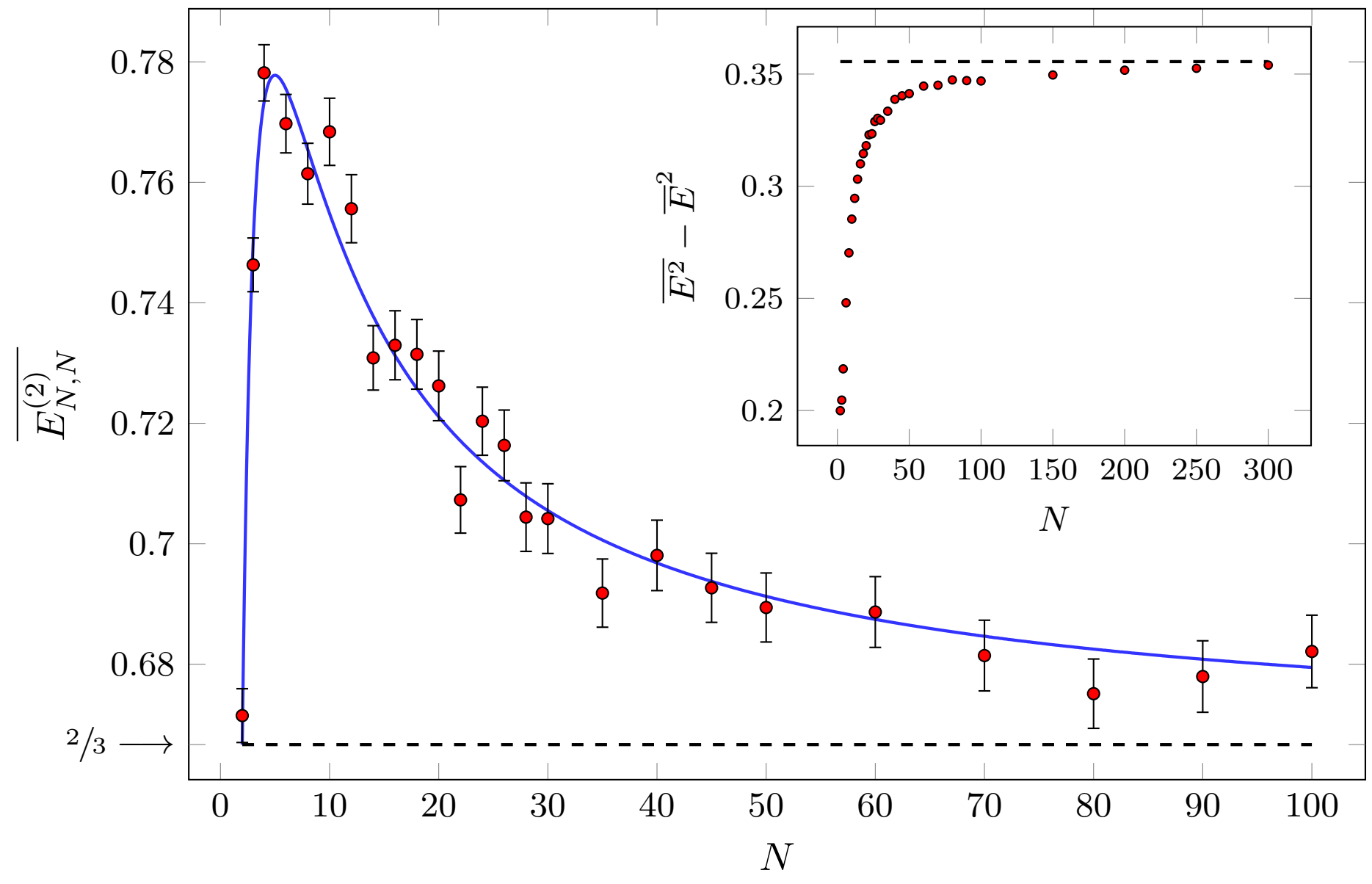

**Figure 6.3.** Numerical results for $\overline{E^{(2)}_{N,N}}$ for several values of $N$. The continuous line represents the exact prediction given in (6.40) and the dashed line gives the value for infinitely large $N$. For every $N$ we have used $10^4$ instances. In the inset we show the numerical results for the variance of the cost $E^{(2)}_{N,N}$ obtained using the exact solution provided by (6.16) and (6.17). The dashed line represents the theoretical large $N$ asymptotic value. Error bars are also plotted but they are smaller than the mark size.

where we have denoted with a bar $\overline{\cdot}$ the average over all the instances. By starting at finite $N$ with the representation (6.35), the large $N$ limit can be obtained setting $k = Ns + \frac{1}{2}$ and introducing the variables $\xi$, $\eta$ and $\varphi$ such that

$$x = s + \frac{\xi}{\sqrt{N}}, \quad y = s + \frac{\eta}{\sqrt{N}}, \quad \phi_k = \frac{\varphi(s)}{\sqrt{N}}, \tag{6.47}$$

in such a way that $s$ is kept fixed when $N \to +\infty$. Using the fact that

$$\Phi_\rho^{-1}(x) \approx \Phi_\rho^{-1}\left(s + \frac{\xi}{\sqrt{N}}\right) = \Phi_\rho^{-1}(s) + \frac{\xi}{\sqrt{N}\left(\rho \circ \Phi_\rho^{-1}\right)(s)}, \tag{6.48}$$

we obtain, at the leading order,

$$\begin{aligned} \Pr\left[\varphi(s) \in d\varphi\right] &= d\varphi \int d\xi\, d\eta\, \delta\left(\varphi - \frac{\eta - \xi}{\rho\left(\Phi_\rho^{-1}(s)\right)}\right) \frac{e^{-\frac{\xi^2+\eta^2}{2s(1-s)}}}{2\pi s(1-s)} \\ &= \frac{\left(\rho \circ \Phi_\rho^{-1}\right)(s)}{\sqrt{4\pi s(1-s)}} e^{-\frac{\left[\left(\rho\circ\Phi_\rho^{-1}\right)(s)\right]^2}{4s(1-s)}\varphi^2} d\varphi, \end{aligned} \tag{6.49}$$

that implies that

$$\begin{aligned}\overline{E_p} &= 2\int_0^1 \overline{|\varphi(s)|^p}\,\mathrm{d}s = 2\int_0^1 ds \frac{s^{\frac{p}{2}}(1-s)^{\frac{p}{2}}}{\left[\left(\rho\circ\Phi_\rho^{-1}\right)(s)\right]^p}\int_{-\infty}^{+\infty}\frac{d\varphi}{\sqrt{4\pi}}e^{-\frac{\varphi^2}{4}}|\varphi|^p \\ &= \frac{2^{1+p}}{\sqrt{\pi}}\Gamma\left(\frac{p+1}{2}\right)\int_0^1 ds \frac{s^{\frac{p}{2}}(1-s)^{\frac{p}{2}}}{\left[\left(\rho\circ\Phi_\rho^{-1}\right)(s)\right]^p} \\ &= \frac{2^{1+p}}{\sqrt{\pi}}\Gamma\left(\frac{p+1}{2}\right)\int_0^1 dx \frac{\Phi_\rho^{\frac{p}{2}}(x)(1-\Phi_\rho(x))^{\frac{p}{2}}}{\rho^{p-1}(x)}.\end{aligned} \tag{6.50}$$

In the particular case of a flat distribution the average cost converges to

$$\overline{E_p} = \frac{2^{1+p}}{\sqrt{\pi}}\Gamma\left(\frac{p+1}{2}\right)\int_0^1 ds\,[s(1-s)]^{\frac{p}{2}} = 2\frac{\Gamma\left(\frac{p}{2}+1\right)}{p+1} \tag{6.51}$$

which is two times the value of the optimal matching. For $p=2$ this gives $\overline{E_2}=2/3$, according to exact result (6.40). Formula (6.49) becomes

$$p_s(x) = \overline{\delta(\varphi(s)-x)} = \frac{e^{-\frac{x^2}{4s(1-s)}}}{\sqrt{4\pi s(1-s)}} \tag{6.52}$$

and similarly, see for example [CS14, Appendix A], it can be derived that the joint probability distribution $p_{t,s}(x,y)$ for $\varphi(s)$ is (for $t<s$) a bivariate Gaussian distribution

$$p_{t,s}(x,y) = \overline{\delta(\varphi(t)-x)\,\delta(\varphi(s)-y)} = \frac{e^{-\frac{x^2}{4t}-\frac{(x-y)^2}{4(s-t)}-\frac{y^2}{4(1-s)}}}{4\pi\sqrt{t(s-t)(1-s)}}. \tag{6.53}$$

This allows to compute, for a generic $p>1$, the average of the square of the re-scaled optimal cost

$$\overline{E_p^2} = 4\int_0^1 dt\int_0^1 ds\,\overline{|\varphi(s)|^p\,|\varphi(t)|^p}, \tag{6.54}$$

which is 4 times the corresponding one of a bipartite matching problem. In the case $p=2$, the average in Eq. (6.54) can be evaluated by using the Wick theorem for expectation values in a Gaussian distribution

$$\overline{E_2^2} = 4\int_0^1 ds\int_0^s dt\int_{-\infty}^{\infty} dx\,dy\,p_{t,s}(x,y)\,x^2y^2 = \frac{4}{5}, \tag{6.55}$$

and therefore

$$\overline{E_2^2} - \overline{E_2}^2 = \frac{16}{45} = 0.3\bar{5}. \tag{6.56}$$

This result is in agreement with the numerical simulations (see inset of Fig. 6.3) and proves that the re-scaled optimal cost is not a self-averaging quantity.

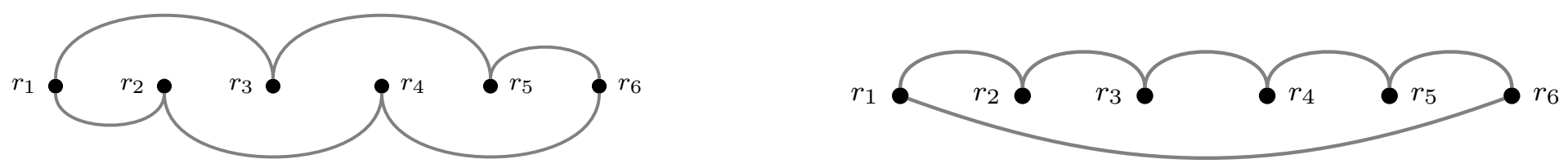

**(a)** $p > 1$ case. **(b)** $0 < p < 1$ case.

**Figure 6.4.** Optimal solutions for $N = 6$.

## 6.4 Optimal cycles on the complete graph

### The $p > 1$ case

We start by proving which is the optimal cycle when $p > 1$, for every realization of the disorder. Let us suppose to have $N$ points $\mathcal{R} = \{r_i\}_{i=1,\dots,N}$ in the interval $[0,1]$. As usual we will assume that the points are ordered, i.e. $r_1 \leq \cdots \leq r_N$. Let us define the following Hamiltonian cycle

$$h^* = h[\tilde{\sigma}] = (r_{\tilde{\sigma}(1)}, r_{\tilde{\sigma}(2)}, \dots, r_{\tilde{\sigma}(N)}, r_{\tilde{\sigma}(1)}) \tag{6.57}$$

with $\tilde{\sigma}$ defined as in (6.16). In Appendix C.2 we prove that

**Proposition 6.4.1.** *The Hamiltonian cycle which provides the optimal cost is $h^*$.*

Here we just sketch the main ideas behind the proof. Consider a generic Hamiltonian cycle, that is a $\sigma \in \mathcal{S}_N$ with $\sigma(1) = 1$ (which corresponds to the irrelevant choice of the starting point of the cycle). We can always introduce a new set of ordered points $\mathcal{B} := \{b_j\}_{j=1,\dots,N} \subset [0,1]$, which we will call blue points, such that

$$b_i = \begin{cases} r_1 & \text{for } \; i = 1 \\ r_{i-1} & \text{otherwise .} \end{cases} \tag{6.58}$$

We consider now an Hamiltonian cycle on the complete bipartite graph with vertex sets $\mathcal{R}$ and $\mathcal{B}$, which only uses links available in our monopartite problem and has the same cost of $\sigma$. In Appendix C.2 we show how to build this cycle. So we have obtained a map between our monopartite problem and a bipartite one, which we know how to solve. Therefore, the solution of the bipartite problem gives us the optimal tour also for the monopartite case, which turns out to be $h^*$. Note also that the same Hamiltonian cycle is obtained using $\tilde{\pi}$ given in (6.17). A graphical representation of the optimal cycle for $p > 1$ and $N = 6$ is given Fig. 6.4a.

### 6.4.1 The $0 < p < 1$ case

We now prove that, given an ordered sequence $\mathcal{R} = \{r_i\}_{i=1,\dots,N}$ of $N$ points in the interval $[0,1]$, with $r_1 \leq \cdots \leq r_N$, if $0 < p < 1$ and if

$$h^* = h[\mathbb{1}] = (r_{\mathbb{1}(1)}, r_{\mathbb{1}\mathbb{1}(2)}, \dots, r_{\mathbb{1}(N)}, r_{\mathbb{1}(1)}) \tag{6.59}$$

where $\mathbb{1}(1)$ is the identity permutation, i.e.:

$$\mathbb{1}(j) = j \tag{6.60}$$

then

**Proposition 6.4.2.** *The Hamiltonian cycle which provides the optimal cost is $h^*$.*

The idea behind this result is that we can define a crossing in the cycle by drawing all the links as arcs in the upper half-plain in such a way that all the crossing between arcs cannot be eliminated by drawing the arcs in another way. An example of crossing is in the following figure

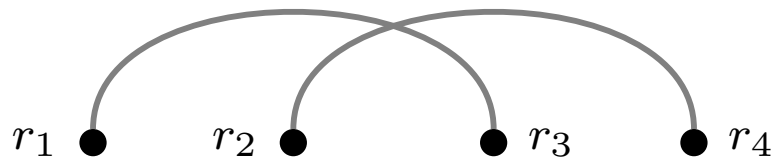

where we have not drawn the arcs which close the cycle to emphasize the crossing. Now, as shown in [BCS14], if we are able to swap two crossing arcs with two non-crossing ones, the difference between the cost of the original cycle and the new one simply consists in the difference between a crossing matching and a non-crossing one, that is positive when $0 < p < 1$. Therefore the proof of Proposition 6.4.2, which is given in Appendix C.2, consists in showing how to remove a crossing (without breaking the cycle into multiple ones) and in proving that $h^*$ is the only Hamiltonian cycle without crossings (see Fig. 6.4b.).

## The $p < 0$ case

Here we study the properties of the solution for $p < 0$. Our analysis is based, again, on the properties of the $p < 0$ optimal matching solution. In [CDS17] it is shown that the optimal matching solution maximizes the total number of crossings, since the cost difference of a non-crossing and a crossing matching is always positive for $p < 0$. This means that the optimal matching solution of $2N$ points on an interval is given by connecting the $i$-th point to the $(i + N)$-th one with $i = 1, \ldots, N$; in this way every edge crosses the remaining $N - 1$. Similarly to the $0 < p < 1$ case, suppose now to have a generic oriented Hamiltonian cycle and draw the connections between the vertices in the upper half plain (as before, eliminating all the crossings which depend on the way we draw the arcs). Suppose it is possible to identify a matching that is non-crossing, then the possible situations are the following two (we draw only the points and arcs involved in the non-crossing matching):

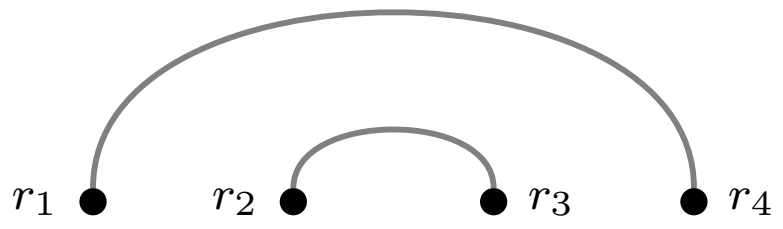

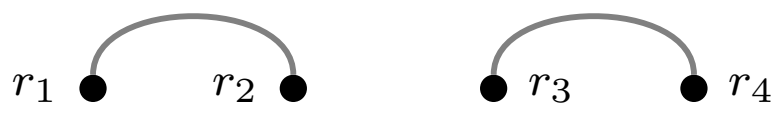

In Appendix C.2, we prove that is not always possible to replace a non-crossing matching by a crossing one keeping unaltered the property of Hamiltonian cycle. This move is such that the cost of the new configuration is lower than the cost of the old one, since the cost gain is the difference between the costs of a non-crossing and a crossing matching, which is always positive for $p < 0$.

In this manner the proof for $p < 0$ goes on the same line of $0 < p < 1$, but instead of finding the cycle with no crossings, now we look for the one or ones that maximize them. However, as we will see in the following, one must distinguish between the $N$ odd and even case. In fact, in the $N$ odd case, only one cycle maximizes the total number of crossings, i.e. we have only one possible solution. In the $N$ even case, on the contrary, the number of Hamiltonian cycles that maximize the total number of crossings are $\frac{N}{2}$.

**N odd case** Given an ordered sequence $\mathcal{R} = \{r_i\}_{i=1,\dots,N}$ of $N$ points, with $N$ odd, in the interval $[0, 1]$, with $r_1 \leq \cdots \leq r_N$, consider the permutation $\sigma$ defined as:

$$\sigma(i) = \begin{cases} 1 & \text{for } i = 1 \\ \frac{N-i+3}{2} & \text{for even } i > 1 \\ \frac{2N-i+3}{2} & \text{for odd } i > 1 \end{cases} \tag{6.61}$$

This permutation defines the following Hamiltonian cycle:

$$h^* := h[\sigma] = (r_{\sigma(1)}, r_{\sigma(2)}, \dots, r_{\sigma(N)}). \tag{6.62}$$

**Proposition 6.4.3.** *The Hamiltonian cycle which provides the optimal cost is $h^*$.*

The proof consist in showing that the only Hamiltonian cycle with the maximum number of crossings is $h^*$. As we discuss in Appendix C.2, the maximum possible number of crossings an edge can have is $N - 3$. The Hamiltonian cycle under exam has $N(N - 3)/2$ crossings, i.e. every edge in $h^*$ has the maximum possible number of crossings. Indeed, the vertex $a$ is connected with the vertices $a + \frac{N-1}{2} \pmod N$ and $a + \frac{N+1}{2} \pmod N$. The edge $(a, a + \frac{N-1}{2} \pmod N)$ has $2\frac{N-3}{2} = N - 3$ crossings due to the $\frac{N-3}{2}$ vertices $a + 1 \pmod N, a + 2 \pmod N, \dots, a + \frac{N-1}{2} - 1 \pmod N$ that contribute with 2 edges each. This holds also for the edge $(a, a + \frac{N+1}{2} \pmod N)$ and for each $a \in [N]$. As shown in Appendix C.2 there is only one cycle with this number of crossings.

Now, notice that an Hamiltonian cycle is a particular loop covering. However, if we search for a loop covering in the $p < 0$ case, we need again to find the one which maximizes the number of crossings. Since the procedure of swapping two non-crossing with two crossing arcs can be applied to each loop covering but $h^*$ it follows that:

**Corollary 1.** *$h^*$ provides also the optimal 2-factor problem solution.*

An example of an Hamiltonian cycle discussed here is given in Fig. 6.5.

**N even case** In this situation, differently from the above case, the solution is not the same irrespectively of the disorder instance. More specifically, there is a set of possible solutions, and at a given instance the optimal is the one among that set with a lower cost. We will show how these solutions can be found and how they are related. In this section we will use the results obtained in Appendix C.3 regarding the Monopartite Euclidean 2-factor for $p < 0$.
Given the usual sequence of points $\mathcal{R} = \{r_i\}_{i=1,\dots,N}$ of $N$ points, with $N$ even, in the interval $[0, 1]$, with $r_1 \leq \cdots \leq r_N$, if $p < 0$, consider the permutation $\sigma$ such that:

$$\sigma(i) = \begin{cases} 1 & \text{for } i = 1 \\ \frac{N}{2} - i + 3 & \text{for even } i \leq \frac{N}{2} + 1 \\ N - i + 3 & \text{for odd } i \leq \frac{N}{2} + 1 \\ i - \frac{N}{2} & \text{for even } i > \frac{N}{2} + 1 \\ i & \text{for odd } i > \frac{N}{2} + 1 \end{cases} \tag{6.63}$$

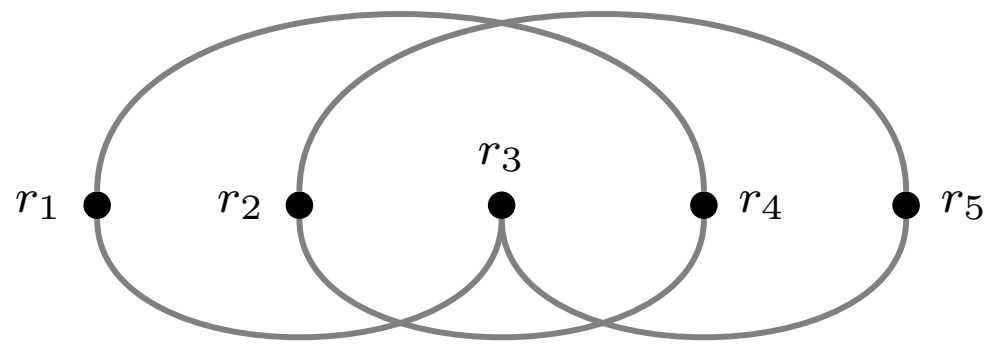

**Figure 6.5.** This is the optimal TSP and 2-factor problem solution for $N = 5$ and $p < 0$. Notice that there are no couples of edges which do not cross and which can be changed in a crossing couple.

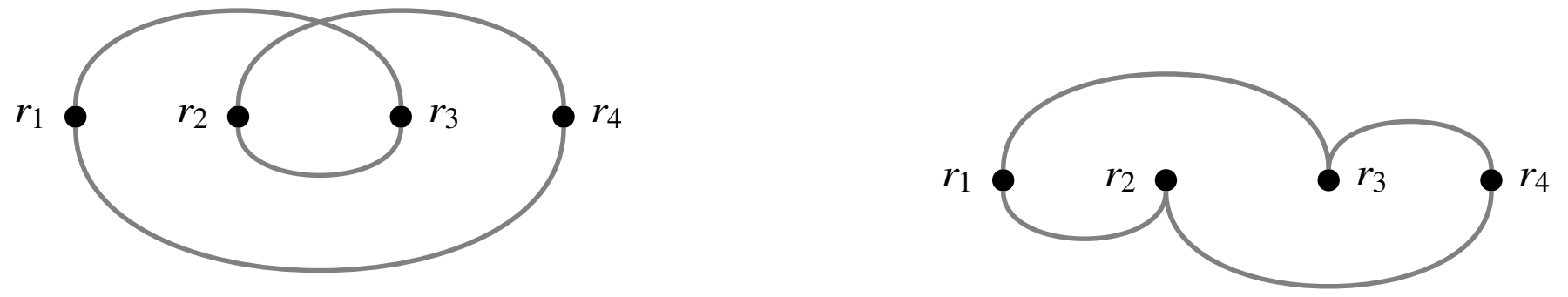

**Figure 6.6.** The two possible optimal Hamiltonian cycles for $p < 0$, $N = 4$. For each specific instance one of them has a lower cost than the other, but differently from all the other cases ($p > 0$ or $N$ odd) the optimal cycle is not the same for each disorder instance.

Given $\tau \in \mathcal{S}_N$ defined by $\tau(i) = i + 1$ for $i \in [N - 1]$ and $\tau(N) = 1$, we call $\Sigma$ the set of permutations $\sigma_k, k = 1, \dots, N$ defined as:

$$\sigma_k(i) = \tau^k(\sigma(i)) \tag{6.64}$$

where $\tau^k = \tau \circ \tau^{k-1}$. Thus we have the following result:

**Proposition 6.4.4.** *The set of Hamiltonian cycles that provides the optimal cost is*

$$h_k^* := h[\sigma_k] = (r_{\sigma_k(1)}, r_{\sigma_k(2)}, \dots, r_{\sigma_k(N)}). \tag{6.65}$$

An example with $N = 4$ points is shown in Fig. 6.6. In Appendix C.3 the optimal solution for the Euclidean 2-factor in obtained. In particular, we show how the solution is composed of a loop-covering of the graph. The idea for the proof of the TSP is to show how to join the loops in the optimal way in order to obtain the optimal TSP. The complete proof of the Proposition 6.4.4 is given in Appendix C.3.

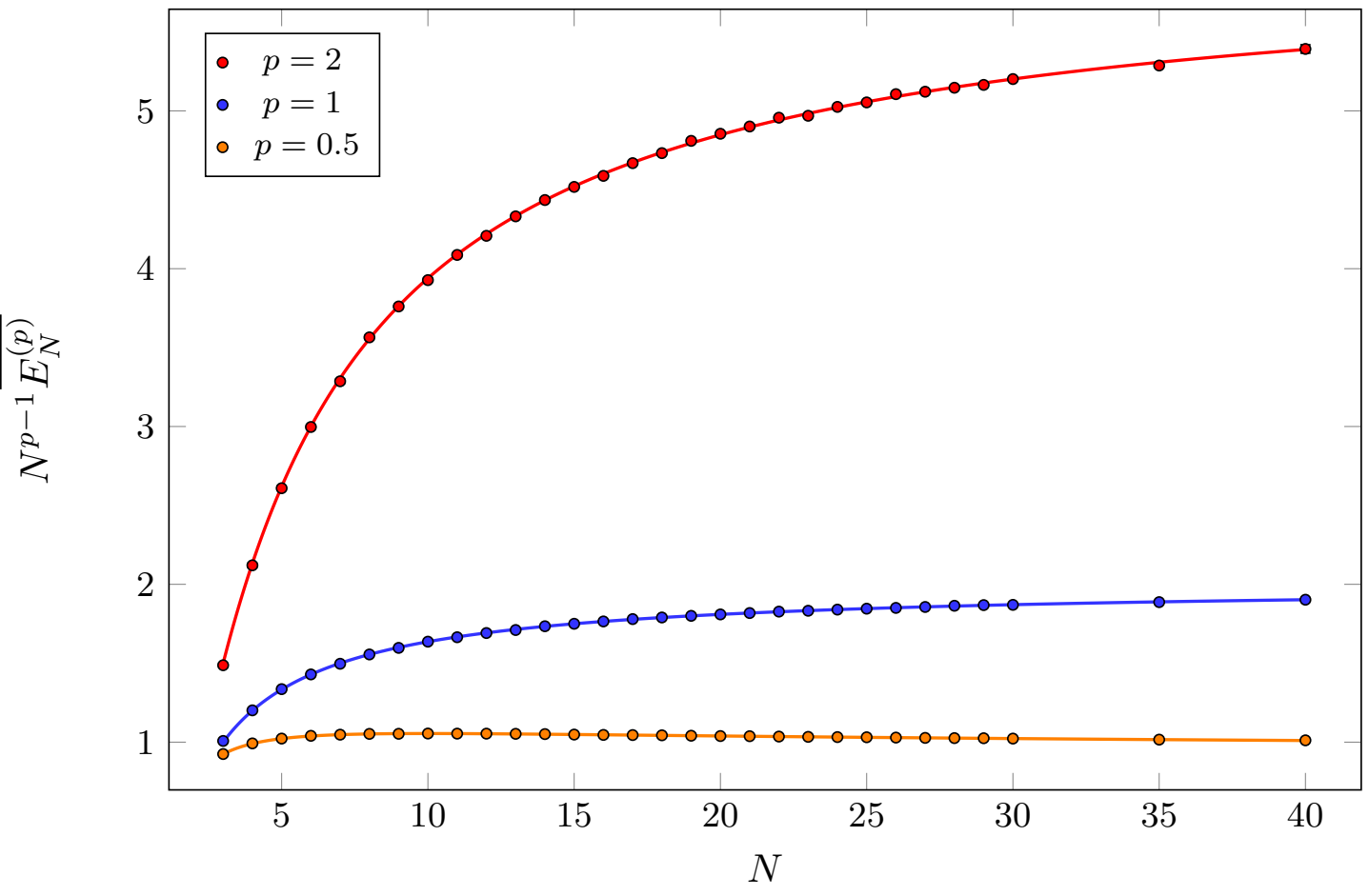

**Figure 6.7.** Rescaled average optimal cost for various values of $p > 0$. The points are the result of a numerical simulation whereas lines are theoretical predictions.

### 6.4.2 Evaluation of the average costs and numerical results

If we have $N$ random points chosen with flat distribution in the interval $[0, 1]$ and we order them in increasing position, the probability for finding the $k$-th point in $x$ is given by (6.36) while the probability for finding the $l$-th point in $x$ and the $s$-th point in $y$ is given, for $s > l$ by

$$p_{l,s}(x,y) = \frac{\Gamma(N+1)}{\Gamma(l)\,\Gamma(s-l)\,\Gamma(N-s+1)}\,x^{l-1} \times (y-x)^{s-l-1}(1-y)^{N-s}\,\theta(y-x)\,, \tag{6.66}$$

see for example [CS14, App. A]. It follows that

$$\int dx\,dy\,(y-x)^{\alpha}\,p_{l,l+k}(x,y) = \frac{\Gamma(N+1)\,\Gamma(k+\alpha)}{\Gamma(N+\alpha+1)\,\Gamma(k)} \tag{6.67}$$

independently from $l$, and, therefore, in the case $p > 1$ we obtain soon

$$\overline{E_N[h^*]} = [(N-2)(p+1)+2]\,\frac{\Gamma(N+1)\,\Gamma(p+1)}{\Gamma(N+p+1)} \tag{6.68}$$

and in particular for $p = 2$

$$\overline{E_N[h^*]} = \frac{2\,(3N-4)}{(N+1)(N+2)}\,, \tag{6.69}$$

and for $p = 1$ we get

$$\overline{E_N[h^*]} = \frac{2\,(N-1)}{N+1}\,. \tag{6.70}$$

In the same way one can evaluate the average optimal cost when $0 < p < 1$, obtaining

$$\overline{E_N[h^*]} = \left[(N-1)\,\Gamma(p+1) + \frac{\Gamma(N+p-1)}{\Gamma(N-1)}\right]\frac{\Gamma(N+1)}{\Gamma(N+p+1)} \tag{6.71}$$

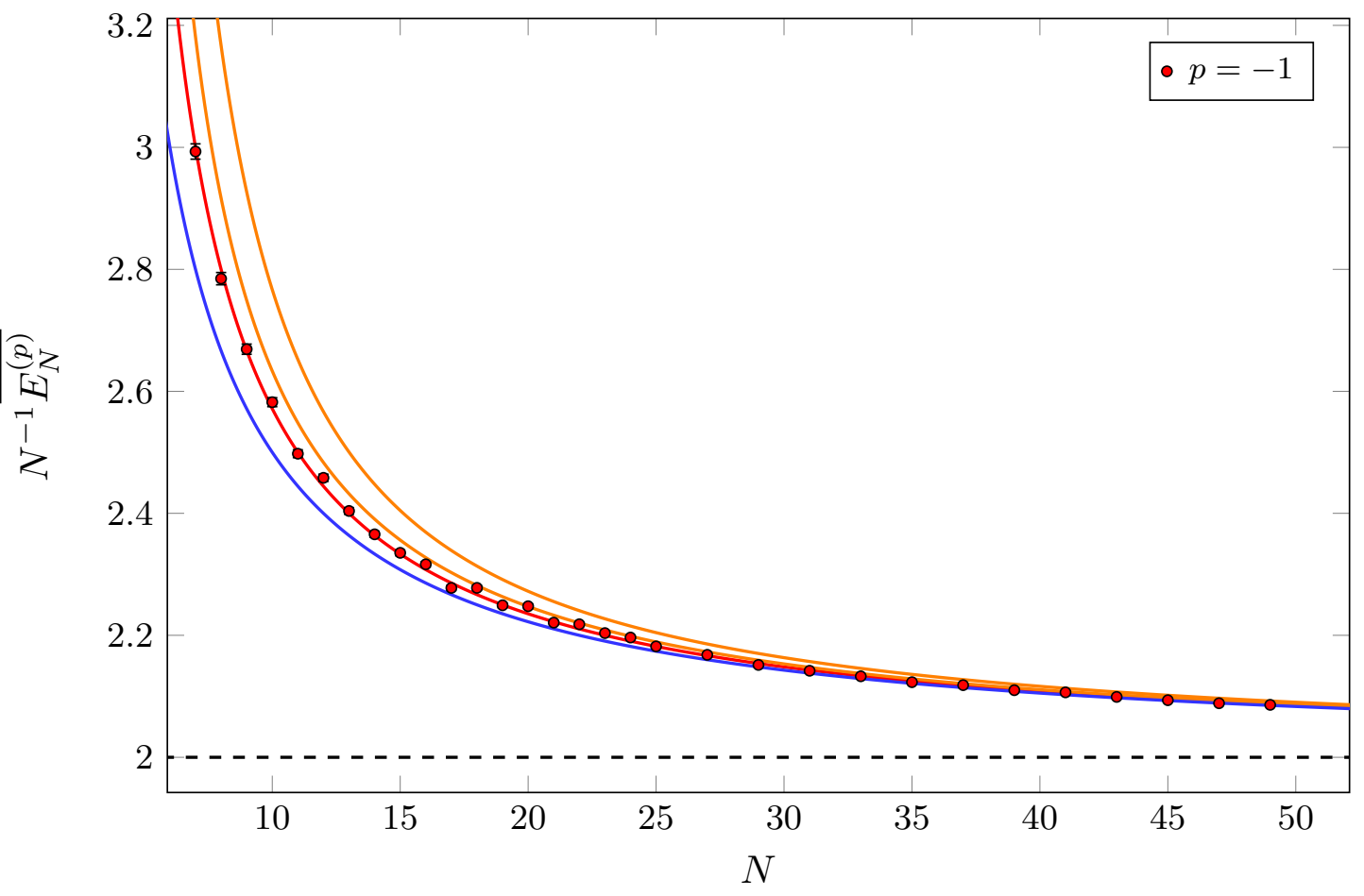

**Figure 6.8.** Rescaled average optimal cost in the $p = -1$ case. The red points and line are respectively the result of a numerical simulation and the theoretical prediction in the odd $N$ case. The blue line is the 2 times the theoretic value of the optimal matching. The orange lines (from top to bottom) are the average costs $\overline{E_N[h_1]}$ and $\overline{E_N[h_2]}$ defined in equation (6.75) and (6.76) respectively. The dashed black line is the large $N$ limit of all the curves.

which coincides at $p = 1$ with (6.70) and, at $p = 0$, provides $\overline{E_N[h^*]} = N$. For large $N$, we get

$$\lim_{N\to\infty} N^{p-1}\overline{E_N[h^*]} = \begin{cases} \Gamma(p+2) & \text{for } p \geq 1 \\ \Gamma(p+1) & \text{for } 0 < p < 1\,. \end{cases} \tag{6.72}$$

The asymptotic cost for large $N$ and $p > 1$ is $2(p+1)$ times the average optimal cost of the matching problem on the complete graph $\mathcal{K}_N$ as can be checked in [CDS17]. We report in Appendix C.4 the computation of the average costs when the points are extracted in the interval $[0, 1]$ using a general probability distribution.

For $p < 0$ and $N$ odd we have only one possible solution, so that the average optimal cost is

$$\overline{E_N[h^*]} = \left[(N-1)\frac{\Gamma\left(\frac{N+1}{2}+p\right)}{\Gamma\left(\frac{N+1}{2}\right)} + (N+1)\frac{\Gamma\left(\frac{N-1}{2}+p\right)}{\Gamma\left(\frac{N-1}{2}\right)}\right]\frac{\Gamma(N+1)}{2\Gamma(N+p+1)}\,. \tag{6.73}$$

For large $N$ it behaves as

$$\lim_{N\to\infty}\frac{\overline{E_N[h^*]}}{N} = \frac{1}{2^p}\,, \tag{6.74}$$

which coincides with the scaling derived before for $p = 0$. Note that for large $N$ the average optimal cost of the TSP problem is two times the one of the corresponding matching problem for $p < 0$. For $N$ even, instead, there are $N/2$ possible solutions. One can see $N/2 - 1$ of these share the same average energy, since they have the same number of links with the same $k$ of equation (6.67). These solutions, in particular have 2 links with $k = N/2$, $N/2$ links with $k = N/2 + 1$ and $N/2 - 2$ links with $k = N/2 + 1$. We denote this set of configurations with $h_1$ (although they are many different configurations, we use only the label $h_1$ to stress that all of them share the same average optimal cost) and its average cost is

$$\overline{E_N[h_1]} = \frac{\Gamma(N+1)}{\Gamma(N+p+1)} \left[ \frac{N}{2} \frac{\Gamma\left(\frac{N}{2}+p-1\right)}{\Gamma\left(\frac{N}{2}-1\right)} + \left(\frac{N}{2}-2\right) \frac{\Gamma\left(\frac{N}{2}+p+1\right)}{\Gamma\left(\frac{N}{2}+1\right)} + 2\frac{\Gamma\left(\frac{N}{2}+p\right)}{\Gamma\left(\frac{N}{2}\right)} \right] \tag{6.75}$$

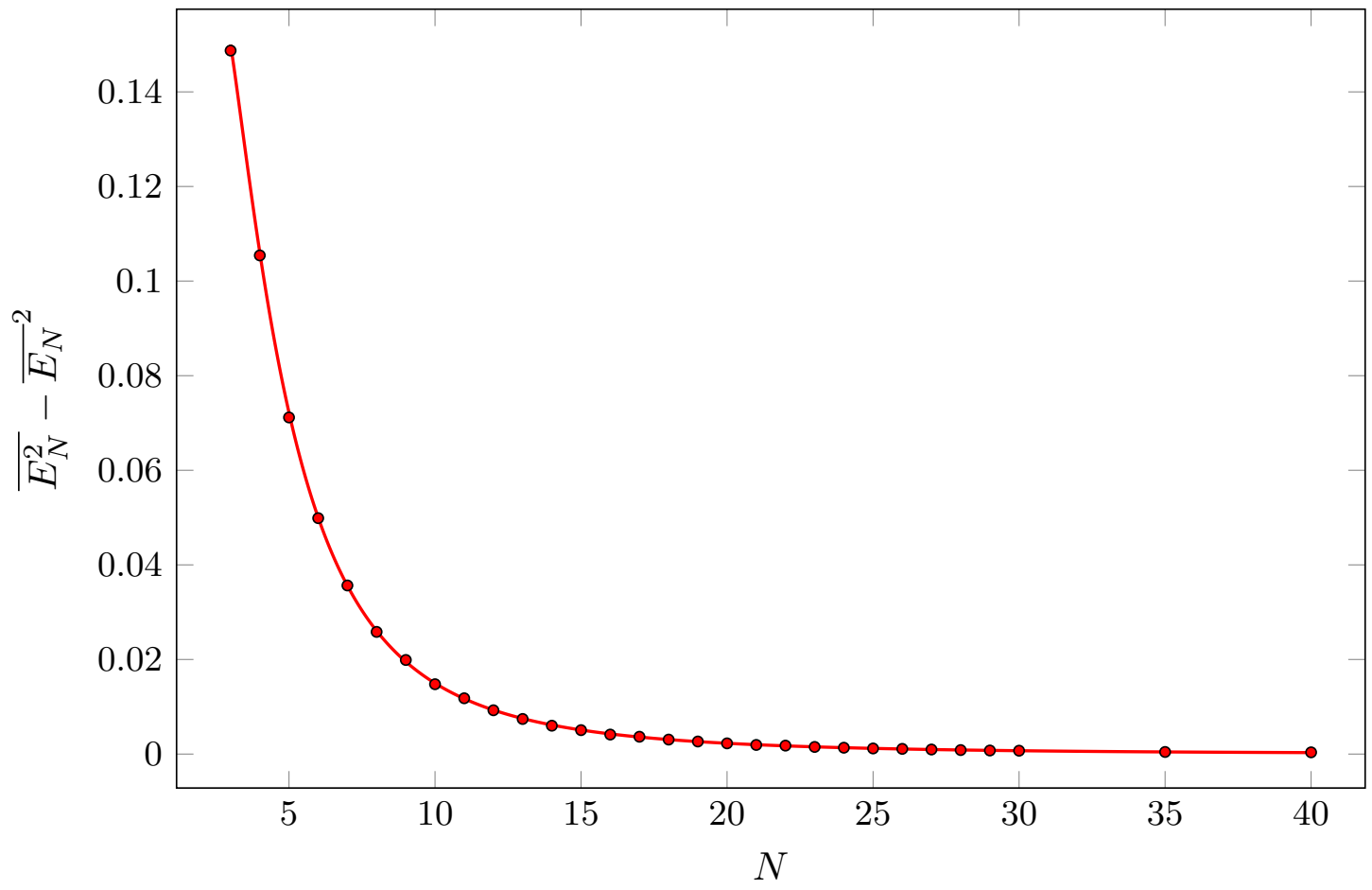

**Figure 6.9.** Variance of the optimal cost in the $p = 2$ case. The red points and line are respectively the result of a numerical simulation and the theoretical prediction as given in (6.78).

The other possible solution, that we will call with $h_2$ has 2 links with $k = N/2 - 1$, $N/2$ links with $k = N/2 + 1$ and $N/2 - 1$ links with $k = N/2 + 1$ and its average cost will be

$$\overline{E_N[h_2]} = \frac{\Gamma(N+1)}{\Gamma(N+p+1)} \left[ \left(\frac{N}{2}-1\right) \frac{\Gamma\left(\frac{N}{2}+p-1\right)}{\Gamma\left(\frac{N}{2}-1\right)} + \left(\frac{N}{2}-1\right) \frac{\Gamma\left(\frac{N}{2}+p+1\right)}{\Gamma\left(\frac{N}{2}+1\right)} + 2\frac{\Gamma\left(\frac{N}{2}+p\right)}{\Gamma\left(\frac{N}{2}\right)} \right] \tag{6.76}$$

In Fig. 6.7 and 6.8 we compare analytical and numerical results respectively for $p = 0.5, 1, 2$ and for $p = -1$. In particular, since $\overline{E_N[h_1]} > \overline{E_N[h_2]}$, $\overline{E_N[h_2]}$ provides our best upper bound for the average optimal cost of the $p = -1$, $N$ even case.

### 6.4.3 Self-averaging property for $p > 1$

An interesting question is whether the average optimal cost is a self-averaging quantity. Previous investigation regarding the matching problem [Ste97, Yuk98] showed that indeed the average optimal cost is self-averaging in every dimensions when the graph is the complete one. This is the case, at least in the one dimensional case, also for the random Euclidean TSP. Again we collect in Appendix C.5 all the technical details concerning the evaluation of the second moment of the optimal cost distribution $\overline{E_N^2}$. Here we only state the main results. $\overline{E_N^2}$ has been computed for all number of points $N$ and, for simplicity, in the case $p > 1$ and it is given in equation (C.42). In the large $N$ limit it goes like

$$\lim_{N\to\infty} N^{2(p-1)}\overline{E_N^2[h^*]} = \Gamma^2(p+2) \tag{6.77}$$

i.e. tends to the square of the rescaled average optimal cost. This proves that the cost is a self-averaging quantity. Using (C.42) together with equation (6.68) one gets the variance of the optimal cost. In particular for $p = 2$ we get

$$\sigma^2_{E_N} = \frac{4(N(5N(N+13)+66)-288)}{(N+1)^2(N+2)^2(N+3)(N+4)}, \tag{6.78}$$

which goes to zero as $\sigma^2_{E_N} \simeq 20/N^3$. In Figure 6.9 we compare the theoretical result with the numerical ones for the variance of the optimal cost for $p = 2$.

# Chapter 7

# The Euclidean 2-factor problem in one dimension

In this chapter we shall consider the 2-factor (or 2-matching) problem which consists, given an undirected graph, in finding the minimum spanning subgraph that contains only disjoint cycles. This problem has been previously considered on a generic graph $\mathcal{G}$ by Bayati et al. [Bay+11] using the BP algorithm. In Statistical Mechanics models on loops have been considered [Bax16]. In particular in two dimension loop coverings have been studied also in connection to conformal field theories (CFT), Schramm-Loewner evolution (SLE) and integrable models [MDJ17, JRS04].

The 2-factor problem can be seen as a relaxation of the TSP, in which one has the additional constraint that there must be a unique cycle. For this reason, if $\mathcal{H}$ is the set of Hamiltonian cycles for the graph $\mathcal{G}$, of course $\mathcal{H} \subset \mathcal{M}_2$. Therefore if $\nu^* \in \mathcal{M}_2$ and $h^* \in \mathcal{H}$ are respectively the optimal 2-factor and the optimal Hamiltonian cycle, we have

$$E[h^*] \geq E[\nu^*] \,. \tag{7.1}$$

In infinite dimensions, where the 2-factor can be studied using replica and cavity method, one finds that, for large number of points, its average optimal cost is the same of the RTSP, that is inequality (7.1) is saturated on average. This result is valid both on the complete and on the complete bipartite graph.

Here we study the 2-factor problem in one dimension, both on the complete graph bipartitioning two sets of $N$ points and on the complete graph of $N$ vertices, throwing the points independently and uniformly in the compact interval $[0, 1]$. The weights on the edges are chosen as a convex function of the Euclidean distance between adjacent vertices. Despite the fact that it is a one-dimensional problem, it is not a trivial one. In the following we show that, while almost for every instance of the disorder there is only one solution, by looking at the whole ensemble of instances there appears an exponential number of possible solutions scaling as $\mathsf{p}^N$, where $\mathsf{p}$ is the plastic constant. This is at variance with what happens for other random combinatorial optimization problems, like the matching problem and the TSP that were studied so far [BCS14, CS14, CDS17, Car+18b]. In both cases we know that, for every realization of the disorder, the configuration that solves the problem is unique.

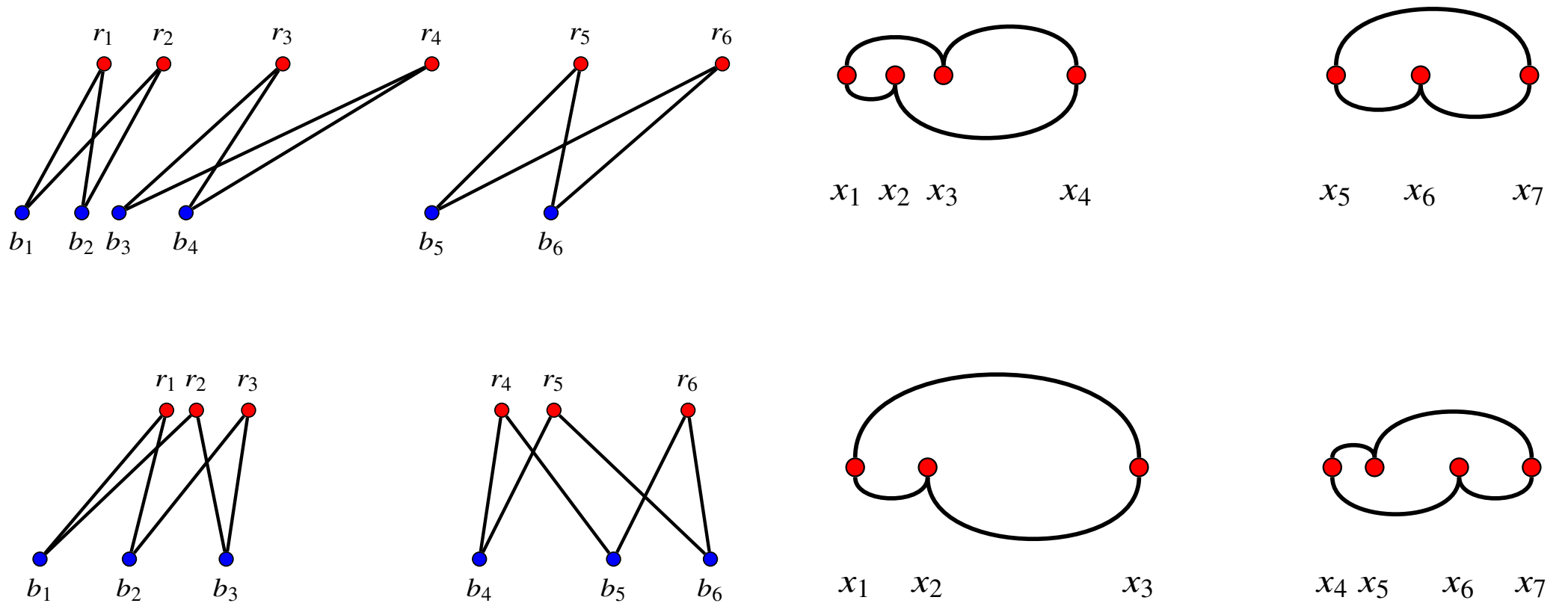

**(a)** Two instances whose optimal solutions are the two possible $\nu^*$ for $N = 6$ on the complete bipartite graph $\mathcal{K}_{N,N}$. For each instance the blue and red points are chosen in the unit interval and sorted in increasing order, then plotted on parallel lines to improve visualization.

**(b)** Two instances whose optimal solutions are the two possible $\nu^*$ for $N = 7$ on the complete graph $\mathcal{K}_N$. For each instance the points are chosen in the unit interval and sorted in increasing order.

**Figure 7.1.** Optimal solutions for small $N$ cases.

The rest of the paper is organized as follows: in Sect. 6.2 we give some definitions and we present our model in more detail. In Sect. 7.1 we write the cost of the 2-factor in terms of permutations and for every number of points we compare its cost with that of matching and TSP. We argue that, in the thermodynamic limit and in the bipartite case, its cost is twice the cost of the optimal matching. In Sect. 7.2 we characterize, for every number of points, the properties of the optimal solution. We compare it with the corresponding one of the TSP problem and we conclude that the number of possible solutions grows exponentially with $N$. In Sect. 7.3 we derive some upper bounds on the average optimal cost and in Sect. 7.4 we compare them with numerical simulations, describing briefly the algorithm we have used to find numerically the solution. We study the finite-size corrections to the asymptotic average cost in the complete bipartite case, and the leading order in the complete case.

## 7.1 The Euclidean 2-factor problem

Let us start by making some considerations when the problem is defined on the complete bipartite graph $\mathcal{K}_{N,N}$, where each cycle must have an even length.

Let $\mathcal{S}_N$ be the symmetric group of order $N$ and consider two permutations $\sigma, \pi \in \mathcal{S}_N$. If for every $i \in [N]$ we have that $\sigma(i) \neq \pi(i)$, then the two permutations define the 2-factor $\nu(\sigma, \pi)$ with edges

$$e_{2i-1} := (r_i, b_{\sigma(i)}) \tag{7.2}$$

$$e_{2i} := (r_i, b_{\pi(i)}) \tag{7.3}$$

for $i \in [N]$. And, vice versa, for any 2-factor $\nu$ there is a couple of permutations

$\sigma, \pi \in \mathcal{S}_N$, such that for every $i \in [N]$ we have that $\sigma(i) \neq \pi(i)$. It will have total cost

$$E[\nu(\sigma,\pi)] = \sum_{i\in[N]} \left[ |r_i - b_{\sigma(i)}|^p + |r_i - b_{\pi(i)}|^p \right] . \tag{7.4}$$

By construction, if we denote by $\mu[\sigma]$ the matching associated to the permutation $\sigma$ and by

$$E[\mu(\sigma)] := \sum_{i\in[N]} |r_i - b_{\sigma(i)}|^p \tag{7.5}$$

its cost, we soon have that

$$E[\nu(\sigma,\pi)] = E[\mu(\sigma)] + E[\mu(\pi)] \tag{7.6}$$

and we recover that

$$E[\nu^*] \geq 2\, E[\mu^*] \tag{7.7}$$

the cost of the optimal 2-factor is necessarily greater or equal to twice the optimal 1-factor. Together with inequality (7.1), which is valid for any graph, we obtain that

$$E[h^*] \geq E[\nu^*] \geq 2\, E[\mu^*] . \tag{7.8}$$

In Chap. 6 we have seen that in the limit of infinitely large $N$, in one dimension and with $p > 1$, the average cost of the optimal Hamiltonian cycle is equal to twice the average cost of the optimal matching (1-factor). We conclude that the average cost of the 2-factor must be the same. In the following we will denote with $\overline{E^{(p)}_{N,N}[\nu^*]}$ the average optimal cost of the 2-factor problem on the complete bipartite graph. Its scaling for large $N$ will be the same of the TSP and the matching problem, that is the limit

$$\lim_{N\to\infty} \frac{\overline{E^{(p)}_{N,N}[\nu^*]}}{N^{1-p/2}} = E^{(p)}_B , \tag{7.9}$$

is finite. An explicit evaluation in the case $p = 2$ is presented in Sec. 7.3.

On the complete graph $\mathcal{K}_N$ inequality (7.7) does not hold, since a general 2-factor configuration cannot always be written as a sum of two disjoint matchings, due to the presence of odd-length loops. Every 2-factor configuration on the complete graph can be determined by only one permutation $\pi$, satisfying $\pi(i) \neq i$ and $\pi(\pi(i)) \neq i$ for every $i \in [N]$. The cost can be written as

$$E[\nu(\pi)] = \sum_{i\in[N]} |x_i - x_{\pi(i)}|^p . \tag{7.10}$$

The two constraints on $\pi$ assure that the permutation does not contain fixed points and cycles of length 2. In the following we will denote with $\overline{E^{(p)}_N[\nu^*]}$ the average optimal cost of the 2-factor problem on the complete graph. Even though inequality (7.7) does not hold, we expect that for large $N$, the average optimal cost scales in the same way as the TSP and the matching problem, i.e. as

$$\lim_{N\to\infty} \frac{\overline{E^{(p)}_N[\nu^*]}}{N^{1-p}} = E^{(p)}_M . \tag{7.11}$$

In Sect. 7.4 we give numerical evidence for this scaling.

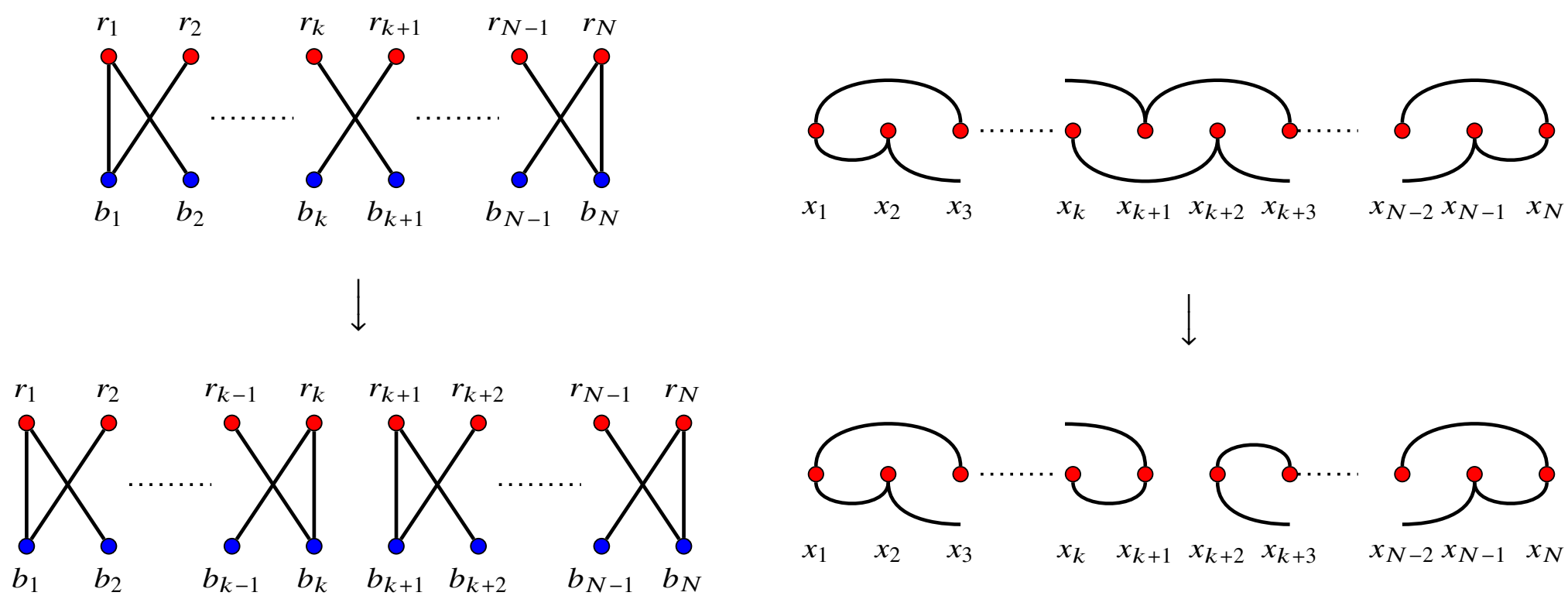

**(a)** $\mathcal{K}_{N,N}$ case **(b)** $\mathcal{K}_N$ case

**Figure 7.2.** Result of one cut of the shoelace in two smaller ones for both the complete bipartite and complete graph cases. The cost gained is exactly the difference between an unordered matching and an ordered one.

## 7.2 Properties of the solution for $d = 1$

We restrict here to the particular case in which the parameter $p$ appearing in the definition of the cost (1.93) is such that $p > 1$, that is the weight associated to an edge is a convex and increasing function of the Euclidean distance between its two vertices. In such a case we know exactly, for every number of points, the optimal solution of the matching problem both on the bipartite [BCS14, CS14, CDS17] and the complete graph [CDS17] and of the TSP problem, again on both its bipartiteand complete graph version as examined in Chap. 6. The knowledge of the optimal configuration of those problems permits to write down several properties of the solution of the 2-factor.

### Bipartite Case

We remind that the adjacency matrix of a bipartite graph with the same cardinality $N$ of red and blue points can always be written in block form (1.3) which defines the $N \times N$ matrix $B$. Now, suppose that both blue and red points are labeled in increasing order, that is if $i > j$ with $i, j \in [N]$, then $r_i > r_j$ and $b_i > b_j$, the permutation which minimizes the cost of the matching is necessarily the identity permutation $\mu^*(i) = i$ for $i \in [N]$, so that

$$E[\mu^*] = \sum_{i \in [N]} |r_i - b_i|^p \,. \tag{7.12}$$

The optimal matching corresponds to

$$B = \begin{pmatrix} 1 & \cdots & \cdots & 0 \\ \vdots & \ddots & & \vdots \\ \vdots & & \ddots & \vdots \\ 0 & \cdots & \cdots & 1 \end{pmatrix} . \tag{7.13}$$

Since the total adjacency matrix is of the form (1.3), $B$ has to satisfy the following constraints

$$\sum_{i=1}^{N} B_{ij} = 1\,, \qquad j \in [N] \tag{7.14a}$$

$$\sum_{j=1}^{N} B_{ij} = 1\,, \qquad i \in [N] \tag{7.14b}$$

$$B_{ij} \in \{0,1\}\,. \tag{7.14c}$$

The first two constraints impose that only one edge must depart from each blue and each red vertex respectively. For the TSP, the optimal Hamiltonian cycle $h^*$ is identified by the two permutations $\tilde{\sigma}$ and $\tilde{\pi}$ defined in equations (6.16) and (6.17) with optimal cost (6.32). The optimal Hamiltonian cycle corresponds to the adjacency matrix

$$B = \begin{pmatrix} 1 & 1 & 0 & 0 & \cdots & 0 \\ 1 & 0 & 1 & 0 & \cdots & 0 \\ 0 & 1 & 0 & 1 & \cdots & 0 \\ \vdots & \ddots & \ddots & \ddots & \ddots & \vdots \\ 0 & \cdots & 0 & 1 & 0 & 1 \\ 0 & \dots & 0 & 0 & 1 & 1 \end{pmatrix}. \tag{7.15}$$

Let us now look for the optimal solutions for the 2-factor. The adjacency matrix of a valid 2-factor must satisfy constraints analogous to those of the matching problem, i.e.

$$\sum_{i=1}^{N} B_{ij} = 2\,, \qquad j \in [N] \tag{7.16a}$$

$$\sum_{j=1}^{N} B_{ij} = 2\,, \qquad i \in [N] \tag{7.16b}$$

$$B_{ij} \in \{0,1\}\,. \tag{7.16c}$$

The only difference with (7.14) is that from every blue or red vertex must depart two edges. For $N = 2$ there is only one configuration. It can be defined by the adjacency matrix

$$B_2 = \begin{pmatrix} 1 & 1 \\ 1 & 1 \end{pmatrix}. \tag{7.17}$$

For $N = 3$ the solution is the same as in the TSP

$$B_3 = \begin{pmatrix} 1 & 1 & 0 \\ 1 & 0 & 1 \\ 0 & 1 & 1 \end{pmatrix}. \tag{7.18}$$

For $N = 4$ the solution has two simple cycles

$$B_2^{(2)} = \begin{pmatrix} 1 & 1 & 0 & 0 \\ 1 & 1 & 0 & 0 \\ 0 & 0 & 1 & 1 \\ 0 & 0 & 1 & 1 \end{pmatrix}. \tag{7.19}$$

For $N = 5$ there are two symmetric possible solutions

$$B_{2,3} = \begin{pmatrix} 1 & 1 & 0 & 0 & 0 \\ 1 & 1 & 0 & 0 & 0 \\ 0 & 0 & 1 & 1 & 0 \\ 0 & 0 & 1 & 0 & 1 \\ 0 & 0 & 0 & 1 & 1 \end{pmatrix} \qquad B_{3,2} = \begin{pmatrix} 1 & 1 & 0 & 0 & 0 \\ 1 & 0 & 1 & 0 & 0 \\ 0 & 1 & 1 & 0 & 0 \\ 0 & 0 & 0 & 1 & 1 \\ 0 & 0 & 0 & 1 & 1 \end{pmatrix} . \tag{7.20}$$

For $N = 6$ there are two possible solutions (not related by symmetry)

$$B_2^{(3)} = \begin{pmatrix} 1 & 1 & 0 & 0 & 0 & 0 \\ 1 & 1 & 0 & 0 & 0 & 0 \\ 0 & 0 & 1 & 1 & 0 & 0 \\ 0 & 0 & 1 & 1 & 0 & 0 \\ 0 & 0 & 0 & 0 & 1 & 1 \\ 0 & 0 & 0 & 0 & 1 & 1 \end{pmatrix} \qquad B_3^{(2)} = \begin{pmatrix} 1 & 1 & 0 & 0 & 0 & 0 \\ 1 & 0 & 1 & 0 & 0 & 0 \\ 0 & 1 & 1 & 0 & 0 & 0 \\ 0 & 0 & 0 & 1 & 1 & 0 \\ 0 & 0 & 0 & 1 & 0 & 1 \\ 0 & 0 & 0 & 0 & 1 & 1 \end{pmatrix} . \tag{7.21}$$

The possible solutions for $N = 6$ and are represented schematically in Fig. 7.1a. For $N = 7$ there are three solutions and so on.

**Lemma 3.** *In any optimal 2-factor $\nu^*$ all the loops must be in the shoelace configuration.*

*Proof.* In each loop there is the same number of red and blue points. Our general result for the TSP of Sect. 6.3 shows indeed that the shoelace loop is always optimal when restricted to one loop. ∎

**Lemma 4.** *In any optimal 2-factor $\nu^*$ there are no loops with more than 3 red points.*

*Proof.* As soon as the number of red points (and therefore blue points) in a loop is larger than 3, a more convenient 2-factor is obtained by considering a 2-factor with two loops. In fact, as can be seen in Fig. 7.2a, the cost gain is exactly equal to the difference between an ordered and an unordered matching which we know is always negative for $p > 1$ [BCS14]. ∎

It follows that

**Proposition 7.2.1.** *In any optimal bipartite 2-factor $\nu^*$ there are only shoelaces loops with 2 or 3 red points.*

In different words to the optimal bipartite 2-factor solution $\nu^*$ is associated an adjacency matrix which is a block matrix built with the sub-matrices $B_2$ and $B_3$. Two different 2-factors in this class are not comparable, that is all of them can be optimal in particular instances.

**Proposition 7.2.2.** *At given number N of both red and blue points there are at most* $\mathrm{Pad}(N-2)$ *optimal 2-factor $\nu^*$.*

$\mathrm{Pad}(N)$ is the $N$-th *Padovan* number, see the D.1, where it is also shown in (D.7) that for large $N$

$$\mathrm{Pad}(N) \sim \mathrm{p}^N \tag{7.22}$$

with p the *plastic* number (D.3) (see Appendix D.3 for a discussion on this constant). Actually, for values of $N$ which we could explore numerically, we saw that all $\mathrm{Pad}(N-2)$ possible solutions appear as optimal solutions in the ensemble of instances.

### Complete Case

Similar conclusions can be derived in the case of the complete graph $\mathcal{K}_N$ since, as we have said, both the analytical solution for the matching [CDS17] and the TSP are known. Let us order the points in increasing order, i.e. $x_i > x_j$ if $i > j$ with $i, j \in [N]$. In the matching problem on the complete graph the number of points must be even, and with $p > 1$ the solution is very simple: if $j > i$ then the point $x_i$ will be matched to $x_j$ if and only if $i$ is odd and $j = i + 1$ that is

$$E[\mu^*] = \sum_{i\in[N]} |x_{2i} - x_{2i-1}|^p \,. \tag{7.23}$$

The corresponding adjacency matrix assumes the block diagonal form

$$A = \begin{pmatrix} \mathbf{a} & \mathbf{0} & \cdots & \cdots & \mathbf{0} \\ \mathbf{0} & \mathbf{a} & \cdots & \cdots & \mathbf{0} \\ \vdots & \vdots & \ddots & & \vdots \\ \mathbf{0} & \mathbf{0} & & \mathbf{a} & \mathbf{0} \\ \mathbf{0} & \mathbf{0} & \cdots & \mathbf{0} & \mathbf{a} \end{pmatrix} , \tag{7.24}$$

where

$$\mathbf{a} = \begin{pmatrix} 0 & 1 \\ 1 & 0 \end{pmatrix} . \tag{7.25}$$

The adjacency matrix (7.24) satisfies constraints (1.4), with $k = 1$. In the case of the TSP on the complete graph, where the number of points can also be odd, the optimal permutation is the same $\tilde{\sigma}$ defined in (6.16). With a slightly abuse of language we will call "shoelace" also the optimal loop configuration for the TSP problem on complete graph given in (6.57). The adjacency matrix is

$$A = \begin{pmatrix} 0 & 1 & 1 & 0 & \cdots & \cdots & 0 \\ 1 & 0 & 0 & 1 & \cdots & \cdots & 0 \\ 1 & 0 & 0 & 0 & 1 & \cdots & 0 \\ \vdots & \ddots & \ddots & \ddots & \ddots & \ddots & \vdots \\ 0 & \cdots & 1 & 0 & 0 & 0 & 1 \\ 0 & \cdots & 0 & 1 & 0 & 0 & 1 \\ 0 & \cdots & 0 & 0 & 1 & 1 & 0 \end{pmatrix} . \tag{7.26}$$

The possible solutions for the 2-factor on complete graph can be constructed by cutting in a similar way the corresponding TSP solution into smaller loops as can be seen pictorially in Fig. 7.2b. Note that one cannot have a loop with two points. Analogously to the bipartite case we have analyzed before, each loop that form the 2-factor configuration must be a shoelace. However the length of allowed loops will be different, since one cannot cut, on a complete graph, a TSP of 4 and 5 points in two smaller sub-tours. It follows that

**Proposition 7.2.3.** *On the complete graph, in the optimal 2-factor $\nu^*$ there are only loops with 3, 4 or 5 points.*

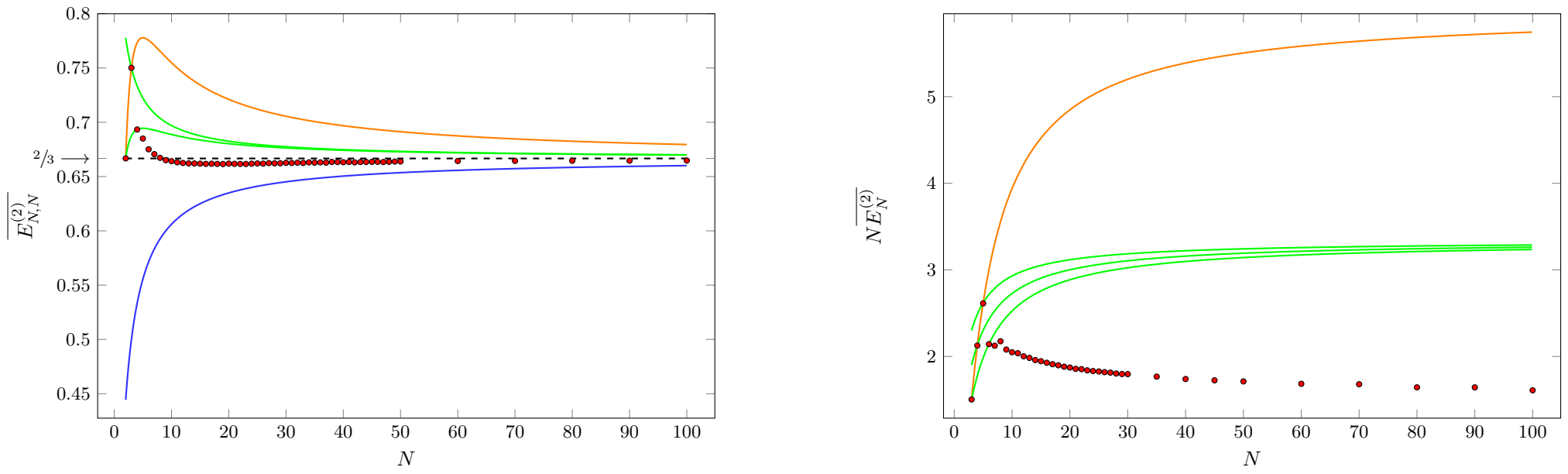

**(a)** $\mathcal{K}_{N,N}$ case. The orange line is the cost of the TSP given in (7.29); the green lines are, from above, the cost of the optimal fixed 2-factor $\nu_{(2,2,\dots,2,3)}$ given in (7.38) and $\nu_{(2,2,\dots,2)}$ given in (7.37). The dashed black line is the asymptotic value $\frac{2}{3}$ and the blue continuous one is twice the cost of the optimal 1-matching $\frac{2}{3}\frac{N}{N+1}$. Red points are the results of a 2-factor numerical simulation, in which we have averaged over $10^7$ instances.

**(b)** $\mathcal{K}_N$ case. Here the average cost is rescaled with $N$. The orange line is the cost of the TSP given in (7.39). The green lines are from above the cost of the fixed 2-factor $\nu_{(3,3,\dots,3,5)}$ given in (7.45), $\nu_{(3,3,\dots,4)}$ given in (7.44) and $\nu_{(3,3,\dots,3)}$ given in (7.43). Red points are the results of a numerical simulation for the 2-factor, in which we have averaged over $10^5$ instances for $N \leq 30$, $10^4$ for $30 < N \leq 50$ and $10^3$ for $N > 50$.

**Figure 7.3.** Average optimal costs for various $N$ and for $p = 2$.

In other terms the optimal configurations are composed by the adjacency matrices $A_1$, $A_2$, $A_3$ that are

$$A_3 = \begin{pmatrix} 0 & 1 & 1 \\ 1 & 0 & 1 \\ 1 & 1 & 0 \end{pmatrix}, \quad A_4 = \begin{pmatrix} 0 & 1 & 1 & 0 \\ 1 & 0 & 0 & 1 \\ 1 & 0 & 0 & 1 \\ 0 & 1 & 1 & 0 \end{pmatrix}, \quad A_5 = \begin{pmatrix} 0 & 1 & 1 & 0 & 0 \\ 1 & 0 & 0 & 1 & 0 \\ 1 & 0 & 0 & 0 & 1 \\ 0 & 1 & 0 & 0 & 1 \\ 0 & 0 & 1 & 1 & 0 \end{pmatrix}. \tag{7.27}$$

In Fig. 7.1b we represent the two solutions when $N = 7$. In Appendix D.2 we prove that, similarly to the bipartite case, the number of 2-factor solutions is at most $g_N$ on the complete graph, which for large $N$ grows according to

$$g_N \sim \mathsf{p}^N . \tag{7.28}$$

Also in this case we verified numerically, for accessible $N$, that the set of possible solutions that we have identified is actually realized by some instance of the problem.

## 7.3 Bounds on the cost

Here we will derive the consequences of the results of the previous section, obtaining explicitly some upper bounds on the average optimal cost of the 2-factor problem. We will examine the complete bipartite case first, where we consider, for simplicity, the $p = 2$ case. Indeed the calculation we perform below can be done also for general $p > 1$, but it is much more involved and we sketch it in Appendix E.3. Then we will examine the complete graph case, where we have obtained a very simple expression of the average optimal cost for every $N$, and for every $p > 1$.

### Bipartite Case

Let us analyze the problem on the complete bipartite graph $\mathcal{K}_{N,N}$. In Sect. 6.3 we derived for $p = 2$ the exact result for all $N$ of the TSP when all the points are chosen with a flat distribution in the interval $[0, 1]$

$$\overline{E^{(2)}_{N,N}[h^*]} = \frac{2}{3}\frac{N^2 + 4N - 3}{(N+1)^2} \tag{7.29}$$

from which we soon obtain that

$$\overline{E^{(2)}_{N,N}[\nu^*]} = \begin{cases} \frac{2}{3} & \text{for } N = 2 \\ \frac{3}{4} & \text{for } N = 3\,, \end{cases} \tag{7.30}$$

since in the cases $N = 2$ and $N = 3$ the solutions are the same as in the TSP. For $N = 4$ we have still only one solution, which corresponds to two cycles on the first and the last 2 red points. Both cycles have the same cost and we easily get that

$$\overline{E^{(2)}_{4,4}[\nu^*]} = \frac{52}{75}\,. \tag{7.31}$$

This result can be obtained also in a different way. We first remark that

$$\overline{(r_k - b_k)^2} + \overline{(r_{k+1} - b_{k+1})^2} - \overline{(r_k - b_{k+1})^2} - \overline{(r_{k+1} - b_k)^2} = -\frac{2}{(N+1)^2} \tag{7.32}$$

irrespectively from the choice of $1 \leq k \leq N - 1$. This is exactly the cost gained by cutting a longer cycle into two smaller ones at position $k$, see Fig. 7.2a. Therefore the cost for the optimal 2-factor for $N = 4$ is the cost for the optimal Hamiltonian cycle, which from (7.29) is $\frac{58}{75}$, decreased because of a cut, that is by $-\frac{2}{25}$.

For $N = 5$ there are two possible optimal solutions that we will denote by $\nu_{(2,3)}$ and $\nu_{(3,2)}$. For both of them

$$\overline{E^{(2)}_{5,5}[\nu_{(2,3)}]} = \overline{E^{(2)}_{5,5}[\nu_{(3,2)}]} = \frac{13}{18} \tag{7.33}$$

and therefore

$$\overline{E^{(2)}_{5,5}[\nu^*]} = \overline{\min\left\{E^{(2)}_{5,5}[\nu_{(2,3)}], E^{(2)}_{5,5}[\nu_{(3,2)}]\right\}} \leq \frac{13}{18}\,. \tag{7.34}$$

For $N = 6$ there are still two possible optimal solutions, that is $\nu_{(3,3)}$ and $\nu_{(2,2,2)}$, but this time they have not the same average cost, indeed

$$\overline{E^{(2)}_{6,6}[\nu_{(3,3)}]} = \frac{36}{49} = \frac{38}{49} - \frac{2}{49} \tag{7.35}$$

$$\overline{E^{(2)}_{6,6}[\nu_{(2,2,2)}]} = \frac{34}{49} = \frac{38}{49} - \frac{4}{49} \tag{7.36}$$

that we have written as the TSP value from (7.29) decreased, respectively, by one and two cuts (7.32) for $N = 6$.

Now it is clear that when $N$ is even the 2-factor with lowest average energy is $\nu_{(2,2,\dots,2)}$ and that

$$\overline{E^{(2)}_{N,N}[\nu_{(2,2,\dots,2)}]} = \frac{2}{3}\frac{N^2+4N-3}{(N+1)^2} - \frac{N-2}{(N+1)^2} = \frac{1}{3}\frac{N(2N+5)}{(N+1)^2}\,, \tag{7.37}$$

which is an upper bound for the optimal average cost since, even though this configuration has the minimum average cost, for every fixed instance of disorder there can be another one which is optimal. For $N$ odd one of the 2-factors with lowest average energy is $\nu_{(2,2,\dots,2,3)}$ and

$$\overline{E^{(2)}_{N,N}[\nu_{(2,2,\dots,2,3)}]} = \frac{2}{3}\frac{N^2+4N-3}{(N+1)^2} - \frac{N-3}{(N+1)^2} = \frac{1}{3}\frac{2N^2+5N+3}{(N+1)^2}\,, \tag{7.38}$$

a result which shows that essentially the upper bound for the optimal average cost for even and odd large $N$ is the same. In Appendix E.3 we generalize these computations to generic $p$.

### Complete Case

Let us now turn to the problem on the complete graph. In Sect. 6.4 it is shown that, for every $N$ and every $p > 1$, the average optimal cost of the TSP has the expression

$$\overline{E^{(p)}_N[h^*]} = [(N-2)(p+1)+2]\,\frac{\Gamma(N+1)\,\Gamma(p+1)}{\Gamma(N+p+1)}\,. \tag{7.39}$$

An analogous expression is present in the case of the matching problem [CDS17], where the number of points $N$ is even

$$\overline{E^{(p)}_N[\mu^*]} = \frac{N\,\Gamma(N+1)\,\Gamma(p+1)}{2\,\Gamma(N+p+1)}\,. \tag{7.40}$$

Let us now turn to the evaluation of the cost gain when we cut the cycle in two shoelaces sub-cycles. For $p > 1$ the cost gain doing one cut can be written as

$$\overline{(x_{k+1}-x_k)^p} + \overline{(x_{k+3}-x_{k+2})^p} - \overline{(x_{k+3}-x_{k+1})^p} - \overline{(x_{k+2}-x_k)^p} = -\frac{2\,p\,\Gamma(N+1)\,\Gamma(p+1)}{\Gamma(N+p+1)}\,. \tag{7.41}$$

For example for $N = 6$ (in which the solution is unique since 6 can be written as a sum of 3, 4 and 5 in an unique way as 3+3) and $p = 2$ we have

$$\overline{E^{(2)}_6} = \frac{1}{2} - \frac{1}{7} = \frac{5}{14}\,. \tag{7.42}$$

If $N$ is multiple of 3, the lowest 2-factor is, on average, the one with the largest number of cuts i.e. $\nu_{(3,3,\dots,3)}$. The number of cuts is $(N-3)/3$ so that the average cost of this configuration is

$$\overline{E^{(p)}_N[\nu_{(3,3,\dots,3)}]} = N\left(\frac{p}{3}+1\right)\frac{\Gamma(N+1)\,\Gamma(p+1)}{\Gamma(N+p+1)}\,. \tag{7.43}$$

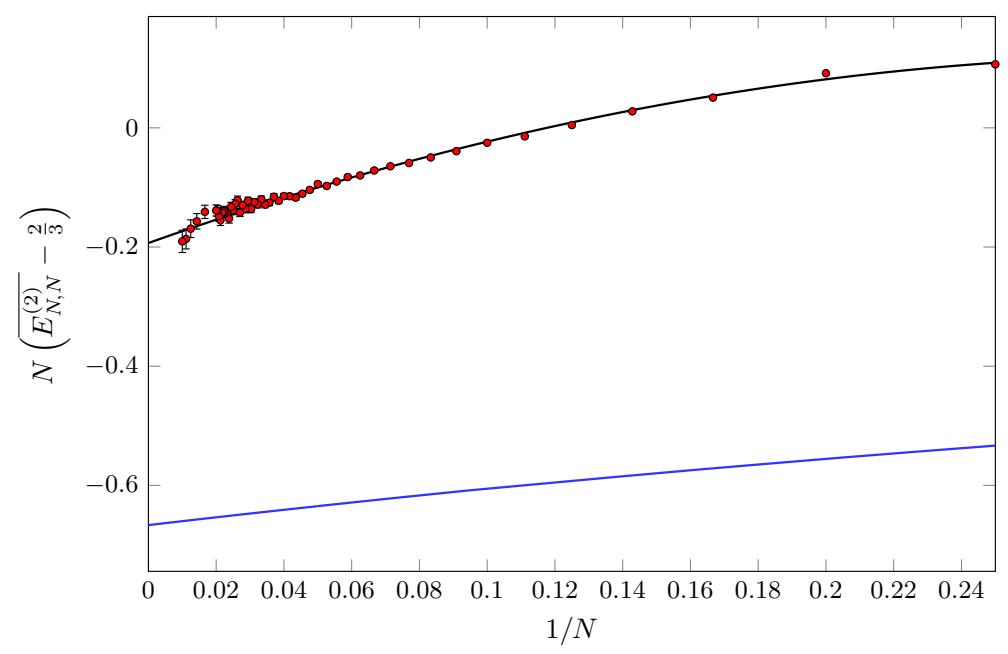

**(a)** Numerical results of $N\left(\overline{E^{(2)}_{N,N}} - \frac{2}{3}\right)$ (red points) in the complete bipartite case; the black line is the fitting function (7.48). The blue line is two times the matching. The values of $a_1$, $a_2$ and $a_3$ are reported in Table 7.1.

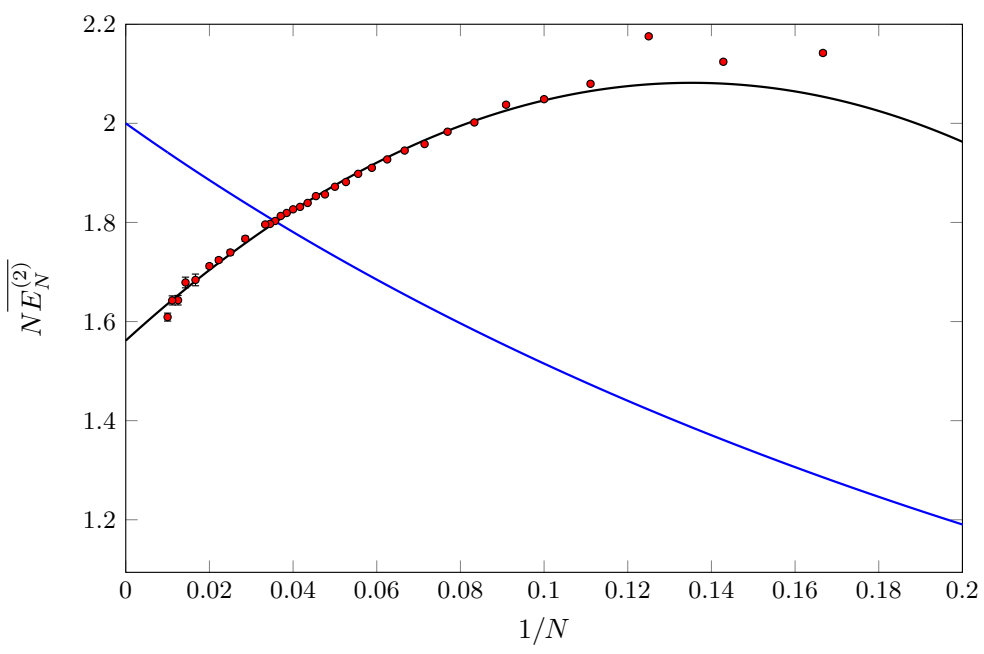

**(b)** Numerical results of $N\overline{E^{(2)}_N}$ (red points) in the complete graph case; the black line is the fitting function (7.49). The blue line is two times the value of the optimal matching as given in equation (7.40). The values of $b_0$, $b_1$ and $b_2$ are reported in Table 7.2.

**Figure 7.4**

| $a_1$ | $a_2$ | $a_3$ |
|---|---|---|
| $-0.193 \pm 0.003$ | $2.03 \pm 0.05$ | $-3.3 \pm 0.2$ |

**Table 7.1.** Numerical estimates of the parameters $a_1$, $a_2$ and $a_3$ defined in (7.48).

Instead if $N$ can be written as a multiple of 3 plus 1, the minimum average energy configuration is $\nu_{(3,3,\dots,3,4)}$, which has $(N-4)/3$ cuts and

$$\overline{E^{(p)}_N}[\nu_{(3,3,\dots,4)}] = \left[N\left(\frac{p}{3}+1\right) + \frac{2}{3}p\right]\frac{\Gamma(N+1)\,\Gamma(p+1)}{\Gamma(N+p+1)}\,. \tag{7.44}$$

The last possibility is when $N$ is a multiple of 3 plus 2, so the minimum average energy configuration is $\nu_{(3,3,\dots,3,5)}$, with $(N-4)/3$ cuts and

$$\overline{E^{(p)}_N}[\nu_{(3,3,\dots,5)}] = \left[N\left(\frac{p}{3}+1\right) + \frac{4}{3}p\right]\frac{\Gamma(N+1)\,\Gamma(p+1)}{\Gamma(N+p+1)}\,. \tag{7.45}$$

In the limit of large $N$ all those three upper bounds behave in the same way. For example

$$\lim_{N\to\infty} \overline{E^{(p)}_N}[\nu_{(3,3,\dots,3)}] = N^{1-p}\left(1+\frac{p}{3}\right)\Gamma(p+1)\,. \tag{7.46}$$

Note that the scaling of those upper bounds for large $N$ is the same of those of matching and TSP.

## 7.4 Numerical Results

In this section we present our numerical simulations describing briefly the algorithm we have used to find the solution for every instance of the problem. The 2-factor problem has an integer programming formulation. Given a generic simple graph $\mathcal{G} = (\mathcal{V}, \mathcal{E})$, the solution can be uniquely identified by a $|\mathcal{V}| \times |\mathcal{V}|$ matrix of occupation numbers $A_{ij}$ which can assume values 0 or 1. In particular $A_{ij}$ assumes value 0 if node $i$ is not connected to node $j$ in the 2-factor solution and 1 otherwise. The problem can be stated as the minimization of the energy function

$$E(A) = \frac{1}{2} \sum_{i=1}^{|\mathcal{V}|} \sum_{j=1}^{|\mathcal{V}|} A_{ij} w_{ij} , \tag{7.47}$$

subject to the constrain (1.4) with $k = 2$. We have performed some numerical simulations using a `C++` code and the open source `GLPK` package, a library that solves general large scale linear programming problems. In Fig. 7.3a and 7.3b we plot the results of some numerical simulations for $p = 2$ respectively for the complete bipartite and complete graph case and we compare them with some exact results. In the complete graph case we plot $N\overline{E_N^{(2)}}$ revealing that the scaling of the cost is the same of the TSP and the matching problem. However the two situations are completely different, since in the complete case the bound estimate only gets worse when $N$ increases.

| $b_0$ | $b_1$ | $b_2$ |
|---|---|---|
| $1.562 \pm 0.005$ | $7.7 \pm 0.2$ | $-28 \pm 2$ |

**Table 7.2.** Numerical estimates of the parameters $b_0$, $b_1$ and $b_2$ defined in (7.49).

In order to understand the analytic form of the finite-size correction, we have also performed a parametric fit of the quantity $N\left(\overline{E_{N,N}^{(2)}} - \frac{2}{3}\right)$ using a fitting function of the type

$$f_B(N) = a_1 + \frac{a_2}{N} + \frac{a_3}{N^2} . \tag{7.48}$$

In Fig. 7.4a we plot the numerical data and $f_B(N)$. The estimate of the parameters is reported in Table 7.1.

In the complete graph case, we have performed a fit of the rescaled cost $N\overline{E_N^{(2)}}$ in order to evaluate numerically the asymptotic value of the cost. The fitting function was chosen to be

$$f_M(N) = b_0 + \frac{b_1}{N} + \frac{b_2}{N^2} . \tag{7.49}$$

In Fig. 7.4b we report the plot of the numerical data together with $f_M(N)$. Remember that in the complete graph case, the cost of the 2-factor cannot be bounded from below by two times the cost of the optimal matching as happens on the complete bipartite graph. For this reason in Fig. 7.4b we have added the plot of the theoretical value of the optimal matching (given in equation (7.40)) multiplied by two. The numerical values of the parameters are reported in Table 7.2. Note also how in the complete graph case, the first finite-size correction $b_1$ is not only positive but its magnitude is much greater than $a_1$, its bipartite counterpart.

# Chapter 8

# Going to higher dimensions

In previous chapters we have focused mainly on one-dimensional problems, showing how, in most of the cases under consideration, the solution is always of the same "shape", irrespectively of the positions of the points. This permits us to find the solution of these problems efficiently (i.e. polynomially), their time complexity scaling as the best algorithm for sorting the positions of $N$ points in increasing order. Therefore, even the Euclidean TSP problem, which is in the worst case scenario NP-complete, is polynomial when considering the subspace of one-dimensional instances. This property is obviously lost in 2 dimension where it is not possible to find the features we found in 1 dimension. However the study of one-dimensional problems can indeed be an important starting point since it may give an indication or a suggestion of what can possibly happen in higher dimensional cases. In this final chapter we study the two-dimensional version of the bipartite TSP. Indeed in previous chapters we have found, in one dimension, a very strong connection with the assignment problem. We list here what we consider the three important progresses that helped us to reassemble the puzzle in two dimensions:

- for other optimization problems similar to the TSP, the monopartite and bipartite versions have different optimal cost properties. For example for the matching, 1-factor and 2-factor (or loop-covering) problems, the optimal cost is expected to be a self-averaging quantity whose average scales according to
$$\overline{E_N^{(p,d)}} \sim N^{1-\frac{p}{d}} \tag{8.1}$$
(see [BHH59] for a proof in the case $p = 1$). On the other hand, in the bipartite version [AKT84, Car+14, BCS14] it is expected that
$$\overline{E_N^{(p,d)}} \sim \begin{cases} N^{1-\frac{p}{2}} & \text{for} \quad d = 1 \\ N^{1-\frac{p}{2}}(\log N)^{\frac{p}{2}} & \text{for} \quad d = 2 \\ N^{1-\frac{p}{d}} & \text{for} \quad d > 2 \end{cases} \tag{8.2}$$
that is a larger cost with respect to the monopartite case when $d \leq 2$. Moreover, this quantity is expected to be not self-averaging. This anomalous scaling is due to the fact that in low dimensions density fluctuations, due to the local difference between the number of red and blue points, become relevant;

- in the bipartite case it is always true (see Chap. 7) that the total optimal cost of the TSP $E^*_{\mathcal{H}}$ is larger than the total optimal cost of the 2-factor problem (loop covering) $E^*_{\mathcal{M}_2}$, which is larger than twice the total optimal cost of the corresponding matching problem (assignment) $E^*_{\mathcal{M}_1}$

$$E^*_{\mathcal{H}} \geq E^*_{\mathcal{M}_2} \geq 2E^*_{\mathcal{M}_1} \,. \tag{8.3}$$

In Chap. 6 it has been shown that in $d = 1$, in the asymptotic limit of an infinitely large number points, this bound is saturated, that is the total optimal cost of the TSP, rescaled with $N^{1-\frac{p}{2}}$, is exactly twice the total rescaled optimal cost of the assignment problem, and, therefore, all the three quantities coincide.

- in the bipartite case, in $d = 2$ and $p = 2$, thanks to a deep connection with the continuum version of the problem, that is the well known *transport problem*, it has been possible to compute, exactly, the total optimal cost of the assignment problem in the asymptotic limit of an infinitely large number points [Car+14, CS15b, CS15a, AST18]:

$$\overline{E^{(2,2)}_{N,N}} = \frac{1}{2\pi} \log N \,. \tag{8.4}$$

We considered, therefore, the possibility that also in $d = 2$, and $p > 1$, exactly, thanks to the logarithmic violation present in the bipartite case, the asymptotic total cost of the TSP can be exactly twice the one of the assignment, that for $p = 2$ is also exactly known. Indeed, this is the case!

The paper is organized as follows. In Section 8.1 we introduce a scale argument that justifies our claims and allows to find sub-optimal TSP and 2-factor problem solutions whose difference in cost with the optimal solution goes to zero as the number of cities goes to infinity. In Section 8.2 we provide evidence of our results by extensive numerical simulations. We also examined the case $p = 1$, which is the most largely considered in the literature.

## 8.1 Scaling argument

In this section we will provide a scaling argument to support our claim, that is, also in two dimensions, for any given choice of the positions of the points, in the asymptotic limit of large $N$, the cost of the bipartite TSP converges to twice the cost of the assignment.

Given an instance, let us consider the optimal assignment $\mu^*$ on them. Let us now consider $N$ points which are taken between the red an blue point of each edge in $\mu^*$ and call $\mathcal{T}^*$ the optimal *monopartite* TSP solution on these points. For simplicity, as these $N$ points we take the blue points.

We shall use $\mathcal{T}^*$ to provide an ordering among the red and blue points. Given two consecutive points in $\mathcal{T}^*$, for example $b_1$ and $b_2$, let us denote by $(r_1, b_1)$ and $(r_2, b_2)$ the two edges in $\mu^*$ involving the blue points $b_1$ and $b_2$ and let us consider also the new edge $(r_1, b_2)$.We know that, in the asymptotic limit of large $N$, the typical distance between two matched points in $\mu^*$ scales as $(\log N/N)^{1/2}$ while the typical distance between two points matched in the monopartite case scales only as $1/N^{1/2}$, that is (for all points but a fraction which goes to zero with $N$)

$$\begin{aligned} w_{(b_1,r_1)} &= \left(\alpha_{11}\frac{\log N}{N}\right)^{\frac{p}{2}}, \\ w_{(b_2,r_1)} &= \left[\beta_{22}\frac{1}{N} + \alpha_{11}\frac{\log N}{N} - \gamma\frac{\sqrt{\log N}}{N}\right]^{\frac{p}{2}}, \end{aligned} \tag{8.5}$$

where $(\alpha_{11}\log N/N)^{1/2}$ is the length of the edge $(r_1, b_1)$ of $\mu^*$, $(\beta_{22}/N)^{1/2}$ is the length of the edge $(b_1, b_2)$ of $\mathcal{T}^*$ and $\gamma = 2\sqrt{\alpha_{11}\beta_{22}}\cos\theta$, where $\theta$ is the angle between the edges $(r_1, b_1)$ of $\mu^*$ and $(b_1, b_2)$ of $\mathcal{T}^*$. This means that, typically, the difference in cost

$$\Delta E = w_{(b_2,r_1)} - w_{(b_1,r_1)} \sim \frac{(\log N)^{\frac{p-1}{2}}}{N^{\frac{p}{2}}} \tag{8.6}$$

is small as compared to the typical cost $(\log N/N)^{\frac{p}{2}}$ of one edge in the bipartite case. To obtain a valid TSP solution, which we call $h^A$, we add to the edges $\mu^* = \{(r_1, b_1), \dots, (r_N, b_N)\}$ the edges $\{(r_1, b_2), \dots, (r_{N-1}, b_N), (r_N, b_1)\}$, see Figure 8.1.

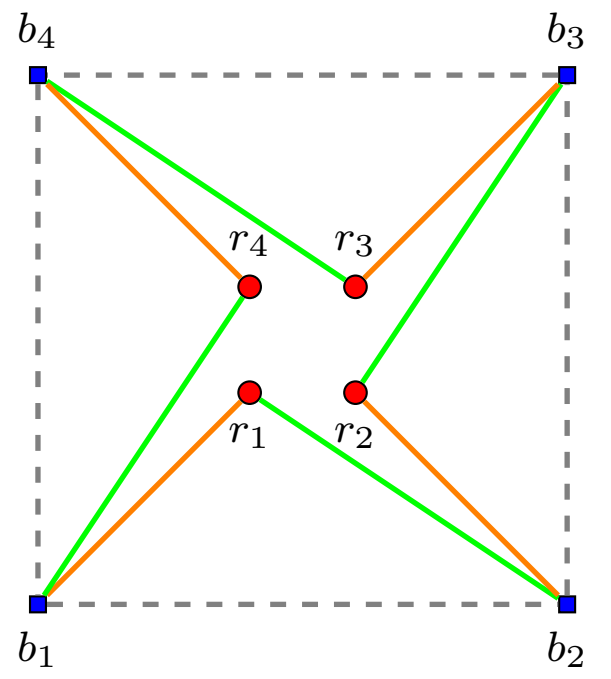

**Figure 8.1.** The optimal assignment $\mu^*$ is given by the orange edges $\{(r_1, b_1), (r_2, b_2), (r_3, b_3), (r_4, b_4)\}$. The monopartite TSP (gray dashed edges) among blue points provides the necessary ordering. In order to obtain the TSP $b_1, r_1, b_2, r_2, b_3, r_3, b_4, r_4, b_1$ in the bipartite graph we have to add the green edges $\{((r_1, b_2), (r_2, b_3), (r_3, b_4), (r_4, b_1)\}$.

Of course $h^A$ is not, in general, the optimal solution of the TSP. However, because of Eq. (8.3), we have that

$$E_{\mathcal{H}}[h^A] \geq E^*_{\mathcal{H}} \geq E^*_{\mathcal{M}_2} \geq 2\,E^*_{\mathcal{M}_1} \tag{8.7}$$

and we have shown that, for large $N$, $E_{\mathcal{H}}[h^A]$ goes to $2\,E^*_{\mathcal{M}_1}$ and therefore also $E^*_{\mathcal{H}}$ must behave in the same way. As a byproduct of our analysis also $E^*_{\mathcal{M}_1}$ for the loop covering problem has the same optimal average cost. Note also that our argument is purely local and therefore it does not depend in any way on the type of boundary conditions adopted. Since in the case of periodic boundary

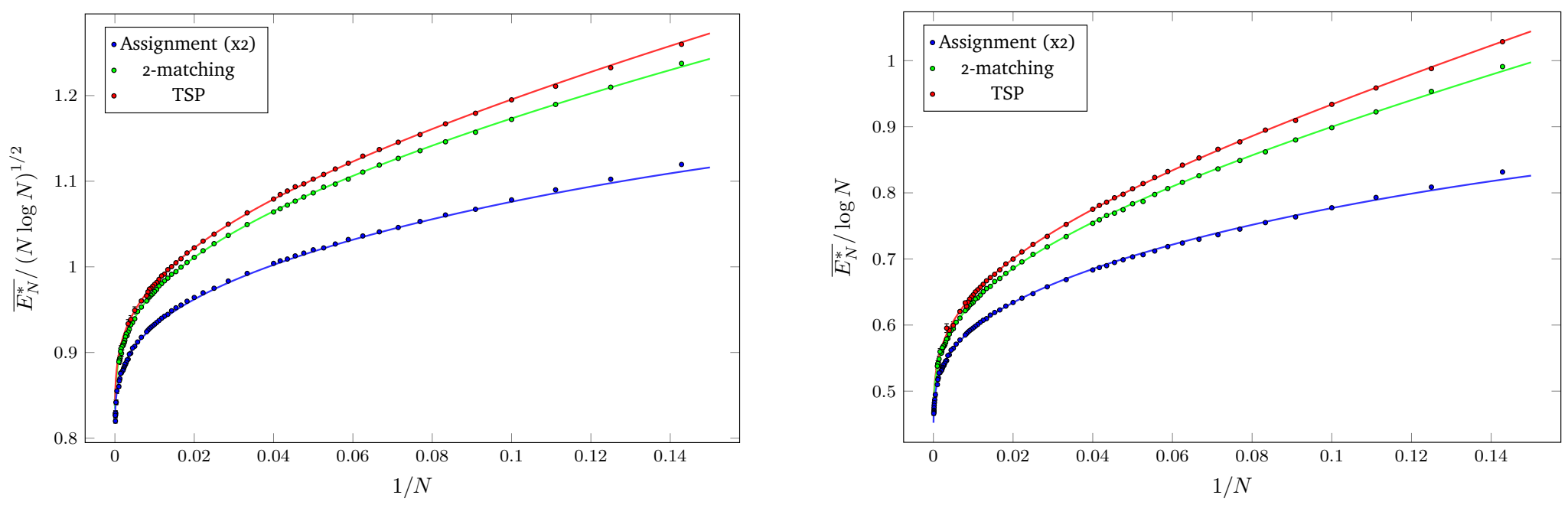

**(a)** $p = 1$ case. **(b)** $p = 2$ case.

**Figure 8.2.** Numerical results for $p = 1$ and $p = 2$ for the TSP (red points), the 2-factor (green points) and 2 times the assignment problem (blue points) in the open boundary condition case. Continuous lines are fit to the data.

conditions, as shown in [CS15a], it holds (8.4), we get that the average optimal cost of both the TSP and 2-factor goes for large $N$ to 2 times the optimal assignment.

Notice that an analogous construction can be used in any number of dimensions. However, the success of the procedure lies in the fact that the typical distance between two points in $\mu^*$ goes to zero slower than the typical distance between two consecutive points in the monopartite TSP. This is true only in one and two dimensions, and it is related to the importance of fluctuations in the number of points of different kinds in a small volume.

This approach allowed us to find also an approximated solution of the TSP which improves as $N \to \infty$. However, this approximation requires the solution of a *monopartite* TSP on $N/2$ points, corroborating the fact that the bipartite TSP is a hard problem (from the point of view of complexity theory).

A similar construction can be used to achieve an approximated solution also for the 2-factor problem. In this case, instead of solving the monopartite TSP on the mean points of each edge of $\mu^*$, one should solve the monopartite matching problem on the same set of points, obtaining a matching $\mathcal{M}^*$. Once more let us denote by $(r_1, b_1)$ and $(r_2, b_2)$ the two edges in $\mu^*$ which give rise to two matched points in $\mathcal{M}^*$, and collect them together with the edges $(r_1, b_2)$ and $(r_2, b_1)$. Repeating the above procedure for each couple of points matched in $\mathcal{M}^*$, the union of the edges obtained gives a valid 2-factor whose cost tends, in the limit of large $N$, to twice the cost of the optimal assignment in one and two dimensions. Notice that, in this case, the procedure is much more efficient because the solution of the matching problem is polynomial in time.

## 8.2 Numerical Results

We have confirmed our theoretical predictions performing numerical simulations on all the three models previously presented: assignment, bipartite 2-factor, and bipartite TSP. We have considered the case of open boundary conditions.

For what concerns the assignment problem, many polynomial-time algorithms are available in the literature, as the famous Hungarian algorithm [Kuh55]. We have implemented an in-house assignment solver based on the `LEMON` optimization library [DJK11], which is based on the Edmonds' blossom algorithm [Edm65]. In the case of the 2-factor and TSP, the most efficient way to tackle numerically those problems is to exploit their *linear* or *integer programming* formulation.

| $p = 1$ | $a_1$ | $a_2$ | $a_3$ |
|---|---|---|---|
| TSP | 0.717(2) | 1.32(1) | $-0.509(1)$ |
| 2-factor | 0.714(2) | 1.31(1) | $-0.58(2)$ |
| Assignment | 0.715(2) | 1.16(2) | $-0.757(3)$ |
| $p = 2$ | $a_1$ | $a_2$ | $a_3$ |
| TSP | 0.321(5) | 1.603(2) | $-0.428(6)$ |
| 2-factor | 0.319(4) | 1.577(2) | $-0.547(7)$ |
| Assignment | 0.31831 | 1.502(2) | $-1.0(2)$ |

**Table 8.1.** Comparison between fit factors in assignment and TSP, for $p = 1, 2$. We have doubled the factors for the assignment to verify our hypothesis. For $p = 2$, we have reported the theoretical value of $a_1$ which is $1/\pi$.

To validate our argument, we solved for both assignment and 2-factor problem (with $p = 1, 2$), $10^5$ independent instances for $2 \leq N \leq 125$, $10^4$ independent instances for $150 \leq N \leq 500$, and $10^3$ independent instances for $600 \leq N \leq 1000$. In the TSP case, the computational cost is dramatically larger; for this reason the maximum number of points we were able to achieve with a good numerical precision using integer programming was $N = 300$, also reducing the total number of instances.

An estimate of the asymptotic average optimal cost and finite-size corrections has been obtained using the fitting function for $p = 1$

$$f^{(p=1)}(N) = \sqrt{N \log N} \left( a_1 + \frac{a_2}{\log N} + \frac{a_3}{\log^2 N} \right) \tag{8.8}$$

while, for $p = 2$

$$f^{(p=2)}(N) = \log N \left( a_1 + \frac{a_2}{\log N} + \frac{a_3}{\log^2 N} \right) . \tag{8.9}$$

These are the first 3 terms of the asymptotic behavior of the cost of the assignment problem [AKT84, Car+14]. Parameters $a_2$ and $a_3$ for $p = 2$ were obtained fixing $a_1$ to $1/\pi$. In Figure 8.2 we plot the data and fit in the case of open boundary conditions. Results are reported in Table 8.1.

To better confirm the behavior of the average optimal cost of the TSP, we also performed some numerical simulations using a much more efficient solver, that is the Concorde TSP solver [App+06], which is based on an implementation of the Branch-and-cut algorithm proposed by Padberg and Rinaldi [PR91]. The results for the leading term of the asymptotic average optimal cost are confirmed while a small systematic error due to the integer implementation of the solver is observed in the finite-size corrections.

## 8.3 Conclusions

In this chapter we have considered three combinatorial optimization problems, the matching problem, 2-factor problem and TSP, where the cost is a convex increasing function of the point distances. Previous investigations have been performed in the one-dimensional case, by means of exact solutions. Here we analyzed the bipartite version of these problems in two dimensions, showing that, as already obtained in one dimension:

$$\lim_{N\to\infty} \frac{\overline{E^*_{\mathcal{H}}}}{\overline{E^*_{\mathcal{M}_1}}} = \lim_{N\to\infty} \frac{\overline{E^*_{\mathcal{M}_2}}}{\overline{E^*_{\mathcal{M}_1}}} = 2\,. \tag{8.10}$$

This implies, for the special case $p = 2$, by using (8.4), our main exact result, that is

$$\lim_{N\to\infty} \frac{\overline{E^*_{\mathcal{H}}}}{\log N} = 1/\pi\,. \tag{8.11}$$

In general, the evaluation of $\overline{E^*_{\mathcal{H}}}$ and $\overline{E^*_{\mathcal{M}_2}}$ for large $N$ is reduced to the solution of the matching problem which requires only polynomial time. This seems to be a peculiar feature of the bipartite problem: the monopartite TSP cannot be approached in a similar way. As a byproduct of our analysis, we provided in Sec. 8.1 two approximate algorithms, for the bipartite TSP and the bipartite 2-factor: both are guaranteed to give a solution with optimal cost for large $N$. The first algorithm allows to solve the bipartite TSP on $N$ points solving the monopartite TSP with $N$ points (notice that, on principle, the bipartite version consists of $2N$ points). The second allows to exploit the fast Hungarian algorithm to obtain an approximate solution of the 2-factor problem.

# Part IV

# Conclusions

# Chapter 9

# Conclusions and perspectives

In this thesis work we have analyzed several combinatorial optimization problems studying both the behavior of average quantities (such as the average optimal cost) and finite-size corrections to them.

In Part II we have started by analyzing mean field models, where the weights are independent identically distributed random variables. These were the first problems that has been studied since 1985 by means of techniques borrowed from the theory of disordered systems, such as replica and cavity method (belief propagation in modern language) that we have reviewed in Part I together with spin glass theory concepts. In Part II we have mainly focused on finite-size corrections to the average optimal cost of several problem, such as the RAP, the RMP and the RFMP. In particular, taking into consideration a generic class of disorder distributions, we have showed how the Maclaurin coefficients of its series expansion near the origin affect both the coefficients and the scaling exponent of the first finite-size correction. In the case of the pure power law probability distribution (where all the Maclaurin coefficients of the probability distribution vanish except the first one) the corrections are only analytical, that is in inverse powers of the number of points. When the first Maclaurin coefficient does not vanish, as in the gamma probability distribution case, the scaling of the first correction is non-analytical and always larger than the power law case. We have also characterized the scaling of the corrections to all orders of perturbation theory. For the RMP an additional correction, due to the fluctuation around the saddle point, is present and it is reminiscent in form of finite-size corrections in many mean-field diluted models. It can in fact be interpreted as the free energy contributions of odd loops of the graph. This correction disappears not only when the problem is defined on the bipartite graph (where there can be only loops of even length), but also when one allows, on the complete graph, the presence of loops in any feasible configuration as happens for the RFMP. In the “loopy” version of the RFMP, obtained by allowing the presence of self-loops, the finite-size correction of order $1/N$ vanishes, coherently with rigorous results. It is an argument of future work to extend these results to locally tree-like random graph which mimic better finite-dimensional results. One expects also in this case that the free energy contributions of even loop in the graph does not contribute to the finite-size correction. We have also pointed out that still, there are no analogous expression for finite-size corrections for the random-link 2-factor problem and for the RTSP since in the context of replica method, the limit of small temperatures is tricky. Using cavity method, instead, one can obtain the average optimal cost in the limit of large number of points. We have proved the key connection between quantities appearing in the replica approach (the RS multi-overlap), with those of the cavity method (the distribution of cavity fields). Interestingly this relation is valid also for the RMP. It is an argument of future work how to use this relation to efficiently compute finite-size corrections the random-link 2-factor problem and the RTSP.

In Part III we have focused on combinatorial optimization problems in finite dimension. In that case, the points are extracted uniformly in an hypercubic space and the weight on an edge is chosen to be a power of the Euclidean distance between them. As a consequence, the weights will be correlated. As a manner of fact the methods developed in mean-field simply fail and one has to resort to combinatorial methods or to the possibility of constructing a continuum field-theory of the problem under investigation. In this part we have mainly analyzed the random Euclidean 2-factor and TSP both on the complete and the bipartite graph. We have started by analyzing one-dimensional models whose solution has often enabled progress in the study of higher-dimensional cases. The one-dimensional analysis has unveiled an important connection with the matching problem, especially when considering the complete bipartite case for convex increasing weights, where not only one obtains the same scalings, but also the average cost itself of the 2-factor and the TSP tends, for large number of points, to 2 times the average optimal cost of the assignment. We proved, by using a scale argument and extensive numerical simulation, that this result is also valid in 2 dimensions. It is desirable therefore, to develop a continuum version approach that is capable to re-derive the previous results and hopefully it is able to compute finite-size corrections, as happens in the random Euclidean assignment problem.

It will be matter of future work also the analysis of large deviation properties of combinatorial optimization problems [MPS19], both in mean-field and in finite dimensional cases.

# Appendices

# Appendix A

# Details of replica computation of the RAP

## A.1 Evaluation of $z[Q]$ on the saddle-point and analytic continuation for $n \to 0$

Let us evaluate now the quantity $z[Q]$ on the RS saddle-point. Using the fact that, for any analytic function $f$

$$\int_0^{2\pi} \frac{d\lambda}{2\pi} e^{i\lambda} f\left(e^{-i\lambda}\right) = \oint \frac{d\xi}{2\pi i} \frac{f(\xi)}{\xi^2} = \left.\frac{df}{d\xi}\right|_{\xi=0}, \tag{A.1}$$

we can write

$$\begin{aligned} &\left[\prod_{a=1}^{n} \int_0^{2\pi} \frac{d\lambda^a}{2\pi} e^{i\lambda^a}\right] \exp\left\{ {\sum_{\alpha\in\mathcal{P}([n])}}' q_{|\alpha|} e^{-i\sum_{b\in\alpha}\lambda^b} \right\} \\ &= \left.\frac{\partial^n}{\partial\xi_1\cdots\partial\xi_n}\right|_{\xi_1=\cdots=\xi_n=0} \exp\left\{ {\sum_{\alpha\in\mathcal{P}([n])}}' q_{|\alpha|} \prod_{b\in\alpha} \xi_b \right\} = \sum_{\boldsymbol{\alpha}} \prod_{\alpha_i\in\boldsymbol{\alpha}} q_{|\alpha_i|}, \end{aligned} \tag{A.2}$$

where $\boldsymbol{\alpha} = \{\alpha_i\}_i$ and $\alpha_i \in \mathcal{P}([n])$ are disjoint subsets whose union is $[n]$; however

$$\begin{aligned} \sum_{\boldsymbol{\alpha}} \prod_{\alpha_i\in\boldsymbol{\alpha}} q_{|\alpha_i|} &= \sum_{m=1}^{n} \sum_{\substack{k_1,\dots,k_m \\ k_1+\cdots+k_m=n}} \binom{n}{k_1\dots k_m} \frac{q_{k_1}\cdots q_{k_m}}{m!} \\ &= \left(\frac{d}{dt}\right)^n \sum_{m=0}^{\infty} \frac{1}{m!} \sum_{k_1,\dots,k_m} \left.\frac{q_{k_1}\cdots q_{k_m}}{k_1!\dots k_m!} t^{k_1+\cdots+k_m}\right|_{t=0} \\ &= \left(\frac{d}{dt}\right)^n \sum_{m=0}^{\infty} \frac{1}{m!} \left.\left(\sum_{k=1}^{\infty} q_k \frac{t^k}{k!}\right)^m\right|_{t=0} \\ &= \left(\frac{d}{dt}\right)^n \exp\left.\left(\sum_{k=1}^{\infty} q_k \frac{t^k}{k!}\right)\right|_{t=0}. \end{aligned} \tag{A.3}$$

To perform the analytic prolongation, we prove now that, if $f(0) = 1$, then

$$\lim_{n\to 0} \frac{1}{n} \ln\left[\left(\frac{d}{dt}\right)^n f(t)\Big|_{t=0}\right] = \int_{-\infty}^{+\infty} dl \left[e^{-e^l} - f(-e^l)\right]. \tag{A.4}$$

This fact can be seen observing that, for $n \to 0$

$$\begin{aligned} \left(\frac{d}{dt}\right)^n f(t) &= f\left(\frac{\partial}{\partial J}\right) J^n e^{Jt}\Big|_{J=0} \approx f(t) + n\, f\left(\frac{\partial}{\partial J}\right) \ln J\, e^{Jt}\Big|_{J=0} \\ &= f(t) + n\, f\left(\frac{\partial}{\partial J}\right) \int_0^\infty \frac{ds}{s} \left(e^{-s} - e^{-sJ}\right) e^{Jt}\Big|_{J=0} \\ &= f(t) + n \int_0^\infty \frac{ds}{s} \left[e^{-s} f(t) - f(t-s)\right] . \end{aligned} \tag{A.5}$$

By the change of variable $s = e^l$, Eq. (A.4) follows.

## A.2 Asymptotic behaviour of the function $B_r$

In this Appendix we study the asymptotic behavior for large $\lambda$ of the function $B_r(\lambda x)$. By definition in Eq. (3.45)

$$\frac{1}{\lambda^r} B_r(\lambda x) \equiv \sum_{k=1}^{\infty} (-1)^{k-1} \frac{e^{\lambda x k}}{(\lambda k)^r (k!)^2}, \tag{A.6}$$

so that

$$\begin{aligned} \frac{B_r(\lambda x)}{\lambda^r} &= -\frac{1}{\Gamma(r)} \int_0^\infty t^{r-1} \sum_{k=1}^{\infty} \frac{(-1)^k}{(k!)^2} e^{\lambda(x-t)k} dt \\ &= -\frac{1}{\Gamma(r)} \int_0^\infty t^{r-1} \left\{ J_0\left[2e^{\frac{1}{2}\lambda(x-t)}\right] - 1 \right\} dt \\ &\xrightarrow{\lambda \to +\infty} \frac{1}{\Gamma(r)} \int_0^\infty t^{r-1} \theta(x-t) dt = \frac{x^r}{\Gamma(r+1)} \theta(x), \end{aligned} \tag{A.7}$$

where we have used the fact that

$$J_0(x) \equiv \sum_{m=0}^{\infty} \frac{(-1)^m}{(m!)^2} \left(\frac{x}{2}\right)^{2m} = \begin{cases} 1 & \text{when } x \to 0, \\ 0 & \text{when } x \to +\infty \end{cases} \tag{A.8}$$

is the Bessel function of zeroth order of the first kind.

## A.3 Some properties of the function $\hat{G}_r$

In this Appendix we give some properties of the function $\hat{G}_r$, defined by the integral equation (3.9c). From the definition, we have that, for $0 \le \alpha < \beta + 1$ and $r > -1$,

$$\hat{G}_r^{(\alpha)}(l) \equiv \mathcal{D}_l^\alpha \, \hat{G}_r(l) = \int_{-\infty}^{+\infty} \frac{(l+y)^{r-\alpha}}{\Gamma(r-\alpha+1)} e^{-\hat{G}_r(y)} \theta(l+y) dy. \tag{A.9}$$

Observe that

$$\hat{G}_r^{(\alpha)}(l) \ge 0 \quad \text{for} \quad 0 \le \alpha < r+1. \tag{A.10}$$

In this equation we have used the fact that, for $0 \le \alpha < \beta + 1$, we have [Pod99]

$$\mathcal{D}_t^\alpha \left[ \frac{t^\beta}{\Gamma(\beta+1)} \theta(t) \right] = \frac{t^{\beta-\alpha}}{\Gamma(\beta-\alpha+1)} \theta(t). \tag{A.11}$$

In particular, for $\alpha = r$ we have the simple relation

$$\hat{G}_r^{(r)}(l) \equiv \mathcal{D}_l^r \, \hat{G}_r(l) = \int_{-\infty}^{\infty} dy \, e^{-\hat{G}_r(y)} \theta(y+l). \tag{A.12}$$

Moreover, for $0 \le \alpha < r + 1$,

$$\lim_{l \to -\infty} \hat{G}_r^{(\alpha)}(l) = 0. \tag{A.13}$$

From Eq. (A.12)

$$\hat{G}_r^{(r+1)}(l) = e^{-\hat{G}_r(-l)} \ge 0 \Rightarrow \lim_{l \to +\infty} \hat{G}_r^{(r+1)}(l) = 1. \tag{A.14}$$

The relations above imply that

$$J_r^{(\alpha)} \equiv \int_{-\infty}^{+\infty} du \, \hat{G}_r(-u) \, \mathcal{D}_u^\alpha \, \hat{G}_r(u) > 0 \, , \quad 0 \le \alpha < r + 1. \tag{A.15}$$

Similarly, for $0 < k < r + 1$ an integer,

$$\begin{aligned} J_r^{(r+k+1)} &\equiv \int_{-\infty}^{+\infty} du \, \hat{G}_r(-u) \, \mathcal{D}_u^{r+k+1} \, \hat{G}_r(u) = \int_{-\infty}^{+\infty} du \, \hat{G}_r(-u) \frac{d^k}{du^k} e^{-\hat{G}_r(-u)} \\ &= \int_{-\infty}^{+\infty} du \, \hat{G}_r^{(k)}(u) e^{-\hat{G}_r(u)} \ge 0. \end{aligned} \tag{A.16}$$

For large $l$ we have

$$\hat{G}_r(l) \approx \frac{l^{r+1}}{\Gamma(r+2)}, \tag{A.17a}$$

$$\hat{G}_r(-l) \approx e^{-\frac{l^{r+1}}{\Gamma(r+2)}}. \tag{A.17b}$$

As anticipated, an exact solution is available in the $r = 0$ case. In particular, for $r = 0$, the second derivative

$$\hat{G}_0^{(2)}(l) = e^{-\hat{G}_0(-l)} \hat{G}_0^{(1)}(-l) = \hat{G}_0^{(1)}(l) \hat{G}_0^{(1)}(-l) \tag{A.18}$$

is an even function of $l$,

$$\hat{G}_0^{(2)}(l) - \hat{G}_0^{(2)}(-l) = 0 \Rightarrow \hat{G}_0^{(1)}(l) + \hat{G}_0^{(1)}(-l) = c, \tag{A.19}$$

with the constant $c = 1$ by evaluating the left-hand side in the limit of infinite $l$ and

$$\hat{G}_0^{(1)}(0) = e^{-\hat{G}_0(0)} = \frac{1}{2}. \tag{A.20}$$

Then we have that

$$\hat{G}_0(l) - \hat{G}_0(-l) = l \Rightarrow \hat{G}_0^{(1)}(l) = e^{-\hat{G}_0(-l)} = e^{l-\hat{G}_0(l)}, \tag{A.21}$$

which means that

$$\frac{d}{dl} e^{\hat{G}_0(l)} = e^l \Rightarrow e^{\hat{G}_0(x)} - e^{\hat{G}_0(0)} = e^x - 1, \tag{A.22}$$

where we have used the initial condition at $l = 0$, that is, because of Eq. (A.20),

$$e^{\hat{G}_0(x)} = 1 + e^x \Rightarrow \hat{G}_0(x) = \ln(1 + e^x). \tag{A.23}$$

## A.4 Evaluation of the integrals in the $r = 0$ case

To explicitly evaluate some of the integrals above, let us introduce the polygamma function

$$\psi_m(z) \equiv \frac{d^{m+1}}{dz^{m+1}} \ln \Gamma(z) = (-1)^{m+1} \int_0^\infty dt \, \frac{t^m e^{-zt}}{1 - e^{-t}}, \tag{A.24}$$

which satisfies the recursion relation

$$\psi_m(z+1) = \psi_m(z) + (-1)^m \frac{m!}{z^{m+1}}, \tag{A.25}$$

which, for a positive integer argument and assuming $m \geq 1$, leads to

$$\frac{\psi_m(k)}{(-1)^{m+1} m!} = \zeta(m+1) - \sum_{r=1}^{k-1} \frac{1}{r^{m+1}} = \sum_{r=k}^{\infty} \frac{1}{r^{m+1}}. \tag{A.26}$$

For $m = 0$ this implies

$$\psi_0(k) = -\gamma_E + H_{n-1} \Rightarrow \psi_0(1) = -\gamma_E, \tag{A.27}$$

with $\gamma_E$ is Euler's gamma constant and

$$H_n \equiv \sum_{k=1}^{n} \frac{1}{k} \tag{A.28}$$

are the harmonic numbers. With these considerations in mind and using Eq. (A.23), we have that

$$\begin{aligned} J_0^{(1)} &\equiv \int_{-\infty}^{+\infty} dy \, \frac{\ln(1+e^y)}{1+e^y} = \int_0^{+\infty} dt \, \frac{t \, e^{-t}}{1-e^{-t}} \\ &= \psi_1(1) = \zeta(2) = \sum_{k \geq 1} \frac{1}{k^2} = \frac{\pi^2}{6}. \end{aligned} \tag{A.29}$$

Then we compute

$$\begin{aligned} J_0^{(0)} &\equiv \int_{-\infty}^{+\infty} dy \, \frac{1}{1+e^y} \int_{-\infty}^{y} du \, \ln(1+e^u) = \int_0^{+\infty} dt \, \frac{e^{-t}}{1-e^{-t}} \int_0^t dw \, \frac{w}{1-e^{-w}} \\ &= -\int_0^{+\infty} dt \, t \, \frac{\ln(1-e^{-t})}{1-e^{-t}} = \sum_{k=1}^{\infty} \frac{1}{k} \int_0^{+\infty} dt \, \frac{te^{-kt}}{1-e^{-t}} = \sum_{k=1}^{\infty} \frac{1}{k} \psi_1(k). \end{aligned} \tag{A.30}$$

We remark now that

$$\sum_{k\geq 1}\frac{\psi_1(k)}{k}=\sum_{k=1}^{\infty}\sum_{r=0}^{\infty}\frac{1}{k}\frac{1}{(r+k)^2}=\sum_{s\geq 1}\sum_{k=1}^{s}\frac{1}{k}\frac{1}{s^2}=\sum_{s\geq 1}\frac{1}{s^2}H_s. \tag{A.31}$$

Applying now the identity

$$\sum_{s=1}^{\infty}\frac{H_s}{s^2}=2\zeta(3), \tag{A.32}$$

discovered by Euler, we obtain

$$J_0^{(0)}=2\zeta(3)=-\psi_2(1). \tag{A.33}$$

To finally evaluate $J_0^{(3)}$, we remark now that

$$\int_{-\infty}^{+\infty}\frac{dy}{1+e^y}\frac{d}{dy}\ln(1+e^y)=-\int_{-\infty}^{+\infty}dy\frac{d}{dy}\frac{1}{1+e^y}=1. \tag{A.34}$$

Then, as

$$\frac{d^2}{dy^2}\ln(1+e^y)=\frac{d}{dy}\ln(1+e^y)-\left[\frac{d}{dy}\ln(1+e^y)\right]^2, \tag{A.35}$$

we have

$$\begin{aligned}J_0^{(3)}&=\int_{-\infty}^{+\infty}dy\,\frac{1}{1+e^y}\frac{d^2}{dy^2}\ln(1+e^y)=-\int_{-\infty}^{+\infty}dy\,\left(\frac{d}{dy}\frac{1}{1+e^y}\right)\frac{d}{dy}\ln(1+e^y)\\&=\int_{-\infty}^{+\infty}dy\,\frac{1}{1+e^y}\left[\frac{d}{dy}\ln(1+e^y)\right]^2=\frac{1}{2}.\end{aligned} \tag{A.36}$$

## A.5 Calculation of $\Delta F_r^T$

To evaluate explicitly $\Delta F_r^T$, let us start from Eq. (3.60),

$$\begin{aligned}\Delta\hat{F}_r^T&=\frac{1}{2\hat{\beta}N}\sum_{s=1}^{\infty}\sum_{t=1}^{\infty}(-1)^{s+t-1}\frac{(s+t-1)!}{s!t!}\frac{\hat{g}_s\hat{g}_t}{\hat{g}_{s+t}^2}q_{s+t}^2\\&=\frac{\eta_0\Gamma(r+1)}{2\hat{\beta}^{r+2}N}\sum_{s=1}^{\infty}\sum_{t=1}^{\infty}\frac{(-1)^{s+t-1}}{s!t!}\left(\frac{t+s}{s\,t}\right)^{r+1}\int_{-\infty}^{+\infty}dy\,e^{-G_r(y)}e^{y(s+t)}q_{s+t}\\&=\frac{\eta_0\Gamma(r+1)}{2\hat{\beta}^{r+2}N}\sum_{k=2}^{\infty}\sum_{s=1}^{k-1}\frac{(-1)^{k-1}}{s!\,(k-s)!}\frac{k^{r+1}}{s^{r+1}(k-s)^{r+1}}\int_{-\infty}^{+\infty}dy\,e^{-G_r(y)}e^{yk}q_k\,,\end{aligned} \tag{A.37}$$

and, in order to perform the sum over $s$, we introduce integral representations

$$\begin{aligned}\sum_{s=1}^{k-1}\frac{1}{s!(k-s)!}\frac{1}{s^{r+1}(k-s)^{r+1}}&=\sum_{s=1}^{k-1}\binom{k}{s}\int_0^{+\infty}du\int_0^{+\infty}dv\,\frac{u^rv^re^{-su}e^{-(k-s)v}}{k!\Gamma^2(r+1)}\\&=\int_0^{+\infty}du\int_0^{+\infty}dv\,u^rv^r\frac{(e^{-u}+e^{-v})^k-e^{-uk}-e^{-vk}}{k!\Gamma^2(r+1)}.\end{aligned} \tag{A.38}$$

Observing now that the value $k = 1$ can be included in the sum over $k$ and defining

$$h \equiv \frac{\hat{\beta}}{[\eta_0 \Gamma(r+1)]^{1/(r+1)}}, \tag{A.39}$$

we can write

$$\begin{aligned}
&\frac{2\hat{\beta}^{r+2}\Gamma(r+1)N}{\eta_0}\Delta\hat{F}_r^T = \sum_{k=1}^{\infty}\frac{(-1)^{k-1}}{k!}q_k \int_{-\infty}^{+\infty} dy\, e^{-G_r(y)}\, \mathcal{D}_y^{r+1}\, e^{yk} \\
&\times \int_0^{+\infty} dudv\, u^r v^r \left[\left(e^{-u}+e^{-v}\right)^k - e^{-uk} - e^{-vk}\right] = \int_{-\infty}^{+\infty} dy\, e^{-G_r(y)}\, \mathcal{D}_y^{r+1} \\
&\times \left\{\int_0^{+\infty} du \int_0^{+\infty} dv\, u^r v^r \left[G_r\left(y + \ln\left(e^{-u}+e^{-v}\right)\right) - G_r(y-u) - G_r(y-v)\right]\right\} \\
&= 2h^r \int_{-\infty}^{+\infty} dy\, e^{-\hat{G}_r(y)}\, \mathcal{D}_y^{r+1} \int_0^{+\infty} du \int_0^{u} dv\, u^r v^r \\
&\times \left[G_r\left(h(y-v) + \ln\left(e^{-h(u-v)}+1\right)\right) - \hat{G}_r(y-u) - \hat{G}_r(y-v)\right].
\end{aligned} \tag{A.40}$$

This implies that, for $h \to \infty$,

$$\begin{aligned}
\Delta\hat{F}_r^T &= \frac{1}{N\eta_0^{\frac{1}{r+1}}\Gamma^{2+\frac{1}{r+1}}(r+1)} \int_{-\infty}^{+\infty} dy\, e^{-\hat{G}_r(y)}\, \mathcal{D}_y^{r+1} \left[\int_0^{+\infty} du \int_0^{u} dv\, u^r v^r \hat{G}_r(y-u)\right] \\
&= -\frac{1}{N\eta_0^{\frac{1}{r+1}}\Gamma^{2+\frac{1}{r+1}}(r+1)} \frac{1}{r+1} \int_{-\infty}^{+\infty} dy\, e^{-\hat{G}_r(y)}\, \mathcal{D}_y^{r+1} \left[\int_{-\infty}^{y} du\, (y-u)^{2r+1}\hat{G}_r(u)\right] \\
&= -\frac{\Gamma(2r+2)}{N\eta_0^{\frac{1}{r+1}}\Gamma^{3+\frac{1}{r+1}}(r+1)} \frac{1}{r+1} \int_{-\infty}^{+\infty} dy\, e^{-\hat{G}_r(y)} \int_{-\infty}^{y} du\, (y-u)^r \hat{G}_r(u) \\
&= -\frac{\Gamma(2r+2)}{N\eta_0^{\frac{1}{r+1}}\Gamma^{2+\frac{1}{r+1}}(r+1)} \frac{1}{r+1} \int_{-\infty}^{+\infty} du\, \hat{G}_r(-u)\hat{G}_r(u),
\end{aligned} \tag{A.41}$$

which is exactly Eq. (3.61).

## A.6 General approach for computing $I_p$

In the following we will write formulas for the quantity $I_p$ defined in (3.114). Using identities (A.13) and (A.14) for the generating function $\hat{G}_r$ we can rewrite $I_p$ as

$$\begin{aligned}
I_p = &\frac{(-1)^p}{\Gamma^p(r+1)} \int_0^{\infty} dt \int_{-\infty}^{+\infty} dx_1 \dots dx_p \partial_1 \hat{G}_r(-x_1 - t) \dots \partial_p \hat{G}_r(-x_p - t) \\
&\times \left[\mathcal{D}_{x_1}^r \dots \mathcal{D}_{x_p}^r (x_1+x_2)^r \dots (x_p+x_1)^r\right] \theta(x_1+x_2) \dots \theta(x_p+x_1)
\end{aligned} \tag{A.42}$$

Since $(x_1+x_2)^r \dots (x_p+x_1)^r$ is an homogeneous polynomial of degree $pr$ the $pr$ derivatives in square bracket give the coefficient of the monomial $x_1^r \dots x_p^r$. This coefficient can be evaluated simply using $p$ times the binomial expansion

$$(x_1+x_2)^r \ldots (x_p+x_1)^r = \prod_{i=1}^{p} \sum_{k_i=0}^{r} \prod_{i=1}^{p} \binom{r}{k_i} x_1^{r+k_1-k_p} x_2^{r+k_2-k_1} \ldots x_p^{r+k_p-k_{p-1}} . \quad (A.43)$$

Next the coefficient of the monomial $x_1^r \ldots x_p^r$ is selected imposing $k_1 = k_2 = \cdots = k_p$, so that

$$\frac{1}{\Gamma^p(r+1)} \mathcal{D}^r_{x_1} \ldots \mathcal{D}^r_{x_p} \left[ (x_1+x_2)^r \ldots (x_p+x_1)^r \right] = \sum_{k=0}^{r} \binom{r}{k}^p \equiv C_{r,p} . \quad (A.44)$$

$$I_{r,p} = (-1)^p C_{r,p} \int_0^\infty dt \left[ \prod_{j=1}^{p} \int_{-\infty}^{+\infty} dx_j \right] \prod_{i=1}^{p} \partial_i \hat{G}_r(-x_i - t)\, \theta(x_i + x_{i+1}) \quad (A.45)$$

where $x_{p+1} = x_1$. Note that

$$\begin{aligned} C_{0,p} &= 1 \\ C_{r,1} &= 2^r \\ C_{r,2} &= \frac{\Gamma(2r+1)}{\Gamma^2(r+1)} \\ C_{r,p} &= {}_pF_{p-1}\left[\{-r,\ldots,-r\},\{1,\ldots,1\};-1\right] , \qquad p \geq 3 \end{aligned} \quad (A.46)$$

where ${}_pF_q$ is the hypergeometric function. Next we perform an integration by parts on all the variables with even index, that is $x_2, x_4, \ldots x_{2[\frac{p}{2}]}$. There is no contribution from the boundaries. We use

$$\begin{aligned} &\partial_{2k}\theta(x_{2k-1}+x_{2k})\theta(x_{2k}+x_{2k+1}) \\ &\qquad = \delta(x_{2k-1}+x_{2k})\theta(x_{2k}+x_{2k+1}) + \theta(x_{2k-1}+x_{2k})\delta(x_{2k}+x_{2k+1}) \end{aligned} \quad (A.47)$$

and then we perform the trivial integration of $\delta$-function. Remark that within this procedure no $\delta'$ appears, so we can also first perform all the derivatives and afterward all the integrations. We will end up with $2^{2[\frac{p}{2}]}$ terms and the integration on only the variables with odd label. Afterwards for each of the remaining variables, say going from the one with highest index we perform an integration by parts of the term $\partial_i \hat{G}_r(-x_i - t)$ only whenever it does not multiply $\hat{G}_r(-x_i - t)$ or $\hat{G}_r(x_i - t)$, that is it appears only in $\theta$-function and perform the integration on $dx_i$ by resolving the $\delta$-functions which have $x_i$ as an argument.

The cases $p = 1$ and $p = 2$ are exceptional. For example in he case with $p = 1$ we have

$$\begin{aligned} I_{r,1} &= -2^r \int_0^\infty dt \int_{-\infty}^{+\infty} dx_1\, \partial_1 \hat{G}_r(-x_1 - t)\theta(x_1) \\ &= 2^r \int_0^\infty dt \int_{-\infty}^{+\infty} dx_1 \hat{G}_r(-x_1 - t)\delta(x_1) = 2^r \int_0^\infty dt\, \hat{G}_r(-t) \end{aligned} \quad (A.48)$$

For $r=0$ we obtain $I_1 = \frac{\zeta(2)}{2}$ as it should [Rat03]. For $p=2$ the two $\theta$-functions have the same argumentand we can start now our procedure on $x_2$

$$\begin{aligned} I_{r,2} &= -\frac{\Gamma(2r+2)}{\Gamma^2(r+1)}\int_0^\infty dt \int_{-\infty}^{+\infty} dx_1\, dx_2\, \partial_1 \hat{G}_r(-x_1-t)\hat{G}_r(-x_2-t)\delta(x_1+x_2) \\ &= -\frac{\Gamma(2r+2)}{\Gamma^2(r+1)}\int_0^\infty dt \int_{-\infty}^{+\infty} dx_1\, \hat{G}_r(x_1-t)\, \partial_1 \hat{G}_r(-x_1-t)\,. \end{aligned} \tag{A.49}$$

Here we can no more proceed in the standard way. Anyhow we can perform a shift and proceed by using $t$ as a variable

$$I_{r,2} = -\frac{\Gamma(2r+2)}{\Gamma^2(r+1)}\int_{-\infty}^{+\infty} dx\, \hat{G}_r(x) \int_0^\infty dt\, \partial_1 \hat{G}_r(-x-2t) \tag{A.50}$$

$$= \frac{\Gamma(2r+1)}{2\Gamma^2(r+1)}\int_{-\infty}^{+\infty} dx\, \hat{G}_r(-x)\hat{G}_r(x) = \frac{\Gamma(2r+1)}{2\Gamma^2(r+1)} J_r^{(0)} \tag{A.51}$$

For $r=0$ we obtain $I_2 = J_0^{(0)}/2 = \zeta(3)$ as in [Rat03]. For $p=3$ our procedure allows us to obtain

$$I_{r,3} = C_{r,3}\left[3\int_0^\infty dt \int_{-\infty}^{+\infty} dx_2 \hat{G}_r^2(-x_2-t)\partial_2 \hat{G}_r(x_2-t)\theta(x_2) + \int_0^\infty dt\, \hat{G}_r^3(-t)\right] \tag{A.52}$$

For $p=4$ we get

$$I_{r,4} = \frac{C_{r,4}}{2}\int_{-\infty}^{+\infty} dx\, \hat{G}_r^2(x)\, \hat{G}_r^2(-x) \tag{A.53}$$

This procedure can be applied for values of $p$ higher.

# Appendix B

# Details of the replica computation of the RFMP

## B.1 One-site partition function

The evaluation of the one-site partition function in the RFMP follows the same type of arguments adopted in the literature for the RMP [MP85, MP87, PR02] and for the RTSP [MP86a]. The one-site partition function $z$ in the RFMP is

$$\begin{aligned} z[\mathbf{Q}] \equiv & \prod_{a=1}^{n}\left[\int_0^{2\pi}\frac{e^{-2i\lambda^a}d\lambda^a}{2\pi}\right] \\ & \times \exp\left[\sum_{\substack{\alpha\cap\beta=\emptyset\\ \alpha\cup\beta\neq\emptyset}} Q_{\alpha,\beta}\, e^{i\sum_{a\in\alpha}\lambda^a+2i\sum_{b\in\beta}\lambda^b} + \frac{\varepsilon-1}{4N}\sum_{\alpha\neq\emptyset} g_{|\alpha|}\, e^{2i\sum_{a\in\alpha}\lambda^a}\right] \\ \equiv & \; z_0[\mathbf{Q}] + \frac{\varepsilon-1}{2N}\sum_{\alpha\neq\emptyset} g_{|\alpha|}\frac{\partial z_0[\mathbf{Q}]}{\partial Q_{\emptyset,\alpha}}. \end{aligned} \tag{B.1}$$

In the previous equation, we have introduced $z_0[\mathbf{Q}]$, that coincides with the expression given in Eq. (4.21e). It follows that, for $N \gg 1$, $\ln z[\mathbf{Q}]$ provides a contribution both to the leading term $S[\mathbf{Q}]$ and to the finite-size corrections $\Delta S^{\mathrm{T}}[\mathbf{Q}]$, namely, up to $o\left(1/N\right)$ terms,

$$2\ln z[\mathbf{Q}] = 2\ln z_0[Q] - \frac{\varepsilon-1}{2N}\sum_{\alpha\neq\emptyset}\frac{g_{|\alpha|}}{z_0[\mathbf{Q}]}\frac{\partial z_0[\mathbf{Q}]}{\partial Q_{\emptyset,\alpha}}, \tag{B.2}$$

to be compared with the terms appearing in Eqs. (4.21). The evaluation of $z_0[\mathbf{Q}]$ is nontrivial in general. In the replica symmetric hypothesis, $z_0[\mathbf{Q}]$ can be written as

$$z_0[\mathbf{Q}] = \prod_{a=1}^{n}\left[\int_0^{2\pi}\frac{e^{-2i\lambda^a}d\lambda^a}{2\pi}\right]\exp\left[\sum_{\substack{\alpha\cap\beta=\emptyset\\ \alpha\cup\beta\neq\emptyset}} Q_{|\alpha|,|\beta|}\, e^{i\sum_{a\in\alpha}\lambda^a+2i\sum_{b\in\beta}\lambda^b}\right]. \tag{B.3}$$

Denoting now by $\xi_a \equiv e^{i\lambda_a}$, observe now that

$$\begin{split}\sum_{\substack{\alpha\cap\beta=\emptyset\\ \alpha\cup\beta\neq\emptyset}} Q_{|\alpha|,|\beta|}\prod_{a\in\alpha}\xi_a\prod_{b\in\beta}\xi_b^2 &= \sum_{p+q\geq 1} Q_{p,q}\sum_{|\beta|=q}\prod_{b\in\beta}\xi_b^2\sum_{\substack{|\alpha|=p\\ \alpha\cap\beta=\emptyset}}\prod_{a\in\alpha}\xi_a\\ \to \sum_{p+q\geq 1} Q_{p,q}\sum_{|\beta|=q}\prod_{b\in\beta}\xi_b^2\sum_{|\alpha|=p}\prod_{a\in\alpha}\xi_a &\to \sum_{p+q\geq 1}\frac{Q_{p,q}}{q!}\left(\sum_{|\alpha|=p}\prod_{a\in\alpha}\xi_a\right)\left(\sum_b \xi_b^2\right)^q, \end{split} \tag{B.4}$$

where each substitution is justified because of the overall constraint that allows us to neglect powers $\xi_b^k$ with $k\geq 3$. It can be seen that

$$\sum_{|\alpha|=p}\prod_{a\in\alpha}\xi_a = \frac{(\sum_a \xi_a^2)^{\frac{p}{2}}}{p!}\mathrm{He}_p\left(\frac{\sum_a \xi_a}{\sqrt{\sum_a \xi_a^2}}\right). \tag{B.5}$$

Here $\mathrm{He}_p(x)$ is the probabilists' Hermite polynomial [AS72]. Substituting the previous identity in the expression for $z_0$, we obtain

$$z_0[\mathbf{Q}] = \int_{-\infty}^{+\infty}\frac{dx dk_x dy dk_y}{(2\pi)^2} e^{\sum_{p+q\geq 1}\frac{Q_{p,q}}{p!q!}y^{\frac{p}{2}+q}\mathrm{He}_p\left(\frac{x}{\sqrt{y}}\right)+ik_x x+ik_y y}\Phi(k_x,k_y), \tag{B.6}$$

where

$$\Phi(k_x,k_y) = \left[\int_0^{2\pi}\frac{d\lambda}{2\pi}\, e^{-2i\lambda-ik_x e^{i\lambda}-ik_y e^{2i\lambda}}\right]^n = \left(-\frac{k_x^2}{2}-ik_y\right)^n. \tag{B.7}$$

Using now the identity $a^n = \partial_t^n e^{at}\big|_{t=0}$, we can write [LPS17]

$$\begin{split}\frac{\ln z_0[\mathbf{Q}]}{n} &= \frac{1}{n}\ln\left[\frac{\partial^n}{\partial t^n}\int_{-\infty}^{+\infty} Dz\, e^{\sum_{p+q\geq 1}\frac{Q_{p,q}}{p!q!}t^{\frac{p}{2}+q}\mathrm{He}_p(z)}\bigg|_{t=0}\right]\\ &= \int_0^{+\infty}\frac{dt}{t}\left[e^{-t}-\int_{-\infty}^{+\infty} Dz\, e^{\sum_{p+q\geq 1}\frac{Q_{p,q}}{p!q!}(-t)^{\frac{p}{2}+q}\mathrm{He}_p(z)}\right]+O(n).\end{split} \tag{B.8}$$

At this point, some considerations are in order. Let us first observe that, on the matching saddle-point in Eq. (4.28), $Q^{\mathrm{sp}}_{p,q} = \delta_{p,0}Q_q$, the previous expression becomes

$$\lim_{n\to 0}\frac{\ln z_0[Q]}{n} = \int_0^{+\infty}\frac{dt}{t}\left[e^{-t}-e^{\sum_{q=1}^{\infty}\frac{Q_q}{q!}(-t)^q}\right]. \tag{B.9}$$

The expression above coincides with the one-site partition function in the RAP and RMP as can be seen in Appendix A.1. In particular, Eqs. (4.29a) can be obtained introducing the function $G$ in Eq. (4.29b). On the other hand, if we consider the RTSP saddle-point solution, $\tilde{Q}^{\mathrm{sp}}_{q,p} = \delta_{q,0}\tilde{Q}_p$, the RTSP one-site partition function is recovered, as it can be easily seen comparing Eq. (B.8) with the results in Ref. [MP86a].

## B.2 Evaluation of $\Delta S^{\mathrm{T}}$ on the matching saddle-point

To evaluate the $\Delta S^{\mathrm{T}}$ contribution (4.21c) on the matching saddle-point, we follow the approach in Ref. [MP87, PR02]. In particular,

$$\frac{1}{n}\Delta S^{\mathrm{T}}[\mathbf{Q}^{\mathrm{sp}}] = \frac{1}{Nn}\sum_{\alpha,\beta|\alpha,\hat{\beta}}{}' \frac{g_{|\alpha|+2|\beta|}g_{|\alpha|+2|\hat{\beta}|}}{4g^2_{2|\beta|+2|\hat{\beta}|+2|\alpha|}} Q^2_{|\beta|+|\hat{\beta}|+|\alpha|} - \frac{\varepsilon-1}{4nN}\sum_{\alpha\neq\emptyset}\frac{g_{|\alpha|}}{g_{2|\alpha|}}Q_{0,|\alpha|}$$
$$= \frac{1}{N}\sum_{\substack{s+p\geq 1\\ s+q\geq 1}} \frac{(-1)^{p+q+s-1}\Gamma(s+p+q)}{p!q!s!}\frac{g_{s+2p}g_{s+2q}}{4g^2_{2p+2q+2s}}Q^2_{p+q+s} - \frac{\varepsilon-1}{4N}\sum_{p=1}^{\infty}\frac{(-1)^{p-1}}{p}\frac{g_p}{g_{2p}}Q_p. \tag{B.10}$$

The asymptotic value of the last sum is given by

$$\begin{aligned}\lim_{\beta\to+\infty}\frac{\varepsilon-1}{4N\beta}\sum_{p=1}^{\infty}\frac{(-1)^{p-1}}{p}\frac{g_p}{g_{2p}}Q_p &= \lim_{\beta\to+\infty}\frac{\varepsilon-1}{2N\beta}\sum_{p=1}^{\infty}\frac{(-1)^{p-1}}{pp!}\int_{-\infty}^{+\infty}dx\, e^{\beta p x-\hat{G}(x)}\\ &= \frac{\varepsilon-1}{2N}\lim_{\beta\to+\infty}\int_{-\infty}^{+\infty}dx\,\frac{\gamma_{\mathrm{E}}+\beta x+\Gamma(0,e^{\beta x})}{\beta}e^{-\hat{G}(x)}\\ &= \frac{\varepsilon-1}{2N}\int_0^{+\infty}xe^{-\hat{G}(x)}dx = \frac{\varepsilon-1}{4N}\zeta(2).\end{aligned} \tag{B.11}$$

In the previous expression, $\gamma_E$ is the Euler-Mascheroni constant, whereas $\Gamma(a,z) \equiv \int_z^{\infty}e^{-t}t^{a-1}dt$ is the incomplete gamma function [AS72]. In the contribution from $\Delta S^{\mathrm{T}}$ we also have

$$\begin{aligned}&\frac{1}{n}\sum_{\substack{s+p\geq 1\\ s+q\geq 1}}^{\infty}\binom{n}{s\ p\ q}\frac{g_{s+2p}g_{s+2q}}{4g_{2p+2q+2s}}Q^2_{p+q+s}\\ &= \sum_{\substack{s+p\geq 1\\ s+q\geq 1}}\frac{(-1)^{p+q+s-1}\Gamma(s+p+q)}{p!q!s!}\frac{g_{s+2p}g_{s+2q}}{4g^2_{2p+2q+2s}}Q^2_{p+q+s}\\ &= \sum_{\substack{s+p\geq 1\\ s+q\geq 1}}\frac{(-1)^{p+q+s-1}g_{s+2p}g_{s+2q}}{2p!q!s!g_{2p+2q+2s}}Q_{p+q+s}\int_{-\infty}^{+\infty}e^{(p+q+s)x-G(x)}dy = \frac{1}{\beta}\int_{-\infty}^{+\infty}dx\, e^{-G(x)}\\ &\times\sum_{k=1}^{\infty}(-1)^{k-1}e^{kx}Q_k\left[\sum_{p=1}^{k-1}\frac{1}{2pp!(k-p)!}+\sum_{s=1}^{k}\sum_{p=0}^{k-s}\frac{1}{s!(s+2p)p!(k-s-p)!}\right].\end{aligned} \tag{B.12}$$

In Ref. [MP87] it has been proved that

$$\lim_{\beta\to+\infty}\frac{1}{\beta^2}\int_{-\infty}^{+\infty}dx\, e^{-G(x)}\sum_{k=1}^{\infty}(-1)^{k-1}e^{kx}Q_k\sum_{p=1}^{k-1}\frac{1}{2pp!(k-p)!} = -\frac{\zeta(3)}{N}. \tag{B.13}$$

The remaining contribution is zero. To prove this fact, let us start from

$$\begin{aligned}
&\sum_{s=1}^{k}\sum_{p=0}^{k-s}\frac{1}{s!(s+2p)p!(k-s-p)!}\\
&=\sum_{s=1}^{k}\frac{i}{2\pi s!}\int_0^{+\infty}dt\oint_{\gamma_{\mathrm{H}}}dz\sum_{p=0}^{\infty}(-z)^{p+s-k-1}\frac{e^{-(s+2p)t-z}}{p!}\\
&=\sum_{s=1}^{k}\frac{i}{2\pi s!}\int_0^{+\infty}dt\,e^{-st}\oint_{\gamma_{\mathrm{H}}}dz\,e^{-(1+e^{-2t})z}(-z)^{s-k-1}\\
&=\sum_{s=1}^{k}\int_0^{+\infty}dt\,\frac{\left(1+e^{-2t}\right)^{k-s-1}e^{-st}}{\Gamma(k-s+1)s!}=\int_0^1 d\tau\sum_{s=1}^{k}\frac{(1+\tau)^{k-s}\,\tau^{s/2-1}}{2s!(k-s)!}\\
&=\int_0^1 d\tau\frac{\left(\tau+\sqrt{\tau}+1\right)^k-(1+\tau)^k}{2k!\tau}.
\end{aligned} \tag{B.14}$$

In the expression above, $\gamma_{\mathrm{H}}$ is the Hankel contour in the complex plane. Summing over $k$ we get

$$\begin{aligned}
&\frac{1}{2\beta}\int_{-\infty}^{+\infty}e^{-G(x)}dx\int_0^1 d\tau\frac{G\left(x+\ln\left(1+\sqrt{\tau}+\tau\right)\right)-G\left(x+\ln(1+\tau)\right)}{\tau}\\
&=\frac{\beta}{2}\int dx\,e^{-\hat{G}(x)}\int_0^{+\infty}du\left[\hat{G}\left(x+\frac{\ln\left(1+e^{-\frac{\beta u}{2}}+e^{-\beta u}\right)}{\beta}\right)-\hat{G}\left(x+\frac{\ln(1+e^{-\beta u})}{\beta}\right)\right],
\end{aligned} \tag{B.15}$$

whose corresponding contribution goes to zero as $\beta\to\infty$. We can finally write

$$\lim_{\beta\to+\infty}\lim_{n\to0}\frac{\Delta S^{\mathrm{T}}[\mathbf{Q}^{\mathrm{sp}}]}{n\beta}=-\frac{1}{N}\left(\frac{\varepsilon-1}{4}\zeta(2)+\zeta(3)\right). \tag{B.16}$$

## B.3 On the evaluation of $\ln\det\Omega$ on the matching saddle-point

In this Appendix we will give some details about the evaluation of the logarithm of the determinant of the Hessian matrix $\mathbf{\Omega}$ on the matching saddle-point using the replica approach, showing that it is equal to the RMP contribution, plus an additional contribution that we expect to be opposite from Wästlund's formula.

We start from its general expression in Eq. (4.24),

$$\begin{aligned}\Omega_{\alpha\beta,\hat\alpha\hat\beta}[\mathbf{Q}^{\text{sp}}] &\equiv \sqrt{g_{|\alpha|+2|\beta|}g_{|\hat\alpha|+2|\hat\beta|}}\,\frac{\partial^2 S[\mathbf{Q}]}{\partial Q_{\alpha,\beta}\partial Q_{\hat\alpha,\hat\beta}}\Bigg|_{\mathbf{Q}=\mathbf{Q}^{\text{sp}}}\\
&= \delta_{\alpha,\hat\alpha}\,\delta_{\beta,\hat\beta}\,\mathbb{I}(\alpha\cap\beta=\emptyset) - 2\sqrt{g_{|\alpha|+2|\beta|}g_{|\hat\alpha|+2|\hat\beta|}}\left\langle e^{i\sum_{a\in\alpha\cup\hat\alpha}\lambda^a+2i\sum_{b\in\beta\cup\hat\beta}\lambda^b}\right\rangle_{z_0}\\
&+2\sqrt{g_{|\alpha|+2|\beta|}g_{|\hat\alpha|+2|\hat\beta|}}\left\langle e^{i\sum_{a\in\alpha}\lambda^a+2i\sum_{b\in\beta}\lambda^b}\right\rangle_{z_0}\left\langle e^{i\sum_{a\in\hat\alpha}\lambda^a+2i\sum_{b\in\hat\beta}\lambda^b}\right\rangle_{z_0}\\
&= \delta_{\alpha,\hat\alpha}\delta_{\beta,\hat\beta}+\frac{1}{2}\frac{Q_{\alpha,\beta}}{\sqrt{g_{|\alpha|+2|\beta|}}}\frac{Q_{\hat\alpha,\hat\beta}}{\sqrt{g_{|\hat\alpha|+2|\hat\beta|}}}\\
&-\frac{\sqrt{g_{|\alpha|+2|\beta|}g_{|\hat\alpha|+2|\hat\beta|}}\,Q_{\alpha\triangle\hat\alpha,\beta\cup\hat\beta\cup(\alpha\cap\hat\alpha)}}{g_{|\alpha\triangle\hat\alpha|+2|\beta\cup\hat\beta\cup(\alpha\cap\hat\alpha)|}}\mathbb{I}(\beta\cap\hat\beta=\emptyset)\mathbb{I}((\alpha\cup\hat\alpha)\cap(\beta\cup\hat\beta)=\emptyset).\end{aligned} \tag{B.17}$$

In the replica symmetric ansatz on the matching saddle-point solution the expression greatly simplifies, becoming

$$\begin{aligned}\Omega_{\alpha\beta,\hat\alpha\hat\beta} &= \delta_{\alpha,\hat\alpha}\delta_{\beta,\hat\beta}\mathbb{I}(\beta\cap\alpha=\emptyset)+\frac{\delta_{|\alpha|,0}\delta_{|\hat\alpha|,0}}{2}\frac{Q_{|\beta|}}{\sqrt{g_{2|\beta|}}}\frac{Q_{|\hat\beta|}}{\sqrt{g_{2|\hat\beta|}}}\\
&-\delta_{\alpha,\hat\alpha}\frac{Q_{|\beta\cup\hat\beta\cup\alpha|}\sqrt{g_{|\alpha|+2|\beta|}g_{|\alpha|+2|\hat\beta|}}\,\mathbb{I}(\beta\cap\hat\beta=\emptyset)\mathbb{I}((\beta\cup\hat\beta)\cap\alpha=\emptyset)}{g_{2|\beta\cup\hat\beta\cup\alpha|}}.\end{aligned} \tag{B.18}$$

Observe first that the quantity above is diagonal respect to the index $\alpha$. In particular

$$\begin{aligned}\ln\det\mathbf{\Omega}[\mathbf{Q}^{\text{sp}}] &= \ln\det\mathbf{\Omega}^{(0)}[\mathbf{Q}^{\text{sp}}]+\sum_{|\alpha|=1}^{\infty}\binom{n}{|\alpha|}\ln\det\mathbf{\Omega}^{(|\alpha|)}[\mathbf{Q}^{\text{sp}}]\\
&= \ln\det\mathbf{\Omega}^{(0)}[\mathbf{Q}^{\text{sp}}]+n\sum_{s=1}^{\infty}\frac{(-1)^{s-1}}{s}\ln\det\mathbf{\Omega}^{(s)}[\mathbf{Q}^{\text{sp}}]+o(n),\end{aligned} \tag{B.19}$$

where we have separated the contributions for different values of $s=|\alpha|$ and introduced

$$\Omega^{(s)}_{\beta\hat\beta}\equiv\delta_{\beta,\hat\beta}-Q_{|\beta|+|\hat\beta|+s}\mathbb{I}(\beta\cap\hat\beta=\emptyset)\frac{\sqrt{g_{2|\beta|+s}g_{2|\hat\beta|+s}}}{g_{2|\beta|+2|\hat\beta|+2s}}+\frac{\delta_{s,0}}{2}\frac{Q_{|\beta|}Q_{|\hat\beta|}}{\sqrt{g_{2|\beta|}g_{2|\hat\beta|}}}, \tag{B.20}$$

where $\beta$ is a set of $n-s$ replica indices. A vector $\mathbf{q}$ is eigenvector of $\mathbf{\Omega}^{(s)}$ with eigenvalue $\lambda$ if

$$q_\beta-\sum_{\hat\beta:\,\beta\cap\hat\beta=\emptyset}\frac{\sqrt{g_{s+2|\beta|}g_{s+2|\hat\beta|}}}{g_{2|\beta|+2|\hat\beta|+2s}}Q_{|\beta|+|\hat\beta|+s}q_{\hat\beta}+\frac{\delta_{s,0}}{2}\sum_{|\hat\beta|\neq\emptyset}\frac{Q_{|\beta|}Q_{|\hat\beta|}q_{\hat\beta}}{\sqrt{g_{2|\beta|}g_{2|\hat\beta|}}}=\lambda q_\beta. \tag{B.21}$$

A first step is to diagonalize $\mathbf{\Omega}^{(s)}$ according to the irreducible representations of the permutation group [WM13] in the space of $n-s$ replica indices. In the spirit of the strategy of De Almeida and Thouless [AT78], and strictly following Refs. [MP87, PR02], we observe that an eigenvector $\mathbf{q}^{(c)}$ with $c$ distinguished replicas, is such that

$$q_\beta^{(c)} \equiv \begin{cases} 0 & \text{if } |\beta| < c, \\ \omega^i_{|\beta|} & \text{if } \beta \text{ contains } c-i \text{ of the } c \text{ distinguished indices, with } i = 0, 1, \dots, c. \end{cases} \tag{B.22}$$

For $c = 0$, than we only have $n - s$ possible eigenvectors in the form $q_\beta^{(0)} \equiv q_{|\beta|}^{(0)}$. The eigenvalue equation can be written down for the $(n-s) \times (n-s)$ matrix $\mathbf{N}^{(s,0)}$ given by

$$N_{pq}^{(s,0)} = \delta_{pq} - \binom{n-s-p}{q} \frac{\sqrt{g_{2p+s}g_{2q+s}}Q_{p+q+s}}{g_{2p+2q+2s}} + \frac{\delta_{s,0}}{2}\binom{n}{q}\frac{Q_pQ_q}{\sqrt{g_{2p}g_{2q}}}, \tag{B.23}$$

whose eigenvalues have multiplicity 1 in the set of eigenvalues of $\mathbf{\Omega}^{(s)}$. For $c \geq 1$, imposing the ortogonality relation between $\mathbf{q}^{(c)}$ and $\mathbf{q}^{(c-1)}$, we obtain

$$\begin{aligned} \sum_\beta q_\beta^{(c)} q_\beta^{(c-1)} &= \sum_{|\beta| \geq c} q_\beta^{(c)} q_\beta^{(c-1)} \\ &= \sum_{p \geq c} \sum_{j=0}^{c-1} \sum_{r=0}^{c-j} \binom{n-s-c}{p-j-r}\binom{c}{j}\binom{c-j}{r} \omega_p^{c-(r+j)} \omega_p^{c-1-j} = 0 \\ &\Longrightarrow \sum_{r=0}^{c-j} \binom{n-s-c}{p-j-r}\binom{c-j}{r}\omega_p^{c-(r+j)} = 0 \text{ for } p \geq c \text{ and } j = 0, 1, \dots, c-1, \end{aligned} \tag{B.24a}$$

that for $n \to 0$ becomes, for $p \geq c$ and $j = 0, 1, \dots, c-1$,

$$\sum_{r=0}^{c-j} (-1)^r \binom{c-j}{r} \frac{\Gamma(s+c+p-r-j)}{\Gamma(p-r-j+1)} \omega_p^{c-(r+j)} = 0, \tag{B.24b}$$

where we have used the property

$$\lim_{n \to 0} \binom{n-a}{b} = \frac{(-1)^b \Gamma(a+b)}{\Gamma(a)\Gamma(b+1)}. \tag{B.25}$$

Eq. (B.24a) allows us to keep $\omega_p^0$ as independent only. In particular, for $c = 1$ we have $p\omega_p^0 + (n-s-p)\omega_p^1 = 0$ and therefore the diagonalization of $\mathbf{\Omega}$ in the subspace $c = 1$ can be reduced to the diagonalization of the $(n-s-1) \times (n-s-1)$ matrix

$$\begin{aligned} N_{pq}^{(s,1)} = \delta_{pq} &- \binom{n-s-p}{q} \frac{q}{q+s-n} \frac{\sqrt{g_{2p+s}g_{2q+s}}Q_{p+q+s}}{g_{2p+2q+2s}} \\ &+ \frac{\delta_{s,0}}{2}\left[\binom{n-1}{q}\frac{q}{q-n} + \binom{n-1}{q-1}\right]\frac{Q_pQ_q}{\sqrt{g_{2p}g_{2q}}}, \end{aligned} \tag{B.26}$$

with eigenvalue multiplicity $n-s-1$ respect to the original matrix $\mathbf{\Omega}^{(s)}$. Before proceeding further, some considerations are in order. The matrices $\mathbf{N}^{(0,0)}$ and $\mathbf{N}^{(0,1)}$ have the same limit as $n\to 0$, in particular $\lim_{n\to 0}\mathbf{N}^{(0,1)}=\mathbf{N}^{(0,0)}$. The calculation of the contribution of these two matrices requires some care, but it can be proved that it is eventually zero for $\beta\to+\infty$ [MP87, PR02]. For $c\geq 1$ Eq. (B.24b) implies

$$\frac{\omega_p^0}{\prod_{u=0}^{c-1}(s+p+u)}=\frac{\omega_p^1}{(p-c+1)\prod_{u=1}^{c-1}(s+p+u)}=\cdots$$
$$=\cdots=\frac{\omega_p^l}{\prod_{v=1}^{l}(p-c+v)\prod_{u=l}^{c-1}(s+p+u)}=\cdots=\frac{\omega_p^c}{\prod_{v=1}^{c}(p-c+v)}$$
$$\implies \omega_p^l=\frac{\Gamma(p+l-c+1)\Gamma(s+p)}{\Gamma(p-c+1)\Gamma(s+p+l)}\omega_p^0. \tag{B.27}$$

Using the previous result, Eq. (B.21) for the $|\beta|=0$ component of an eigenvector with $c\geq 1$ becomes

$$\begin{aligned}\lambda\omega_p^0 &= \frac{\delta_{s,0}}{2}\sum_{q=1}^{\infty}\sum_{i=0}^{c}\binom{n-c}{q+i-c}\binom{c}{c-i}\frac{Q_pQ_q\omega_q^i}{\sqrt{g_{2p}g_{2q}}}\\ &\quad+\omega_p^0-\sum_{q=1}^{\infty}\binom{n-s-p}{q}\frac{\sqrt{g_{s+2p}g_{s+2q}}}{g_{2s+2p+2q}}Q_{p+q+s}\omega_q^c\\ &=\omega_p^0-\sum_{q=1}^{\infty}(-1)^q\frac{\Gamma(s+q)\Gamma(s+p+q)\sqrt{g_{s+2p}g_{s+2q}}}{\Gamma(s+p)\Gamma(s+q+c)\Gamma(q-c+1)}\frac{Q_{p+q+s}}{g_{2s+2p+2q}}\omega_q^0+o(n)\end{aligned} \tag{B.28}$$

(observe that the quadratic term is zero because of Eq. (B.24a)). We can finally write for each value of $s$

$$\ln\det\mathbf{\Omega}^{(s)}=\sum_{c=0}^{\infty}\left[\binom{n-s}{c}-\binom{n-s}{c-1}\right]\ln\det\mathbf{N}^{(s,c)} \tag{B.29}$$

being the $(n-s-c)\times(n-s-c)$ matrix $\mathbf{N}^{(s,c)}$ in the $n\to 0$ limit

$$N_{pq}^{(s,c)}=\delta_{pq}-(-1)^q\frac{\Gamma(s+q)\Gamma(p+q+s)\sqrt{g_{2p+s}g_{2q+s}}}{\Gamma(s+p)\Gamma(q-c+1)\Gamma(s+q+c)}\frac{Q_{p+q+s}}{g_{2p+2q+2s}}. \tag{B.30}$$

Shifting by $c$ the indices in the expression above, transposing and then multiplying by

$$(-1)^{p+q}\sqrt{\frac{g_{2q+2c+s}}{g_{2p+2c+s}}}\frac{\Gamma(q+c+s)\Gamma(p+1)}{\Gamma(p+c+s)\Gamma(q+1)}$$

we get a new matrix $\mathbf{M}^{(s,c)}$ with the same spectrum of $\mathbf{N}^{(s,c)}$, namely

$$M_{pq}^{(s,c)}=\delta_{pq}-(-1)^{q+c}\frac{\Gamma(p+q+s+2c)}{\Gamma(p+s+2c)\Gamma(q+1)}\frac{g_{2q+2c+s}}{g_{2(p+q+s+2c)}}Q_{p+q+s+2c}. \tag{B.31}$$

To evaluate the correction to the free-energy we need to compute therefore

$$\begin{aligned}\ln\det\mathbf{\Omega} &= \sum_{c=2}^{\infty}\left[\binom{n}{c}-\binom{n}{c-1}\right]\ln\det\mathbf{M}^{(0,c)}\\&+\sum_{s=1}^{\infty}\binom{n}{s}\sum_{c=0}^{\infty}\left[\binom{n-s}{c}-\binom{n-s}{c-1}\right]\ln\det\mathbf{M}^{(s,c)}\\&= n\sum_{c=2}^{\infty}(-1)^{c-1}\frac{2c-1}{c(c-1)}\ln\det\mathbf{M}^{(0,c)}\\&-n\sum_{s=1}^{\infty}\sum_{c=0}^{\infty}(-1)^{c+s}(2c+s-1)\frac{\Gamma(s+c-1)}{s!c!}\ln\det\mathbf{M}^{(s,c)}+o(n).\end{aligned}\tag{B.32}$$

At this point is important to observe that, in the expression above, the $s=0$ contribution exactly coincides with the fluctuation contribution appearing in the finite-size corrections of RMP. Its evaluation has been performed in Refs. [MP87, PR02], but no closed formula is known for it. The $s\geq 1$ contribution is instead absent in the RMP. Generalizing therefore the analysis of Refs. [MP87, PR02], we note that eigenvalues of $\mathbf{M}^{(s,c)}$ are the same as the eigenvalues of the operator

$$\mathcal{M}^{(s,c)}(x,y)=\delta(x-y)-(-1)^c\mathcal{A}^{(c+s/2)}(x,y)\tag{B.33}$$

where the operator $\mathcal{A}^{(k)}(x,y)$ was the one introduced in Ref. [PR02] and it is defined as

$$\mathcal{A}^{(k)}(x,y)=2e^{-\frac{G(x)+G(y)}{2}}\sum_{q=0}^{\infty}\frac{(-1)^q e^{(q+k)(x+y)}}{\Gamma(2k+q)\Gamma(q+1)}g_{2(q+k)}.\tag{B.34}$$

Indeed, if $\psi_p$ is an eigenvector of $\mathbf{M}^{(s,c)}$ with corresponding eigenvalue $\lambda$, then, by a straightforward computation, it can be verified that

$$\phi(x)\equiv e^{\left(c+\frac{s}{2}\right)x-\frac{G(x)}{2}}\sum_{p=0}^{+\infty}\frac{(-1)^p}{p!}g_{2p+2c+s}\psi_p e^{px}\tag{B.35}$$

is an eigenvector of $\mathcal{M}^{(s,c)}(x,y)$ with the same eigenvalue. To evaluate the $\beta\to\infty$ limit of $\mathcal{A}^{(k)}(x,y)$ we observe that it has the same eigenvalues of the operator

$$\mathcal{H}^{(k)}(x,y)\equiv\beta\mathcal{A}^{(k)}(\beta x,\beta y)=e^{-\frac{\hat{G}(x)+\hat{G}(y)}{2}}\sum_{q=0}^{\infty}\frac{(-1)^q e^{(q+k)\beta(x+y)}}{\Gamma(2k+q)\Gamma(q+1)}\frac{1}{q+k}.\tag{B.36}$$

In Ref. [MP87, PR02] it has been shown that, for $\beta\to+\infty$, if we impose $\ln k=\beta t$ with $t$ fixed, the $\beta\to+\infty$ limit exists. In particular

$$\mathcal{H}^{(k)}(x,y)\xrightarrow[t\text{ fixed}]{\ln k=\beta t}\mathcal{H}_t(x,y)=e^{-\frac{\hat{G}(x)+\hat{G}(y)}{2}}\theta\left(x+y-2t\right).\tag{B.37}$$

This result suggests that the evaluation of the sums in Eq. (B.32) must be performed scaling $c$, $s$ and $\beta$ in a proper way. Given the known result for the RMP, we distinguish now between the contributions with $s\geq 1$ and the contribution obtained for $s=0$. We know indeed that we can obtain a finite limit for the $s=0$, that is the corresponding fluctuation correction to the aoc in the RMP. We can write

$$\begin{aligned}
&n\sum_{c=2}^{\infty}(-1)^{c-1}\frac{2c-1}{c(c-1)}\ln\det\mathbf{M}^{(0,c)} = n\sum_{E=1}^{\infty}\frac{1}{E}\sum_{c=2}^{\infty}(-1)^{(E+1)c}\frac{2c-1}{c(c-1)}\mathrm{tr}\left[\left(\mathcal{A}^{(c)}\right)^{E}\right]\\
&= n\sum_{E=1}^{\infty}\frac{1}{E}\sum_{c=1}^{\infty}\frac{(4c-1)\mathrm{tr}\left[\left(\mathcal{A}^{(2c)}\right)^{E}\right]}{2c(2c-1)} - n\sum_{E=1}^{\infty}\frac{(-1)^{E}}{E}\sum_{c=1}^{\infty}\frac{(4c+1)\mathrm{tr}\left[\left(\mathcal{A}^{(2c+1)}\right)^{E}\right]}{2c(2c+1)}\\
&\xrightarrow{\beta\to+\infty} 2n\beta\sum_{E\text{ odd}}\frac{1}{E}\int_0^{+\infty}\mathrm{tr}\left[\mathcal{H}_t^{E}\right]dt.
\end{aligned} \tag{B.38}$$

A numerical estimation of the quantity above can be found in Refs. [PR02, LPS17]. The evaluation of the $s\geq 1$ terms is more complicated and we will present here a non-rigorous treatment. We have

$$\begin{aligned}
&n\sum_{s=1}^{\infty}\sum_{c=0}^{\infty}(-1)^{c+s-1}(2c+s-1)\frac{\Gamma(s+c-1)}{s!c!}\ln\det\mathbf{M}^{(s,c)}\\
&= n\sum_{E=1}^{\infty}\frac{1}{E}\sum_{s=1}^{\infty}\sum_{c=0}^{\infty}(-1)^{(E+1)c+s}(2c+s-1)\frac{\Gamma(s+c-1)}{s!c!}\mathrm{tr}\left[\left(\mathcal{A}^{(c+\frac{s}{2})}\right)^{E}\right]+o(n)\\
&= n\sum_{E=1}^{\infty}\frac{1}{E}\sum_{s=1}^{\infty}\sum_{c=0}^{\infty}(-1)^{s}(4c+s-1)\frac{\Gamma(s+2c-1)\mathrm{tr}\left[\left(\mathcal{A}^{(2c+\frac{s}{2})}\right)^{E}\right]}{s!(2c)!}\\
&- n(-1)^{E}\sum_{E=1}^{\infty}\frac{1}{E}\sum_{s=1}^{\infty}\sum_{c=0}^{\infty}(-1)^{s}(4c+s+1)\frac{\Gamma(s+2c)\mathrm{tr}\left[\left(\mathcal{A}^{(2c+1+\frac{s}{2})}\right)^{E}\right]}{s!(2c+1)!}+o(n).
\end{aligned} \tag{B.39}$$

Introducing $4c+s=z$, the first sum in the last line of Eq. (B.39) becomes

$$\begin{aligned}
&\sum_{s=1}^{\infty}\sum_{c=0}^{\infty}(-1)^{s}(4c+s-1)\frac{\Gamma(s+2c-1)\mathrm{tr}\left[\left(\mathcal{A}^{(2c+\frac{s}{2})}\right)^{E}\right]}{s!(2c)!}\\
&= \sum_{z=1}^{\infty}(z-1)\mathrm{tr}\left[\left(\mathcal{A}^{(z/2)}\right)^{E}\right]\sum_{s=1}^{\infty}\sum_{c=0}^{\infty}\frac{(-1)^{s}\Gamma(s+2c-1)\,\mathbb{I}\,(s+4c=z)}{\Gamma(s+1)\Gamma(2c+1)}\\
&= \mathrm{tr}\left[\left(\mathcal{A}^{(1)}\right)^{E}\right]+\sum_{z=3}^{\infty}(z-1)\mathrm{tr}\left[\left(\mathcal{A}^{(z/2)}\right)^{E}\right]\\
&\sum_{s=1}^{\infty}\sum_{c=0}^{\infty}\frac{(-1)^{s}\Gamma(s+2c-1)\,\mathbb{I}\,(s+4c=z)}{\Gamma(s+1)\Gamma(2c+1)}.
\end{aligned} \tag{B.40a}$$

Similarly, after the change of variable $z = 4c + s + 2$, the second sum becomes

$$\sum_{s=1}^{\infty}\sum_{c=0}^{\infty}(-1)^s(4c+s+1)\frac{\Gamma(s+2c)}{s!\Gamma(2c+2)}\operatorname{tr}\left[\left(\mathcal{A}^{(2c+\frac{s}{2}+1)}\right)^E\right]$$
$$= \sum_{z=3}^{\infty}(z-1)\operatorname{tr}\left[\left(\mathcal{A}^{(z/2)}\right)^E\right]\sum_{s=1}^{\infty}\sum_{c=0}^{\infty}\frac{(-1)^s\Gamma(s+2c)\mathbb{I}\left(2+s+4c=z\right)}{\Gamma(s+1)\Gamma(2c+2)}. \quad \text{(B.40b)}$$

In other words, we can write

$$\sum_{s=1}^{\infty}\sum_{c=0}^{\infty}(-1)^{c+s-1}(2c+s-1)\frac{\Gamma(s+c-1)}{s!c!}\ln\det\mathbf{M}^{(s,c)}$$
$$= \sum_{E=1}^{\infty}\frac{1}{E}\sum_{z=3}^{\infty}(z-1)\operatorname{tr}\left[\left(\mathcal{A}^{(z/2)}\right)^E\right]\left[h_1(z)-(-1)^E h_2(z)\right] \quad \text{(B.41)}$$

The main difficulty in the evaluation of the quantities above is that the coefficients

$$h_1(z) := \sum_{s=1}^{\infty}\sum_{c=0}^{\infty}\frac{(-1)^s\Gamma\left(s+2c-1\right)\mathbb{I}\left(s+4c=z\right)}{\Gamma(s+1)\Gamma\left(2c+1\right)} \quad \text{(B.42)}$$

$$h_2(z) := \sum_{s=1}^{\infty}\sum_{c=0}^{\infty}\frac{(-1)^s\Gamma(s+2c)\mathbb{I}\left(2+s+4c=z\right)}{\Gamma(s+1)\Gamma(2c+2)} \quad \text{(B.43)}$$

are oscillating with diverging amplitude in $z$ for $z \to +\infty$, and therefore the large $z$ estimation is not straightforward. It is possible, however, that a different rearrangement of the contributions appearing in the sums might lead to a simpler asymptotic evaluation.

In the present work, we avoided this estimation using Wästlund's formula, but it is interesting to observe that Eq. (4.2c) implies that the contribution from Eq. (B.39) is equal and opposite to the one in Eq. (B.38), that coincides with the fluctuation finite-size correction to the aoc in the RMP. A more comprehensive study of these quantities can be matter of future investigations.

# Appendix C

# Appendix to Chapter 6

## C.1 The case $N = 3$ of the Euclidean bipartite TSP

In the case $N = 2$ there is only one Hamiltonian cycle, that is $\tilde{h}$. The first nontrivial case is $N = 3$. There are 6 Hamiltonian cycles. If we fix the starting point to be $r_1$ there are only two possibilities for the permutation $\sigma$ of the red points, that is $(1,2,3)$ and $(1,3,2)$. One is the dual of the other. We can restrict to the $(1,3,2)$ by removing the degeneracy in the orientation of the cycles. Indeed $\tilde{\sigma}$ is exactly $(1,3,2)$ according to (6.16). With this choice the 6 cycles are in correspondence with the permutations $\pi \in \mathcal{S}_3$ of the blue points. We sort in increasing order both the blue and red points. We have

$$\begin{aligned} E(\pi) = {} & |r_1 - b_{\pi(1)}|^p + |r_1 - b_{\pi(3)}|^p + |r_3 - b_{\pi(2)}|^p \\ & + |r_3 - b_{\pi(1)}|^p + |r_2 - b_{\pi(3)}|^p + |r_2 - b_{\pi(2)}|^p \,. \end{aligned} \tag{C.1}$$

The optimal solution is $\tilde{\pi} = (2,3,1)$. The permutations $(1,3,2)$ and $(3,2,1)$ have always a grater cost than $\tilde{\pi}$, indeed the corresponding cycles are $(r_1 b_1 r_3 b_3 r_2 b_2)$ and $(r_1 b_3 r_3 b_2 r_2 b_1)$, where we have colored in orange the path that, according to Lemma 1, can be reversed to lower the total cost. Doing this we obtain the optimal cycle in both cases. Notice that, since we can label each cycle using only the $\pi$ permutation, we can restrict ourself to moves that only involve blue points. Since there are three blue points, these moves will always reverse paths of the form $b_i r_j b_k$, so they correspond simply to a swap in the permutation $\pi$. Therefore our moves cannot be used to reach the optimal cycle from every starting cycle. A diagram showing all the possible moves is shown in Fig. C.1. In conclusion, the cost function makes $\mathcal{S}_3$ a *poset* with an absolute minimum and an absolute maximum. The permutation $(2,3,1)$ is preceded by both $(1,3,2)$ and $(3,2,1)$, which cannot be compared between them, but both precede $(1,2,3)$ and $(3,1,2)$, which cannot be compared between them. $(2,1,3)$ is the greatest element.

We compute the average costs for all the permutations. Using the same tech-

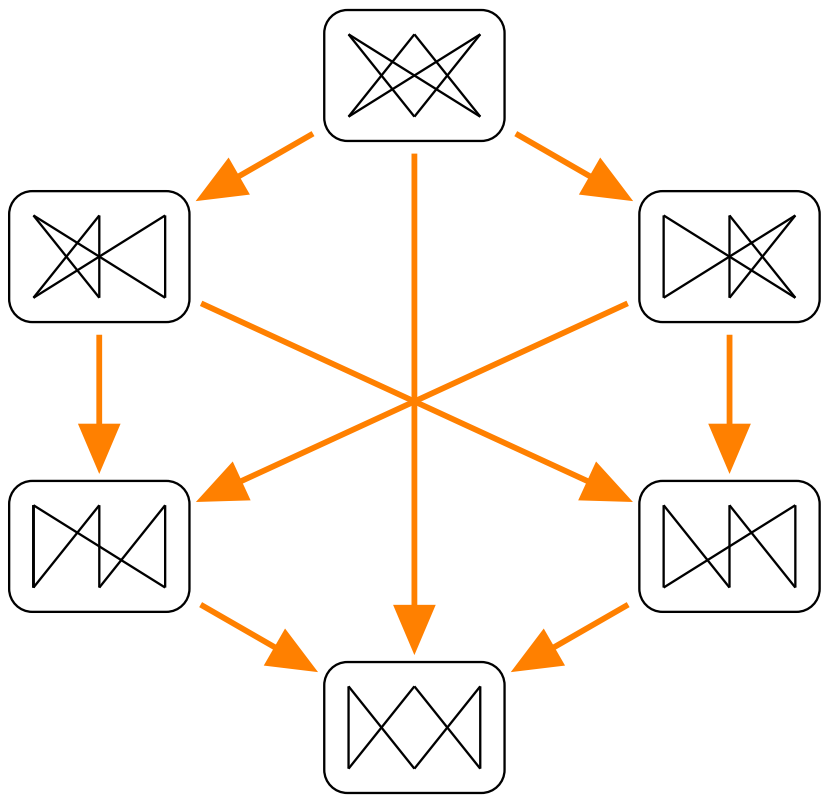

**Figure C.1.** The whole diagram describing the $N = 3$ case. In the squared boxes the various cycle configurations are represented. Lower boxes correspond to lower costs. All the possible moves suggested in Lemma 1 are represented by orange arrows.

niques used in section 6.3.3, we get that, for the $p = 2$ case

$$\begin{aligned}\overline{E[(2,3,1)]} = \frac{3}{4} &< \overline{E[(1,3,2)]} = \overline{E[(3,2,1)]} = \frac{7}{8} \\ &< \overline{E[(1,2,3)]} = \overline{E[(3,1,2)]} = \frac{9}{8} \\ &< \overline{E[(2,1,3)]} = \frac{5}{4}\,.\end{aligned} \tag{C.2}$$

## C.2 Proofs

In this appendix we prove various propositions stated in the main text.

### C.2.1 Proof of Proposition 6.4.1

Consider a $\sigma \in \mathcal{S}_N$ with $\sigma(1) = 1$. As we said before, taking $\sigma(1) = 1$ correspond to the irrelevant choice of the starting point of the cycle. Let us introduce now a new set of ordered points $\mathcal{B} := \{b_j\}_{j=1,\dots,N} \subset [0,1]$ such that

$$b_i = \begin{cases} r_1 & \text{for} \quad i = 1 \\ r_{i-1} & \text{otherwise} \end{cases} \tag{C.3}$$

and consider the Hamiltonian cycle on the complete bipartite graph with vertex sets $\mathcal{R}$ and $\mathcal{B}$

$$\begin{aligned}&h[(\sigma, \pi_\sigma)] \\ &:= (r_1, b_{\pi_\sigma(1)}, r_{\sigma(2)}, b_{\pi_\sigma(2)}, \dots, r_{\sigma(N)}, b_{\pi_\sigma(N)}, r_{\sigma(1)})\end{aligned} \tag{C.4}$$

so that

$$\pi_\sigma(i) = \begin{cases} 2 & \text{for} \quad i = 1 \\ \sigma(i) + 1 & \text{for} \quad i < k \\ \sigma(i+1) + 1 & \text{for} \quad i \geq k \\ 1 & \text{for} \quad i = N \end{cases} \tag{C.5}$$

where $k$ is such that $\sigma(k) = N$. We have therefore

$$\begin{aligned}(b_{\pi_\sigma(1)}, b_{\pi_\sigma(2)}, \dots, b_{\pi_\sigma(k-1)}, b_{\pi_\sigma(k)}, \dots, b_{\pi_\sigma(N-1)}, b_{\pi_\sigma(N)}) \\ = (r_1, r_{\sigma(2)}, \dots, r_{\sigma(k-1)}, r_{\sigma(k+1)}, \dots, r_{\sigma(N)}, r_1).\end{aligned} \tag{C.6}$$

In other words we are introducing a set of blue points such that we can find a bipartite Hamiltonian tour which only use link available in our monopartite problem and has the same cost of $\sigma$. Therefore, by construction (using (C.6)):

$$\begin{aligned}E_N(h[\sigma]) = E_N(h[(\sigma, \pi_\sigma)]) \geq E_N(h[(\tilde{\sigma}, \tilde{\pi})]) \\ = E_N(h[(\tilde{\sigma}, \pi_{\tilde{\sigma}})]) = E_N(h[\tilde{\sigma}]),\end{aligned} \tag{C.7}$$

where the fact that $\tilde{\pi} = \pi_{\tilde{\sigma}}$ can be checked using (6.16) and (6.17) and (C.5).

### C.2.2 Proof of Proposition 6.4.2

Before proving Proposition 6.4.2, we enunciate and demonstrate two lemmas that will be useful for the proof. The first one will help us in understand how to remove two crossing arcs without breaking the TSP cycle into multiple ones. The second one, instead will prove that removing a crossing between two arcs will always lower the total number of crossing in the TSP cycle.

**Lemma 5.** *Given an Hamiltonian cycle with its edges drawn as arcs in the upper half-plain, let us consider two of the arcs that cannot be drawn without crossing each other. Then, this crossing can be removed only in one way without splitting the original cycle into two disjoint cycles; moreover, this new configuration has a lower cost than the original one.*

*Proof.* Let us consider a generic oriented Hamiltonian cycle and let us suppose it contains a matching as in figure:

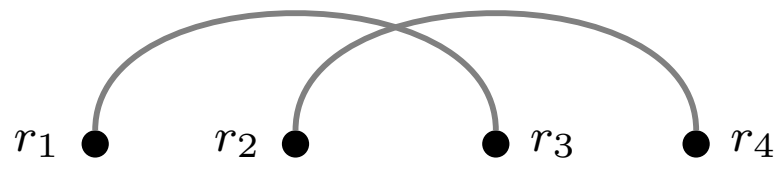

There are two possible orientations for the matching that correspond to this two oriented Hamiltonian cycles:

1. $(C_1 r_1 r_3 C_2 r_2 r_4 C_3)$,
2. $(C_1 r_1 r_3 C_2 r_4 r_2 C_3)$,

where $C_1$, $C_2$ and $C_3$ are paths (possibly visiting other points of our set). The other possibilities are the dual of this two, and thus they are equivalent. In both cases, a priori, there are two choices to replace this crossing matching $(r_1, r_3)$, $(r_2, r_4)$ with a non-crossing one: $(r_1, r_2)$, $(r_3, r_4)$ or $(r_1, r_4)$, $(r_2, r_3)$. We now show, for the two possible prototypes of Hamiltonian cycles, which is the right choice for the non-crossing matching, giving a general rule. Let us consider case 1: here, if we replace the crossing matching with $(r_1, r_4)$, $(r_2, r_3)$, the cycle will split; in fact we would have two cycles: $(C_1 r_1 r_4 C_3)$ and $(r_3 C_2 r_2)$. Instead, if we use the other non-crossing matching, we would have: $(C_1 r_1 r_2 [C_2]^{-1} r_3 r_4 C_3)$. This way we have removed the crossing without splitting the cycle. Let us consider now case 2: in this situation, using $(r_1, r_4)$, $(r_2, r_3)$ as the new matching, we would have: $(C_1 r_1 r_4 [C_2]^{-1} r_3 r_2 C_3)$; the other matching, on the contrary, gives: $(C_1 r_1 r_2 C_3)$ and $(r_3 C_2 r_4)$.

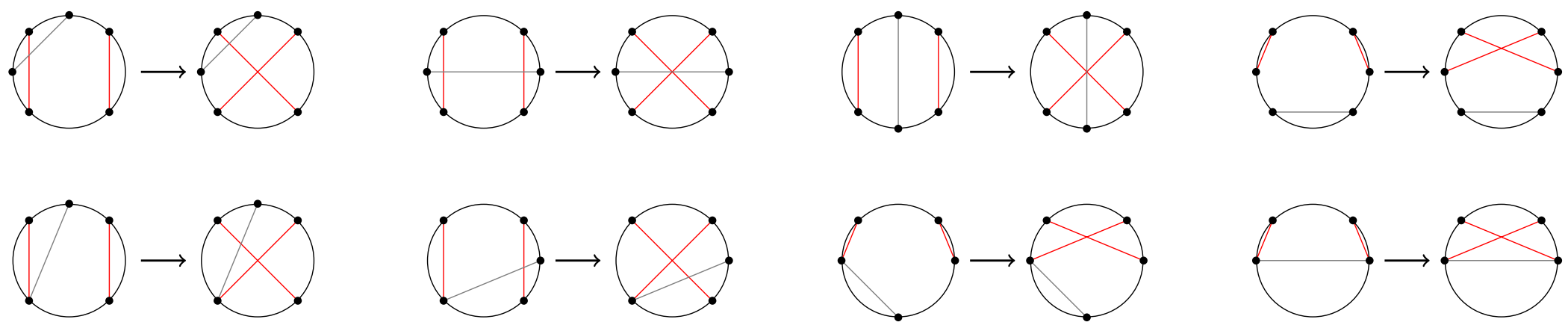

**Figure C.2.** Replacing a non-crossing matching with a crossing one in an Hamiltonian cycle always increase the number of crossings. Here we list all the possible topological configurations one can have.

The general rule is the following: given the oriented matching, consider the four oriented lines going inward and outward the node. Then, the right choice for the non-crossing matching is obtained joining the two couples of lines with opposite orientation.

Since the difference between the cost of the original cycle and the new one simply consists in the difference between a crossing matching and a non-crossing one, this is positive when $0 < p < 1$, as shown in [BCS14]. ∎

Now we deal with the second point: given an Hamiltonian cycle, in general it is not obvious that replacing non-crossing arcs with a crossing one, the total number of intersections increases. Indeed there could be the chance that one or more nodes are removed in the operation of substituting the matching we are interested in. However, we now show that it holds the following

**Lemma 6.** *Given an Hamiltonian cycle with a matching that is non-crossing, if it is replaced by a crossing one, the total number of intersections always increases. Vice versa, if a crossing matching is replaced by a non-crossing one, the total number of crossings always decreases.*

*Proof.* This is a topological property we will prove for cases, using the representation on the circle, since crossings are easier to visualize. All the possibilities are displayed in Fig. C.2, where we have represented with red lines the edges involved in the matching, while the other lines span all the possible topological configurations.

∎

Now we can prove Proposition 6.4.2:

*Proof.* Consider a generic Hamiltonian cycle and draw the connections between the points in the upper half-plain. Suppose to have an Hamiltonian cycle where there are, let us say, $n$ intersections between edges. Thanks to Lemma 5, we can swap two crossing arcs with a non-crossing one without splitting the Hamiltonian cycle. As shown in Lemma 6, this operation lowers always the total number of crossings between the edges, and the cost of the new cycle is smaller than the cost of the starting one. Iterating this procedure, it follows that one can find a cycle with no crossings. Now we prove that there are no other cycles out of $h^*$ and its dual with no crossings. This can be easily seen, since $h^*$ is the only cycle that visits all the points, starting from the first, in order. This means that all the other cycles do not visit the points in order and, thus, they have a crossing, due to the fact that the point that is not visited in a first time, must be visited next, creating a crossing. ■

### C.2.3 Proof of Proposition 6.4.3

To complete the proof given in the main text, we need to discuss two points. Firstly, we address which is the correct move that swap a non-crossing matching with a crossing one; thanks to Lemma 6, by performing such a move one always increases the total number of crossings. Secondly we prove that there is only one Hamiltonian cycle to which this move cannot be applied (and so it is the optimal solution).

We start with the first point: consider an Hamiltonian cycle with a matching that is non-crossing, then the possible situations are the following two:

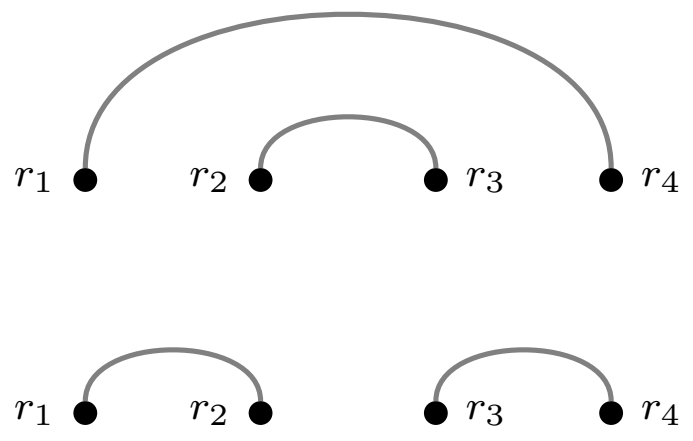

For the first case there are two possible independent orientations:

1. $(r_1 r_4 C_2 r_2 r_3 C_3)$,
2. $(r_1 r_4 C_2 r_3 r_2 C_3)$.

If we try to cross the matchings in the first cycle, we obtain $(r_1 r_3 C_3)(r_2 [C_2]^{-1} r_4)$, and this is not anymore an Hamiltonian cycle. On the other hand, in the second cycle, the non-crossing matching can be replaced by a crossing one without breaking the cycle: $(r_1 r_3 [C_2]^{-1} r_4 r_2 C_3)$. For the second case the possible orientations are:

1. $(r_1 r_2 C_2 r_4 r_3 C_3)$,
2. $(r_1 r_2 C_2 r_3 r_4 C_3)$.

By means of the same procedure used in the first case, one finds that the non-crossing matching in the second cycle can be replaced by a crossing one without splitting the cycle, while in the first case the cycle is divided by this operation.

The last step is the proof that the Hamiltonian cycle given in Proposition 6.4.3 has the maximum number of crossings.

Let us consider an Hamiltonian cycle $h[\sigma] = \left(r_{\sigma(1)}, \ldots, r_{\sigma(N)}\right)$ on the complete graph $\mathcal{K}_N$. We now want to evaluate what is the maximum number of crossings an edge can have depending on the permutation $\sigma$. Consider the edge connecting two vertices $r_{\sigma(i)}$ and $r_{\sigma(i+1)}$: obviously both the edges $(r_{\sigma(i-1)}, r_{\sigma(i)})$ and $(r_{\sigma(i+1)}, r_{\sigma(i+2)})$ share a common vertex with $(r_{\sigma(i)}, r_{\sigma(i+1)})$, therefore they can never cross it. So, if we have $N$ vertices, each edge has $N-3$ other edges that can cross it. Let us denote with $\mathcal{N}[\sigma(i)]$ the number of edges that cross the edge $(r_{\sigma(i)}, r_{\sigma(i+1)})$ and let us define the sets:

$$A_j := \begin{cases} \{r_k\}_{k=\sigma(i)+1 \pmod N, \ldots, \sigma(i+1)-1 \pmod N} & \text{for} \quad j=1 \\ \{r_k\}_{k=\sigma(i+1)+1 \pmod N, \ldots, \sigma(i)-1 \pmod N} & \text{for} \quad j=2 \end{cases} \tag{C.8}$$

These two sets contain the points between $r_{\sigma(i)}$ and $r_{\sigma(i+1)}$. In particular, the maximum number of crossings an edge can have is given by:

$$\max(\mathcal{N}[\sigma(i)]) = \begin{cases} 2\min_j |A_j| & \text{for} \quad |A_1| \neq |A_2| \\ 2|A_1| - 1 & \text{for} \quad |A_1| = |A_2| \end{cases} \tag{C.9}$$

This is easily seen, since the maximum number of crossings an edge can have is obtained when all the points belonging to the smaller between $A_1$ and $A_2$ contributes with two crossings. This cannot happen when the cardinality of $A_1$ and $A_2$ is the same because at least one of the edges departing from the nodes in $A_1$ for example, must be connected to one of the ends of the edge $(r_{\sigma(i)}, r_{\sigma(i+1)})$, in order to have an Hamiltonian cycle. Note that this case, i.e. $|A_1| = |A_2|$ can happen only if $N$ is even.

Consider the particular case such that $\sigma(i) = a$ and $\sigma(i+1) = a + \frac{N-1}{2} \pmod N$ or $\sigma(i+1) = a + \frac{N+1}{2} \pmod N$. Then (C.9) in this cases is exactly equal to $N-3$, which means that the edges $(r_a, r_{a+\frac{N-1}{2} \pmod N})$ and $(r_a, r_{a+\frac{N+1}{2} \pmod N})$ can have the maximum number of crossings if the right configuration is chosen. Moreover, if there is a cycle such that every edge has $N-3$ crossings, such a cycle is unique, because the only way of obtaining it is connecting the vertex $r_a$ with $r_{a+\frac{N-1}{2} \pmod N}$ and $r_{a+\frac{N+1}{2} \pmod N}$, $\forall a$.

## C.3 The 2-factor and TSP solution for $p < 0$ and even $N$

We start considering here the 2-factor solution for $p < 0$ in the even-$N$ case. After that, we proof Proposition 6.4.4.

In the following we will say that, given a permutation $\sigma \in \mathcal{S}_N$, the edge $(r_{\sigma(i)}, r_{\sigma(i+1)})$ has length $L \in \mathbb{N}$ if:

$$L = \mathcal{L}(i) := \min_j |A_j(i)| \tag{C.10}$$

where $A_j(i)$ was defined in equation (C.8).

### C.3.1 $N$ is a multiple of 4

Let us consider the sequence of points $\mathcal{R} = \{r_i\}_{i=1,\dots,N}$ of $N$ points, with $N$ a multiple of 4, in the interval $[0,1]$, with $r_1 \leq \dots \leq r_N$, consider the permutations $\sigma_j$, $j = 1,2$ defined by the following cyclic decomposition:

$$\begin{aligned}\sigma_1 = &(r_1, r_{\frac{N}{2}+1}, r_2, r_{\frac{N}{2}+2}) \dots \\ &(r_a, r_{a+\frac{N}{2}}, r_{a+1}, r_{a+\frac{N}{2}+1}) \dots (r_{\frac{N}{2}-1}, r_{N-1}, r_{\frac{N}{2}}, r_N)\end{aligned} \tag{C.11a}$$

$$\begin{aligned}\sigma_2 = &(r_1, r_{\frac{N}{2}+1}, r_N, r_{\frac{N}{2}}) \dots (r_a, r_{a+\frac{N}{2}}, r_{a-1}, r_{a+\frac{N}{2}-1}) \\ &\dots (r_{\frac{N}{2}-1}, r_{N-1}, r_{\frac{N}{2}-2}, r_{N-2})\end{aligned} \tag{C.11b}$$

for integer $a = 1, \dots, \frac{N}{2} - 1$. Defined $h_1^* := h[\sigma_1]$ and $h_2^* := h[\sigma_2]$, it holds the following:

**Proposition C.3.1.** *$h_1^*$ and $h_2^*$ are the 2-factors that contain the maximum number of crossings between the arcs.*

*Proof.* An edge can be involved, at most, in $N-3$ crossing matchings. In the even N case, this number is achieved by the edges of the form $(r_a, r_{a+\frac{N}{2} \pmod N})$, i.e. by the edges of length $\frac{N}{2} - 1$. There can be at most $\frac{N}{2}$ edges of this form in a 2-factor. Thus, in order to maximize the number of crossings, the other $\frac{N}{2}$ edges must be of the form $(r_a, r_{a+\frac{N}{2}+1 \pmod N})$ or $(r_a, r_{a+\frac{N}{2}-1 \pmod N})$, i.e. of length $\frac{N}{2} - 2$. It is immediate to verify that both $h_1^*$ and $h_2^*$ have this property; we have to prove they are the only ones with this property.
Consider, then, to have already fixed the $\frac{N}{2}$ edges $(r_a, r_{a+\frac{N}{2} \pmod N}), \forall a \in [N]$. Suppose to have fixed also the edge $(r_1, r_{\frac{N}{2}})$ (the other chance is to fix the edge $(r_1, r_{\frac{N}{2}+2})$: this brings to the other 2-factor). Consider now the point $r_{\frac{N}{2}+1}$: suppose it is not connected to the point $r_N$, but to the point $r_2$, i.e., it has a different edge from the cycle $h_2^*$. We now show that this implies it is not possible to construct all the remaining edges of length $\frac{N}{2} - 2$. Consider, indeed, of having fixed the edges $(r_1, r_{\frac{N}{2}})$ and $(r_2, r_{\frac{N}{2}+1})$ and focus on the vertex $r_{\frac{N}{2}+2}$: in order to have an edge of length $\frac{N}{2} - 2$, this vertex must be connected either with $r_1$ or with $r_3$, but $r_1$ already has two edges, thus, necessarily, there must be the edge $(r_{\frac{N}{2}+2}, r_3)$. By the same reasoning, there must be the edges $(r_{\frac{N}{2}+3}, r_4)$, $(r_{\frac{N}{2}+4}, r_5), \dots, (r_{N-2}, r_{\frac{N}{2}-1})$. Proceeding this way, we have constructed $N-1$ edges; the remaining one is uniquely determined, and it is $(r_{N-1}, r_N)$, which has null length.
Therefore the edge $(r_2, r_{\frac{N}{2}+1})$ cannot be present in the optimal 2-factor and so, necessarily, there is the edge $(r_{\frac{N}{2}+1}, r_N)$; this creates the cycle $(r_1, r_{\frac{N}{2}}, r_N, r_{\frac{N}{2}+1})$. Proceeding the same way on the set of the remaining vertices $\{r_2, r_3, \dots, r_{\frac{N}{2}-1}, r_{\frac{N}{2}+2}, \dots, r_{N-1}\}$, one finds that the only way of obtaining $\frac{N}{2}$ edges of length $\frac{N}{2} - 1$ and $\frac{N}{2}$ edges of length $\frac{N}{2} - 2$ is generating the loop coverings of the graph $h_1^*$ or $h_2^*$. ■

Proposition C.3.1, together with the fact that the optimal 2-factor has the maximum number of crossing matchings, guarantees that the optimal 2-factor is either $h_1^*$ or $h_2^*$.

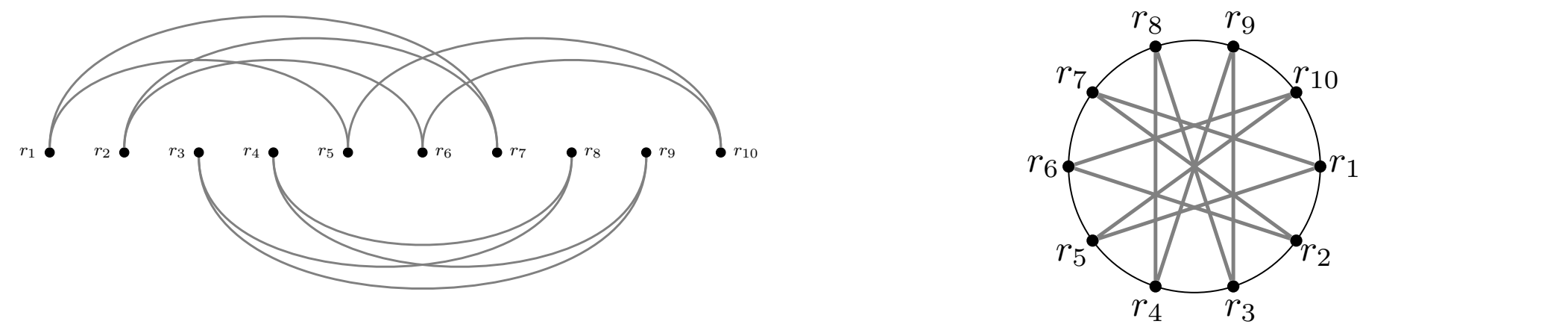

**(a)** One of the optimal 2-factor solutions for $N = 10$ and $p < 0$; the others are obtainable cyclically permuting this configuration

**(b)** The same optimal 2-factor solution, but represented on a circle, where the symmetries of the solutions are more easily seen

**Figure C.3**

### C.3.2 $N$ is not a multiple of 4

Let us consider the usual sequence $\mathcal{R} = \{r_i\}_{i=1,\dots,N}$ of $N$ points, with even $N$ but not a multiple of 4, in the interval $[0,1]$, with $r_1 \le \dots \le r_N$, consider the permutation $\pi$ defined by the following cyclic decomposition:

$$\begin{aligned} \pi = &(r_1, r_{\frac{N}{2}}, r_N, r_{\frac{N}{2}+1}, r_2, r_{\frac{N}{2}+2})(r_3, r_{\frac{N}{2}+3}, r_4, r_{\frac{N}{2}+4}) \\ &\dots (r_{\frac{N}{2}-2}, r_{N-1}, r_{\frac{N}{2}-1}, r_{N-2}) \end{aligned} \tag{C.12}$$

Defined

$$\pi_k(i) := \pi(i) + k \ (\mathrm{mod}\ N),\ k \in [0, N-1] \tag{C.13}$$

and

$$h_k^* := h[\pi_k] \tag{C.14}$$

the following proposition holds:

**Proposition C.3.2.** *$h_k^*$ are the 2-factors that contain the maximum number of crossings between the arcs.*

*Proof.* Also in this case the observations done in the proof of Proposition C.3.1 holds. Thus, in order to maximize the number of crossing matchings, one considers, as in the previous case, the $\frac{N}{2}$ edges of length $\frac{N}{2} - 1$, i.e. of the form $(r_a, r_{a+\frac{N}{2} \ (\mathrm{mod}\ N)})$, and then tries to construct the remaining $\frac{N}{2}$ edges of length $\frac{N}{2} - 2$, likewise the previous case. Again, if one fixes the edge $(r_1, r_{\frac{N}{2}})$, the edge $(r_2, r_{\frac{N}{2}+1})$ cannot be present, by the same reasoning done in the proof of Proposition B.1. The fact that, in this case, $N$ is not a multiple of 4 makes it impossible to have a 2-factor formed by 4-vertices loops, as in the previous case. The first consequence is that, given $\frac{N}{2}$ edges of length $\frac{N}{2} - 1$, it is not possible to have $\frac{N}{2}$ edges of length $\frac{N}{2} - 2$. In order to find the maximum-crossing solution, one has the following options:

- to take a 2-factor with $\frac{N}{2}$ edges of length $\frac{N}{2} - 1$, $\frac{N}{2} - 1$ edges of length $\frac{N}{2} - 2$ and one edge of length $\frac{N}{2} - 2$: in this case the theoretical maximum number of crossing matchings is $\frac{N(N-3)}{2} + (\frac{N}{2} - 1)(N-4) + N - 6 = N^2 - \frac{7N}{2} - 2$;

- to take a 2-factor with $\frac{N}{2} - 1$ edges of length $\frac{N}{2} - 1$, $\frac{N}{2} + 1$ edges of length $\frac{N}{2} - 2$: in this case the theoretical maximum number of crossing matchings is $(\frac{N}{2} - 1)(N-3) + (\frac{N}{2} + 1)(N-4) = N^2 - \frac{7N}{2} - 1$.

Clearly the second option is better, at least in principle, than the first one. The cycles $h_k^*$ belong to the second case and saturate the number of crossing matchings. Suppose, then, to be in this case. Let us fix the $\frac{N}{2} - 1$ edges of length $\frac{N}{2} - 1$; this operation leaves two vertices without any edge, and this vertices are of the form $r_a, r_{a+\frac{N}{2} \pmod N}, a \in [1, N]$ (this is the motivation for the degeneracy of solutions). By the reasoning done above, the edges that link this vertices must be of length $\frac{N}{2} - 2$, and so they are uniquely determined. They form the 6-points loop $(r_a, r_{a-1+\frac{N}{2} \pmod N}, r_{N-1+a \pmod N}, r_{a+\frac{N}{2} \pmod N}, r_{a+1 \pmod N}, r_{a+1+\frac{N}{2} \pmod N})$. The remaining $N-6$ points, since $4|(N-6)$, by the same reasoning done in the proof of Proposition C.3.1, necessarily form the $\frac{N-6}{4}$ 4-points loops given by the permutations (C.13). ∎

Proposition C.3.2, together with the fact that the optimal 2-factor has the maximum number of crossing matchings, guarantees that the optimal 2-factor is such that $h^* \in \{h_k^*\}_{k=1}^N$.

### C.3.3 Proof of Proposition 6.4.4

*Proof.* Let us begin from the permutations that define the optimal solutions for the 2-factor, that is those given in Eqs. C.11 if is $N$ a multiple of 4 and in Eq. C.12 otherwise. In both cases, the optimal solution is formed only by edges of length $\frac{N}{2} - 1$ and of length $\frac{N}{2} - 2$. Since the optimal 2-factor is not a TSP, in order to obtain an Hamiltonian cycle from the 2-factor solution, couples of crossing edges need to became non-crossing, where one of the two edges belongs to one loop of the covering and the other to another loop. Now we show that the optimal way of joining the loops is replacing two edges of length $\frac{N}{2} - 1$ with other two of length $\frac{N}{2} - 2$. Let us consider two adjacent 4-vertices loops, i.e. two loops of the form:

$$(r_a, r_{a+\frac{N}{2}}, r_{a+1}, r_{a+\frac{N}{2}+1}), (r_{a+2}, r_{a+2+\frac{N}{2}}, r_{a+3}, r_{a+\frac{N}{2}+3}) \tag{C.15}$$

and let us analyze the possible cases:

1. to remove two edges of length $\frac{N}{2} - 2$, that can be replaced in two ways:
   - either with an edge of length $\frac{N}{2} - 2$ and one of length $\frac{N}{2} - 4$; in this case the maximum number of crossings decreases by 4;
   - or with two edges of length $\frac{N}{2} - 3$; also in this situation the maximum number of crossings decreases by 4.
2. to remove one edge of length $\frac{N}{2} - 2$ and one of length $\frac{N}{2} - 1$, and also this operation can be done in two ways:
   - either with an edge of length $\frac{N}{2} - 2$ and one of length $\frac{N}{2} - 3$; in this case the maximum number of crossings decreases by 3;

- or with an edge of length $\frac{N}{2}-3$ and one of length $\frac{N}{2}-4$; in this situation the maximum number of crossings decreases by 7.

3. the last chance is to remove two edges of length $\frac{N}{2}-1$, and also this can be done in two ways:

    - either with two edges of length $\frac{N}{2}-3$; here the maximum number of crossings decreases by 6;
    - or with two edges of length $\frac{N}{2}-2$; in this situation the maximum number of crossings decreases by 2. This happens when we substitute two adjacent edges of length $\frac{N}{2}-1$, that is, edges of the form $(r_a, r_{\frac{N}{2}+a \pmod N})$ and $(r_{a+1}, r_{\frac{N}{2}+a+1 \pmod N})$, with the non-crossing edges $(r_a, r_{\frac{N}{2}+a+1 \pmod N})$ and $(r_{a+1}, r_{\frac{N}{2}+a \pmod N})$

The last possibility is the optimal one, since our purpose is to find the TSP with the maximum number of crossings, in order to conclude it has the lower cost. Notice that the cases discussed above holds also for the 6-vertices loop and an adjacent 4-vertices loop when N is not a multiple of 4. We have considered here adjacent loops because, if they were not adjacent, then the difference in maximum crossings would have been even bigger.
Now we have a constructive pattern for building the optimal TSP. Let us call $\mathcal{O}$ the operation described in the second point of (3). Then, starting from the optimal 2-factor solution, if it is formed by $n$ points, $\mathcal{O}$ has to be applied $\frac{N}{4}-1$ times if N is a multiple of 4 and $\frac{N-6}{4}$ times otherwise. In both cases it is easily seen that $\mathcal{O}$ always leaves two adjacent edges of length $\frac{N}{2}-1$ invariant, while all the others have length $\frac{N}{2}-2$. The multiplicity of solutions is given by the $\frac{N}{2}$ ways one can choose the two adjacent edges of length $\frac{N}{2}-1$. In particular, the Hamiltonian cycles $h_k^*$ saturates the maximum number of crossings that can be done, i.e., every time that $\mathcal{O}$ is applied, exactly 2 crossings are lost.
We have proved, then, that $h_k^*$ are the Hamiltonian cycles with the maximum number of crossings. Now we prove that any other Hamiltonian cycle has a lower number of crossings. Indeed any other Hamiltonian cycle must have

- either every edge of length $\frac{N}{2}-2$;
- or at least one edge of length less than or equal to $\frac{N}{2}-3$.

This is easily seen, since it is not possible to build an Hamiltonian cycle with more than two edges or only one edge of length $\frac{N}{2}-1$ and all the others of length $\frac{N}{2}-2$. It is also impossible to build an Hamiltonian cycle with two non-adjacent edges of length $\frac{N}{2}-1$ and all the others of length $\frac{N}{2}-2$: the proof is immediate. Consider then the two cases presented above: in the first case the cycle (let us call it $H$) is clearly not optimal, since it differs from $h_k^*, \forall k$ by a matching that is crossing in $h_k^*$ and non-crossing in $H$. Let us consider, then, the second case and suppose the shortest edge, let us call it $b$, has length $\frac{N}{2}-3$: the following reasoning equally holds if the considered edge is shorter. The shortest edge creates two subsets of vertices: in fact, called $x$ and $y$ the vertices of the edge considered and supposing

$x < y$, there are the subsets defined by:

$$A = \{r \in \mathcal{V} : x < r < y\} \tag{C.16}$$

$$B = \{r \in \mathcal{V} : r < x \vee r > y\} \tag{C.17}$$

Suppose, for simplicity, that $|A| < |B|$: then, necessarily $|A| = \frac{N}{2} - 3$ and $|B| = \frac{N}{2} + 1$. As an immediate consequence, there is a vertex in $B$ whose edges have both vertices in $|B|$. As a consequence, fixed an orientation on the cycle, one of this two edges and $b$ are obviously non-crossing and, moreover, have the right relative orientation so that they can be replaced by two crossing edges without splitting the Hamiltonian cycle. Therefore also in this case the Hamiltonian cycle considered is not optimal. ■

## C.4 General distribution of points

In this section we shall consider a more general distribution of points. Let choose the points in the interval $[0, 1]$ according to the distribution $\rho$, which has no zero in the interval, and let

$$\Phi_\rho(x) = \int_0^x dt\, \rho(t) \tag{C.18}$$

its *cumulative*, which is an increasing function with $\Phi_\rho(0) = 0$ and $\Phi_\rho(1) = 1$.

In this case, the probability of finding the $l$-th point in the interval $[x, x + dx]$ and the $s$-th point in the interval $[y, y + dy]$ is given, for $s > l$ by

$$\begin{aligned} p_{l,s}(x,y)\, d\Phi_\rho(x)\, d\Phi_\rho(y) = &\frac{\Gamma(N+1)}{\Gamma(l)\,\Gamma(s-l)\,\Gamma(N-s+1)} \\ &\times \Phi_\rho^{l-1}(x)\, \left[\Phi_\rho(y) - \Phi_\rho(x)\right]^{s-l-1} \left[1 - \Phi_\rho(y)\right]^{N-s} \\ &\times \theta(y-x)\, d\Phi_\rho(x)\, d\Phi_\rho(y) \end{aligned} \tag{C.19}$$

We have that, in the case $p > 1$

$$\begin{aligned} \overline{E_N[h^*]} = &\int d\Phi_\rho(x)\, d\Phi_\rho(y)\, (y-x)^p \\ &\times \left[ p_{1,2}(x,y) + p_{N-1,N}(x,y) + \sum_{l=1}^{N-2} p_{l,l+2}(x,y) \right] \end{aligned} \tag{C.20}$$

and

$$\begin{aligned} \sum_{l=1}^{N-2} p_{l,l+2}(x,y) = &\frac{\Gamma(N+1)}{\Gamma(N-2)} \left[1 - \Phi_\rho(y) + \Phi_\rho(x)\right]^{N-3} \\ &\times \left[\Phi_\rho(y) - \Phi_\rho(x)\right]\, \theta(y-x) \end{aligned} \tag{C.21}$$

while

$$[p_{1,2}(x,y) + p_{N-1,N}(x,y)] = \frac{\Gamma(N+1)}{\Gamma(N-1)} \left[\left(1 - \Phi_\rho(y)\right)^{N-2} + \Phi_\rho^{N-2}(x)\right] \theta(y-x) \tag{C.22}$$

For large $N$ we can make the approximation

$$\begin{aligned}\overline{E_N[h^*]} \approx N^3 \int & d\Phi_\rho(x)\, d\Phi_\rho(y)\, (y-x)^p \\ & \times \left[1-\Phi_\rho(y)+\Phi_\rho(x)\right]^N \left[\Phi_\rho(y)-\Phi_\rho(x)\right]\, \theta(y-x)\end{aligned} \tag{C.23}$$

and we remark that the maximum of the contribution to the integral comes from the region where $\Phi_\rho(y) \approx \Phi_\rho(x)$ and we make the change of variables

$$\Phi_\rho(y) = \Phi_\rho(x) + \frac{\epsilon}{N} \tag{C.24}$$

so that

$$y = \Phi_\rho^{-1}\left[\Phi_\rho(x) + \frac{\epsilon}{N}\right] \approx x + \frac{\epsilon}{N\rho(x)} \tag{C.25}$$

and we get

$$\overline{E_N[h^*]} \approx N^3 \int d\Phi_\rho(x) \int_0^\infty \frac{d\epsilon}{N} \left[\frac{\epsilon}{N\rho(x)}\right]^p \frac{\epsilon}{N}\, e^{-\epsilon} \frac{\Gamma(p+2)}{N^{p-1}} \int dx\, \rho^{1-p}(x)\,. \tag{C.26}$$

When $0 < p < 1$

$$\overline{E_N[h^*]} = \int d\Phi_\rho(x)\, d\Phi_\rho(y)\, (y-x)^p \left[p_{1,N}(x,y) + \sum_{l=1}^{N-1} p_{l,l+1}(x,y)\right] \tag{C.27}$$

and

$$\sum_{l=1}^{N-1} p_{l,l+1}(x,y) = N(N-1)\, \left[1-\Phi_\rho(y)+\Phi_\rho(x)\right]^{N-2} \theta(y-x) \tag{C.28}$$

while

$$p_{1,N}(x,y) = N(N-1)\, \left[\Phi_\rho(y)-\Phi_\rho(x)\right]^{N-2} \theta(y-x) \tag{C.29}$$

For large $N$ we can make the approximation

$$\begin{aligned}\overline{E_N[h^*]} &\approx N^2 \int d\Phi_\rho(x)\, d\Phi_\rho(y)\, (y-x)^p \left[1-\Phi_\rho(y)+\Phi_\rho(x)\right]^N \theta(y-x) \\ &\approx N^2 \int d\Phi_\rho(x) \int_0^\infty \frac{d\epsilon}{N} \left[\frac{\epsilon}{N\rho(x)}\right]^p e^{-\epsilon} \\ &= \frac{\Gamma(p+1)}{N^{p-1}} \int dx\, \rho^{1-p}(x)\,.\end{aligned} \tag{C.30}$$

Indeed the other term, for large $N$, gives a contribution

$$N^2 \int (y-x)^p \left[\Phi_\rho(y)-\Phi_\rho(x)\right]^N \theta(y-x)\, d\Phi_\rho(x)\, d\Phi_\rho(y) \tag{C.31}$$

so that, we will set

$$\Phi_\rho(y) = 1 - \frac{\epsilon}{N}\,, \qquad \Phi_\rho(x) = \frac{\delta}{N}\,, \qquad y - x \approx 1 \tag{C.32}$$

and therefore we get a contribution

$$\int_0^\infty d\epsilon\, e^{-\epsilon} \int_0^\infty d\delta\, e^{-\delta} = 1 \tag{C.33}$$

which is of the same order of the other term only at $p = 1$.

## C.5 Calculation of the second moment of the optimal cost distribution

In this Appendix we compute the second moment of the optimal cost distribution. We will restrict for simplicity to the $p > 1$ case, where

$$E_N[h^*] = |r_2 - r_1|^p + |r_N - r_{N-1}|^p + \sum_{i=1}^{N-2} |r_{i+2} - r_i|^p \,. \tag{C.34}$$

We begin by writing the probability distribution for $N$ ordered points

$$\rho_N(r_1, \dots, r_N) = N! \prod_{i=0}^{N} \theta(r_{i+1} - r_i) \tag{C.35}$$

where we have defined $r_0 \equiv 0$ and $r_{N+1} \equiv 1$. The joint probability distribution of their spacings

$$\varphi_i \equiv r_{i+1} - r_i \,, \tag{C.36}$$

is, therefore

$$\rho_N(\varphi_0, \dots, \varphi_N) = N!\,\delta\left[\sum_{i=0}^{N} \varphi_i = 1\right] \prod_{i=0}^{N} \theta(\varphi_i)\,. \tag{C.37}$$

If $\{i_1, i_2, \dots, i_k\}$ is a generic subset of $k$ different indices in $\{0, 1, \dots, N\}$, we soon get the marginal distributions

$$\rho_N^{(k)}(\varphi_{i_1}, \dots, \varphi_{i_k}) = \frac{N!}{(N-k)!} \left(1 - \sum_{n=1}^{k} \varphi_{i_n}\right)^{N-k} \theta\left(1 - \sum_{n=1}^{k} \varphi_{i_n}\right) \prod_{n=1}^{k} \theta(\varphi_{i_n})\,. \tag{C.38}$$

Developing the square of (C.34) one obtains $N^2$ terms, each one describing a particular configuration of two arcs connecting some points on the line. We will denote by $\chi_1$ and $\chi_2$ the length of these arcs; they can only be expressed as a sum of 2 spacings or simply as one spacing. Because the distribution (C.38) is independent of $i_1, \dots, i_k$, these terms can be grouped together on the base of their topology on the line with a given multiplicity. All these terms have a weight that can be written as

$$\int_0^1 d\chi_1\, d\chi_2\, \chi_1^p\, \chi_2^p\, \rho(\chi_1, \chi_2) \tag{C.39}$$

where $\rho$ is a joint distribution of $\chi_1$ and $\chi_2$. Depending on the term in the square of (C.34) one is taking into account, the distribution $\rho$ takes different forms, but it can always be expressed as in function of the distribution (C.38). As an example, we show how to calculate $\overline{|r_3 - r_1|^p |r_4 - r_2|^p}$. In this case $\rho(\chi_1, \chi_2)$ takes the form

$$\begin{aligned}
\rho(\chi_1, \chi_2) = \int d\varphi_1\, d\varphi_2\, d\varphi_3\, \rho_N^{(3)}(\varphi_1, \varphi_2, \varphi_3) \\
\delta\left(\chi_1 - \varphi_1 - \varphi_2\right) \delta\left(\chi_2 - \varphi_2 - \varphi_3\right) \\
= N(N-1) \Big[ (1-\chi_1)^{N-2} \theta(\chi_1) \theta(\chi_2 - \chi_1) \theta(1 - \chi_2) \\
+ (1-\chi_2)^{N-2} \theta(\chi_2) \theta(\chi_1 - \chi_2) \theta(1 - \chi_1) \\
- (1 - \chi_1 - \chi_2)^{N-2} \theta(\chi_1) \theta(\chi_2) \theta(1 - \chi_1 - \chi_2) \Big]\,,
\end{aligned} \tag{C.40}$$

that, plugged into (C.39) gives

$$\overline{|r_3 - r_1|^p |r_4 - r_2|^p} = \frac{\Gamma(N+1)\left[\Gamma(2p+3) - \Gamma(p+2)^2\right]}{(p+1)^2\Gamma(N+2p+1)}\,. \tag{C.41}$$

All the other terms contained can be calculated the same way; in particular there are 7 different topological configurations that contribute. After having counted how many times each configuration appears in $(E_N[h^*])^2$, the final expression that one gets is

$$\begin{aligned}\overline{(E_N[h^*])^2} = \frac{\Gamma(N+1)}{\Gamma(N+2p+1)}\Bigg[&4(N-3)\Gamma(p+2)\Gamma(p+1)\\ &+\left((N-4)(N-3)(p+1)^2 - 2N+8\right)\Gamma(p+1)^2+\\ &+\frac{[N(2p+1)(p+5) - 4p(p+5) - 8]\,\Gamma(2p+1)}{(p+1)}\Bigg]\,. \end{aligned} \tag{C.42}$$

# Appendix D

# Possible solutions in Euclidean 2-factor problem

## D.1 The Padovan numbers

According to Proposition 7.2.1, in the optimal 2-factor configuration of the complete bipartite graph there are only loops of length 2 and 3. Here we will count the number of possible optimal solutions for each value of $N$. Let $f_N$ be the number of ways in which the integer $N$ can be written as a sum in which the addenda are only 2 and 3. For example, $f_4 = 1$ because $N = 4$ can be written only as $2+2$, but $f_5 = 2$ because $N = 5$ can be written as $2+3$ and $3+2$. We simply get the recursion relation

$$f_N = f_{N-2} + f_{N-3} \tag{D.1}$$

with the initial conditions $f_2 = f_3 = f_4 = 1$. The $N$-th *Padovan number* $\mathrm{Pad}(N)$ is defined as $f_{N+2}$. Therefore it satisfies the same recursion relation (D.1) but with the initial conditions $\mathrm{Pad}(0) = \mathrm{Pad}(1) = \mathrm{Pad}(2) = 1$.

A generic solution of (D.1) can be written in terms of the roots of the equation

$$x^3 = x + 1\,. \tag{D.2}$$

There is one real root

$$\mathsf{p} = \frac{(9+\sqrt{69})^{\frac{1}{3}} + (9-\sqrt{69})^{\frac{1}{3}}}{2^{\frac{1}{3}}3^{\frac{2}{3}}} \approx 1.324717957244746\ldots \tag{D.3}$$

known as the *plastic* constant and two complex conjugates roots

$$\begin{aligned} z_\pm &= \frac{(-1 \pm i\sqrt{3})(9+\sqrt{69})^{\frac{1}{3}} + (-1 \mp i\sqrt{3})(9-\sqrt{69})^{\frac{1}{3}}}{2^{\frac{4}{3}}3^{\frac{2}{3}}} \\ &\approx -0.662359\ldots \pm i\,0.56228\ldots \end{aligned} \tag{D.4}$$

of modulus less than unity. Therefore

$$\mathrm{Pad}(N) = a\,\mathsf{p}^N + b\,z_+^N + b^*\,z_-^N \tag{D.5}$$

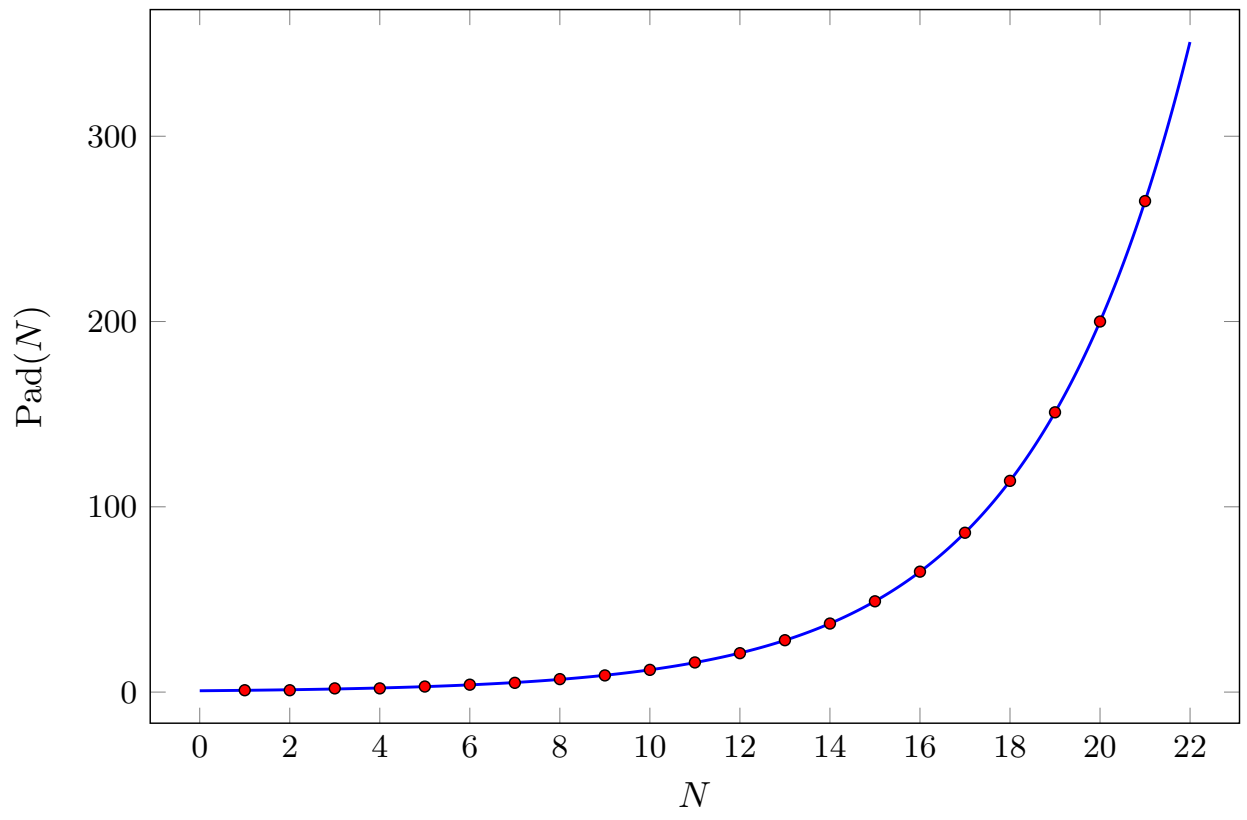

**Figure D.1.** Padovan numbers and their asymptotic expansion.

and by imposing the initial conditions we get

$$\mathrm{Pad}(N) = \frac{(z_+ - 1)(z_- - 1)}{(\mathsf{p} - z_+)(\mathsf{p} - z_-)}\,\mathsf{p}^N + \frac{(\mathsf{p} - 1)(z_- - 1)}{(z_+ - \mathsf{p})(z_+ - z_-)}\,z_+^N + \frac{(\mathsf{p} - 1)(z_+ - 1)}{(z_- - \mathsf{p})(z_- - z_+)}\,z_-^N\,. \tag{D.6}$$

For large $N$ we get

$$\mathrm{Pad}(N) \sim \lambda\,\mathsf{p}^N \tag{D.7}$$

with $\lambda \approx 0.722124\ldots$ the real solution of the cubic equation

$$23\,t^3 - 23\,t^2 + 6\,t - 1 = 0\,. \tag{D.8}$$

In Fig. D.1 we plot the Padovan sequence for a range of values of $N$ and its asymptotic expression.

There is a relation between the Padovan numbers and the Binomial coefficients. If we consider $k$ addenda equal to 3 and $s$ addenda equal to 2, there are $\binom{k+s}{k} = \binom{k+s}{s}$ possible different orderings. If we fix $N = 3\,k + 2\,s$ we easily get that

$$\mathrm{Pad}(N-2) = \sum_{k\geq 0}\sum_{s\geq 0}\delta_{N,3\,k+2\,s}\binom{k+s}{k} = \sum_{m\geq 0}\sum_{k\geq 0}\delta_{N,k+2\,m}\binom{m}{k}\,. \tag{D.9}$$

## D.2 The recursion on the complete graph

A recursion relation analogous to eq. (D.1) can be derived for the number of possible solution of the 2-factor problem on the complete graph $\mathcal{K}_N$. Let $g_N$ be the number of ways in which the integer $N$ can be expressed as a sum of 3, 4 and 5. Then $g_N$ satisfies the recursion relation given by

$$g_N = g_{N-3} + g_{N-4} + g_{N-5}\,, \tag{D.10}$$

with the initial conditions $g_3 = g_4 = g_5 = g_6 = 1$ and $g_7 = 2$. The solution of this recursion relation can be written in function of the roots of the 5-th order polynomial

$$x^5 - x^2 - x - 1 = 0\,. \tag{D.11}$$

This polynomial can be written as $(x^2+1)(x^3-x-1)=0$. Therefore the roots will be the same of the complete bipartite case ($\mathsf{p}$, and $z_\pm$) and in addition

$$y_\pm = \pm i\,. \tag{D.12}$$

$g_N$ can be written as

$$g_N = \alpha_1 \mathsf{p}^N + \alpha_2 z_+^N + \alpha_3 z_-^N + \alpha_4 y_+^N + \alpha_5 y_-^N\,, \tag{D.13}$$

where the constants $\alpha_1$, $\alpha_2$, $\alpha_3$, $\alpha_4$, and $\alpha_5$ are fixed by the initial conditions $g_3 = g_4 = g_5 = g_6 = 1$ and $g_7 = 2$. When $N$ is large the dominant contribution comes from the plastic constant

$$g_N \simeq \alpha_1 \mathsf{p}^N\,. \tag{D.14}$$

with $\alpha_1 \approx 0.262126...$

## D.3 The plastic constant

In 1928, shortly after abandoning his architectural studies and becoming a novice monk of the Benedictine Order, Hans van der Laan discovered a new, unique system of architectural proportions. Its construction is completely based on a single irrational value which he called the plastic number (also known as the plastic constant) [MS12]. This number was originally studied in 1924 by a French engineer, G. Cordonnier, when he was just 17 years old, calling it "radiant number". However, Hans van der Laan was the first who explained how it relates to the human perception of differences in size between three-dimensional objects and demonstrated his discovery in (architectural) design. His main premise was that the plastic number ratio is truly aesthetic in the original Greek sense, i.e. that its concern is not beauty but clarity of perception [Pad02]. The word plastic was not intended, therefore, to refer to a specific substance, but rather in its adjectival sense, meaning something that can be given a three-dimensional shape [Pad02]. The golden ratio or divine proportion

$$\phi = \frac{1+\sqrt{5}}{2} \approx 1.6180339887\,, \tag{D.15}$$

which is a solution of the equation

$$x^2 = x + 1\,, \tag{D.16}$$

has been studied by Euclid, for example for its appearance in the regular pentagon, and has been used to analyze the most aesthetic proportions in the arts. For example, the golden rectangle, of size $(a+b)\times a$ which may be cut into a square of size $a \times a$ and a smaller rectangle of size $b \times a$ with the same aspect ratio

$$\frac{a+b}{a} = \frac{a}{b} = \phi\,. \tag{D.17}$$

This amounts to the subdivision of the interval $AB$ of length $a+b$ into $AC$ of length $a$ and $BC$ of length $b$. By fixing $a+b=1$ we get

$$\frac{1}{a} = \frac{a}{1-a} = \phi\,, \tag{D.18}$$

which implies that $\phi$ is the solution of (D.16). The segments $AC$ and $BC$, of length, respectively $\frac{1}{\phi^2}(\phi, 1)$ are sides of a golden rectangle.

But the golden ratio fails to generate harmonious relations within and between three-dimensional objects. Van der Laan therefore elevates definition of the golden rectangle in terms of space dimension. Van der Laan breaks segment $AB$ in a similar manner, but in three parts. If C and D are points of subdivision, plastic number $\mathsf{p}$ is defined with

$$\frac{AB}{AD} = \frac{AD}{BC} = \frac{BC}{AC} = \frac{AC}{CD} = \frac{CD}{BD} = \mathsf{p} \tag{D.19}$$

and by fixing $AB = 1$, from $AC = 1 - BC$, $BD = 1 - AD$ we get

$$\mathsf{p}^3 = \mathsf{p} + 1\,. \tag{D.20}$$

The segments $AC$, $CD$ and $BD$, of length, respectively, $\frac{1}{(\mathsf{p}+1)\mathsf{p}^2}(\mathsf{p}^2, \mathsf{p}, 1)$ can be interpreted as sides of a cuboid analogous to the golden rectangle.

# Appendix E

# Selberg Integrals

Euler beta integrals

$$B(\alpha,\beta) := \int_0^1 dx\, x^{\alpha-1}(1-x)^{\beta-1} = \frac{\Gamma(\alpha)\Gamma(\beta)}{\Gamma(\alpha+\beta)} \tag{E.1}$$

with $\alpha, \beta \in \mathbb{C}$ and $\Re(\alpha) > 0$, $\Re(\beta) > 0$, have been generalized in the 1940s by Atle Selberg [Sel44]

$$S_n(\alpha,\beta,\gamma) \equiv \left[\prod_{i=1}^{n}\int_0^1 dx_i\, x_i^{\alpha-1}(1-x_i)^{\beta-1}\right]|\Delta(x)|^{2\gamma} \tag{E.2}$$

$$= \prod_{j=1}^{n} \frac{\Gamma(\alpha+(j-1)\gamma)\Gamma(\beta+(j-1)\gamma)\Gamma(1+j\gamma)}{\Gamma(\alpha+\beta+(n+j-2)\gamma)\Gamma(1+\gamma)} \tag{E.3}$$

where

$$\Delta(x) \equiv \prod_{1\leq i<j\leq n}(x_i - x_j) \tag{E.4}$$

is the *Vandermonde determinant*, with $\alpha, \beta, \gamma \in \mathbb{C}$ and $\Re(\alpha) > 0$, $\Re(\beta) > 0$, $\Re(\gamma) > \min(1/n, \Re(\alpha)/(n-1), \Re(\beta)/(n-1))$, see [AAR99, Chap. 8]. Indeed (E.3) reduces to (E.1) when $n = 1$. These integrals have been used by Enrico Bombieri (see [FW08] for the detailed history) to prove what was known as the *Mehta-Dyson conjecture* [MD63, Meh67, Meh74, Meh04] in random matrix theory, but they found many applications also in the context of conformal field theories [DF85] and exactly solvable models, for example in the evaluation of the norm of *Jack polynomials* [Kak98]. In the following we shall also need of an extension of Selberg integrals [AAR99, Sec. 8.3]

$$\begin{aligned} B_n(j,k;\alpha,\beta,\gamma) &\equiv \left[\prod_{i=1}^{n}\int_0^1 dx_i\, x_i^{\alpha-1}(1-x_i)^{\beta-1}\right]\left(\prod_{s=1}^{j} x_s\right)\left(\prod_{s=j+1}^{k}(1-x_s)\right)|\Delta(x)|^{2\gamma} \\ &= S_n(\alpha,\beta,\gamma)\,\frac{\prod_{i=1}^{j}[\alpha+(n-i)\gamma]\prod_{i=1}^{k}[\beta+(n-i)\gamma]}{\prod_{i=1}^{j+k}[\alpha+\beta+(2n-1-i)\gamma]}\,. \end{aligned} \tag{E.5}$$

Here we shall show one more application in the context of Euclidean random combinatorial optimization problems in one dimension [CDGM18b]. In the following we will always assume that both the red and the blue points are extracted with the flat distribution over the interval $[0,1]$ and that we label them in ordered position, i.e. $r_i < r_{i+1}$ and $b_i < b_{i+1}$ for $i = 1, \dots, N-1$. In addition we will consider always the case $p > 1$ in the cost function (1.93).

## E.1 Average cost in the assignment

In the case of the assignment for $p > 1$ the optimal permutation is the identity one [McC99, BCS14], i.e. the cost is

$$E^{(p)}_{N,N}[\mu^*] = \sum_{i=1}^{N} |r_i - b_i|^p \,. \tag{E.6}$$

Using the expression of the probability $P_k(x)$ defined in (6.36), the mean displacement is given by

$$\begin{aligned}\overline{|r_k - b_k|^p} &= \int_0^1 dx\, dy\, P_k(x)\, P_k(y)\; |y-x|^p \\ &= \left[\frac{\Gamma(N+1)}{\Gamma(k)\Gamma(N-k+1)}\right]^2 S_2\left(k, N-k+1, \frac{p}{2}\right) \\ &= \frac{\Gamma^2(N+1)\Gamma\left(k+\frac{p}{2}\right)\Gamma\left(N-k+1+\frac{p}{2}\right)\Gamma(1+p)}{\Gamma(k)\Gamma(N-k+1)\Gamma\left(N+1+\frac{p}{2}\right)\Gamma(N+1+p)\Gamma\left(1+\frac{p}{2}\right)}\,,\end{aligned} \tag{E.7}$$

from which it follows the general $p > 1$ formula for the average optimal cost of the assignment

$$\begin{aligned}\overline{E^{(p)}_{N,N}[\mu^*]} &= \frac{\Gamma^2(N+1)\Gamma(1+p)}{\Gamma\left(N+1+\frac{p}{2}\right)\Gamma(N+1+p)\Gamma\left(1+\frac{p}{2}\right)} \sum_{k=1}^{N} \frac{\Gamma\left(k+\frac{p}{2}\right)\Gamma\left(N-k+1+\frac{p}{2}\right)}{\Gamma(k)\Gamma(N-k+1)} \\ &= \frac{\Gamma\left(1+\frac{p}{2}\right)}{p+1}\, \frac{N\,\Gamma(N+1)}{\Gamma\left(N+1+\frac{p}{2}\right)}\,,\end{aligned} \tag{E.8}$$

where we made repeated use of the duplication and Euler's inversion formula for $\Gamma$-functions. The exact result (E.8) was known only in the cases $p = 2, 4$ where only Euler Beta-functions (3.22) is needed, see [CDS17]. The other cases were known only in the limit of large $N$ [CDS17] at $o(N^{-1})$.

## E.2 Average cost in the TSP

In the bipartite TSP, when $p > 1$, the optimal solution is given by two permutations $\tilde{\sigma}$ and $\tilde{\pi}$ given respectively in (6.16) and (6.17) and the optimal cost is written as in (6.32). We have therefore to solve the following integral

$$\begin{aligned}&\overline{|b_{k+1} - r_k|^p} = \overline{|r_{k+1} - b_k|^p} = \int_0^1 dx\, dy\, P_k(x)\, P_{k+1}(y)\; |x-y|^p \\ &= k(k+1)\binom{N}{k}\binom{N}{k+1} \int_0^1 dx\, dy\, x^{k-1} y^k (1-x)^{N-k} (1-y)^{N-k-1}\, |x-y|^p\,,\end{aligned} \tag{E.9}$$

involving the difference between the $(k+1)$-th point of one color and the $k$-th of the other one. We have solved this integrals in Sec. 6.3 limiting for simplicity to the simple $p=2$ case. This quantity can be evaluated for every $p$ by using the generalized Selberg integral (E.5)

$$\begin{aligned}\overline{|b_{k+1}-r_k|^p} &= \frac{\Gamma^2(N+1)}{\Gamma(k)\,\Gamma(N-k)\,\Gamma(k+1)\,\Gamma(N-k-1)}\,B_2\left(1,1;k,N-k,\frac{p}{2}\right)\\ &= \frac{\Gamma^2(N+1)\,\Gamma(p+1)\,\Gamma\left(k+\frac{p}{2}+1\right)\,\Gamma\left(N-k+\frac{p}{2}+1\right)}{\Gamma(k+1)\,\Gamma(N-k+1)\,\Gamma(N+p+1)\,\Gamma\left(N+\frac{p}{2}+1\right)\,\Gamma\left(1+\frac{p}{2}\right)}\,.\end{aligned} \tag{E.10}$$

from which we obtain

$$\begin{aligned}\sum_{k=1}^{N-1}\overline{|b_{k+1}-r_k|^p} &= 2\,\Gamma(N+1)\,\Gamma(1+p)\\ &\times\left[\frac{(N+p+1)\,\Gamma\left(\frac{p}{2}\right)}{4(p+1)\,\Gamma(p)\,\Gamma\left(N+1+\frac{p}{2}\right)}-\frac{1}{\Gamma(N+1+p)}\right].\end{aligned} \tag{E.11}$$

In addition

$$\begin{aligned}\overline{|r_1-b_1|^p} = \overline{|r_N-b_N|^p} &= N^2\int_0^1 dx\,dy\,(xy)^{N-1}\,|x-y|^p\\ &= N^2 S_2\left(N,1,\frac{p}{2}\right) = \frac{N\,\Gamma(N+1)\,\Gamma(p+1)}{\left(N+\frac{p}{2}\right)\,\Gamma(N+p+1)}\,.\end{aligned} \tag{E.12}$$

Finally, the average optimal cost for every $N$ and every $p>1$ is

$$\begin{aligned}\overline{E^{(p)}_{N,N}}[h^*] &= 2\left[\overline{|r_1-b_1|^p}+\sum_{k=1}^{N-1}\overline{|b_{k+1}-r_k|^p}\right]\\ &= 2\,\Gamma(N+1)\left[\frac{(N+p+1)\,\Gamma\left(1+\frac{p}{2}\right)}{(p+1)\,\Gamma\left(N+1+\frac{p}{2}\right)}-\frac{2\,\Gamma(p+1)}{(2N+p)\,\Gamma(N+p)}\right].\end{aligned} \tag{E.13}$$

For $p=2$ this reduces to

$$\overline{E^{(2)}_{N,N}}[h^*] = \frac{2}{3}\frac{N^2+4N-3}{(N+1)^2}\,, \tag{E.14}$$

which was derived in Sect. 6.3 together with the asymptotic behavior for large $N$ for $p>1$

$$\lim_{N\to\infty} N^{p/2-1}\overline{E^{(p)}_N} = 2\,\frac{\Gamma\left(\frac{p}{2}+1\right)}{p+1}\,. \tag{E.15}$$

## E.3 Cutting shoelaces: the two-factor problem

Selberg integral are also useful to evaluate for generic $p$ and $N$ the cost gained cutting an Hamiltonian cycle into smaller cycles. We denote the cost gain of cutting the TSP shoelace at position $k$ by $E^{(p)}_k$ and it is given by

$$E^{(p)}_k = \overline{|r_{k+1}-b_{k+1}|^p}+\overline{|r_k-b_k|^p}-\overline{|r_k-b_{k+1}|^p}-\overline{|r_{k+1}-b_k|^p}\,. \tag{E.16}$$

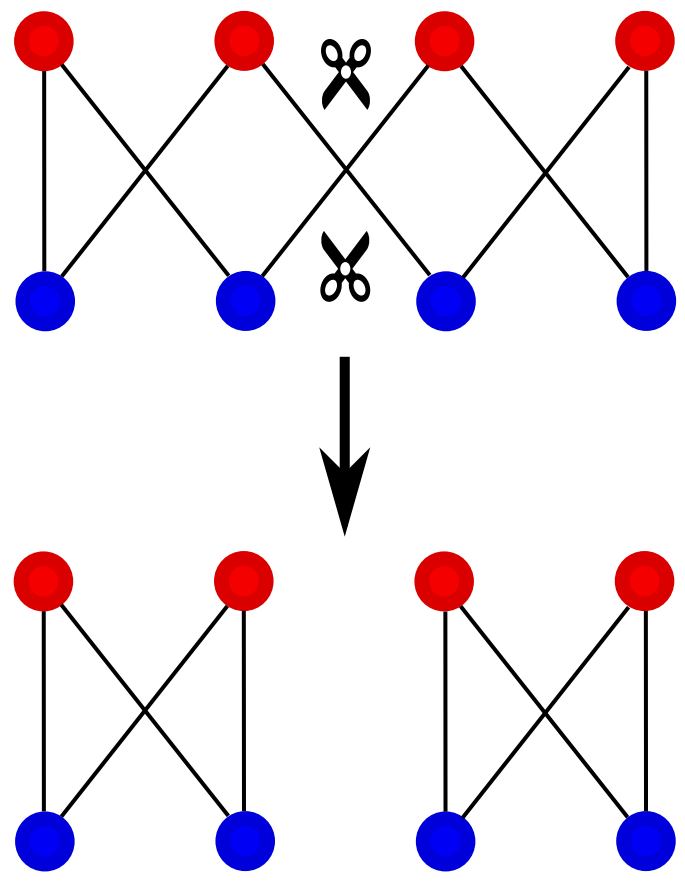

**Figure E.1.** Graphical representation of the cutting operation which brings from the optimal TSP cycle (top) to a possible optimal solution of the 2-factor problem (bottom). Here we have represented the $N = 4$ case, where the cutting operation is unique. Notice that blue and red points are chosen on a interval, but here they are represented equispaced on two parallel lines to improve visualization.

We obtain

$$\begin{aligned}\overline{|r_k - b_k|^p} - \overline{|r_k - b_{k+1}|^p} = \\ = \frac{\Gamma^2(N+1)\,\Gamma(p+1)\,\Gamma\left(k+\frac{p}{2}\right)\,\Gamma\left(N-k+\frac{p}{2}+1\right)}{\Gamma(k)\,\Gamma(N-k+1)\,\Gamma(N+p+1)\,\Gamma\left(N+\frac{p}{2}+1\right)\,\Gamma\left(\frac{p}{2}+1\right)}\left[1-\frac{k+\frac{p}{2}}{k}\right] \\ = -\frac{p}{2}\frac{\Gamma^2(N+1)\,\Gamma(p+1)\,\Gamma\left(k+\frac{p}{2}\right)\,\Gamma\left(N-k+\frac{p}{2}+1\right)}{\Gamma(k+1)\,\Gamma(N-k+1)\,\Gamma(N+p+1)\,\Gamma\left(N+\frac{p}{2}+1\right)\,\Gamma\left(\frac{p}{2}+1\right)},\end{aligned} \tag{E.17}$$

and similarly

$$\begin{aligned}\overline{|r_{k+1} - b_{k+1}|^p} - \overline{|r_{k+1} - b_k|^p} = \\ = \frac{\Gamma^2(N+1)\,\Gamma(p+1)\,\Gamma\left(k+\frac{p}{2}+1\right)\,\Gamma\left(N-k+\frac{p}{2}\right)}{\Gamma(k+1)\,\Gamma(N-k)\,\Gamma(N+p+1)\,\Gamma\left(N+\frac{p}{2}+1\right)\,\Gamma\left(\frac{p}{2}+1\right)}\left[1-\frac{N-k+\frac{p}{2}}{N-k}\right] \\ = -\frac{p}{2}\frac{\Gamma^2(N+1)\,\Gamma(p+1)\,\Gamma\left(k+\frac{p}{2}+1\right)\,\Gamma\left(N-k+\frac{p}{2}\right)}{\Gamma(k+1)\,\Gamma(N-k+1)\,\Gamma(N+p+1)\,\Gamma\left(N+\frac{p}{2}+1\right)\,\Gamma\left(\frac{p}{2}+1\right)}.\end{aligned} \tag{E.18}$$

Their sum is

$$E_k^{(p)} = -\frac{p}{2}\frac{\Gamma^2(N+1)\,\Gamma(p+1)\,\Gamma\left(k+\frac{p}{2}\right)\,\Gamma\left(N-k+\frac{p}{2}\right)}{\Gamma(k+1)\,\Gamma(N-k+1)\,\Gamma(N+p)\,\Gamma\left(N+\frac{p}{2}+1\right)\,\Gamma\left(\frac{p}{2}+1\right)}, \tag{E.19}$$

For $p = 2$ this quantity is in agreement with what we got in Sect. 7.3

$$E_k^{(2)} = -\frac{2}{(N+1)^2}. \tag{E.20}$$

For $p \neq 2$, $E_k$ depends on $k$. In particular, for $1 < p < 2$ the cut near to 0 and 1 are (on average) more convenient than those near the center. For $p > 2$ the reverse is true (see Fig. E.2). Notice that for $p = 2$ we can sum on all the cuts that can be done and obtain the mean total cost gain. For $p \neq 2$, this sum does not give a simple formula.

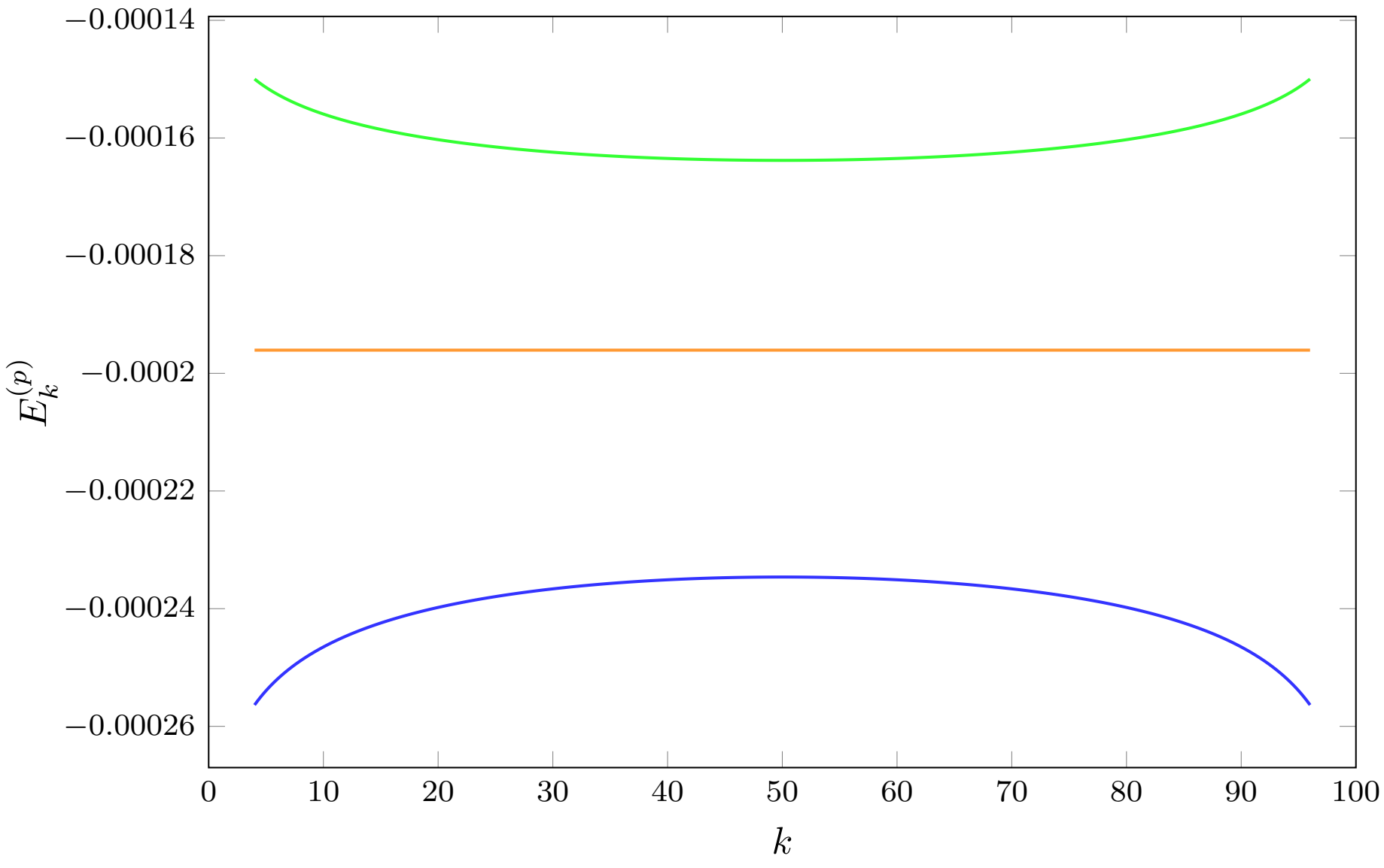

**Figure E.2.** Plot of $E_k^{(p)}$ given in Eq. (E.19) for various values of $p$: the green line is calculated with $p = 2.1$, the orange with $p = 2$ and the blue one with $p = 1.9$; in all cases we take $N = 100$.

# Acronyms

| | |
|---|---|
| **1RSB** | one-step Replica Symmetry Breaking |
| **2RSB** | two-step Replica Symmetry breaking |
| **aoc** | Average Optimal Cost |
| **BP** | Belief Propagation |
| **dAT** | de Almeida Thouless |
| **EA** | Edwards Anderson |
| **ER** | Erdős-Rényi |
| **fRSB** | full Replica Symmetry Breaking |
| **MF** | Mean Field |
| **MST** | Minimum Spanning Tree |
| **RAP** | Random Assignment Problem |
| **RCOP** | Random Combinatorial Optimization Problem |
| **RFMP** | Random Fractional Matching Problem |
| **RKKY** | Ruderman Kittel Kasuya Yosida |
| **RMP** | Random Matching Problem |
| **RRG** | Random Regular Graphs |
| **RS** | Replica Symmetry |
| **RSB** | Replica Symmetry Breaking |

**RTSP** Random Traveling Salesman Problem

**SK** Sherrington Kirkpatrick

**TAP** Thouless Anderson Palmer

**TSP** Traveling Salesman Problem